\documentclass[
    aps,prx,onecolumn,tightenlines,preprintnumbers,
    showpacs,superscriptaddress,notitlepage,nofootinbib,
    floatfix,longbibliography,10pt
    ]{revtex4-2}
\pdfoutput=1

\usepackage{amssymb}   
\usepackage{amsmath}
\usepackage{dsfont}
\usepackage{CJKutf8}   
\usepackage{graphicx}  
\graphicspath{{./figures/}}
\usepackage{dcolumn}   
\usepackage{bm}        
\usepackage{xspace}
\usepackage{standalone}
\usepackage{enumitem}
\usepackage[pdftex]{color}
\usepackage{xcolor}
\usepackage{slashed}
\usepackage{booktabs}
\usepackage{multirow}
\usepackage{stackrel}
\usepackage{rotating}
\usepackage{pifont}
\usepackage{mathtools}
\usepackage{simplewick}
\usepackage{float} 
\usepackage{tikz} 
\usepackage{hyperref}
\hypersetup{
    colorlinks=true,       
    linkcolor=blue,          
    citecolor=blue,        
    filecolor=blue,      
    urlcolor=blue           
}

\def\qutip{\texttt{QuTiP}\xspace}
\def\qiskit{\texttt{QISKIT}\xspace}

\def\eqref#1{{(\ref{#1})}}

\definecolor{kngrey}{HTML}{A6AAA9}
\definecolor{knred}{HTML}{EC5D57}
\definecolor{knorange}{HTML}{F39019}
\definecolor{knyellow}{HTML}{F5D328}
\definecolor{kngreen}{HTML}{70BF41}
\definecolor{knblue}{HTML}{51A7F9}
\definecolor{knpurple}{HTML}{B36AE2}

\def\ket#1{{|#1\rangle}}

\newcount\hour \newcount\hourminute \newcount\minute
\hour=\time \divide \hour by 60
\hourminute=\hour \multiply \hourminute by 60
\minute=\time \advance \minute by -\hourminute

\begin{document}
\preprint{RIKEN-iTHEMS-Report-21, DMUS-MP-21/10}

\title{Matrix Model simulations using Quantum Computing, Deep Learning, and Lattice Monte Carlo}

\author{Enrico Rinaldi}
\affiliation{Physics Department, University of Michigan, Ann Arbor, MI 48109, USA}
\affiliation{Theoretical Quantum Physics Laboratory, Center for Pioneering Research, RIKEN, Wako, Saitama 351-0198, Japan}
\affiliation{Interdisciplinary Theoretical and Mathematical Sciences Program (iTHEMS), RIKEN, Wako, Saitama 351-0198, Japan}

\author{Xizhi Han}
\affiliation{Department of Physics, Stanford University, Stanford, CA 94305-4060, USA}

\author{Mohammad Hassan}
\affiliation{City College of New York, New York, NY 10031, USA}

\author{Yuan Feng}
\affiliation{Pasadena City College, Pasadena, CA 91106, USA}

\author{Franco Nori}
\affiliation{Theoretical Quantum Physics Laboratory, Center for Pioneering Research, RIKEN, Wako, Saitama 351-0198, Japan}
\affiliation{RIKEN Center for Quantum Computing (RQC), Wako, Saitama 351-0198, Japan}
\affiliation{Physics Department, University of Michigan, Ann Arbor, MI 48109, USA}

\author{Michael McGuigan}
\affiliation{Brookhaven National Laboratory, Upton, NY 11973, USA}

\author{Masanori Hanada}
\affiliation{Department of Mathematics, University of Surrey, Guildford, Surrey, GU2 7XH, UK}

\date{\today}
\begin{abstract}
Matrix quantum mechanics plays various important roles in theoretical physics, such as a holographic description of quantum black holes.
Understanding quantum black holes and the role of entanglement in a holographic setup is of paramount importance for the development of better quantum algorithms (quantum error correction codes) and for the realization of a quantum theory of gravity.
Quantum computing and deep learning offer us potentially useful approaches to study the dynamics of matrix quantum mechanics.
In this paper we perform a systematic survey for quantum computing and deep learning approaches to matrix quantum mechanics, comparing them to Lattice Monte Carlo simulations.
In particular, we test the performance of each method by calculating the low-energy spectrum.
\end{abstract}

\maketitle
\tableofcontents

\section{Introduction}\label{sec:introduction}

Gauge/gravity duality~\cite{Maldacena:1997re,Itzhaki:1998dd} translates difficult (or intractable) problems in quantum gravity to well-defined problems in non-gravitational quantum theories.
Although it originated from string/M-theory, connections to various other fields including quantum information theory, condensed matter theory, and cosmology, have been recognized by now with research programs actively pursued.
Quantum mechanics with matrix degrees of freedom (matrix QM, or matrix models in the following) play important roles in gauge/gravity duality.
More specifically, Yang--Mills-type matrix models can provide us with a nonperturbative formulation of superstring/M-theory~\cite{Banks:1996vh,deWit:1988wri,Itzhaki:1998dd,Berenstein:2002jq}.
In certain regions of the parameter space (i.e., strong coupling and large-$N$), weakly-coupled gravity with small stringy corrections can be described by matrix models.
Therefore, matrix models can offer us an ideal setup to study quantum corrections to gravity.
By solving the matrix models, it should be possible to study fascinating problems such as the microscopic mechanism of black hole evaporation.

In the past, Monte Carlo simulations based on the Euclidean path-integral formulation have been used extensively; see e.g., Ref.~\cite{Berkowitz:2016jlq} for a remarkable agreement between the D0-brane matrix model and type-IIA superstring theory.
However, there are certain problems that are hard to be accessed via Monte Carlo methods, such as the Hamiltonian time evolution.
Therefore it is important to develop alternative methods too.
In this paper we consider quantum simulation~\cite{Buluta108,RevModPhys.86.153,Banuls:2019bmf,Buluta_2011} and deep learning as such alternative approaches.
Quantum simulations of lattice gauge theories have recently attracted a lot of attention since the pioneering work of Ref.~\cite{Byrnes:2005qx}, and deep learning is being used to accelerate simulations of lattice gauge theories similar to QCD~\cite{Albergo:2021bna}.
Another interesting approach outside the scope of this paper is bootstrap~\cite{Han:2020bkb}.

We briefly introduce the pros and cons of these numerical approaches to matrix models:
\begin{itemize}
\item
Monte Carlo simulations can be used to study the problems which can be accessed by the Euclidean path integral, such as canonical thermodynamics and Euclidean correlation functions.
Large-scale parallel simulations on High Performance Computing (HPC) systems are doable and actually done regularly.
It is known how to improve the simulation results systematically, by accumulating more chain trajectories and simulating larger systems.

However, Monte Carlo simulation can work only when there is no sign problem.\footnote{
Monte Carlo methods provide us a way to generate a probability distribution.
If the path-integral weight $e^{-S}$ is real and non-negative, the lattice configurations can be generated in such a way that the probability distribution is proportional to $e^{-S}$.
If the path-integral weight is not real and non-negative, it cannot be a `probability distribution', and hence, Monte Carlo methods cannot be applied.
This is a manifestation of the so-called sign problem.
With the Minkowski signature, the path integral weight is $e^{iS}$, which is purely imaginary.}
Due to the sign problem, we cannot study the Hamiltonian time evolution.
The sign problem may appear also at the very low-temperature region of the matrix models. Note also that, even when there is no sign problem, we cannot see the quantum states in the Hilbert space directly.
This might be a disadvantage for several interesting applications, such as decoding the emergent bulk geometry in holography along the line of Ref.~\cite{Hanada:2021ipb}.

\item
On a quantum computer, quantum states are directly realized.
Therefore, quantum computing does not suffer from the sign problem, and the Hamiltonian time evolution can be simulated straightforwardly.
Some research directions that could benefit from direct quantum simulations are the formation and evaporation of a black hole, or a wormhole, and quantum teleportation.

The biggest challenge is that we do not yet have a reliable large-scale quantum computer.
Although we can emulate a quantum computer on a classical computer, these simulations are restricted to systems with only a small number of qubits.
Furthermore, the effects of the regularization needed to map a theory on a quantum device are not well understood yet.

The Variational Quantum Eigensolver (VQE)~\cite{2014NatCo...5.4213P,Moll_2018,Endo_2021} is expected to be a useful quantum-classical hybrid algorithm in the Noisy Intermediate Scale Quantum (NISQ) era.
Being a variational method, the VQE can have systematic errors: \emph{a priori}, we do not know if we can get the right answer, because the simulation might be trapped in a local optimum.
This is a general problem for variational methods.

Quantum computing represents the computation in terms of qubits and quantum gates which can represent larger Hamiltonians and Hilbert spaces on quantum hardware than classical computers that are limited by memory constraints.
Also, quantum computing represents fermions straightforwardly in a similar way to how they represent bosons and, as mentioned above, are not susceptible to the sign problem like classical methods using Monte Carlo algorithms.

Some of the disadvantages of NISQ devices~\cite{Preskill_2018} include the lack of quantum error correction for quantum gates and the presence of noise which can lead to inaccurate quantum gates and decoherence of the qubits during the computation.
For hybrid quantum-classical computations that we discuss here, the advantages are that many parts of a hybrid quantum-classical variational calculation, such as  optimization, can be done efficiently on a classical computer, while expectation values can be computed on a quantum computer.
This leads to a lower depth for the circuits, making them less susceptible to decoherence.
However, these methods include complicated state-preparation protocols and variational ansatze for the wave function that need to be specified to have strong overlaps with the true ground state.
Such wave functions may need a high-depth quantum circuit which is difficult to implement on a NISQ hardware, and it is prone to decoherence.

\item
Deep learning methods have been applied successfully to the study of many-body quantum systems~\cite{Carleo:2019ptp}.
As one such approach, we consider a variational quantum Monte Carlo method with generative autoregressive flows: this method uses a deep neural network to represent the probability distribution given by the system wave function~\cite{Han:2019wue}.
This method may make the calculation on a classical computer much more efficient, and we can get quantum states directly.
This enables us to access a class of physics which cannot be handled via Markov Chain Monte Carlo simulations of the Euclidean path integral.

The biggest disadvantage of this approach is that, \emph{a priori}, we do not know if we can get the right answer, that is, systematic errors are difficult to estimate and the simulation might be trapped in a local optimum.
This is the same problem as the one we mentioned above for the VQE.
However, in practice, this approach has been extensively used across several many-body quantum systems~\cite{Gao_2017,Carleo_2018,Jia_2019,Hermann_2020,Choo_2020,Yoshioka_2021}, with neural quantum states becoming increasingly useful in real-world applications.
\end{itemize}

In this paper, we perform a systematic survey for quantum-simulation and deep-learning approaches.
Specifically, we will focus on two aspects:
\begin{itemize}
    \item
    Quantum simulation requires a truncation of the Hilbert space.
    For SU(2) matrix models, we will study the truncation effects by determining the low-energy states precisely via an exact diagonalization procedure.
    We find that the truncation effects are exponentially small with respect to the truncation level.

    \item
    We use the VQE and deep-learning methods to estimate the low-energy spectrum, and compare them with other methods (exact diagonalization of a truncated Hamiltonian, lattice Monte Carlo simulation and some exact results for supersymmetric models).
    For the deep-learning method, we observe a reasonable agreement.
    As for the VQE, the specific architecture we used did not show a satisfactory performance at strong coupling perhaps due to the variational forms parameterized by the quantum circuits not adequately probing the full gauge-invariant Hilbert space.
\end{itemize}

This paper is organized as follows.
In Sec.~\ref{sec:models}, we define the models: a Yang-Mills type bosonic matrix model, and supersymmetric matrix model which we call ``minimal BMN''.
We can obtain a few exact results for the latter thanks to supersymmetry.
In Sec.~\ref{sec:Hamiltonian-truncation}, we consider the quantum simulation approach.
We introduce a simple scheme of the truncation of the Hilbert space,
and study the truncation effect on the energy spectrum and gauge symmetry.
Then, in Sec.~\ref{sec:VQE}, we study the VQE as an actual example of the application of a quantum computer.
In Sec.~\ref{sec:deep-learning}, we estimate the ground-state energy via the deep learning approach.
In Sec.~\ref{sec:lattice-simulation}, we calculate the ground-state energy of the bosonic matrix model by performing lattice Monte Carlo simulation.
We compare the results in Sec.~\ref{sec:comparison}.

The codes used to generate the data and make the figures are open source and we provide a website with supplementary figures and tables at \url{https://erinaldi.github.io/mm-qc-dl-supplemental}.

\section{Models}\label{sec:models}
In this section, we define the two kinds of matrix models studied in this paper.
The first model contains only bosonic degrees of freedom (dof), while the second contains both bosonic and fermionic dof in a supersymmetric setup.

\subsection{Bosonic Matrix Model}\label{sec:def_bos_MM}
Let us start with the bosonic matrix model, i.e.,~a matrix model consisting of only bosonic degrees of freedom.
We introduce $d$ traceless Hermitian matrices $X_1(t),X_2(t),\cdots,X_d(t)$, where $t$ is time.
We consider the following action (in the Minkowski signature),
\begin{align}
S = N\int dt {\rm Tr}\left(
\frac{1}{2}\sum_I(D_tX_I)^2
-
\frac{m^2}{2}\sum_IX_I^2
+
\frac{\lambda}{4}\sum_{I\neq J}[X_I,X_J]^2
\right) \ ,
\label{bosonic_action_1}
\end{align}
where $\lambda=g^2N$ and $D_t$ is the covariant derivative with gauge field $A_t$, i.e., $D_tX_I=\partial_tX_I-i[A_t,X_I]$.
Via the SU($N$) transformation $X_I(t)\to\Omega(t)X_I(t)\Omega^{-1}(t)$, $D_tX_I$ transforms as $D_tX_I\to\Omega(t)(D_tX_I(t))\Omega^{-1}(t)$, hence this action has the SU($N$) gauge symmetry.
The normalization used in Eq.~\eqref{bosonic_action_1} is convenient when we take the 't Hooft large-$N$ limit.
The number of matrices $d$ can be arbitrary; we consider $d=2$ in this paper.
We can also use different normalization, by rescaling $X_I$ with a factor $\sqrt{N}$:
\begin{align}
S = \int dt {\rm Tr}\left(
\frac{1}{2}(D_tX_I)^2
-
\frac{m^2}{2}X_I^2
+
\frac{g^2}{4}[X_I,X_J]^2
\right) \ .
\label{bosonic_action_2}
\end{align}

In the operator formalism, the Hamiltonian is given by
\begin{align}
\hat{H} = {\rm Tr}\left(
\frac{1}{2}\hat{P}_I^2
+
\frac{m^2}{2}\hat{X}_I^2
-
\frac{g^2}{4}[\hat{X}_I,\hat{X}_J]^2
\right) \ ,
\label{eq:bosonic_Hamiltonian}
\end{align}
where
\begin{align}
\hat{P}_I=\sum_{\alpha=1}^{N^2-1}\hat{P}_I^\alpha\tau_\alpha \ ,
\qquad
\hat{X}_I=\sum_{\alpha=1}^{N^2-1}\hat{X}_I^\alpha\tau_\alpha \ .
\label{bosonic_generator}
\end{align}
$\tau_\alpha$ are the generators of SU($N$) normalized as ${\rm Tr}(\tau_\alpha\tau_\beta)=\delta_{\alpha\beta}$.
The canonical commutation relation is
\begin{align}
[\hat{X}_{I\alpha},\hat{P}_{J\beta}]
=
i\delta_{IJ}\delta_{\alpha\beta} \ .
\label{canonical_commutation_relation_XP}
\end{align}
The Hamiltonian and the canonical commutation relation is invariant under the SU($N$) transformation $\hat{X}_{I,ij}\to (\Omega\hat{X}_{I}\Omega^{-1})_{ij}$, $\hat{P}_{I,ij}\to (\Omega\hat{P}_{I}\Omega^{-1})_{ij}$.
The physical states are restricted to singlets under this SU($N$) transformation (gauge singlets).
We denote the Hilbert space spanned by gauge-singlets by ${\cal H}_{\rm inv}$.
We can also consider a bigger, `extended' Hilbert space ${\cal H}_{\rm ext}$ that contains non-singlets.
Because operators $\hat{X}_I$ and $\hat{P}_I$ are not gauge-invariant, they are defined naturally on ${\cal H}_{\rm ext}$.

It is tedious but straightforward to show the equivalence between the path-integral formalism and operator formalism; see Appendix~\ref{appendix:operator-formalism-vs-path-integral}.

\subsection{Supersymmetric Matrix Model (Minimal BMN)}\label{sec:mini_bmn}
The BMN matrix model~\cite{Berenstein:2002jq} is a one-parameter deformation of the D0-brane quantum mechanics which preserves all supersymmetries.
In Ref.~\cite{Kim:2006wg}, low-SUSY analogues of the BMN matrix model are listed\footnote{
The D0-brane quantum mechanics is the dimensional reduction of 10d ${\cal N}=1$ SYM, which has 16 real supercharges.
By dimensionally reducing 6d, 4d and 3d ${\cal N}=1$ SYM, similar matrix quantum mechanics with 8, 4 and 2 real supercharges can be constructed.
They admit the BMN-like one-parameter deformations.
}.
The one corresponding to 3d ${\cal N}=1$ SYM (Eq.(6.3) in Ref.~\cite{Kim:2006wg}) has the minimal degrees of freedom.
So let us denote this model ``minimal BMN''.

Gamma matrices in three dimensions can be chosen as $\gamma^0=i\sigma_3$, $\gamma^1=\sigma_1$ and $\gamma^2=\sigma_2$.
The charge conjugation matrix $C$, which satisfies $C^{-1}\gamma_\mu C = -\gamma_\mu^{\rm T}$, can be chosen as $C=i\sigma_2$.
The Majorana condition $C\psi=\gamma_0^{\rm T}\psi^\ast$ can be solved by $\psi=\left(\begin{array}{c}\xi\\ i\xi^\ast\end{array}\right)$.
When $\psi$ is a matrix, $\psi=\left(\begin{array}{c}\xi\\ i\xi^\dagger\end{array}\right)$.

Having this convention in mind, the Hamiltonian of minimal BMN is expressed as~\cite{Kim:2006wg}
\begin{align}
\hat{H} &= {\rm Tr}\left(
\frac{1}{2}\hat{P}_I^2
-
\frac{g^2}{4}[\hat{X}_I,\hat{X}_J]^2
+
\frac{g}{2}\hat{\bar{\psi}}\Gamma^I[\hat{X}_I,\hat{\psi}]
-
\frac{3i\mu}{4}\hat{\bar{\psi}}\hat{\psi}
+
\frac{\mu^2}{2}\hat{X}_I^2
\right)
-
(N^2-1)\mu
\nonumber\\
&= {\rm Tr}\left(
\frac{1}{2}\hat{P}_I^2
-
\frac{g^2}{2 }[\hat{X}_1,\hat{X}_2]^2
+
\frac{g}{2}\hat{\xi}[-\hat{X}_1-i\hat{X}_2,\hat{\xi}]
+
\frac{g}{2}\hat{\xi}^\dagger[-\hat{X}_1+i\hat{X}_2,\hat{\xi}^\dagger]
+
\frac{3\mu}{2}\hat{\xi}^\dagger\hat{\xi}
+
\frac{\mu^2}{2}\hat{X}_I^2
\right)
\nonumber\\
& - (N^2-1)\mu \ .
\end{align}
The last term $-(N^2-1)\mu$ makes the ground state energy to be zero\footnote{
For simplicity, we assume $\mu\ge 0$.
This assumption is not essential.}.
Although the fermion number is not conserved because there are interaction terms including two $\hat{\xi}$'s or two $\hat{\xi}^\dagger$'s, fermion parity (i.e., fermion number even or odd) is conserved.

\subsubsection{Symmetry of the model}
The supersymmetry transformation is given by~\cite{Kim:2006wg}
\begin{align}
\delta_{\epsilon}\ \cdot\
=
[\hat{Q}\epsilon^\ast+\hat{Q}^\dagger\epsilon,
\ \cdot\ ] \ ,
\end{align}
where
\begin{align}
\hat{Q} = - \hat{\xi}^\dagger_\alpha
\left[
(\hat{P}_1^\alpha-i\hat{P}_2^\alpha)
-
i\mu
(\hat{X}_1^\alpha-i\hat{X}_2^\alpha)
\right]
-
\frac{ig}{\sqrt{2}}f_{\alpha\beta\gamma} \hat{\xi}^\alpha\hat{X}_1^\beta\hat{X}_2^\gamma \ .
\end{align}
Here $f_{abc}$ is the structure constant of SU($N$), which is $\epsilon_{abc}$ for SU(2).

It is convenient to use
\begin{align}
\hat{Z} = \frac{X_1-iX_2}{\sqrt{2}}, \qquad \hat{P}_Z
=
\frac{\hat{P}_1-i\hat{P}_2}{\sqrt{2}} \ ,
\end{align}
which satisfy
\begin{align}
[\hat{Z},\hat{P}^\dagger_Z] = [\hat{Z}^\dagger,\hat{P}_Z] =i \ , \qquad
[\hat{Z},\hat{P}_Z] = [\hat{Z}^\dagger,\hat{P}^\dagger_Z] =0 \ .
\end{align}

The Hamiltonian is
\begin{align}
\hat{H} &= {\rm Tr}\left(
\hat{P}_Z\hat{P}_Z^\dagger
+
\frac{g^2}{2}[\hat{Z},\hat{Z}^\dagger]^2
-
\frac{g}{\sqrt{2}}\hat{\xi}[\hat{Z}^\dagger,\hat{\xi}]
-
\frac{g}{\sqrt{2}}\hat{\xi}^\dagger[\hat{Z},\hat{\xi}^\dagger]
+
\frac{3\mu}{2}\hat{\xi}^\dagger\hat{\xi}
+
\mu^2\hat{Z}\hat{Z}^\dagger
\right) \ .
\end{align}
Hence the SO(2) rotation is
\begin{align}
\hat{Z} \to e^{i\theta}\hat{Z} \ , \qquad
\hat{\xi} \to e^{i\theta/2}\hat{\xi} \ .
\end{align}
The generator of SO(2) is
\begin{align}
\hat{M} = \sum_\alpha \left(
i(\hat{Z}_\alpha\hat{P}_{Z\alpha}^\dagger-\hat{Z}_\alpha^\dagger\hat{P}_{Z\alpha})
-
\frac{1}{2}\hat{\xi}_\alpha^\dagger\hat{\xi}_\alpha
\right) \ .
\end{align}
Note that $\hat{H}$ and $\hat{M}$ commute.
The supercharge is written as
\begin{align}
\hat{Q} = - \sqrt{2}\hat{\xi}^\dagger_\alpha
\left(
\hat{P}_{Z\alpha}
-
i\mu
\hat{Z}_\alpha
\right)
-
\frac{g}{\sqrt{2}}f_{\alpha\beta\gamma} \hat{\xi}_\alpha\hat{Z}_\beta\hat{Z}^\dagger_\gamma \ .
\end{align}
This satisfies
\begin{align}
\hat{Q}^2 = -i\hat{Z}_\alpha\hat{G}_\alpha \ , \qquad
\hat{Q}^{\dagger 2} = i\hat{Z}^\dagger_\alpha\hat{G}_\alpha \ ,
\end{align}
and
\begin{align}
\{\hat{Q},\hat{Q}^\dagger\} = 2\left(\hat{H} - \mu\hat{M}\right) \ .
\end{align}
This means $\hat{H}-\mu\hat{M}$ is positive semi-definite, and the BPS state should satisfy
\begin{align}
\hat{Q}|{\rm BPS}\rangle = \hat{Q}^\dagger|{\rm BPS}\rangle = \left(\hat{H}-\mu\hat{M}\right)|{\rm BPS}\rangle =0 \ .
\end{align}
When restricted to the gauge-invariant Hilbert space, $\hat{Q}$ and $\hat{Q}^\dagger$ commute with $\hat{H}-\mu\hat{M}$.
Hence any non-BPS states form a pair consisting of states in fermion-number-even and odd sectors.
Therefore, when the coupling is turned on, the BPS states must remain BPS.
In particular, the SO(2)-invariant ground state has to stay at zero energy.
This property is useful for a sanity check of numerical computations.

\section{Hamiltonian truncation and quantum simulation}\label{sec:Quantum-Simulation}
On a quantum computer, the dimension of the Hilbert space is finite, because the number of qubits is finite.
On the other hand, bosons require an infinite-dimensional Hilbert space.
The matrix models contain bosons, and hence, we have to truncate the Hilbert space.
The Hamiltonian is also truncated to a matrix of finite size, and hence, we call this method ``Hamiltonian truncation".
In the Hamiltonian truncation method, it is important to control the truncation effects.
How the full theory is recovered when the truncation level is taken to infinity is a well-known problem in many-body quantum physics~\cite{Cubitt:2015xsa}.

In this section, we test the performance of the Hamiltonian truncation method.
Specifically, we study the truncation effect, to estimate the amount of resources needed for precise quantum simulations.
As a concrete truncation scheme, we use a simple Fock-space truncation.
To confirm the validity of this approach, we implement the regularized Hamiltonian on a classical computer and study its properties numerically.
Numerical results in this section are obtained by using \qutip\footnote{
See \url{https://qutip.org/} for the official web page of \qutip.
For low-dimensional Hilbert spaces we have cross-checked the results using independent codes written in Python and FORTRAN.
} (Quantum Toolbox in Python)~\cite{qutip1,qutip2}.

In Sec.~\ref{sec:Hamiltonian-truncation}, we regularize the Hamiltonian by using the Fock basis, and explain how it can be written in terms of qubits.
Then we discuss the gauge-singlet constraint in Sec.~\ref{sec:singlet-constraint-QC}.
In Sec.~\ref{sec:exact_diagonalization}, we numerically study the regularized Hamiltonian and estimate the truncation effects.
The results in these sections tell us how we can control the truncation effects.
In Sec.~\ref{sec:VQE}, we will use the truncation studied in this section together with the VQE, a hybrid quantum-classical promising tool available for NISQ-era hardware.

\subsection{Regularization of the Hamiltonian}\label{sec:Hamiltonian-truncation}
We start with truncating the infinite-dimensional Hamiltonian to a finite dimension.
Here we closely follow Ref.~\cite{Gharibyan:2020bab} and use the Fock space truncation method.\footnote{
Another common method uses the coordinate basis~\cite{Jordan:2011ne,Jordan:2011ci,Klco:2018zqz,Gharibyan:2020bab} or the conformal truncation~\cite{Anand:2020gnn,Liu:2020eoa}.}
As a concrete example, let us consider the bosonic SU($N$) matrix model~\eqref{eq:bosonic_Hamiltonian}.
We can regularize the minimal BMN in a very similar manner, as we will see in Sec.~\ref{sec:Hamiltonian-truncation-minimal-BMN}.

We write the matrices as~\eqref{bosonic_generator}, where $\tau_\alpha$ in~\eqref{bosonic_generator} is the generator of SU($N$) satisfying ${\rm Tr}(\tau_\alpha\tau_\beta)=\delta_{\alpha\beta}$ and $[\tau_\alpha,\tau_\beta]=if_{\alpha\beta\gamma}\tau_\gamma$.
(The values of the structure constant $f_{\alpha\beta\gamma}$ depend on the detail of the choice of $\tau_\alpha$.)
Then the Hamiltonian~\eqref{eq:bosonic_Hamiltonian} is written as
\begin{align}\label{eq:bos-ham-reg}
\hat{H} = \sum_{\alpha,I}
\left(
\frac{1}{2}\hat{P}_{I\alpha}^2
+
\frac{m^2}{2}\hat{X}_{I\alpha}^2
\right)
+
\frac{g^2}{4}\sum_{\gamma,I,J}\left(\sum_{\alpha,\beta}f_{\alpha\beta\gamma}\hat{X}_I^\alpha\hat{X}_J^\beta\right)^2 \ .
\end{align}
The canonical commutation relation is given by~\eqref{canonical_commutation_relation_XP}.
We introduce the creation and annihilation operators as
\begin{align}
\hat{a}_{I\alpha}^\dagger = \sqrt{\frac{m}{2}} \hat{X}_{I\alpha} - \frac{i\hat{P}_{I\alpha}}{\sqrt{2m}} \ , \qquad
\hat{a}_{I\alpha} = \sqrt{\frac{m}{2}} \hat{X}_{I\alpha} + \frac{i\hat{P}_{I\alpha}}{\sqrt{2m}} \ .
\end{align}
By using the creation and annihilation operators, and the number operator $\hat{n}_{I\alpha}=\hat{a}_{I\alpha}^\dagger\hat{a}_{I\alpha}$, the Hamiltonian can be expressed as
\begin{align}
\hat{H} = m\sum_{\alpha,I} \left(\hat{n}_{I\alpha} + \frac{1}{2} \right)
+
\frac{g^2}{16m^2}\sum_{\gamma,I,J}\left(\sum_{\alpha,\beta}f_{\alpha\beta\gamma}(\hat{a}_{I\alpha}+\hat{a}_{I\alpha}^\dagger)(\hat{a}_{J\beta}+\hat{a}_{J\beta}^\dagger)\right)^2 \ .
\end{align}
The generators of the gauge transformation are
\begin{align}
\hat{G}_\alpha = i\sum_{\beta,\gamma,I}f_{\alpha\beta\gamma} \hat{a}_{I\beta}^\dagger\hat{a}_{I\gamma} \ .
\end{align}
The Hamiltonian is gauge-invariant, or equivalently,
\begin{align}
[\hat{H},\hat{G}_\alpha] = 0 \ .
\end{align}
For each $(I,\alpha)$, we can take the Fock vacuum $\ket{0}_{I\alpha}$  which satisfies
\begin{align}
\hat{a}_{I\alpha}\ket{0}_{I\alpha}=0
\end{align}
and the excited states $\ket{n}_{I\alpha}$ defined by
\begin{align}
\ket{n}_{I\alpha} = \frac{(\hat{a}_{I\alpha}^\dagger)^n}{\sqrt{n!}}\ket{0}_{I\alpha} \ .
\end{align}
We can take the Fock states for the matrix model by taking the tensor product, as
\begin{align}
\ket{\{n_{I\alpha}\}} = \otimes_{I,\alpha}\ket{n}_{I\alpha}.
\end{align}
Note that the Fock vacuum is gauge invariant, i.e.,
\begin{align}
\hat{G}_\alpha\left(\otimes_{I,\beta}\ket{0}_{I\beta}\right)=0
\end{align}
for all gauge generators $\hat{G}_\alpha$.
As a regularization, we truncate the Hilbert space such that the excitation level of each oscillator is below $\Lambda$.
We define the truncated raising, lowering, and number operators to be
\begin{align}\label{raiseoperators-truncated-in-Fock-basis}
\hat{a}_{\rm truncated}^\dag = \sum\limits_{n= 0}^{\Lambda - 2} {\sqrt {n+ 1} } |n+ 1\rangle \langle n| \ ,
\qquad
\hat{a}_{\rm truncated} = \sum\limits_{n= 0}^{\Lambda - 2} {\sqrt {n+ 1} } |n\rangle \langle n+1|
\end{align}
and
\begin{align}\label{numoperator-truncated-in-Fock-basis}
\hat{n}_{\rm truncated} = \sum\limits_{n = 0}^{\Lambda - 1} n  |n \rangle \langle n| \ .
\end{align}
Note that the truncation breaks gauge symmetry manifestly, i.e.,
\begin{align}
[\hat{H}_{\rm truncated},\hat{G}_{\alpha,{\rm truncated}}] \neq 0 \ .
\end{align}

In the truncated Fock basis, the creation and annihilation operators are expressed as simple Pauli strings (i.e., tensor products of Pauli matrices), which are convenient for quantum computations~\cite{Gharibyan:2020bab}.
In Eq.~\eqref{raiseoperators-truncated-in-Fock-basis} and~\eqref{numoperator-truncated-in-Fock-basis}, let $|n\rangle$ ($n=0,1,\cdots,\Lambda-1$) be the $j$-th excited state of an oscillator.
We can write $j$ in terms of binaries as $n = \sum_{l=0}^{K-1}b_l2^l$.
By using $K=\log_2\Lambda$ qubits, we can rewrite the state $\ket{j}$ as
\begin{align}
|n\rangle  = \left| {{b_{0}}} \right\rangle \left| {{b_{1}}} \right\rangle  \ldots \left| {{b_{K-1}}} \right\rangle \ .
\end{align}
Writing $|n\rangle  = \left| {{b_{0}}} \right\rangle \left| {{b_{1}}} \right\rangle  \ldots \left| {{b_{K-1}}} \right\rangle$
and $|n+1\rangle  = \left| {{b'_0}} \right\rangle \left| {{b'_1}} \right\rangle  \ldots \left| {{b'_{K-1}}} \right\rangle$, we can express $|n + 1\rangle \langle n|$ as an operator in this basis as
\begin{align}
|n + 1\rangle \langle n| = \otimes_{l=0}^{K-1} \left(|b'_l\rangle\langle b_l|\right) \ .
\label{adag-in-Fock-basis-2}
\end{align}
Note that each $|b'_l\rangle\langle b_l|$ is a linear combination of the Pauli matrices:
\begin{align}
&
|0\rangle\langle 0|=\frac{\textbf{1}_2-\sigma_z}{2} \ ,
\qquad
|1\rangle\langle 1|=\frac{\textbf{1}_2+\sigma_z}{2} \ ,
\nonumber\\
&
|0\rangle\langle 1|=\frac{\sigma_x+i\sigma_y}{2} \ ,
\qquad
|1\rangle\langle 0|=\frac{\sigma_x-i\sigma_y}{2} \ .
\label{adag-in-Fock-basis-3}
\end{align}
The annihilation operator $\hat{a}_{\rm truncated}$ and the number operator $\hat{n}_{\rm truncated}$ have similar forms.

\subsection{Gauge singlets and non-singlets}\label{sec:singlet-constraint-QC}
The physical states in gauge theory are gauge singlets (gauge-invariant states).
Let us denote the Hilbert space spanned by singlets by $\mathcal{H}_{\rm inv}$.
Our construction involves the extended Hilbert space $\mathcal{H}_{\rm ext}$ which contains gauge non-singlet states.
Depending on the problem under consideration, it may or may not be an issue.
Firstly, it is not a problem as long as one considers an exact time evolution of the exact Hamiltonian (without truncation), for the following reasons:

\begin{itemize}
\item
Time evolution commutes with the gauge transformation, i.e., $[\hat{G}_\alpha,e^{-i\hat{H}t}]=0$.
Therefore, if we take the initial state to be in $\mathcal{H}_{\rm inv}$, it stays in $\mathcal{H}_{\rm inv}$.
\item
From any non-singlet state $|\phi\rangle$, we can obtain a singlet state by projecting\footnote{
Specifically, we take the average over all gauge transformation $\frac{1}{\sqrt{{\cal N}_\phi}}\int dU (\hat{U}|\phi\rangle)$, where the operator $\hat{U}$ acts as the SU($N$) transformation and ${\cal N}_\phi$ is the normalization factor.
} it to $\mathcal{H}_{\rm inv}$.
We can regard $|\phi\rangle$ as a gauge-fixed state.
The projection and time evolution commute with each other.
Often, the gauge-fixed states are easier to handle.
\end{itemize}

The situation can be different when a quantum simulation is considered.
By assuming that the noise is completely removed, there are two sources of errors.
The first one is the error associated with the truncation of the Hilbert space.
The second one is the error associated with the approximation of $e^{-i\hat{H}t}$, such as the Trotter-Suzuki error.
For each algorithm, in principle, it is straightforward to make the second type of error as small as we want.
Hence let us focus on the first type of error.
Namely we assume that $e^{-i\hat{H}_{\rm truncated}t}$ is constructed precisely.
We need to truncate the generators of the gauge transformation $\hat{G}_\alpha$ as well.
Due to the truncation error, neither $\hat{G}_\alpha$ nor $\hat{G}_{\alpha,{\rm truncated}}$ commute with $\hat{H}_{\rm truncated}$.
Therefore, even if the initial state is taken to be (approximately) gauge-invariant, the singlet condition may be broken badly after a long time.

A related problem is that, when we calculate low-lying modes of the Hamiltonian, many non-singlet modes, which are not necessarily of interest, are obtained.
We might be able to circumvent these problems by adding a term proportional to $\hat{G}^2 \equiv \sum_\alpha\hat{G}_\alpha^2$ to the Hamiltonian~\cite{Zohar:2011cw,Zohar:2012ay,Halimeh:2021vzf,Dalmonte:2016alw}:
\begin{align}\label{eq:deformed-Hamiltonian}
\hat{H}' = \hat{H} + c\sum_\alpha\hat{G}_\alpha^2 \ .
\end{align}
Here $c$ is a free parameter.

Let $|E'\rangle$ be an eigenstate of $\hat{H}'$ with eigenvalue $\lambda_{E'}$, i.e., $\hat{H}'|E'\rangle=\lambda_{E'}|E'\rangle$.
Let $E'$ be the expectation value of the original Hamiltonian $\hat{H}$ for eigenstate $|E'\rangle$, i.e., $E'=\langle E'|\hat{H}|E'\rangle$.
Then, by definition, $\lambda_{E'}-E'=c\sum_\alpha\langle E'|\hat{G}_\alpha^2|E'\rangle$.
We will use $c=\Lambda$ later in this paper.
Because gauge-singlet states are zero modes of $\sum_\alpha\hat{G}_\alpha^2$, such a deformation makes the non-singlets heavy and decouple from the low-energy dynamics.
A nontrivial issue here is whether $\lambda_{E'}$ converges to the energy eigenvalue of the original Hamiltonian.
As we will see, this is the case indeed, as $\sum_\alpha\langle E'|\hat{G}_\alpha^2|E'\rangle$ decays exponentially fast as $\Lambda$ becomes large.

If the initial state is taken to be sufficiently low energy in terms of $\hat{H}'_{\rm truncated}$ and $e^{-i\hat{H}'_{\rm truncated}t}$ is calculated precisely, then the singlet condition is preserved with a good precision.
Note that, even with this deformation, the Hamiltonian can be written in a simple form in terms of the Pauli strings.
Therefore, we can study this modified Hamiltonian efficiently on a quantum computer.

\subsection{Energy spectrum via classical computation}\label{sec:exact_diagonalization}

\begin{figure}[htbp]
  \begin{center}
  \begin{tabular}{cc}
      \includegraphics[width=0.30\textwidth]{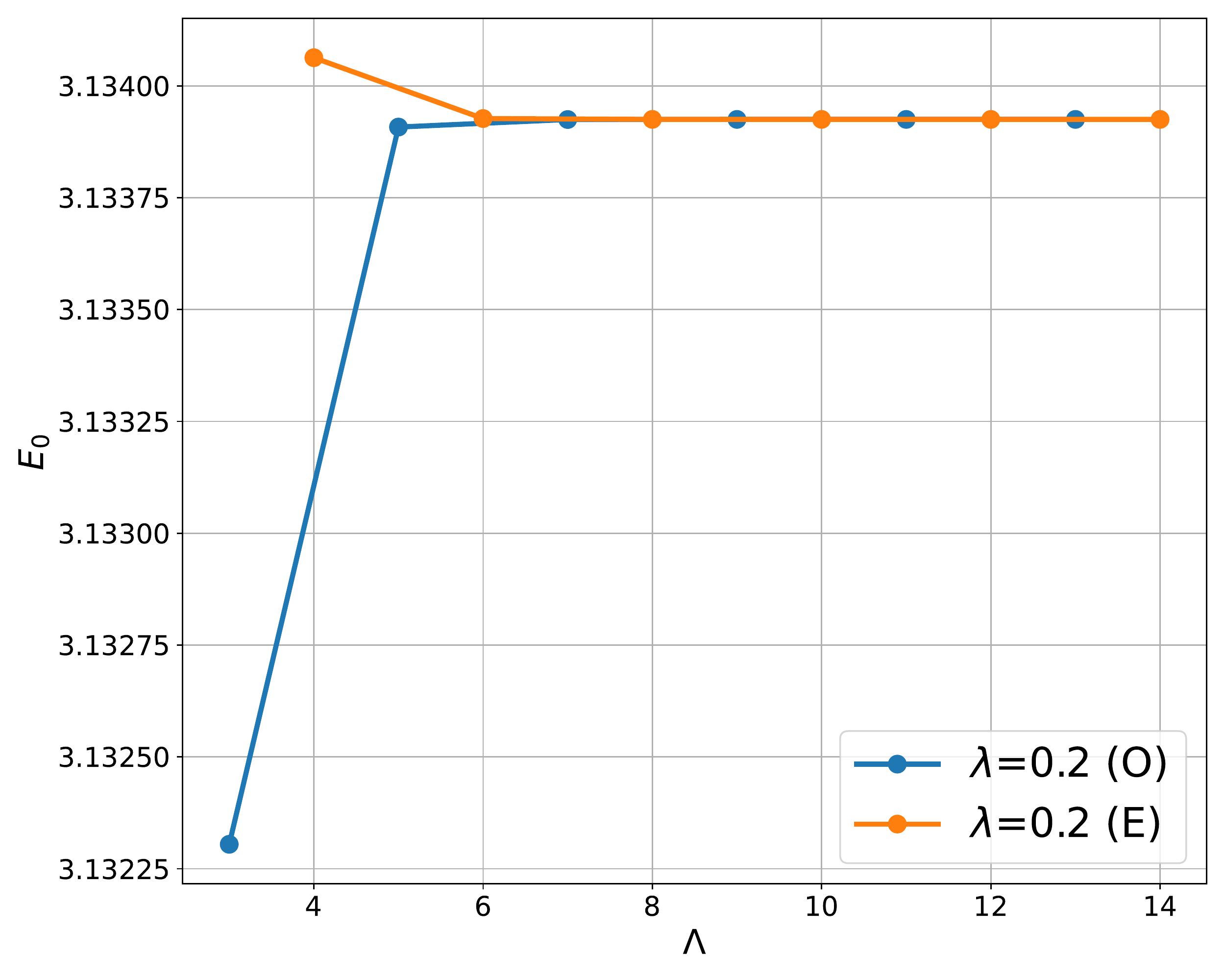} &
      \includegraphics[width=0.30\textwidth]{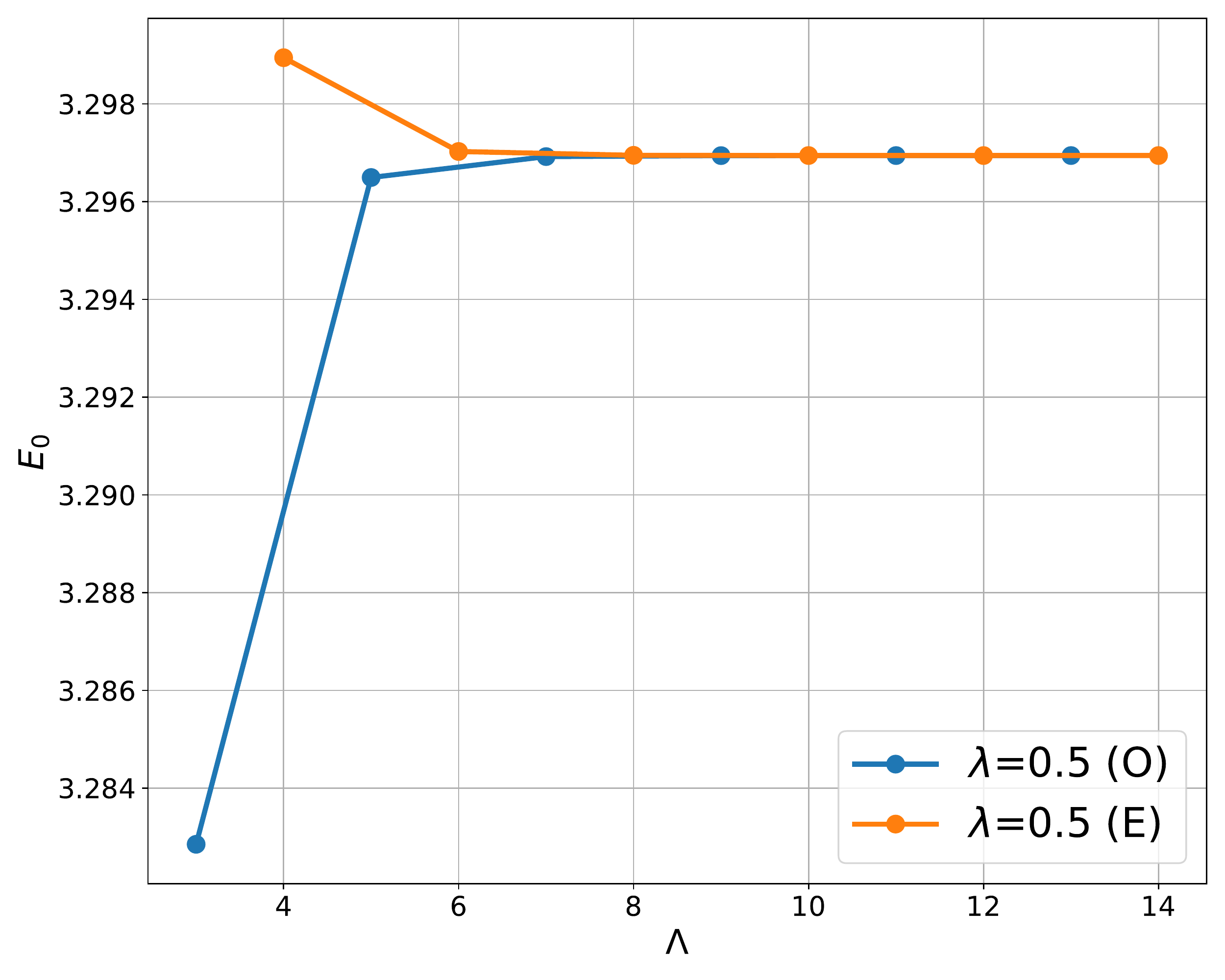} \\
      \includegraphics[width=0.30\textwidth]{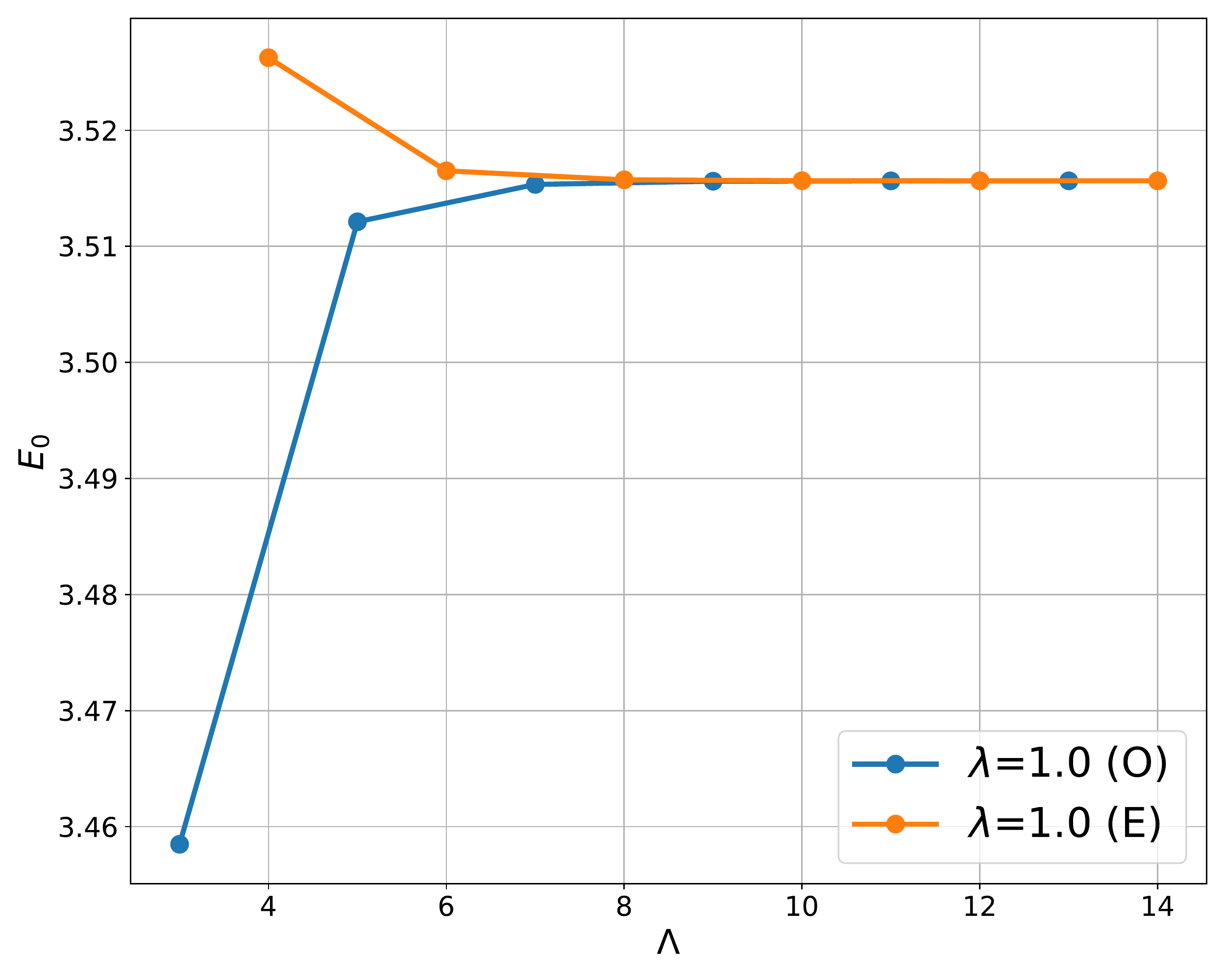} &
      \includegraphics[width=0.30\textwidth]{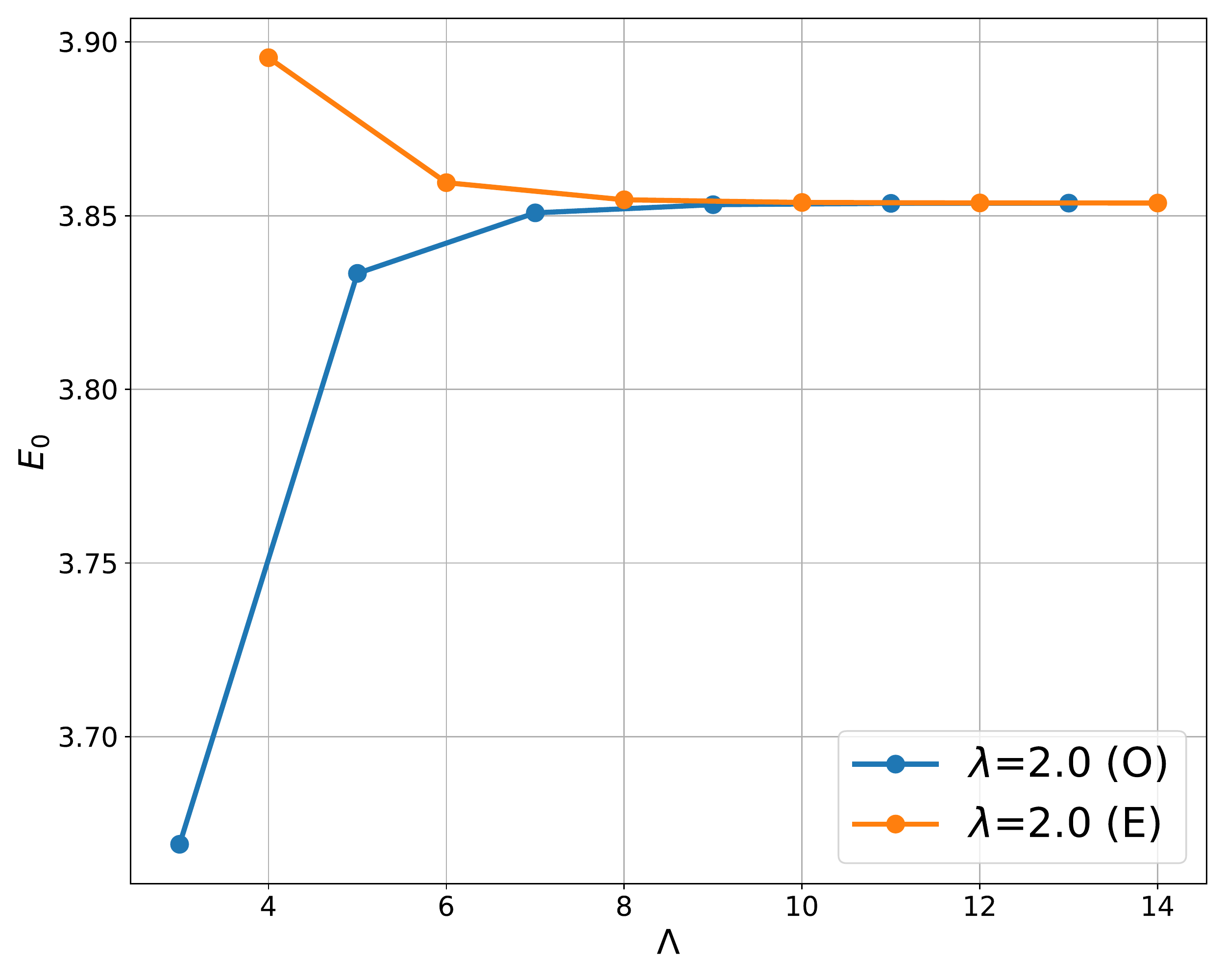}
  \end{tabular}
  \end{center}
  \caption{\label{fig:ham_su2_gs}
  The ground state energy $E_0$ as a function of the cutoff $\Lambda$ for various couplings $\lambda=g^2N=0.2$, 0.5, 1.0, and 2.0 for the SU(2) bosonic model.
  Even (E) and odd (O) values of $\Lambda$ are plotted with different colors.
  Other parameters are $m^2=1$ and $c=0$.
  }
\end{figure}

\subsubsection{SU(2) bosonic matrix model}
We consider the SU(2) two-matrix model.
Following Sec.~\ref{sec:Hamiltonian-truncation}, we can easily write down the Hamiltonian explicitly, by taking $\tau_\alpha=\frac{\sigma_\alpha}{\sqrt{2}}$ and $f_{\alpha\beta\gamma}=\sqrt{2}\epsilon_{\alpha\beta\gamma}$, where $\epsilon_{123}=\epsilon_{231}=\epsilon_{312}=+1$, $\epsilon_{321}=\epsilon_{213}=\epsilon_{132}=-1$.

The Hamiltonian is sparse (only a small fraction of the matrix elements are non-zero) and we can make efficient use of optimized sparse linear algebra solvers from the ARPACK~\cite{arpack} software suite, such as implicitly restarted arnoldi methods~\cite{arpackusers}, to find the eigenvalues and eigenvectors.

First, we use the setup without deformation, i.e., $c=0$ and hence $\hat{H}' = \hat{H}$ (cfr. Eq.~\eqref{eq:deformed-Hamiltonian}).
We obtain the ground state, which we denote by $|E_0\rangle$.
We estimate the ground-state energy $E_0 = \langle E_0 | \hat{H} | E_0 \rangle$ and the violation of the singlet constraint $\sum_\alpha\langle E_0|\hat{G}^2_\alpha|E_0\rangle$ for the cutoff $\Lambda=3,4,\cdots,14$ and coupling $\lambda=g^2N= 0.2$, 0.5, 1.0, and 2.0.
Respectively, we plot them in Fig.~\ref{fig:ham_su2_gs} and Fig.~\ref{fig:ham_su2_gv}.
The tables corresponding to the plots are reported in Appendix~\ref{appendix:Ham_truncation}.

\begin{figure}[htbp]
  \begin{center}
  \begin{tabular}{cc}
      \includegraphics[width=0.30\textwidth]{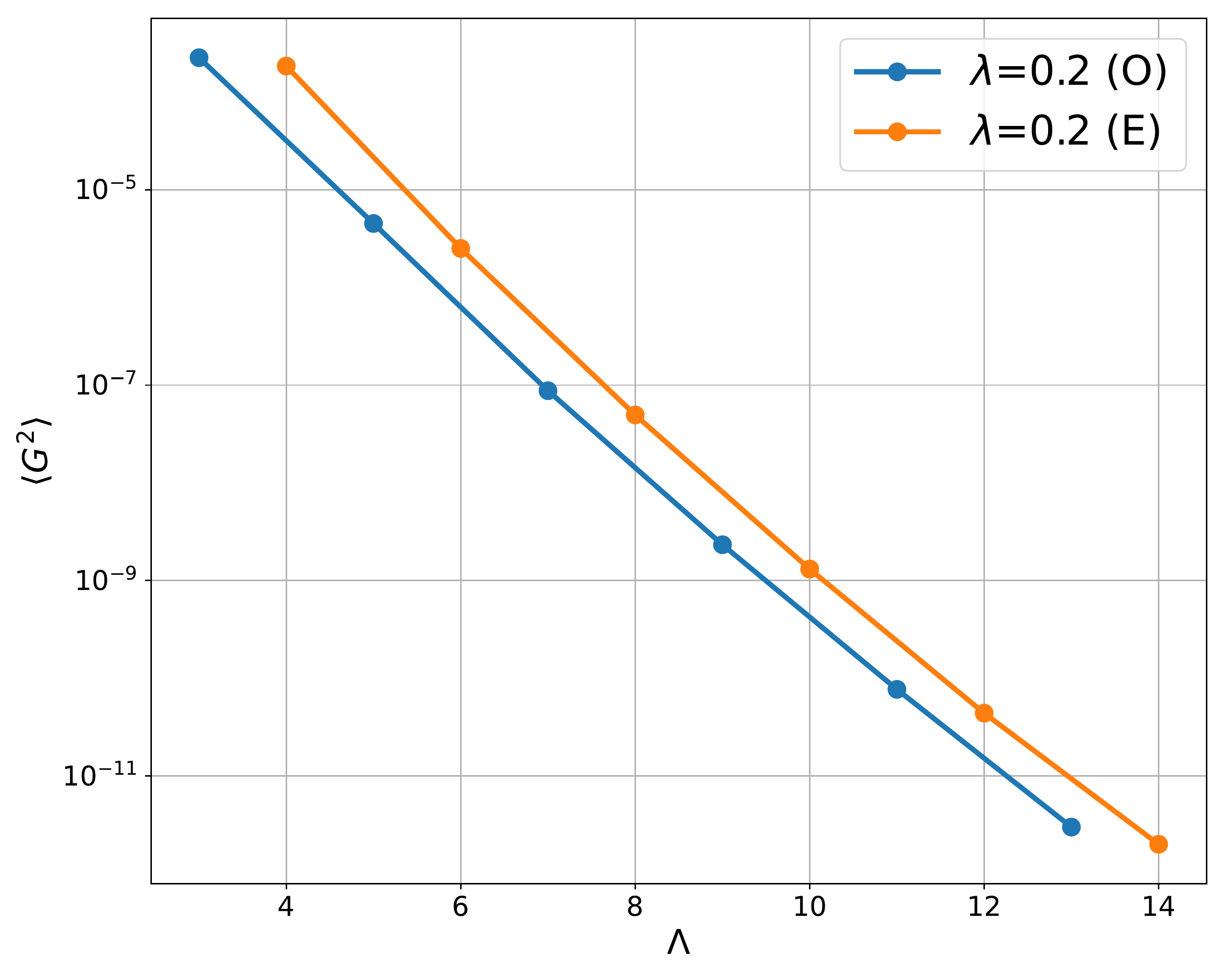} &
      \includegraphics[width=0.30\textwidth]{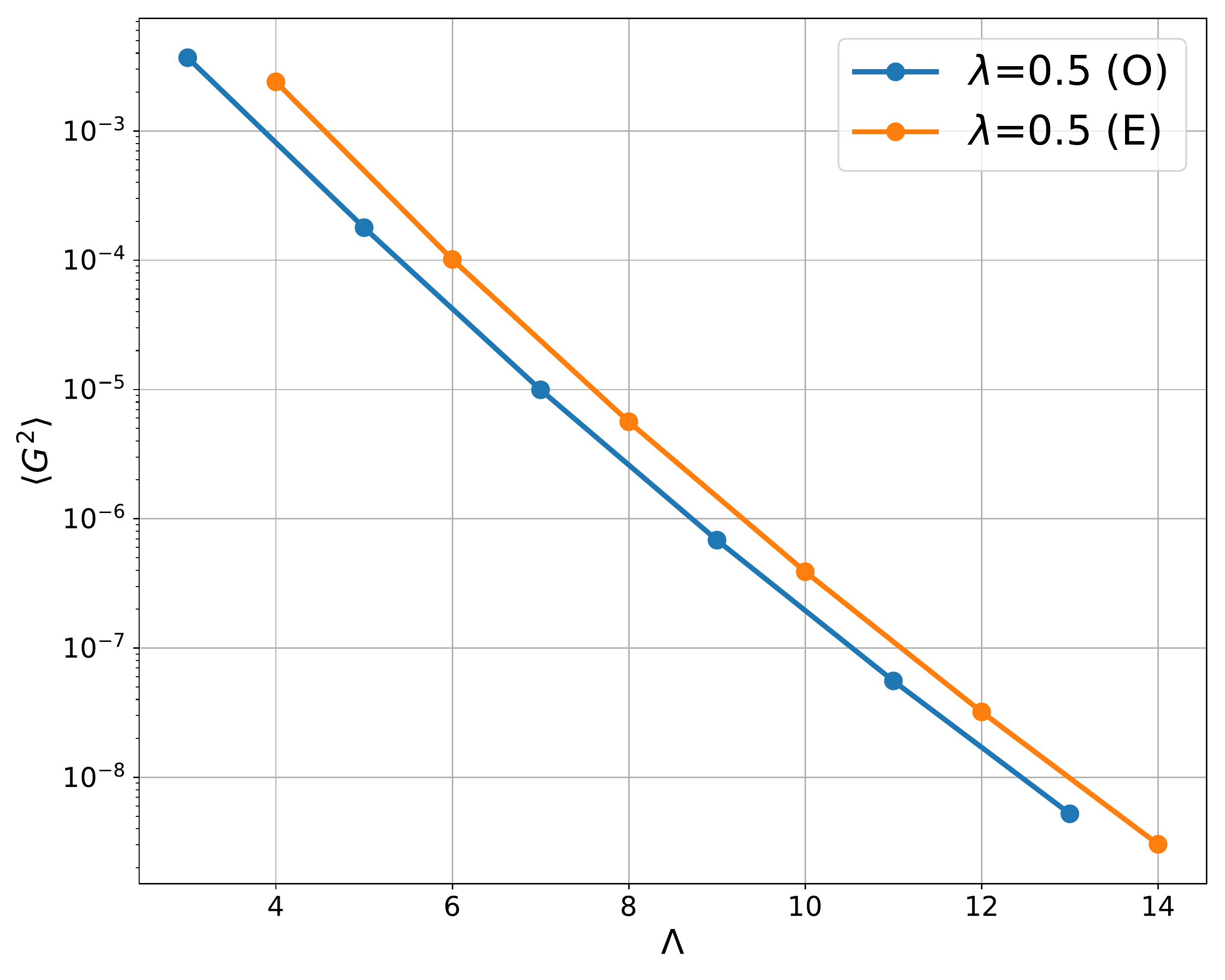} \\
      \includegraphics[width=0.30\textwidth]{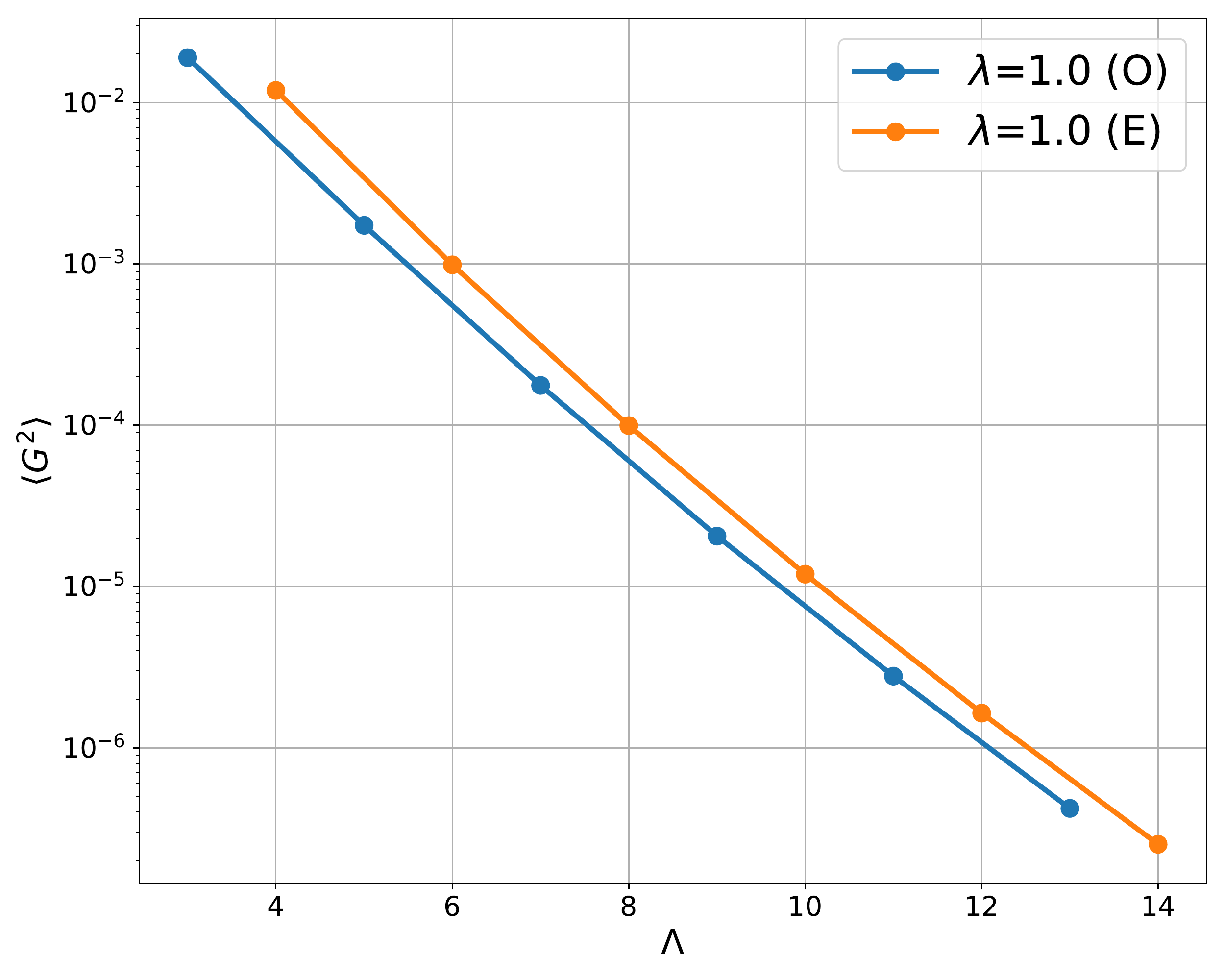} &
      \includegraphics[width=0.30\textwidth]{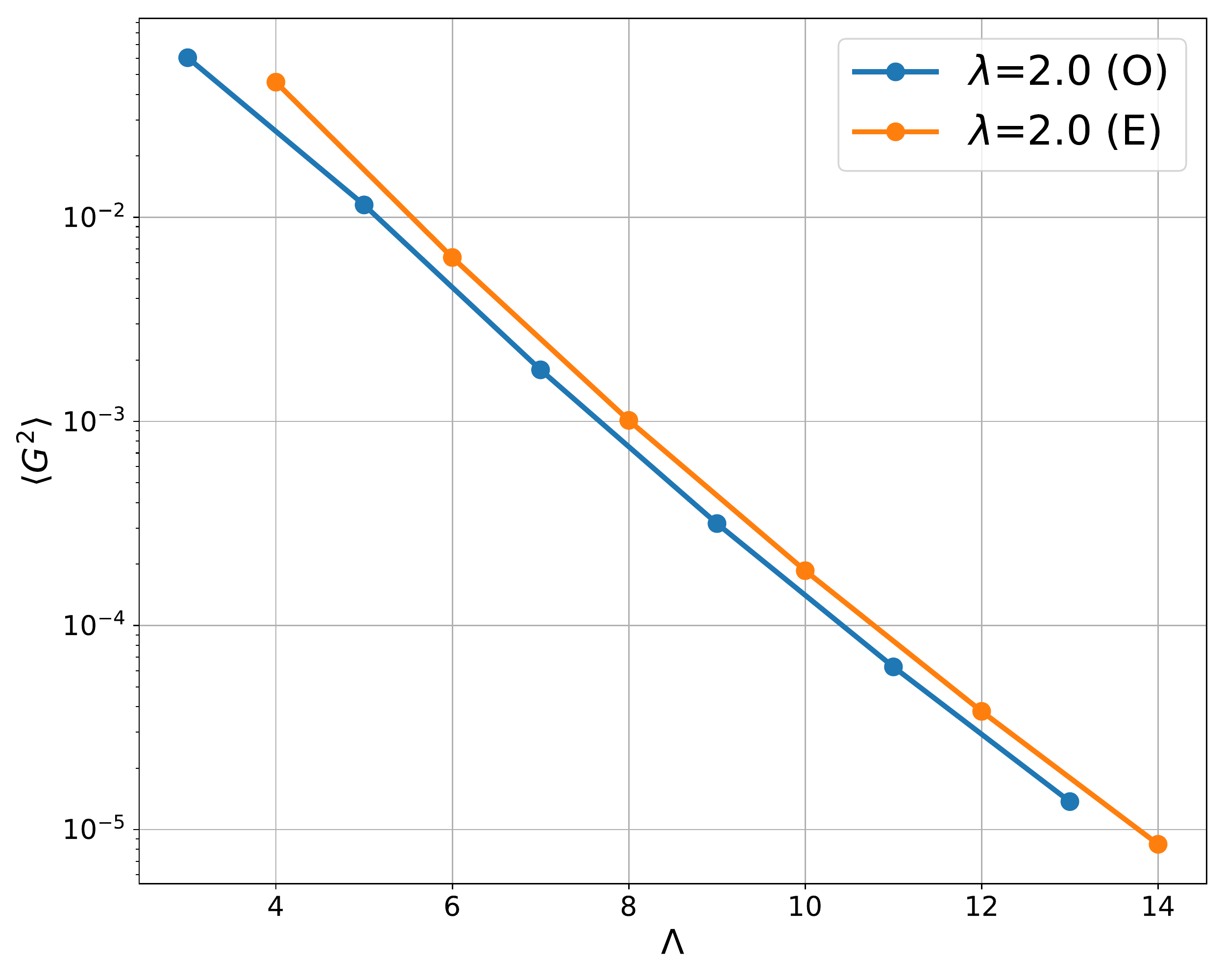}
  \end{tabular}
  \end{center}
  \caption{\label{fig:ham_su2_gv}
  The violation of the singlet constraint $\sum_\alpha\langle E_0|\hat{G}^2_\alpha|E_0\rangle$ as a function of the cutoff $\Lambda$ for various couplings  $\lambda=g^2N=0.2$, 0.5, 1.0, and 2.0 for the SU(2) bosonic model.
  Even (E) and odd (O) values of $\Lambda$ are plotted with different colors in logarithmic scale.
  Other parameters are $m^2=1$ and $c=0$.
  }
\end{figure}

At each value of the coupling constant, we can see that $\sum_\alpha\langle E_0|\hat{G}^2_\alpha|E_0\rangle$ scales as $e^{-a\Lambda}$, with the same exponent $a$ but different overall factor for even and odd $\Lambda$; see Fig.~\ref{fig:ham_su2_gv}.
It suggests that we can take $c$ to be an arbitrary power of $\Lambda$; we expect that, as long as it is smaller that $e^{+a\Lambda}$, it does not affect the ground state, and probably the excited states as well. (We will confirm this shortly, by taking $c=\Lambda$.)

\begin{figure}[htbp]
  \begin{center}
  \begin{tabular}{cc}
      \includegraphics[width=0.30\textwidth]{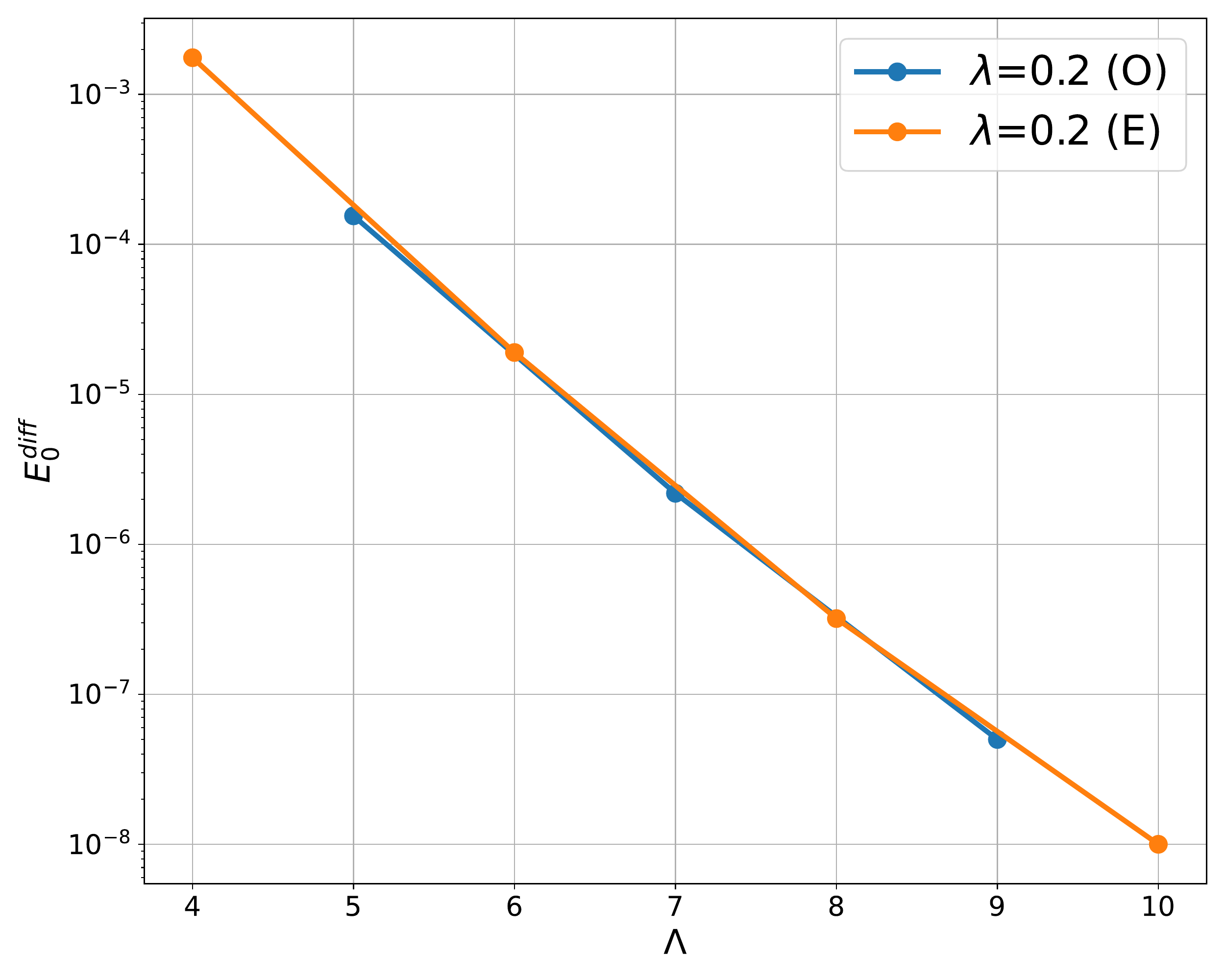} &
      \includegraphics[width=0.30\textwidth]{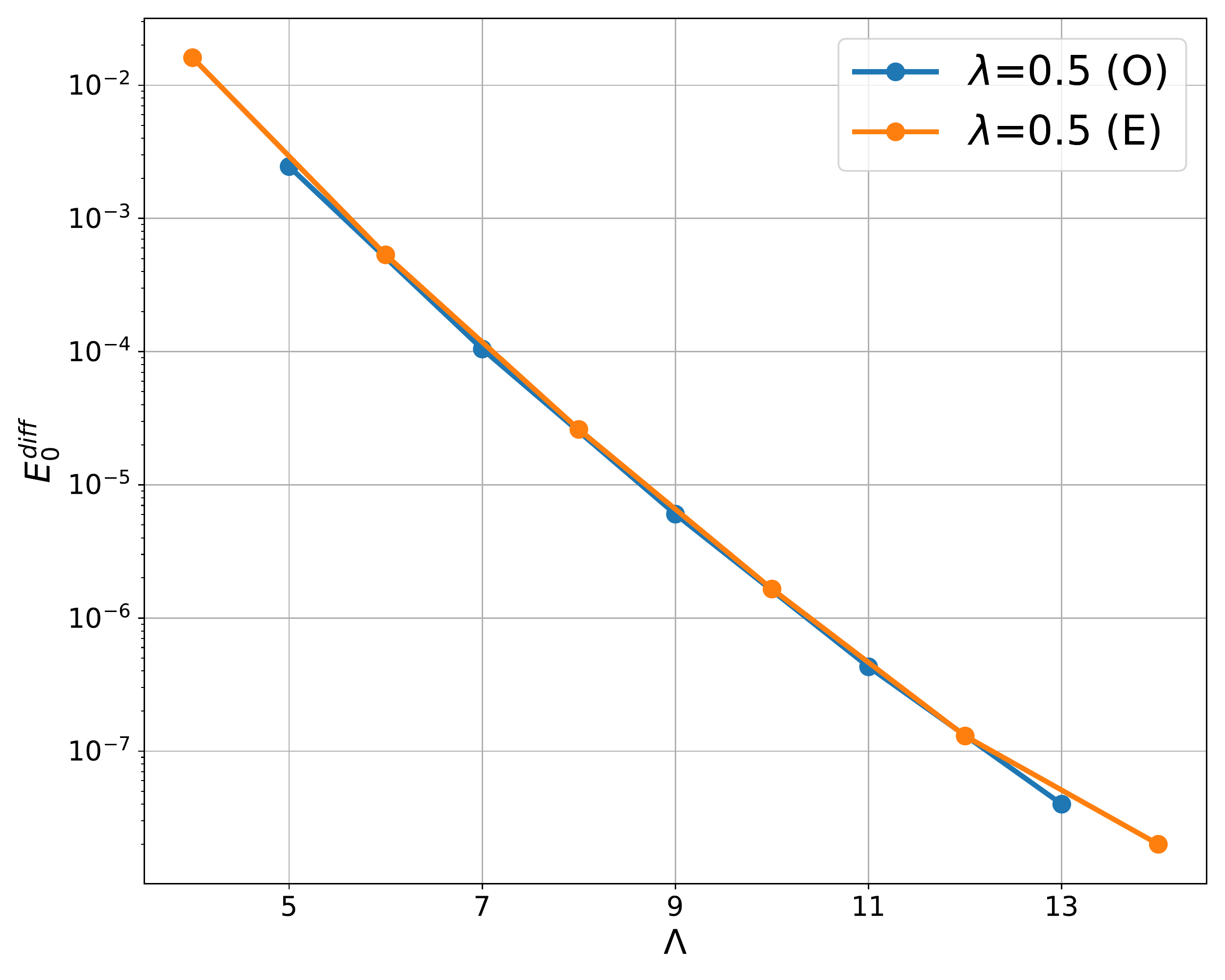} \\
      \includegraphics[width=0.30\textwidth]{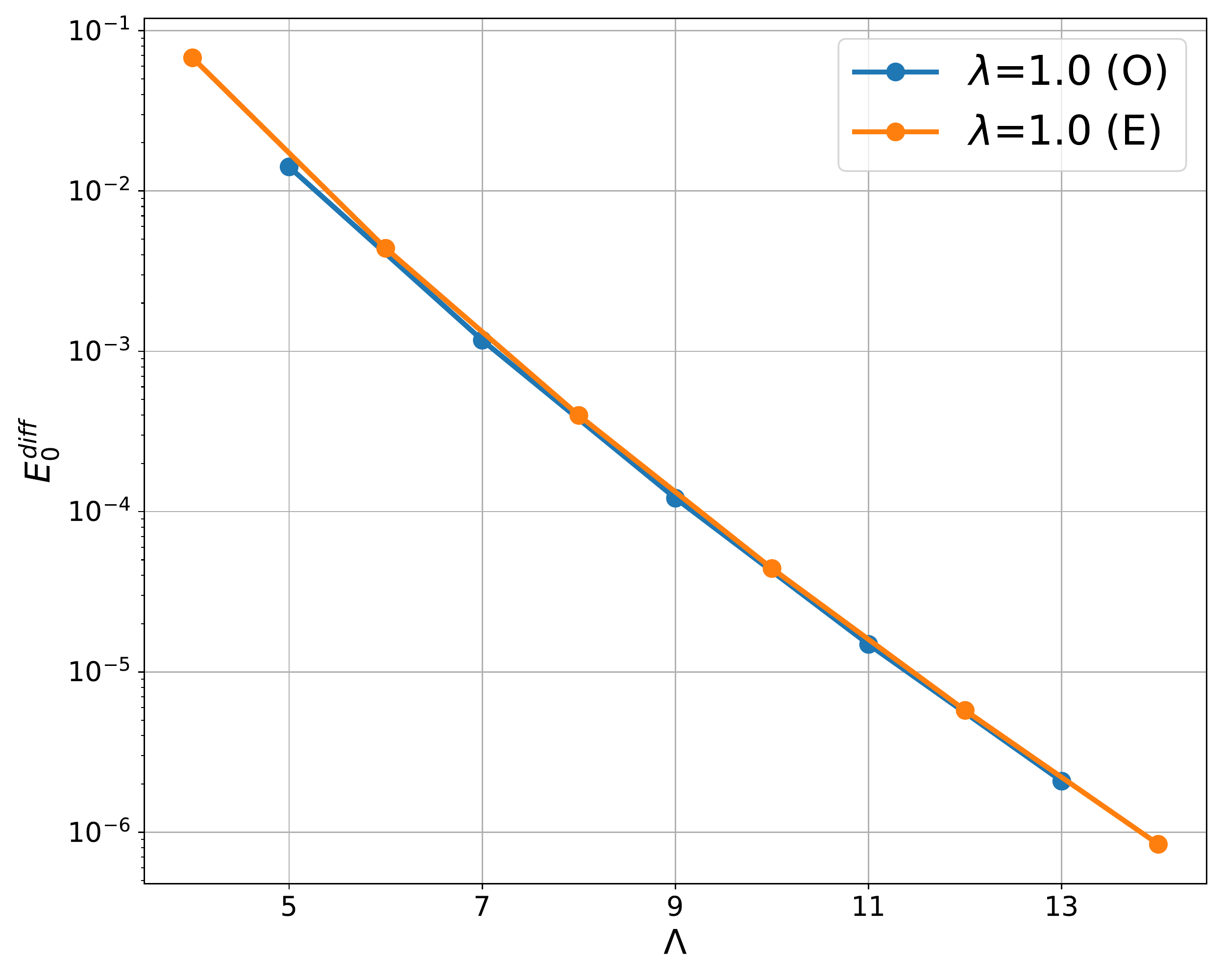} &
      \includegraphics[width=0.30\textwidth]{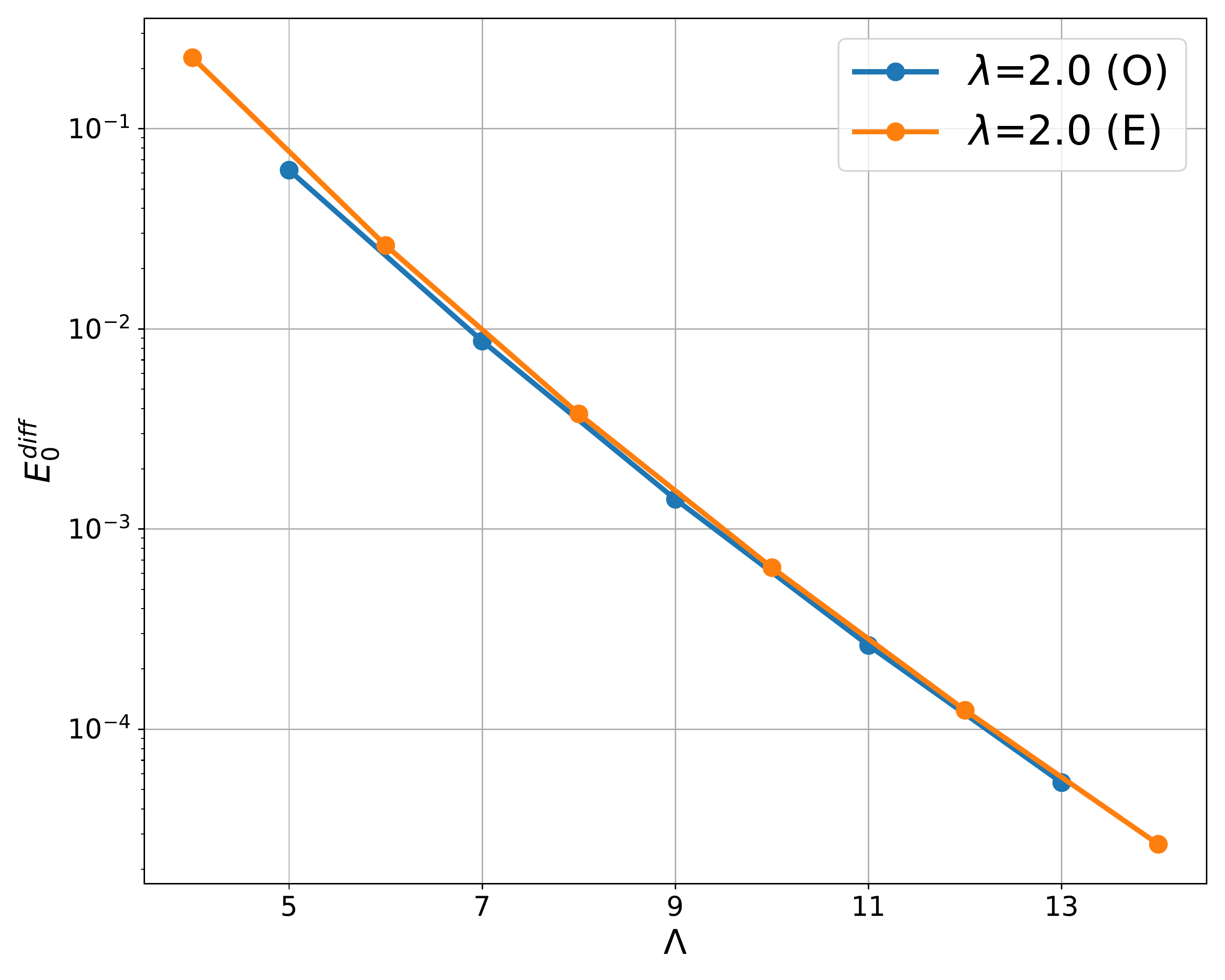}
  \end{tabular}
  \end{center}
  \caption{\label{fig:ham_su2_gs_diff}
  The absolute value of the ground state energy difference between successive values of the cutoff (see text for details) as a function of the cutoff $\Lambda$ for various couplings $\lambda=g^2N=0.2$, 0.5, 1.0, and 2.0 for the SU(2) bosonic model.
  Even (E) and odd (O) values of $\Lambda$ are plotted with different colors and the vertical axis is in logarithmic scale.
  Other parameters are $m^2=1$ and $c=0$.
  }
\end{figure}

In Fig.~\ref{fig:ham_su2_gs}, the approach of the ground-state energy to the large-$\Lambda$ limit appears to be exponentially fast.
To confirm the exponential decay manifestly, we plotted the difference of ground state energy for cutoffs differing by one unit $E_0^{\textrm{diff}}(\Lambda)\equiv \mid E_0(\Lambda)-E_0(\Lambda-1) \mid$ we can see that it decays exponentially with $\Lambda$ in Fig.~\ref{fig:ham_su2_gs_diff}.

Next, we consider the setup with deformation $\hat{H}' = \hat{H} + c \hat{G}^2$ (cfr. Eq.~\eqref{eq:deformed-Hamiltonian}).
We take $c=\Lambda$, such that the non-singlet modes completely decouple in the limit $\Lambda\to\infty$.
For this deformed Hamiltonian $\hat{H}'$ we compute the low-lying eigenstates $|E'\rangle$.
Let us first set the mass and coupling to $m^2=1$ and $\lambda = 0.2$, respectively.

\begin{figure}[htbp]
  \begin{center}
      \includegraphics[width=0.50\textwidth]{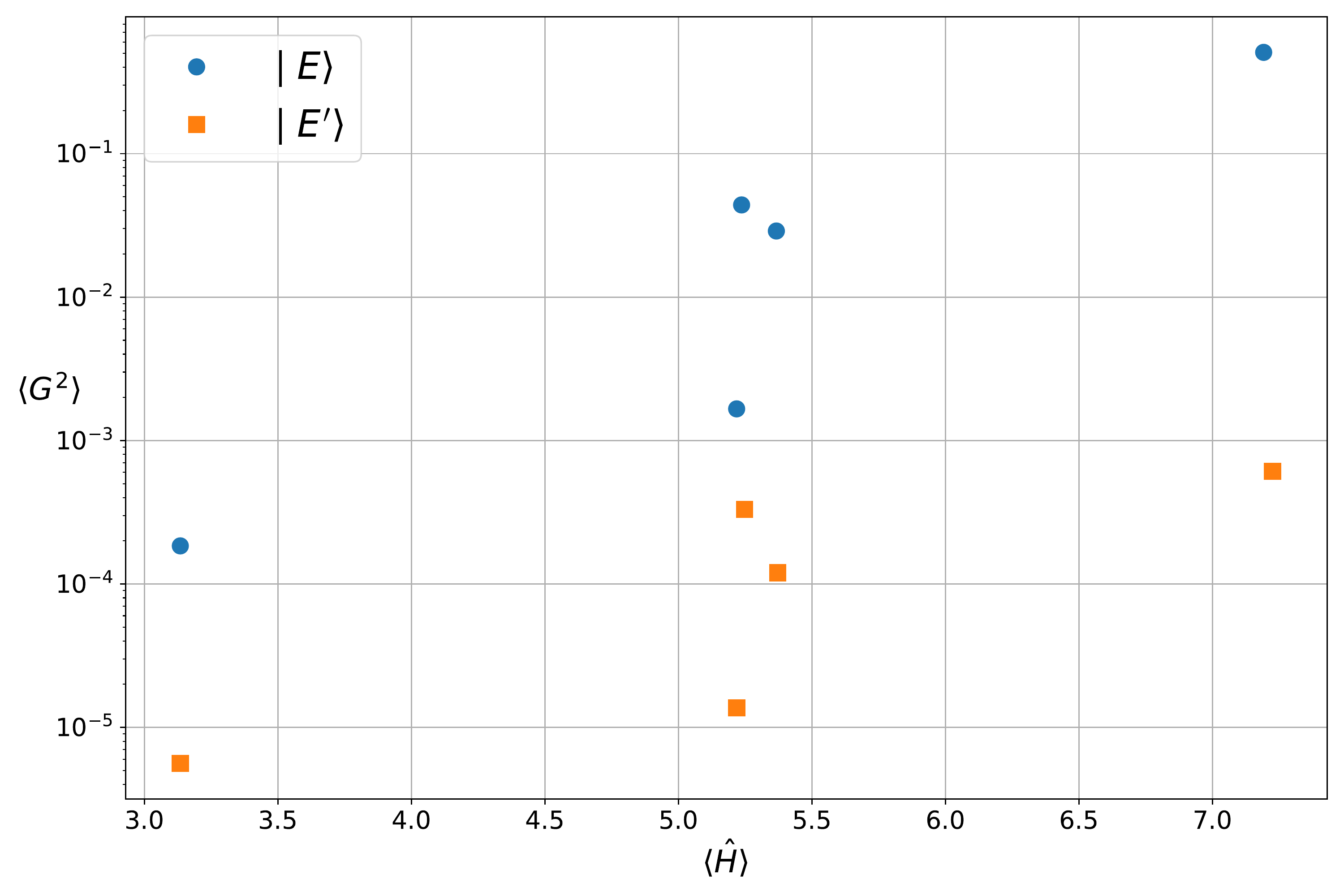}
  \end{center}
  \caption{\label{fig:L4_su2_gs_gv_l02}
  We take the five eigenstates of $\hat{H}$ ($| E_j \rangle$) with the smallest values of $\sum_\alpha\langle E_j|\hat{G}^2_\alpha|E_j\rangle$, and the five lowest eigenstates for $\hat{H}'$ ($| E'_j \rangle$ with $j=0,\ldots,4$) and we plot them as a function of their energy $\langle \hat{H} \rangle$ for $c=\Lambda=4$, $m^2=1$, $\lambda=g^2N=0.2$ in the SU(2) bosonic model.
  We can see the deformation does not affect the low-lying modes, it just removes the non-singlet modes.
  }
\end{figure}

In Fig.~\ref{fig:L4_su2_gs_gv_l02}, we compare five eigenstates of $\hat{H}$ with the smallest values of $\sum_\alpha\langle E|\hat{G}^2_\alpha|E\rangle$, and five eigenstates of $\hat{H}'$ with the smallest eigenvalues $\lambda_{E'} = \langle E' | \hat{H}' | E' \rangle$, for $c=\Lambda=4$, $m^2=1$, $\lambda=0.2$.
We can see the deformation does not affect the low-lying modes, it just removes the non-singlet modes.
\begin{figure}[htbp]
  \begin{center}
  \begin{tabular}{cc}
      \includegraphics[width=0.30\textwidth]{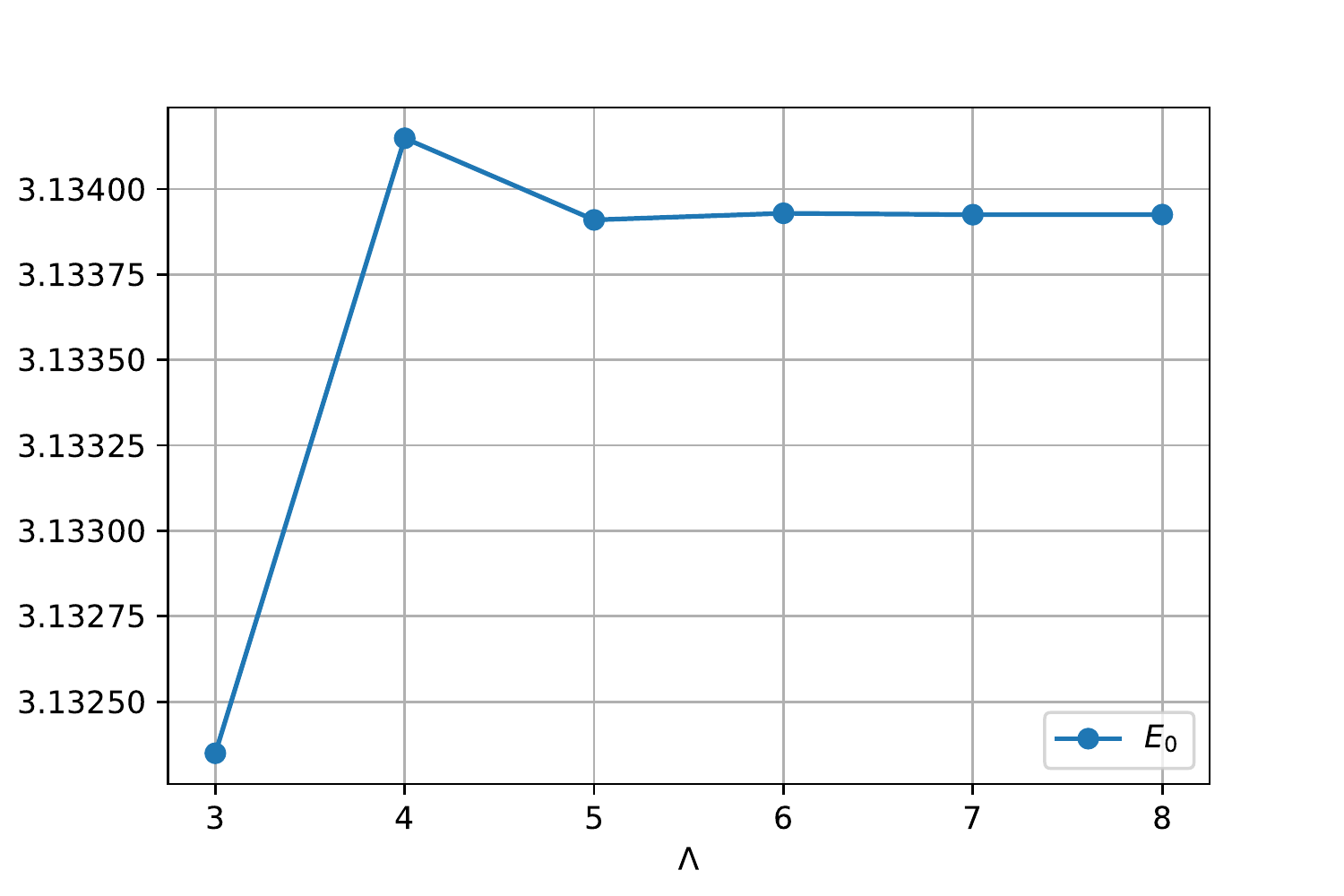} &
      \includegraphics[width=0.30\textwidth]{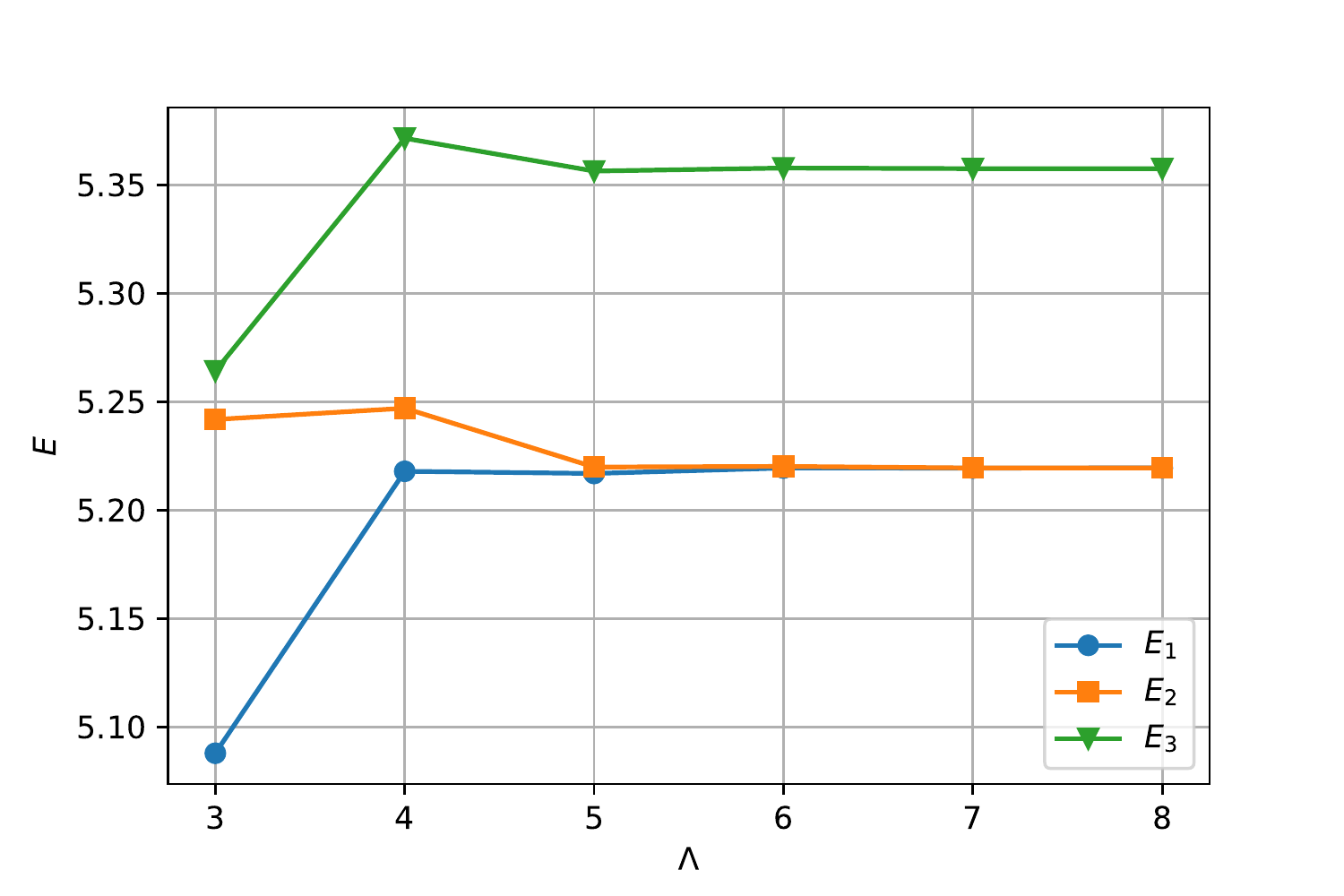} \\
      \includegraphics[width=0.30\textwidth]{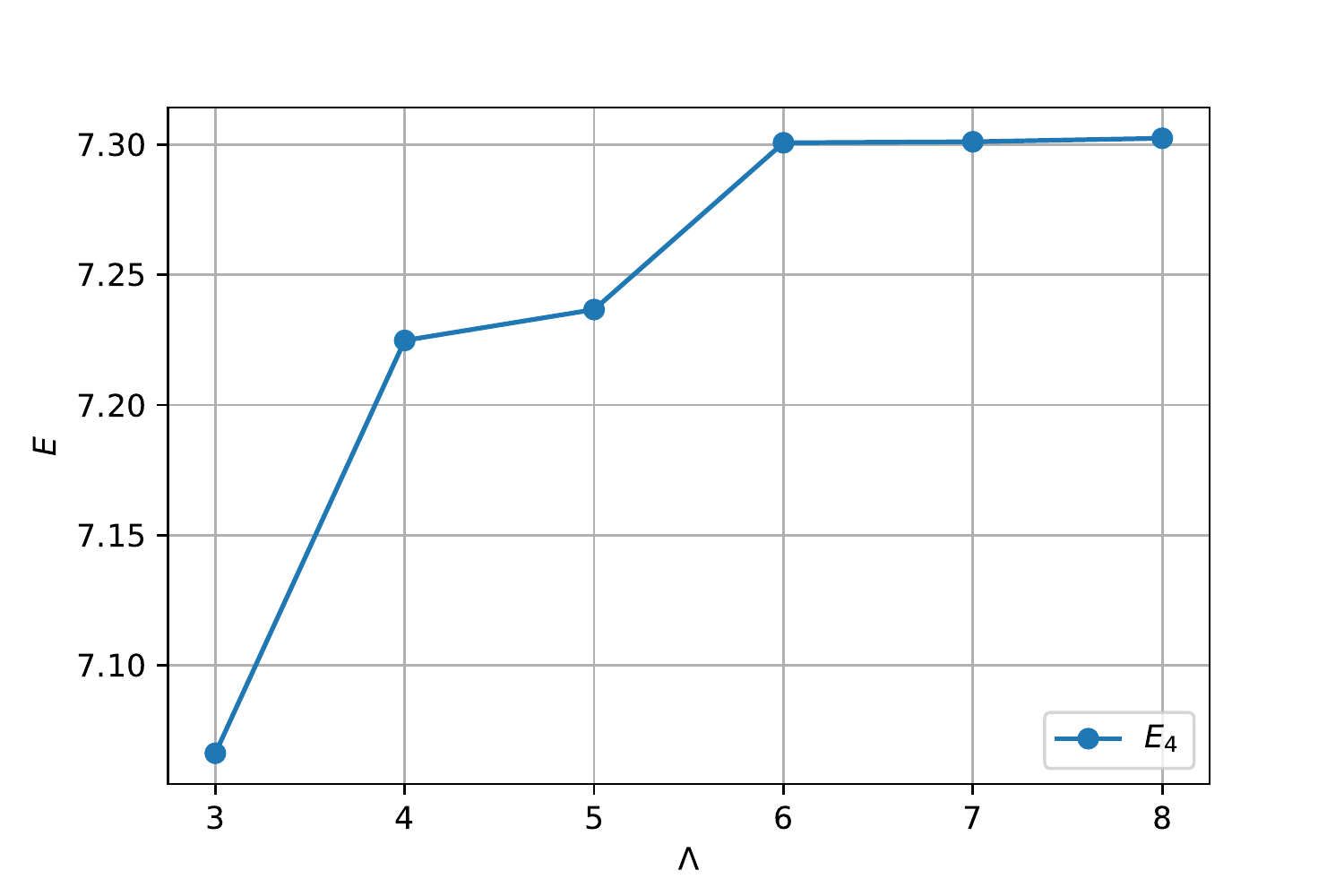} &
      \includegraphics[width=0.30\textwidth]{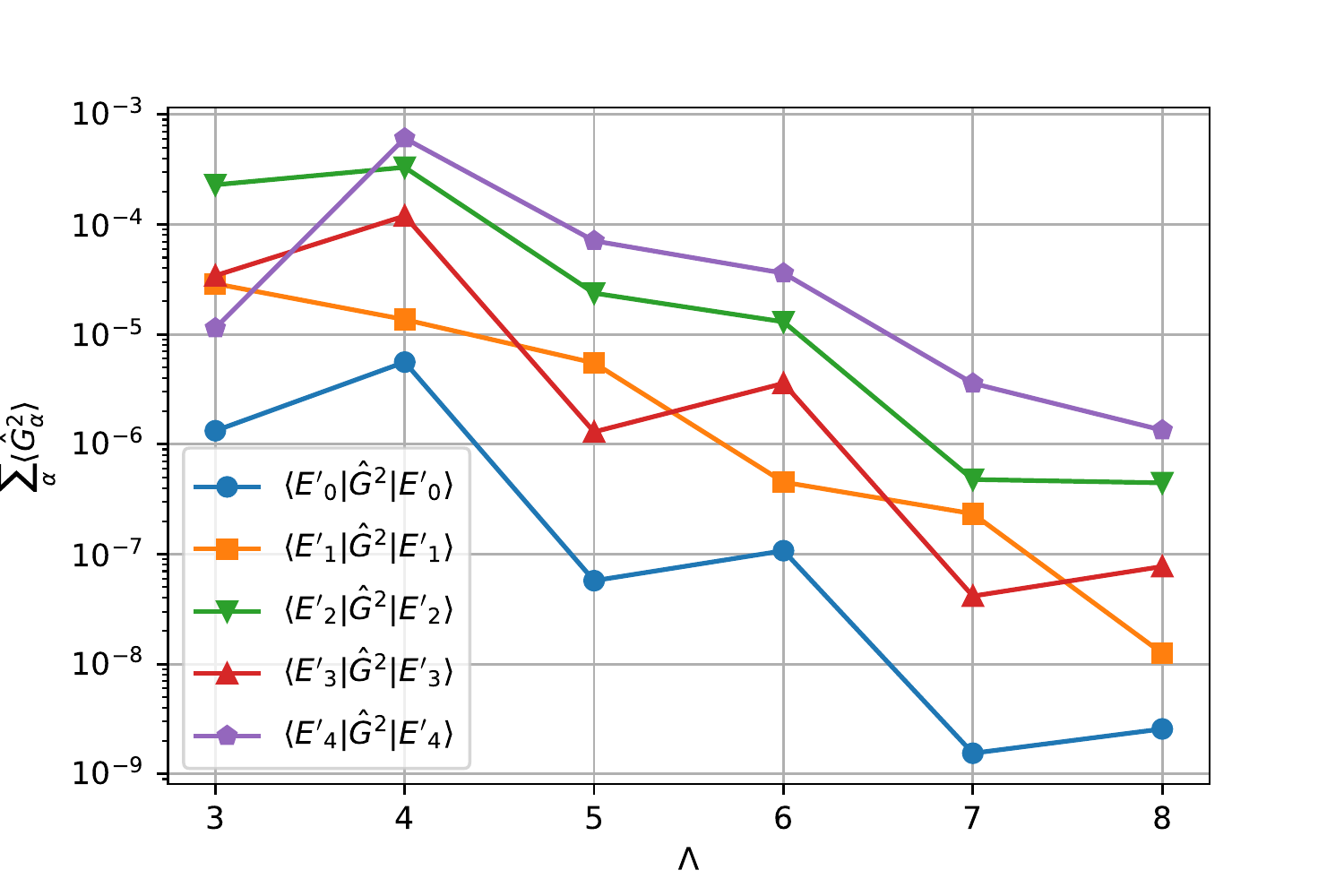}
  \end{tabular}
  \end{center}
  \caption{\label{fig:ham_prime_su2_gs_gv_l02}
  The energy $\langle E'_n|\hat{H}|E'_n\rangle$ and the violation of the singlet constraint $\sum_\alpha\langle E'_n|\hat{G}^2_\alpha|E'_n\rangle$ in the SU(2) bosonic model.
  The parameters are $c=\Lambda$, $m^2=1$, $\lambda=g^2N=0.2$.
  }
\end{figure}
\begin{figure}[htbp]
  \begin{center}
  \begin{tabular}{cc}
      \includegraphics[width=0.30\textwidth]{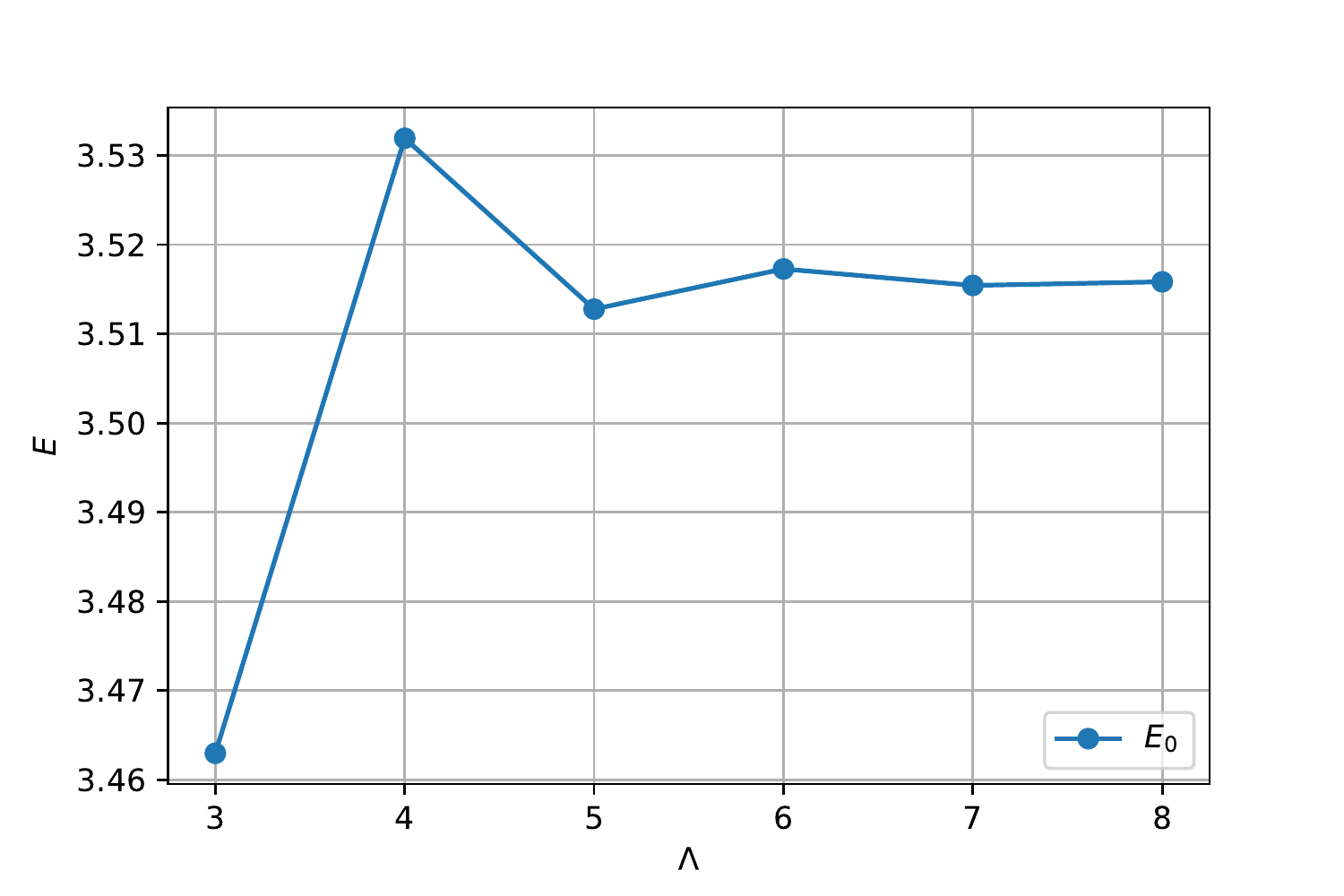} &
      \includegraphics[width=0.30\textwidth]{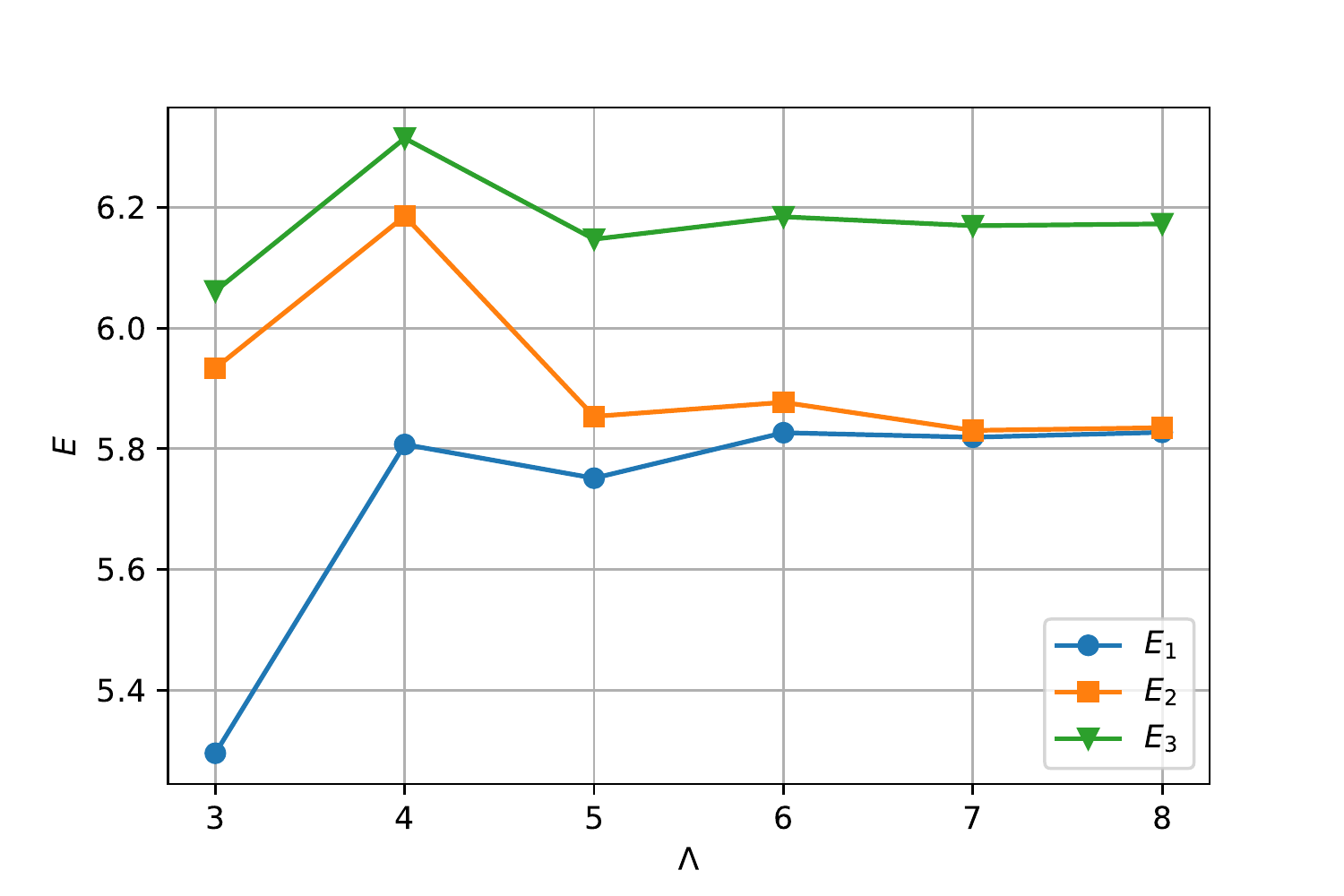} \\
      \includegraphics[width=0.30\textwidth]{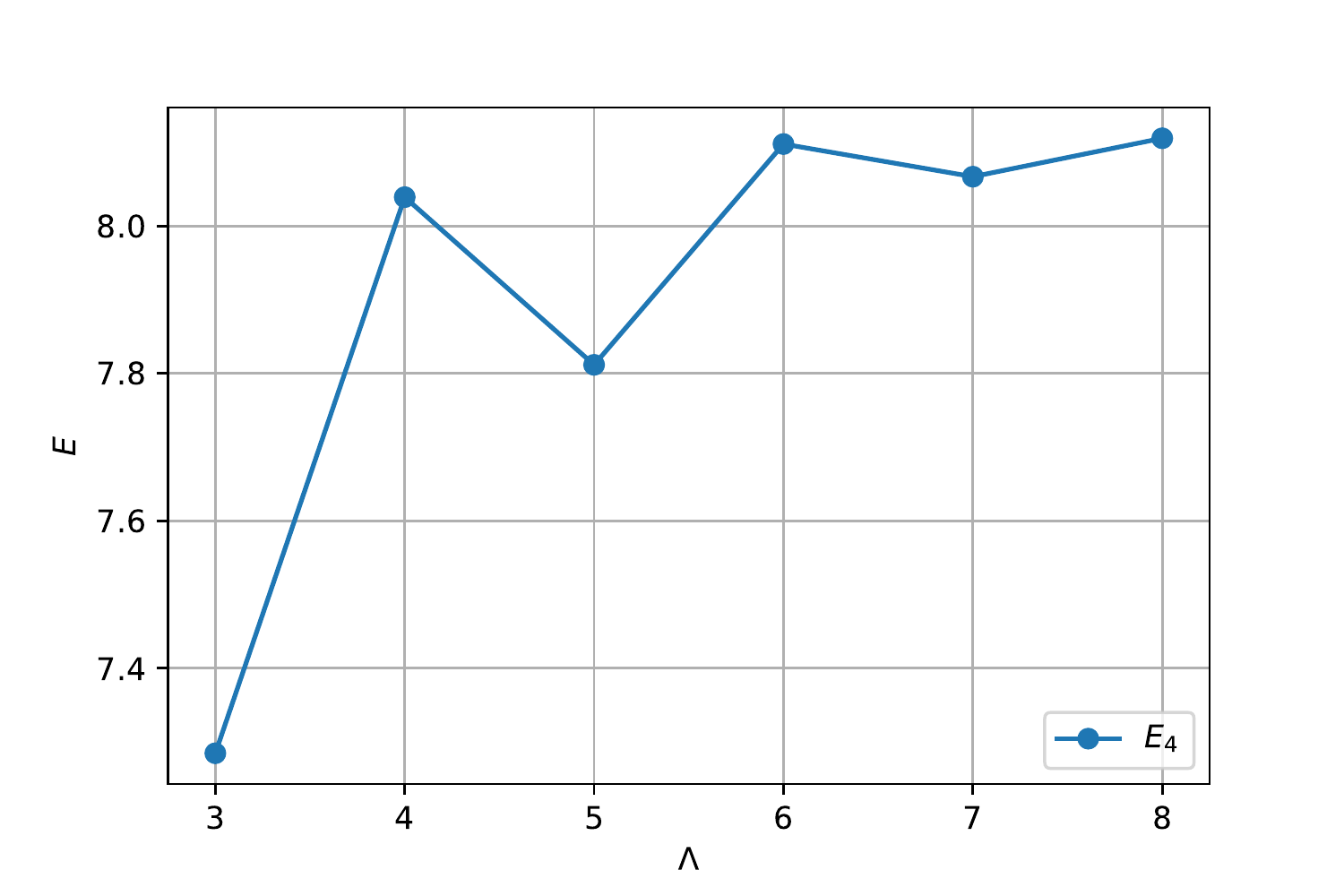} &
      \includegraphics[width=0.30\textwidth]{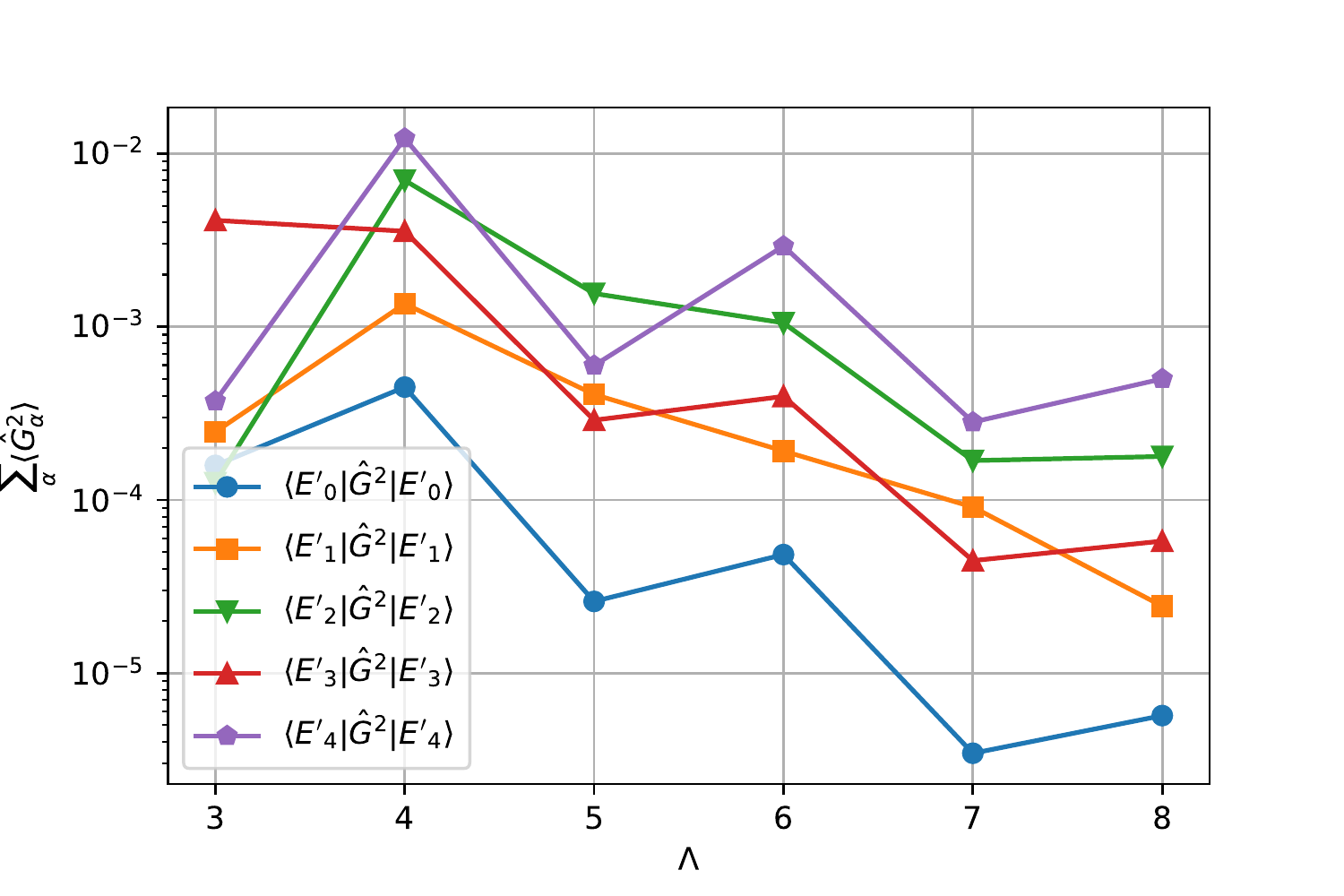}
  \end{tabular}
  \end{center}
  \caption{\label{fig:ham_prime_su2_gs_gv_l10}
  The energy $\langle E'_n|\hat{H}|E'_n\rangle$ and the violation of the singlet constraint $\sum_\alpha\langle E'_n|\hat{G}^2_\alpha|E'_n\rangle$.
  The parameters are $c=\Lambda$, $m^2=1$, $\lambda=g^2N=1.0$ in the SU(2) bosonic model.
  }
\end{figure}

As we can see from Fig.~\ref{fig:ham_prime_su2_gs_gv_l02}, up to the fourth excited level we achieve a reasonably good convergence already at $\Lambda=5$ or $\Lambda=6$.
Higher excited modes receive larger truncation effects, as expected.
We can see that the violation of the singlet constraint $\sum_\alpha\langle E'_n|\hat{G}_\alpha^2|E'_n\rangle$ decays exponentially with $\Lambda$.
We can see the same pattern for $m^2=1$, $\lambda=1.0$ in Fig.~\ref{fig:ham_prime_su2_gs_gv_l10}.
Therefore, both $\lambda_{E'} =  \langle E' | \hat{H}' | E' \rangle$ and $E'\equiv \langle E' | \hat{H} | E' \rangle$ converge to the correct energy eigenvalue.

Because we added $\hat{G}^2$ to the Hamiltonian, the low-lying modes of $\hat{H}'$ are gauge-singlet states.
In the weak-coupling limit ($\lambda \to 0$), the five lowest energy levels in the gauge-singlet sector are $E'_0=3$, $E'_1=E'_2=E'_3=5$ and $E'_4=7$.
Values closer to them are observed at $\lambda=0.2$ (cfr. Fig.~\ref{fig:ham_prime_su2_gs_gv_l02}).

At $\lambda=0$, $|E'_{1,2,3}\rangle$ are linear combinations of ${\rm Tr}(\hat{a}_1^\dagger\hat{a}_1^\dagger)|0\rangle$, ${\rm Tr}(\hat{a}_1^\dagger\hat{a}_2^\dagger)|0\rangle$ and ${\rm Tr}(\hat{a}_2^\dagger\hat{a}_2^\dagger)|0\rangle$.
With respect to the SO(2) symmetry mixing $\hat{a}^\dagger_1$ and $\hat{a}^\dagger_2$, they split to the singlet ${\rm Tr}(\hat{a}_1^\dagger\hat{a}_1^\dagger+\hat{a}_2^\dagger\hat{a}_2^\dagger)|0\rangle$
and the doublet.
In Fig.~\ref{fig:ham_prime_su2_gs_gv_l02} ($\lambda=0.2$) and Fig.~\ref{fig:ham_prime_su2_gs_gv_l10} ($\lambda=1.0$) we can see the degeneracy of $E'_1$ and $E'_2$ as $\Lambda\to\infty$, hence it is natural to expect that the first and second excited levels form the doublet while the third excited level is the singlet.
To check it, we define the generator of the SO(2) rotation $\hat{M}$ which acts on the states as $\hat{M}f(\hat{a}_1^\dagger,\hat{a}_2^\dagger)|0\rangle=f(-\hat{a}_2^\dagger,\hat{a}_1^\dagger)|0\rangle$, and calculate $\langle E'_n|\hat{M}|E'_n\rangle$.
We get $+1, -1, -1, +1, +1$ for $n=0,1,2,3,4$, both for $\lambda=0.2$ and $\lambda=1.0$, with good precision for $\Lambda \geq 4$.

\subsubsection{SU(2) minimal BMN}\label{sec:Hamiltonian-truncation-minimal-BMN}
The SU(2) minimal BMN has 6 bosonic degrees of freedom and 3 fermionic degrees of freedom.
With the truncation scheme for the bosons discussed earlier in this paper, the dimension of the truncated Hilbert space is $8\Lambda^6$, because the dimension of the fermionic Hilbert space is $2^3=8$.
The Hamiltonian is
\begin{align}
\hat{H} &= \sum_{\alpha}
\left(
\frac{\hat{P}_{1\alpha}^2}{2} + \frac{\hat{P}_{2\alpha}^2}{2}
+
\frac{\mu^2\hat{X}_{1\alpha}^2}{2} + \frac{\mu^2\hat{X}_{2\alpha}^2}{2}
+
\frac{3\mu}{2}\hat{\xi}_\alpha^\dagger\hat{\xi}_\alpha
\right)
\nonumber\\
& + g^2 \sum_{\alpha\neq\beta}\hat{X}_{1\alpha}^2\hat{X}_{2\beta}^2
-
2g^2 \sum_{\alpha<\beta} \hat{X}_{1\alpha}\hat{X}_{1\beta} \hat{X}_{2\alpha}\hat{X}_{2\beta}
\nonumber\\
& + \frac{ig}{\sqrt{2}} \sum_{\alpha,\beta,\gamma} \epsilon_{\alpha\beta\gamma}
\left(
(-\hat{X}_{1\alpha}-i\hat{X}_{2\alpha})
\hat{\xi}_\beta^\dagger\hat{\xi}_\gamma^\dagger
+
(-\hat{X}_{1\alpha}+i\hat{X}_{2\alpha})
\hat{\xi}_\beta\hat{\xi}_\gamma
\right)
- 3\mu \ .
\end{align}

The complex fermions $\xi^\alpha$ ($\alpha=1,2,3$) satisfy the anti-commutation relation $\{\xi_\alpha^\dagger,\xi_\beta\}=\delta_{\alpha\beta}$.
Regarding the fermions, we can use the standard Fock basis, and build states from the Fock vacuum $\hat{\xi}_\alpha|0\rangle=0$.
We can define the fermion number by counting how many $\hat{\xi}^\dagger$ operators act on the Fock vacuum.
The Hamiltonian does not mix the fermion number odd and even sectors.

The generators of the gauge transformation are
\begin{align}
\hat{G}_\alpha = i\sum_{\beta,\gamma}\epsilon_{\alpha\beta\gamma}
\left(
\hat{a}_{1\beta}^\dagger\hat{a}_{1\gamma}
+
\hat{a}_{2\beta}^\dagger\hat{a}_{2\gamma}
+
\hat{\xi}_\beta^\dagger\hat{\xi}_\gamma
\right) \ .
\end{align}
As before, we consider $\hat{H}'\equiv\hat{H}+c\sum_\alpha\hat{G}_\alpha^2$, in order to eliminate the non-singlet modes from the dynamics.

We will consider gauge-invariant and fixed-angular-momentum sectors.
Hence, just for numerical purpose, we add $c' (\hat{M}-J)^2$ to the Hamiltonian.
In addition, we keep $c\sum_\alpha\hat{G}_\alpha^2$ to penalize gauge non-singlets.
To summarize, we use a modified Hamiltonian $\hat{H}'$ defined by
\begin{align}\label{eq:deformed-angular-hamiltonian}
\hat{H}' = \hat{H} + c\sum_\alpha\hat{G}_\alpha^2 + c' (\hat{M}-J)^2 \ .
\end{align}
Here $J$ is the angular momentum of the states we want to consider.
We take $c=\Lambda$ as before and several choices of $c'$ including $c' = 10 \Lambda$ for two values of $J=0, \frac{1}{2}$.
If we consider only the low-lying modes of $\hat{H}'$, states with $G^2>0$ or $M\neq J$ are removed because their energies are increased.

\begin{figure}[htbp]
  \begin{center}
  \begin{tabular}{cc}
      \includegraphics[width=0.35\textwidth]{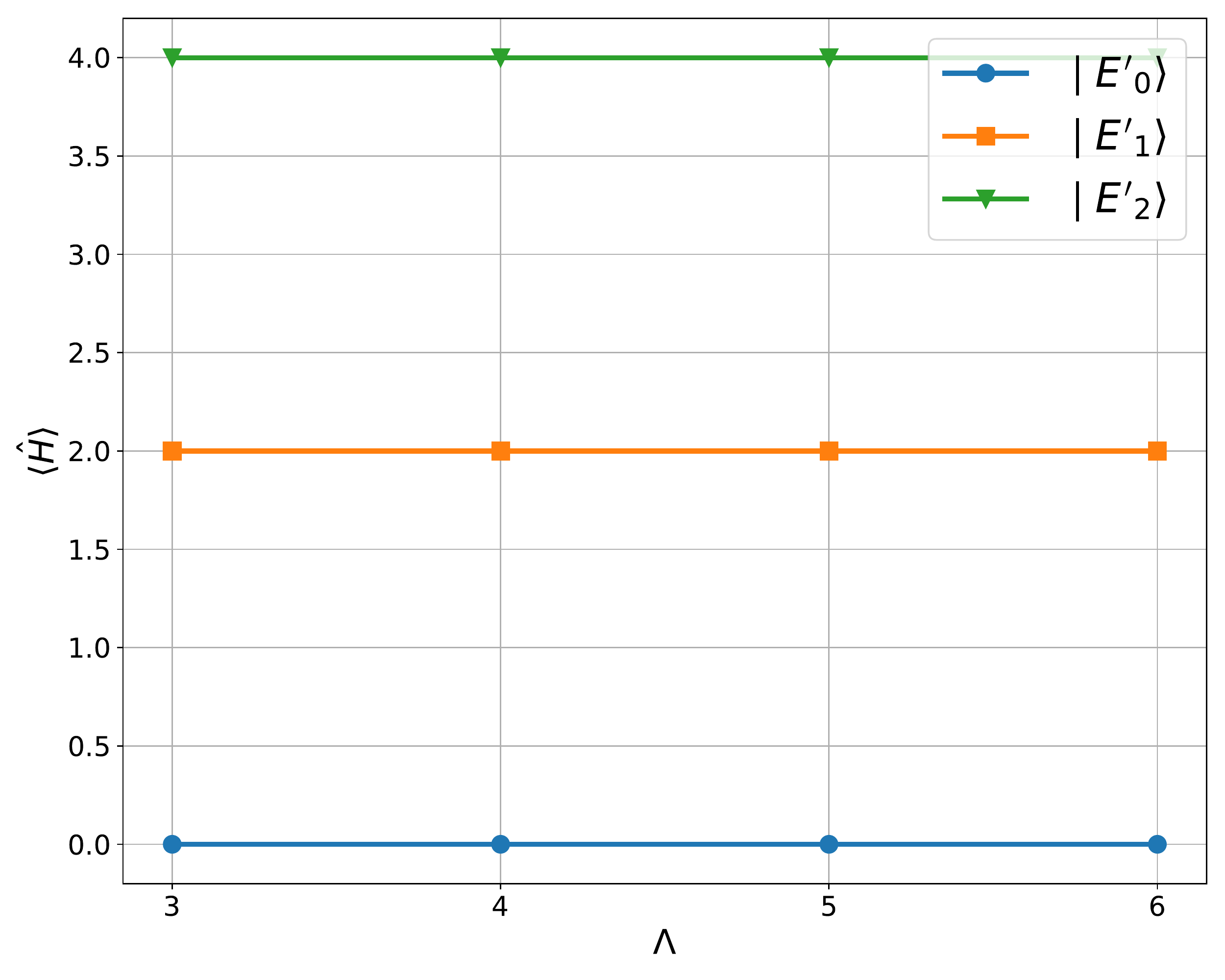} &
      \includegraphics[width=0.35\textwidth]{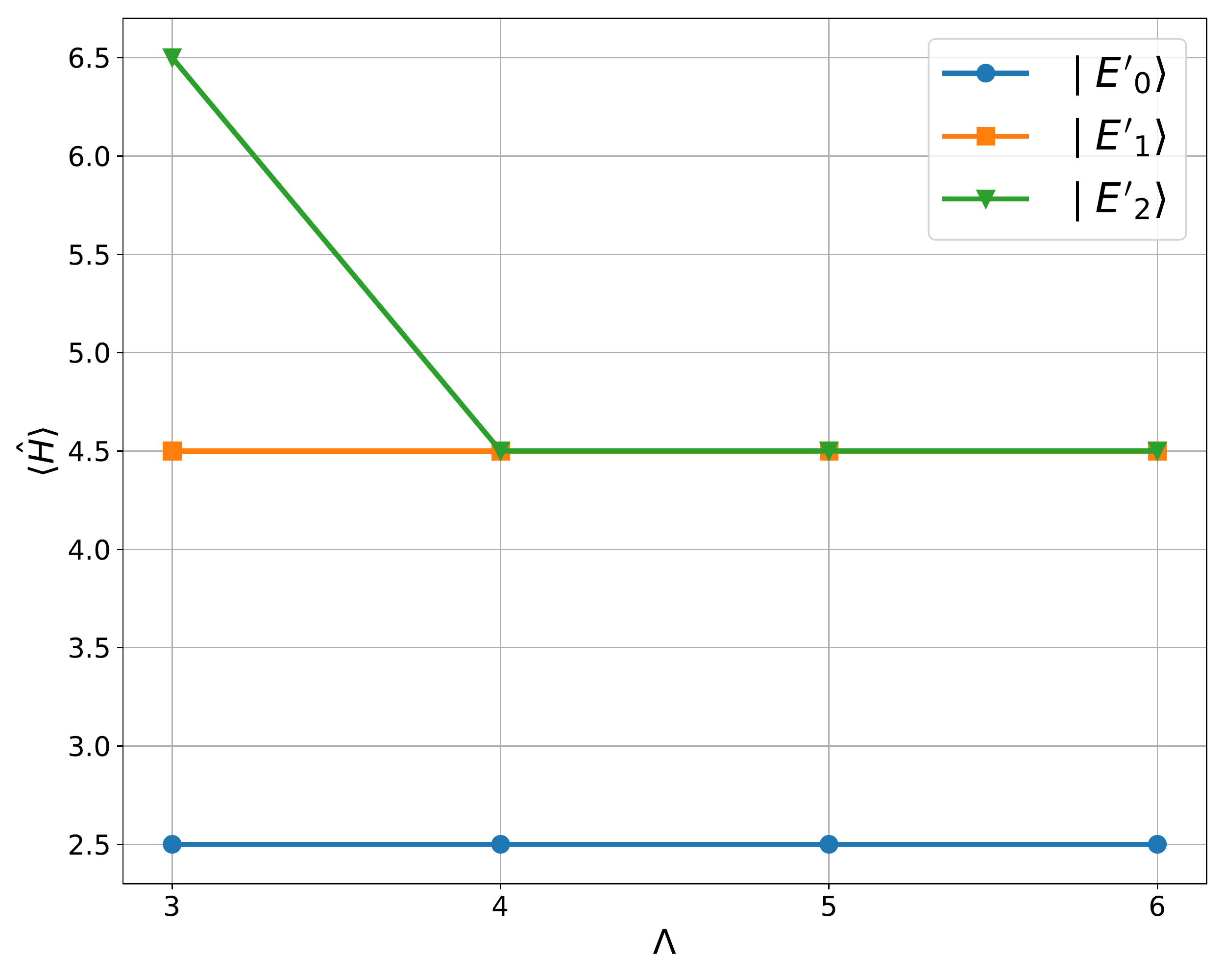} \\
  \end{tabular}
  \end{center}
  \caption{\label{fig:ham_su2_mini_gs_e012}
  $\langle E'_{i,J}|\hat{H}|E'_{i,J}\rangle$ in the SU(2) minimal BMN model at zero coupling, for a few low-lying excited modes.
  The modes are ordered by their $\lambda_{E'}$ eigenvalues of $\hat{H}'$ and their order can change depending on the cutoff $\Lambda$.
  The left panel is for the sector with $J=0$ and the right panel is for $J=1/2$.
  Other parameters are $\lambda=g^2N=0$, $\mu=1$, $c=\Lambda$, and $c'=100$
  (This large value of $c'$ ensures that the low-cutoff states are in the right sector).
  }
\end{figure}

Let us consider the case of zero coupling as a sanity check.
At zero coupling in the $J=0$ sector, the ground state is the Fock vacuum $|0\rangle$, and the first excited mode is ${\rm Tr}(\hat{a}_1^{\dagger 2}+\hat{a}_2^{\dagger 2})|0\rangle$.
The energies are $0$ and $2\mu$ (recall that $\mu$ and $m$ are equivalent).
The ground state is BPS, so the ground state energy has to be zero at any coupling.
On the other hand, at zero coupling in the $J=1/2$ sector, the lowest-energy mode is ${\rm Tr}(\hat{\xi}^\dagger(\hat{a}_1^\dagger+i\hat{a}_2^\dagger))|0\rangle$, whose energy is $5\mu/2$.
The second lightest modes are made of $\hat{\xi}^\dagger$, two $(\hat{a}_1^\dagger+i\hat{a}_2^\dagger)$'s and one $(\hat{a}_1^\dagger-i\hat{a}_2^\dagger)$, such as ${\rm Tr}(\hat{\xi}^\dagger (\hat{a}_1^\dagger+i\hat{a}_2^\dagger)(\hat{a}_1^{\dagger 2}+\hat{a}_2^{\dagger 2}))|0\rangle$, and the energy is $9\mu/2$.
None of them is BPS, hence their energy can change when the interaction is turned on.
We show the zero coupling case explicitly in Fig.~\ref{fig:ham_su2_mini_gs_e012} where the left panel corresponds to $J=0$ and the right panel to $J=1/2$ for $\mu=1$.

\begin{figure}[htbp]
  \begin{center}
  \begin{tabular}{cc}
      \includegraphics[width=0.30\textwidth]{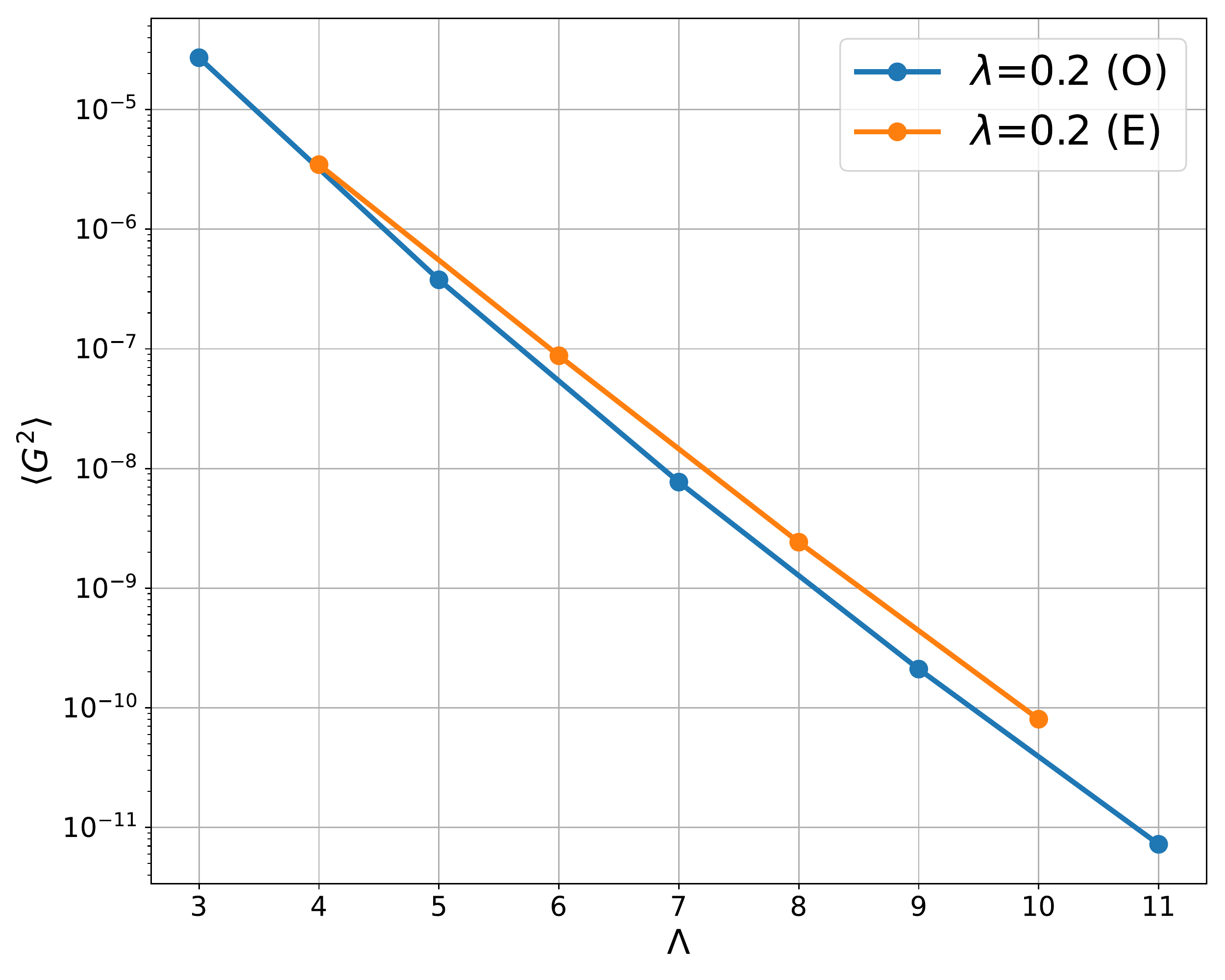} &
      \includegraphics[width=0.30\textwidth]{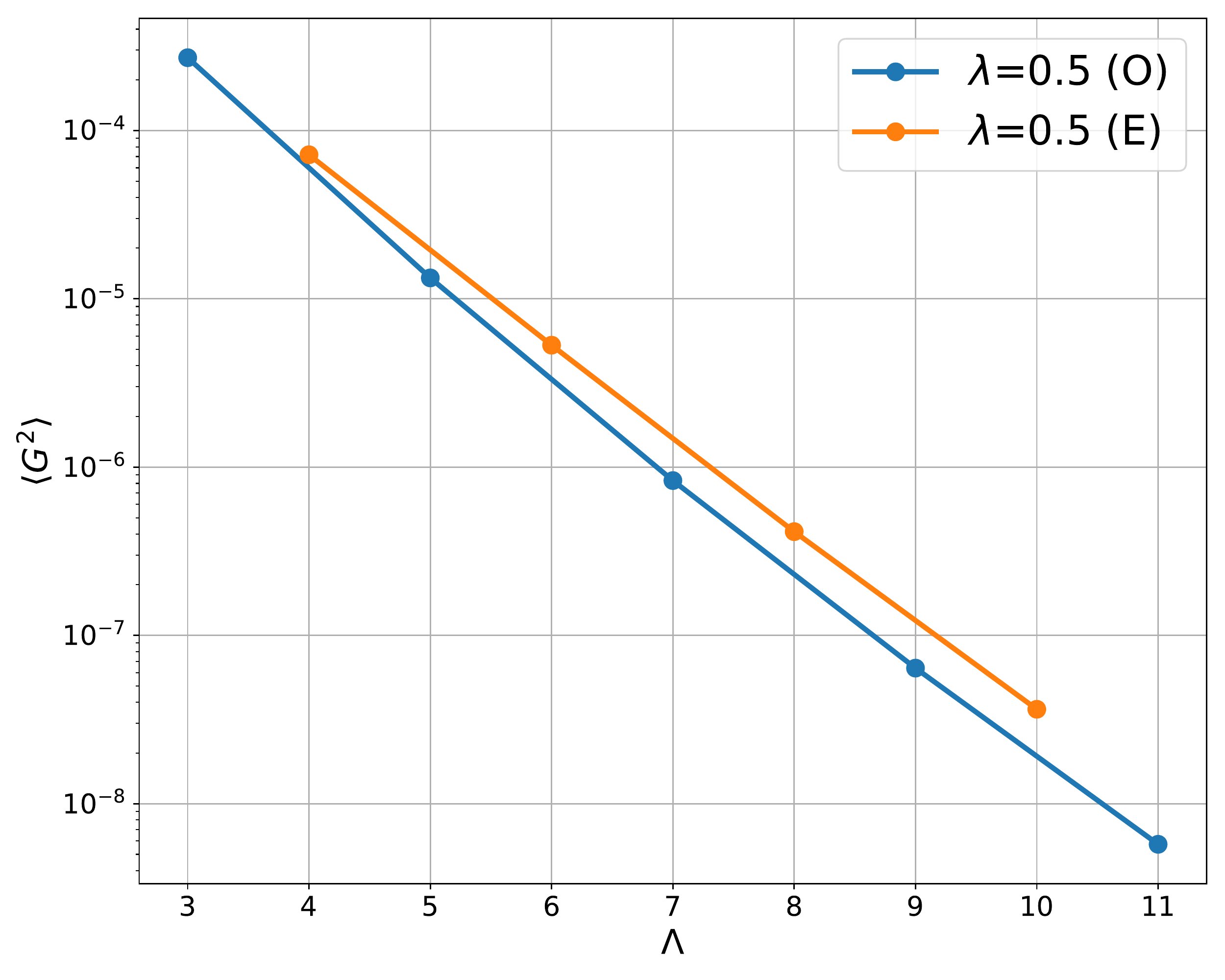} \\
      \includegraphics[width=0.30\textwidth]{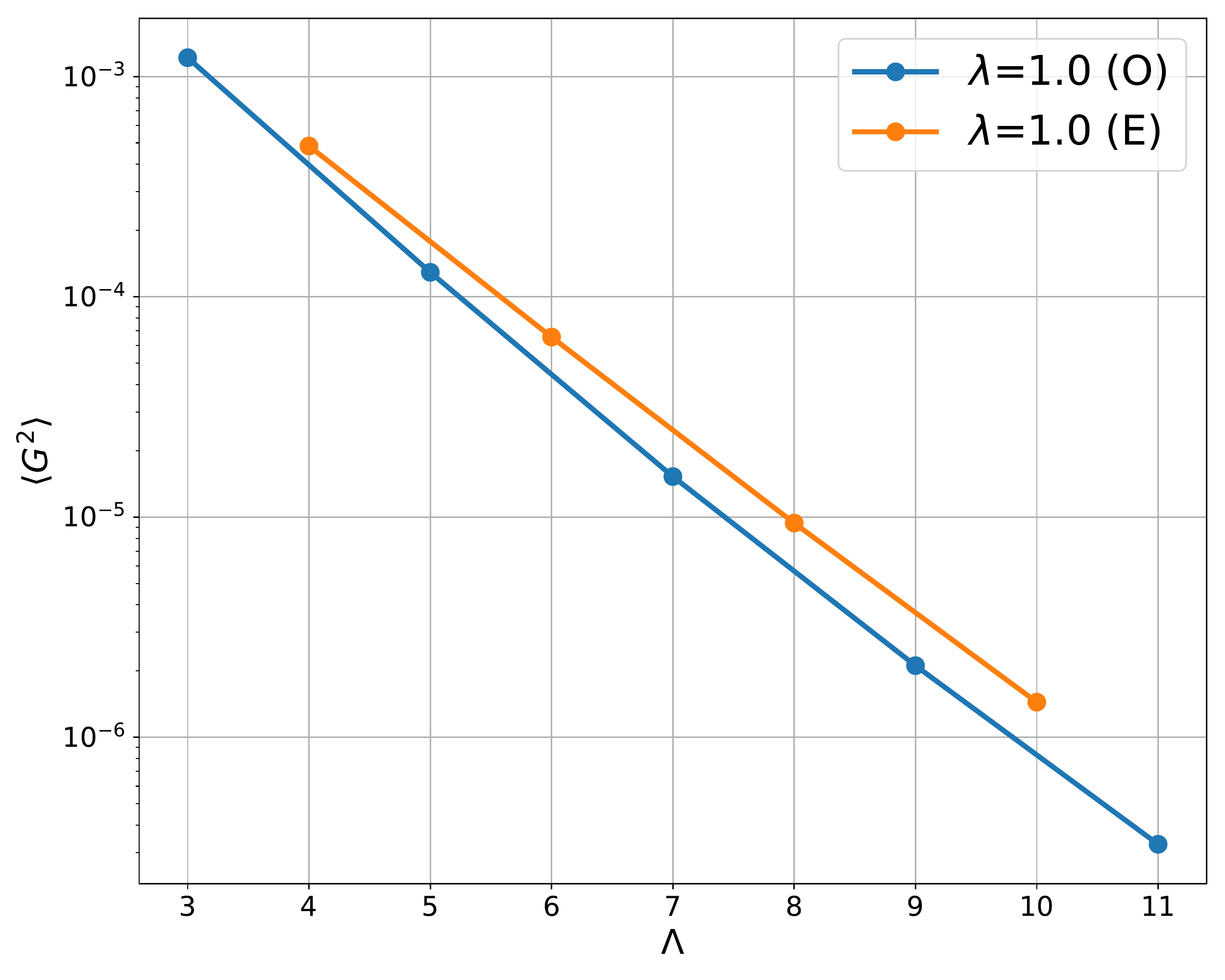} &
      \includegraphics[width=0.30\textwidth]{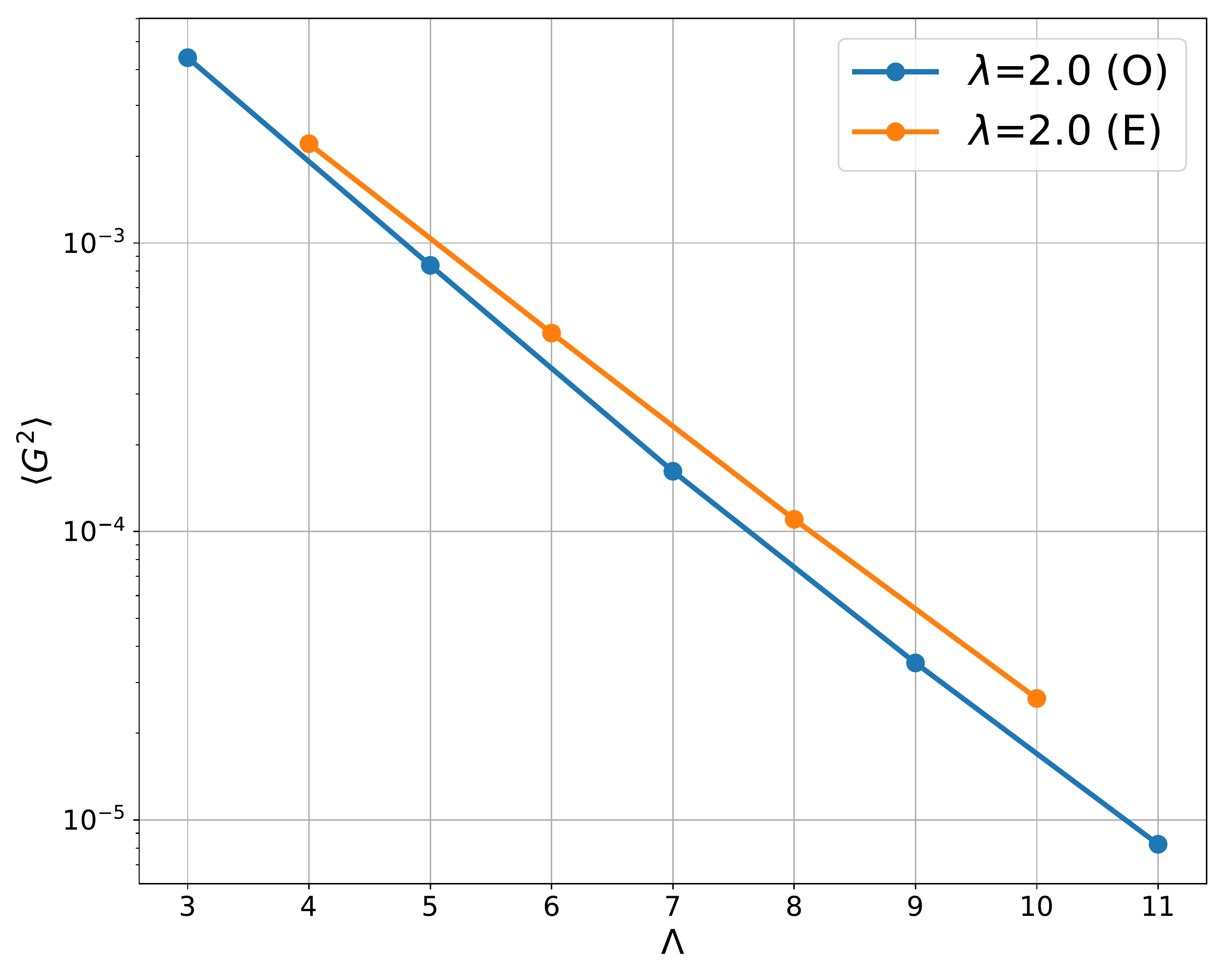}
  \end{tabular}
  \end{center}
  \caption{\label{fig:ham_su2_mini_gv}
  The violation of the singlet constraint for the ground state of the Hamiltonian $\hat{H}'$ as a function of the cutoff $\Lambda$ for various couplings $\lambda=g^2N=0.2$, 0.5, 1.0, and 2.0 in the SU(2) minimal BMN model.
  Even (E) and odd (O) values of $\Lambda$ are plotted with different colors in a logarithmic scale to show that the trend to zero is exponentially fast.
  Other parameters are $\mu=1$, $c=\Lambda$, $c'=1$ and $J=0$ (a low value of $c'$ is enough for the ground state only).
  }
\end{figure}
\begin{figure}[htbp]
  \begin{center}
  \begin{tabular}{cc}
      \includegraphics[width=0.30\textwidth]{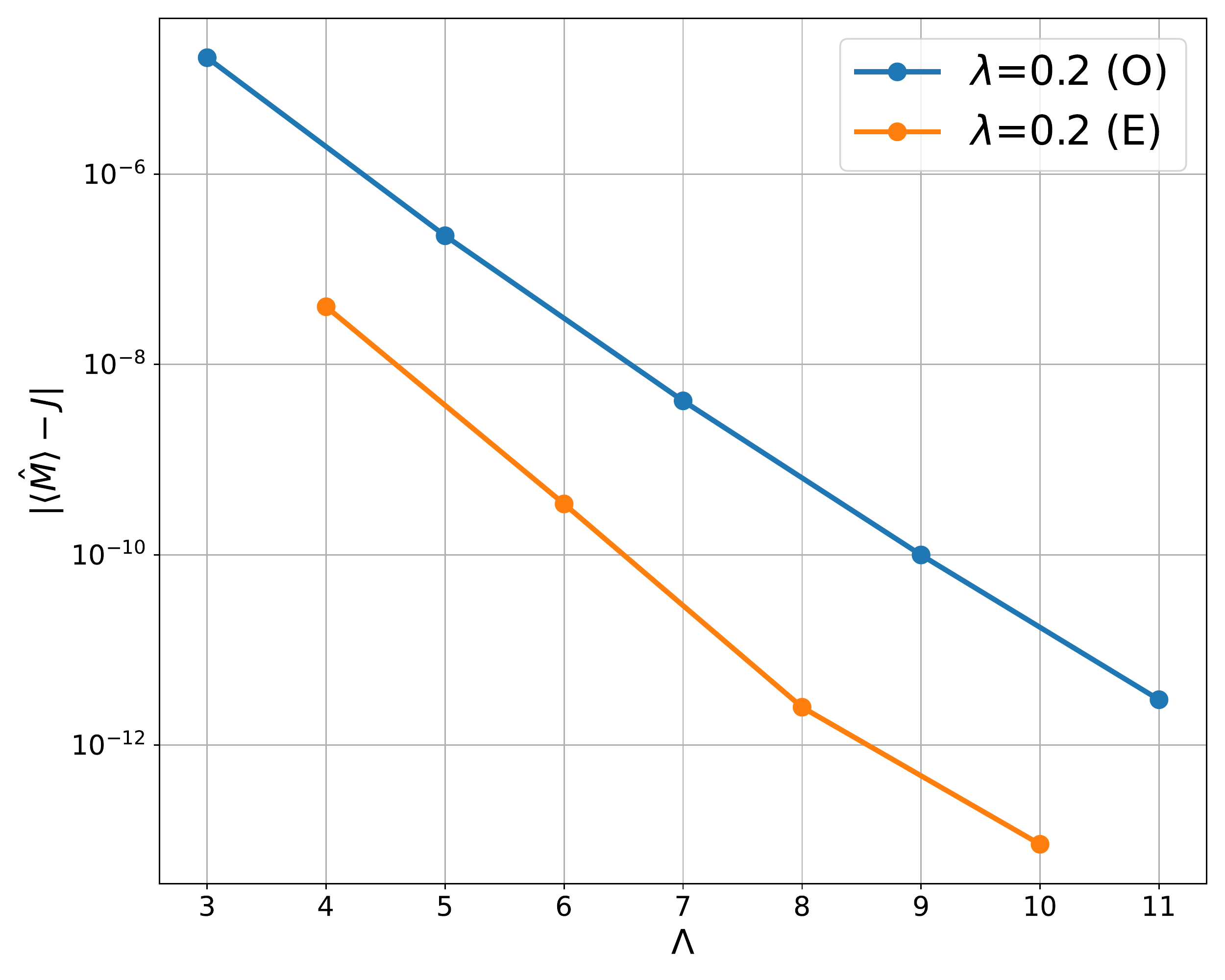} &
      \includegraphics[width=0.30\textwidth]{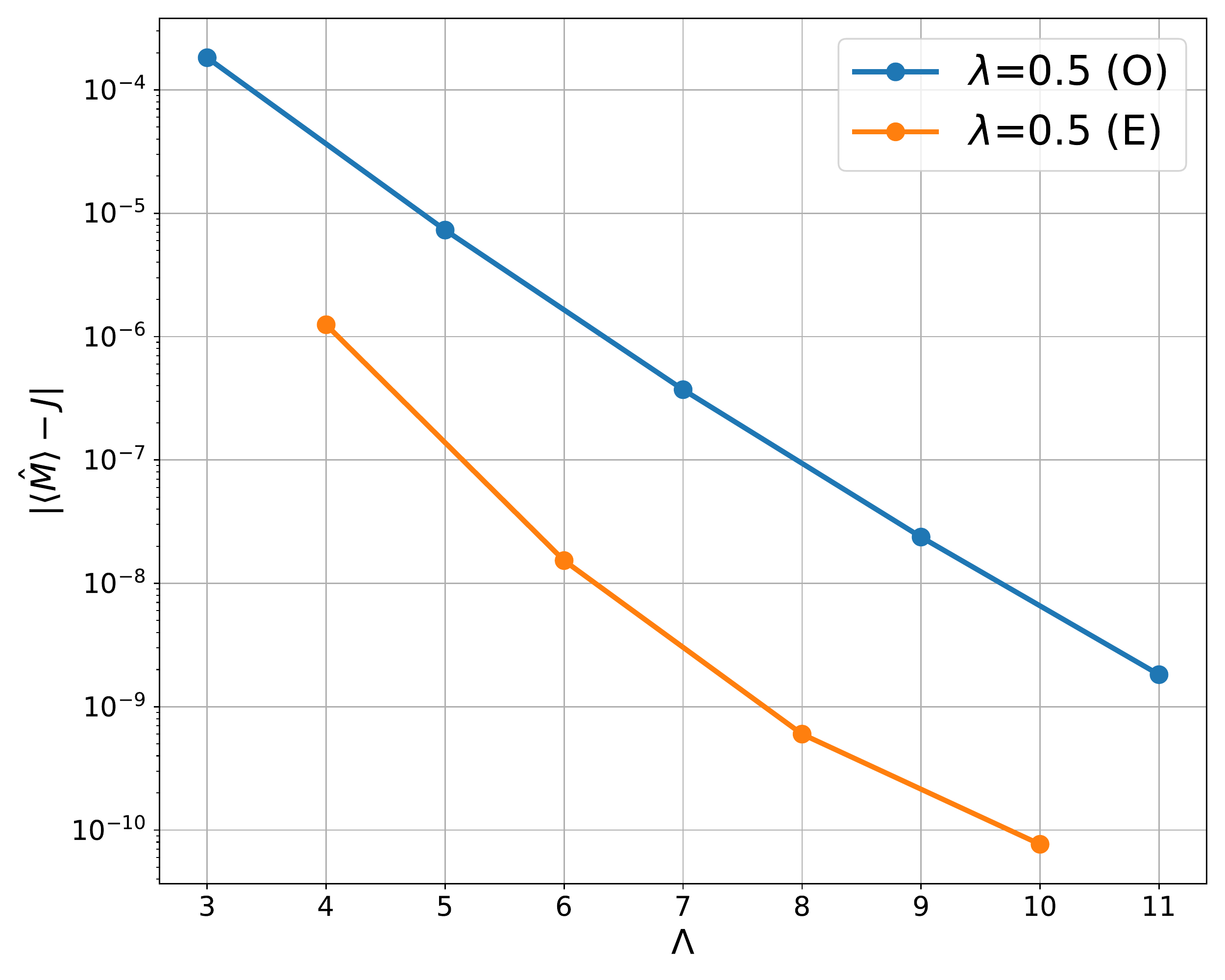} \\
      \includegraphics[width=0.30\textwidth]{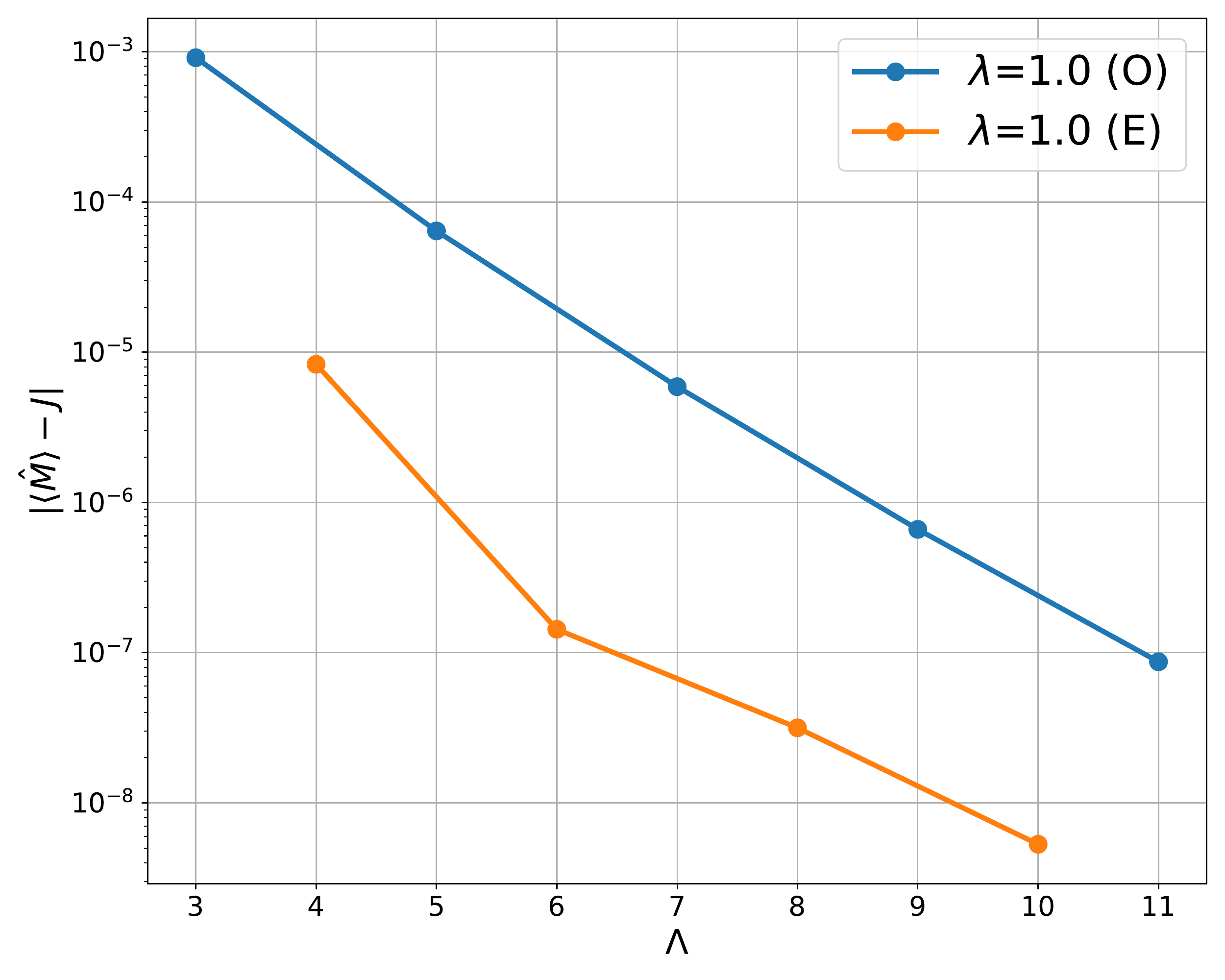} &
      \includegraphics[width=0.30\textwidth]{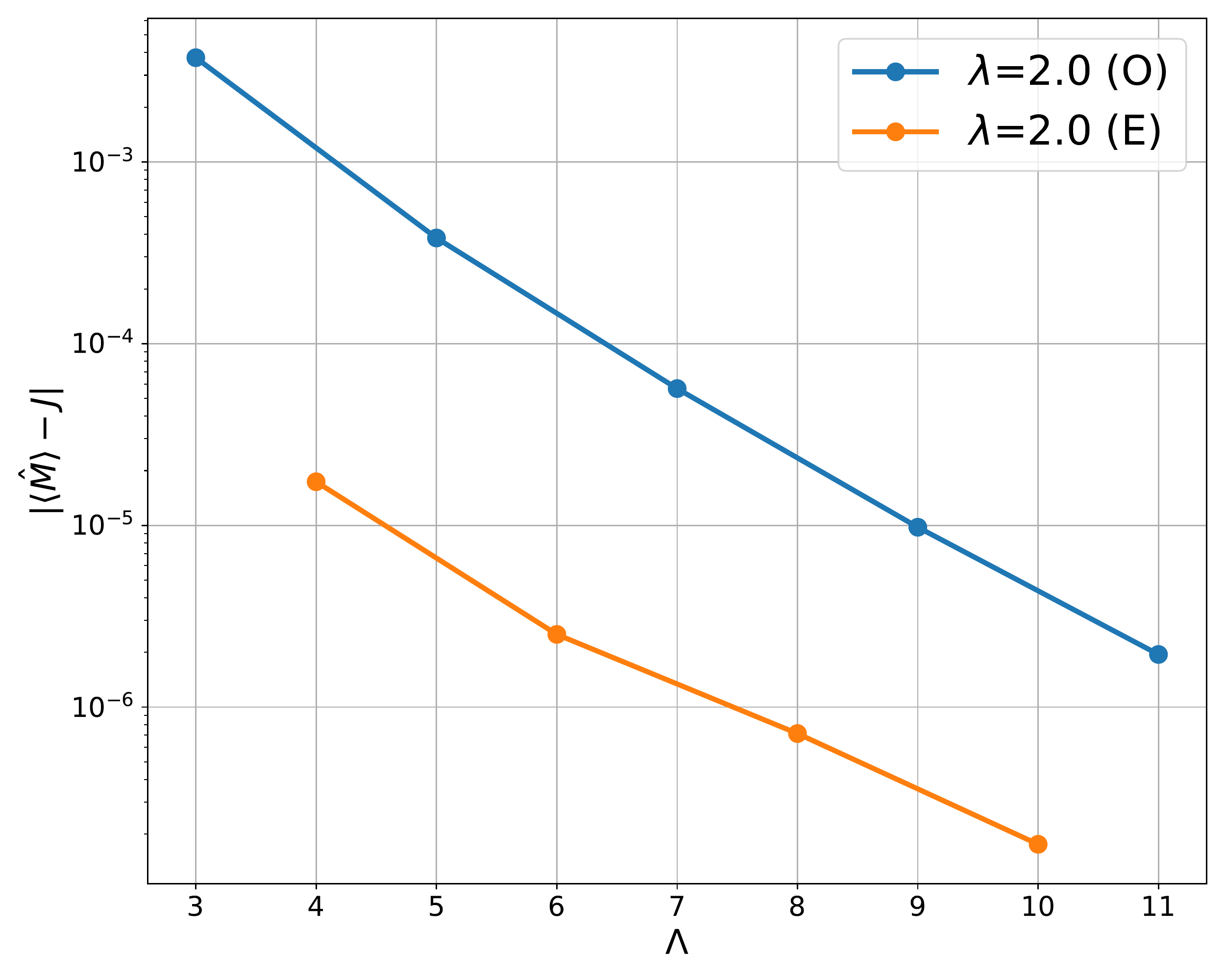}
  \end{tabular}
  \end{center}
  \caption{\label{fig:ham_su2_mini_gm_J0}
  The $|\langle E'_{i,J}|\hat{M}|E'_{i,J}\rangle-J|$ for the ground state of the Hamiltonian $\hat{H}'$ as a function of the cutoff $\Lambda$ for various couplings $\lambda=g^2N=0.2$, 0.5, 1.0, and 2.0 in the SU(2) minimal BMN model.
  Even (E) and odd (O) values of $\Lambda$ are plotted with different colors in logarithmic scale.
  Other parameters are $m^2=1$, $c=\Lambda$, $c'=1$ and $J=0$ (a low value of $c'$ is enough for the ground state only).
  }
\end{figure}

\begin{figure}[htbp]
  \begin{center}
  \begin{tabular}{cc}
      \includegraphics[width=0.30\textwidth]{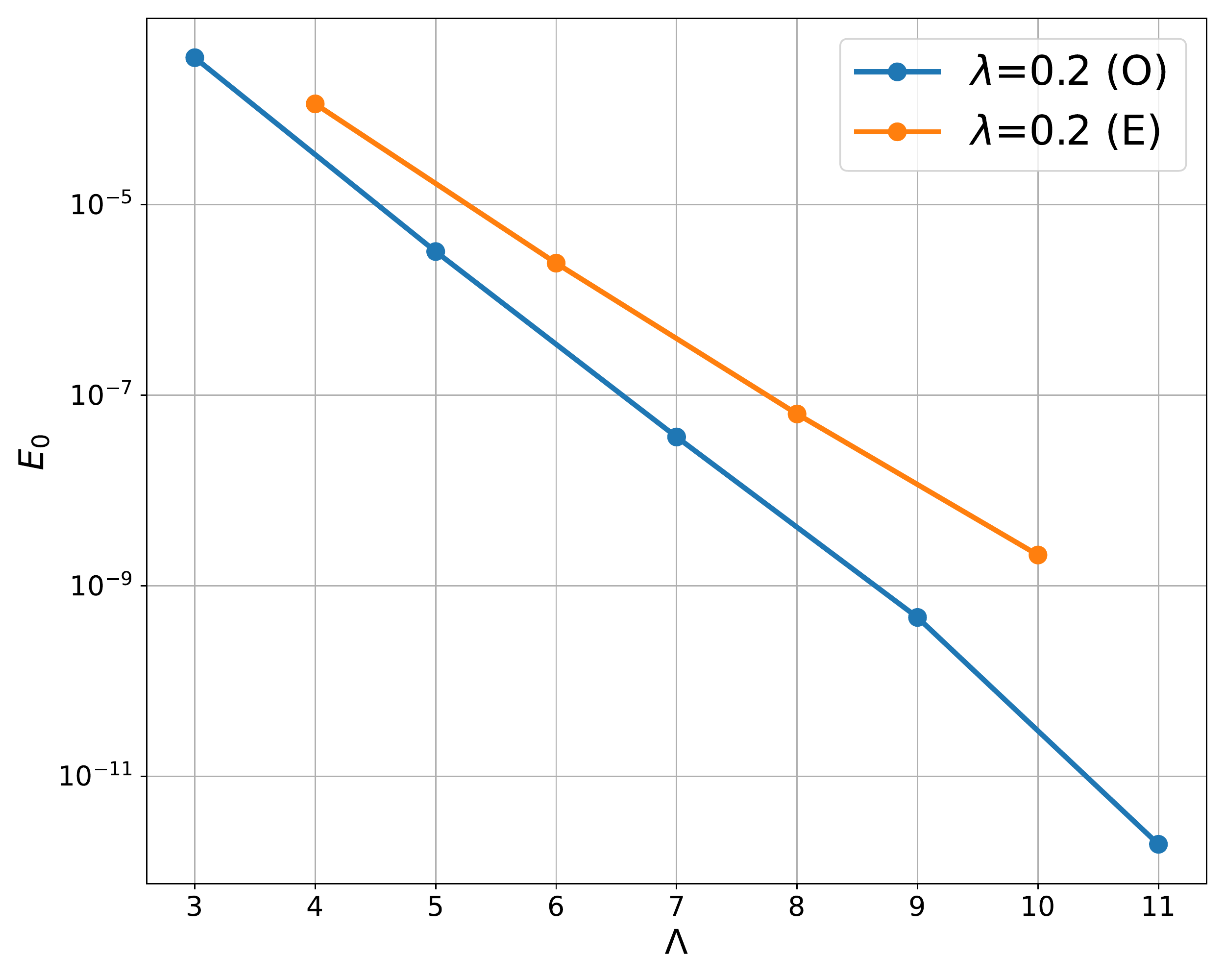} &
      \includegraphics[width=0.30\textwidth]{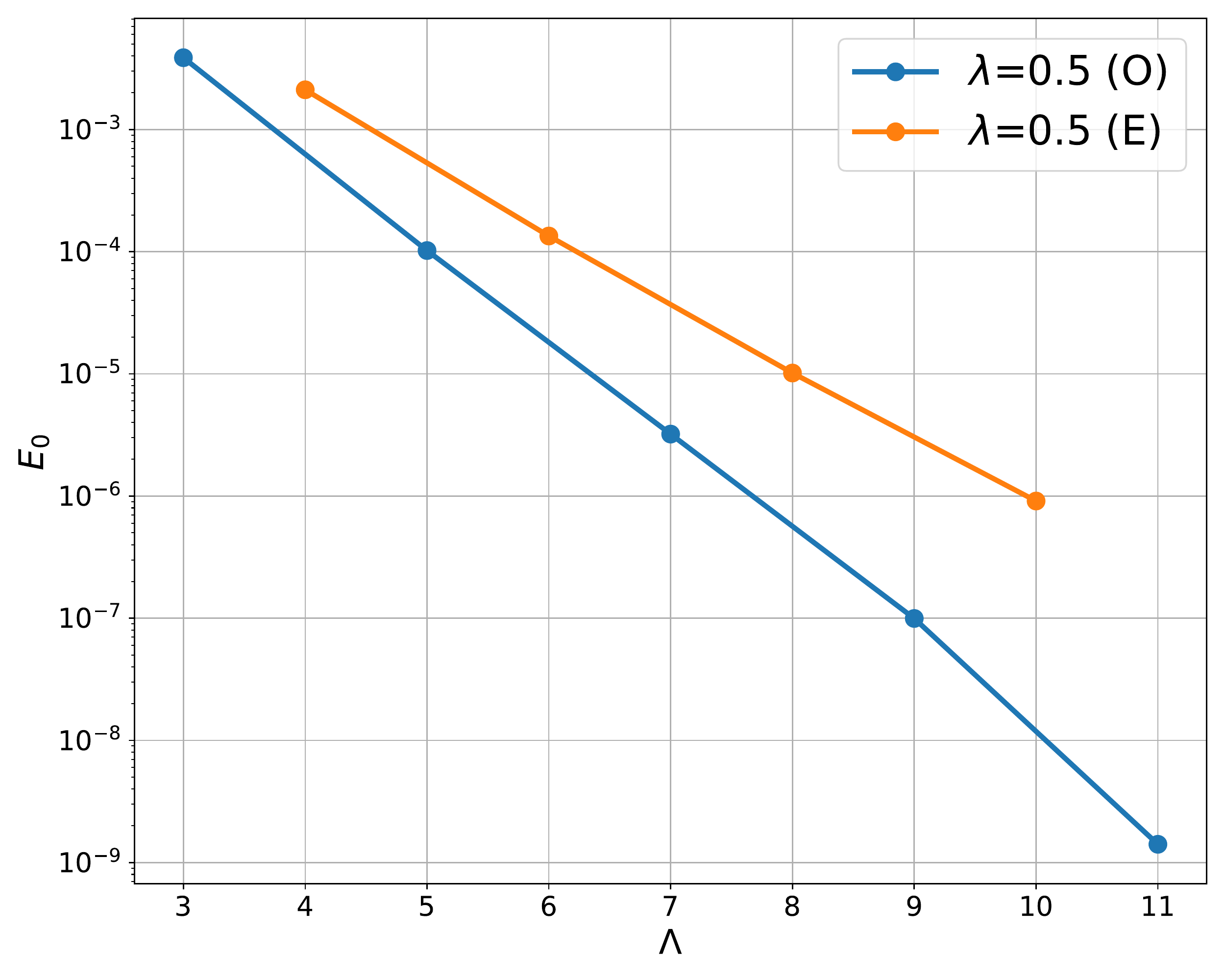} \\
      \includegraphics[width=0.30\textwidth]{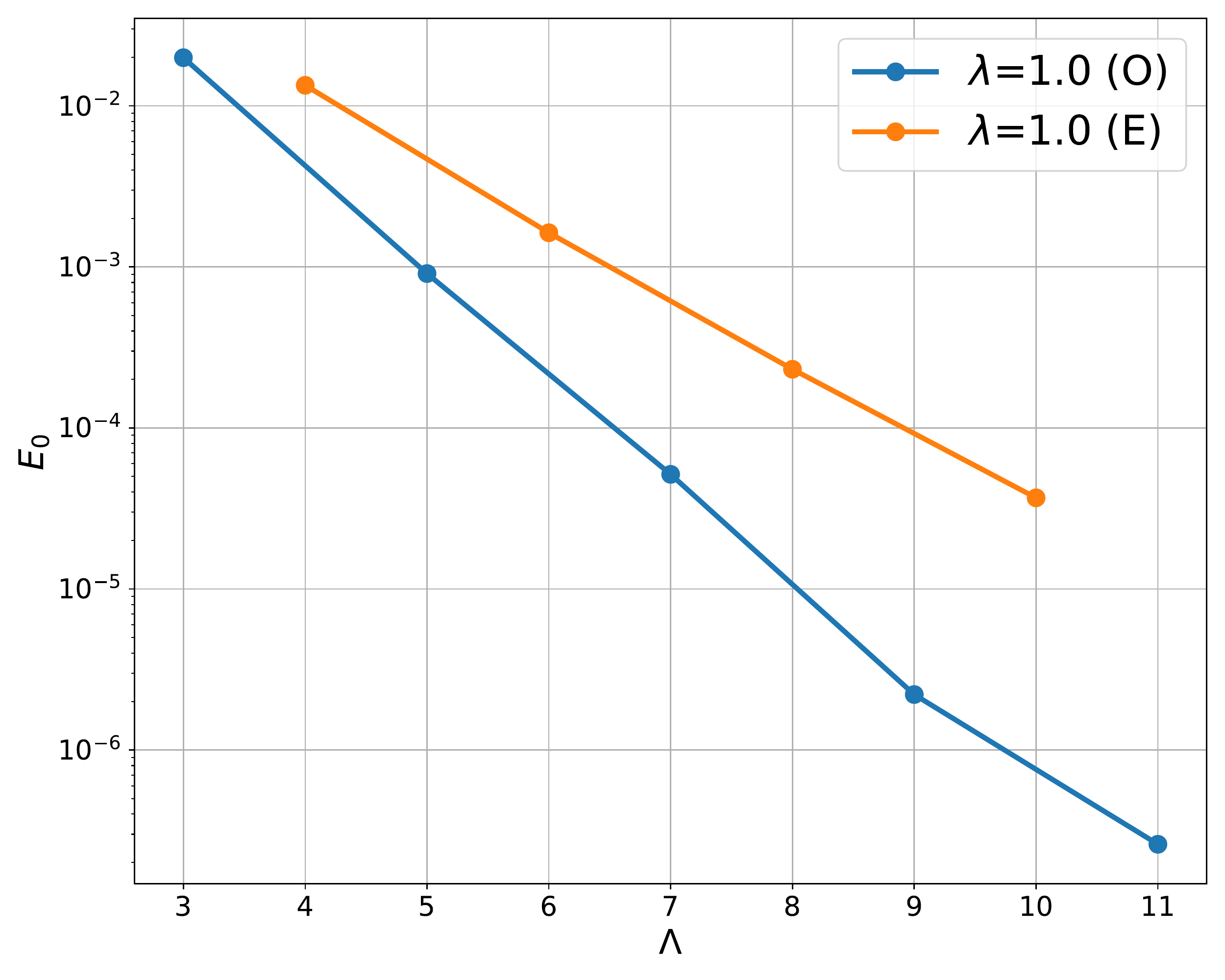} &
      \includegraphics[width=0.30\textwidth]{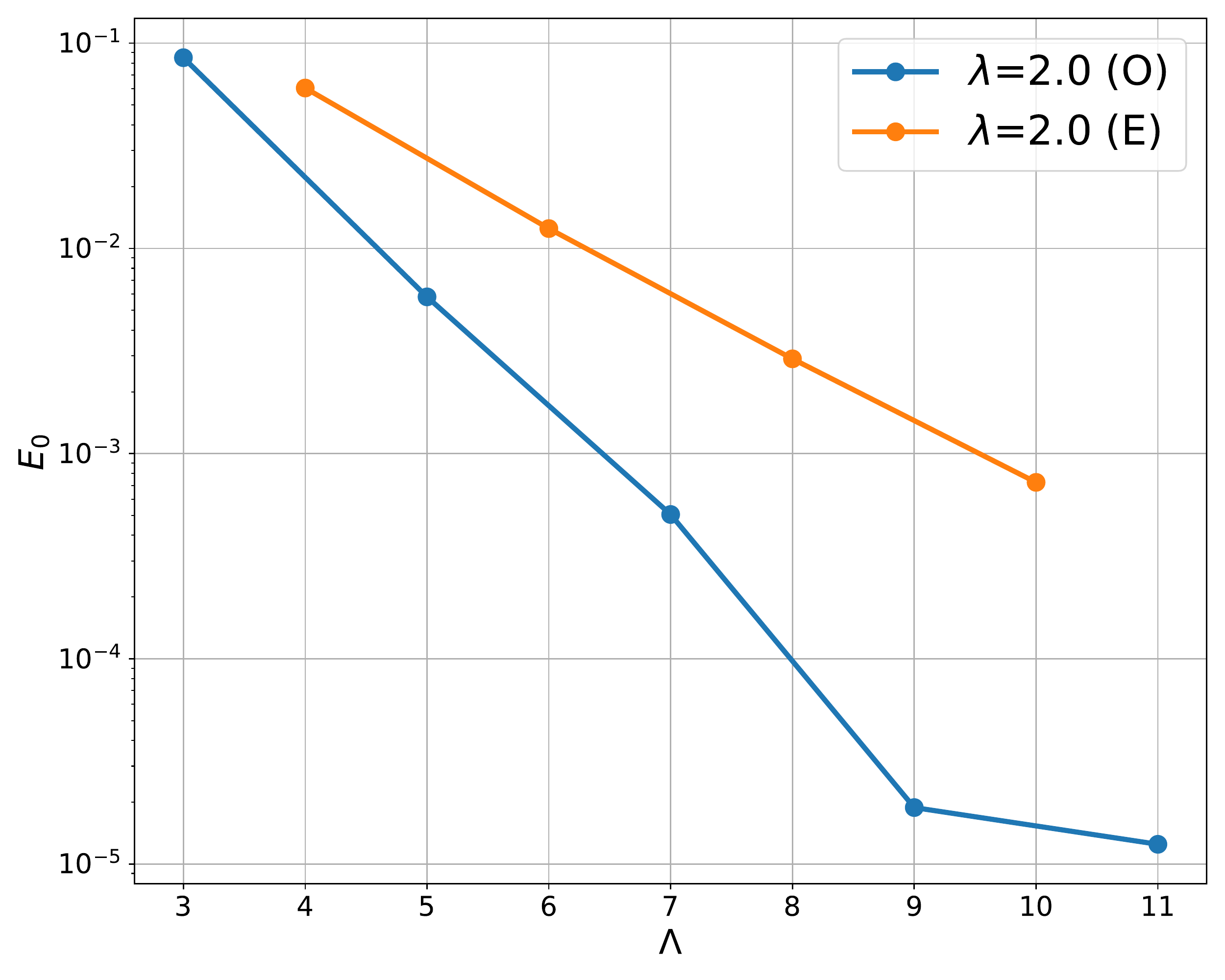}
  \end{tabular}
  \end{center}
  \caption{\label{fig:ham_su2_mini_gs}
  The energy expectation value $|\langle E_{0,J=0}'|\hat{H}|E_{0,J=0}'\rangle|$ of the ground state of the Hamiltonian $\hat{H}'$ as a function of the cutoff $\Lambda$ for various couplings $\lambda=g^2N=0.2$, 0.5, 1.0, and 2.0 in the SU(2) minimal BMN model.
  Even (E) and odd (O) values of $\Lambda$ are plotted with different colors in a logarithmic scale.
  The approach to the $\Lambda \rightarrow \infty$ limit value of zero is exponentially fast.
  Other parameters are $m^2=1$, $c=\Lambda$ and $c'=1$.
  }
\end{figure}

\begin{figure}[htbp]
  \begin{center}
  \begin{tabular}{cc}
      \includegraphics[width=0.30\textwidth]{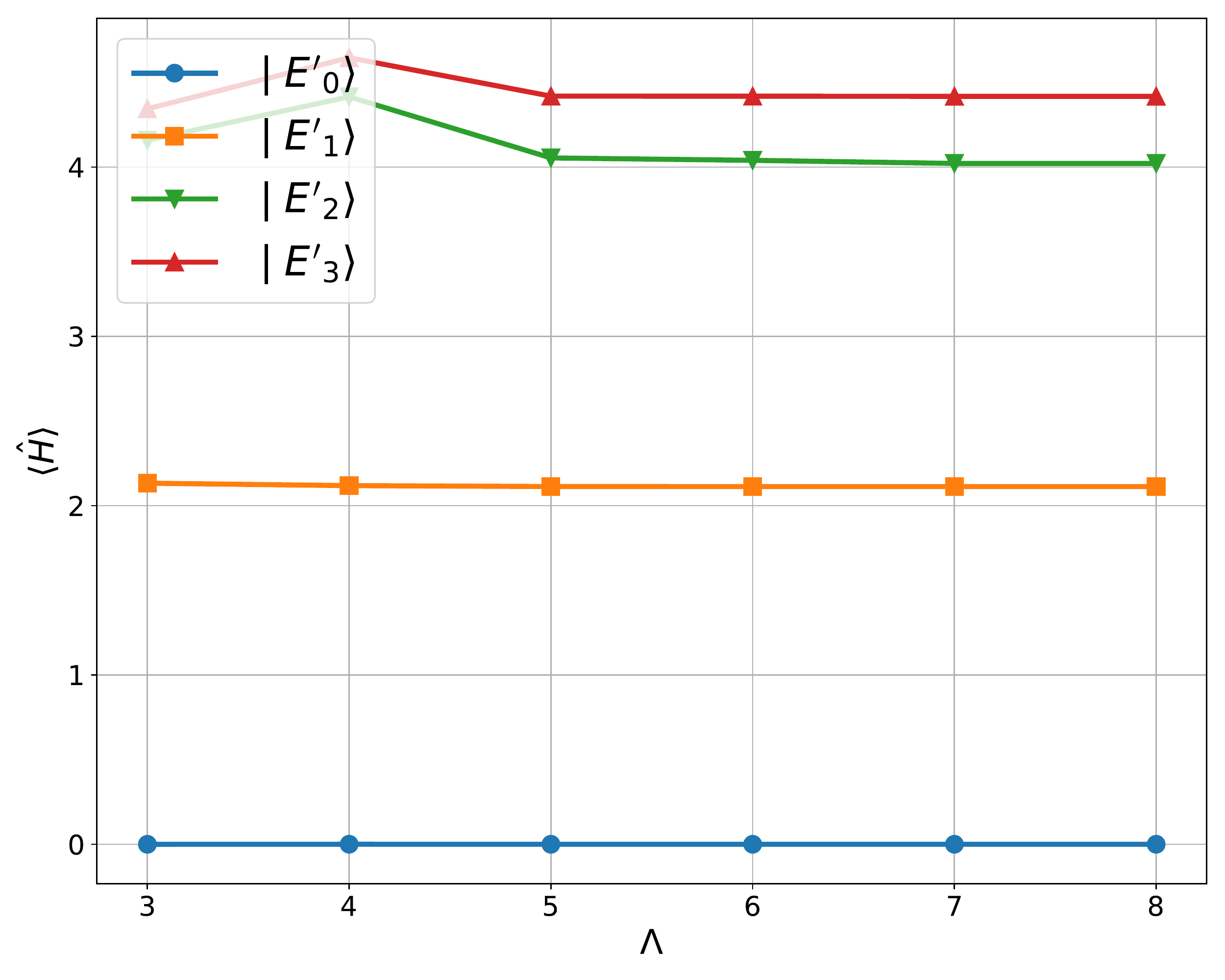} &
      \includegraphics[width=0.30\textwidth]{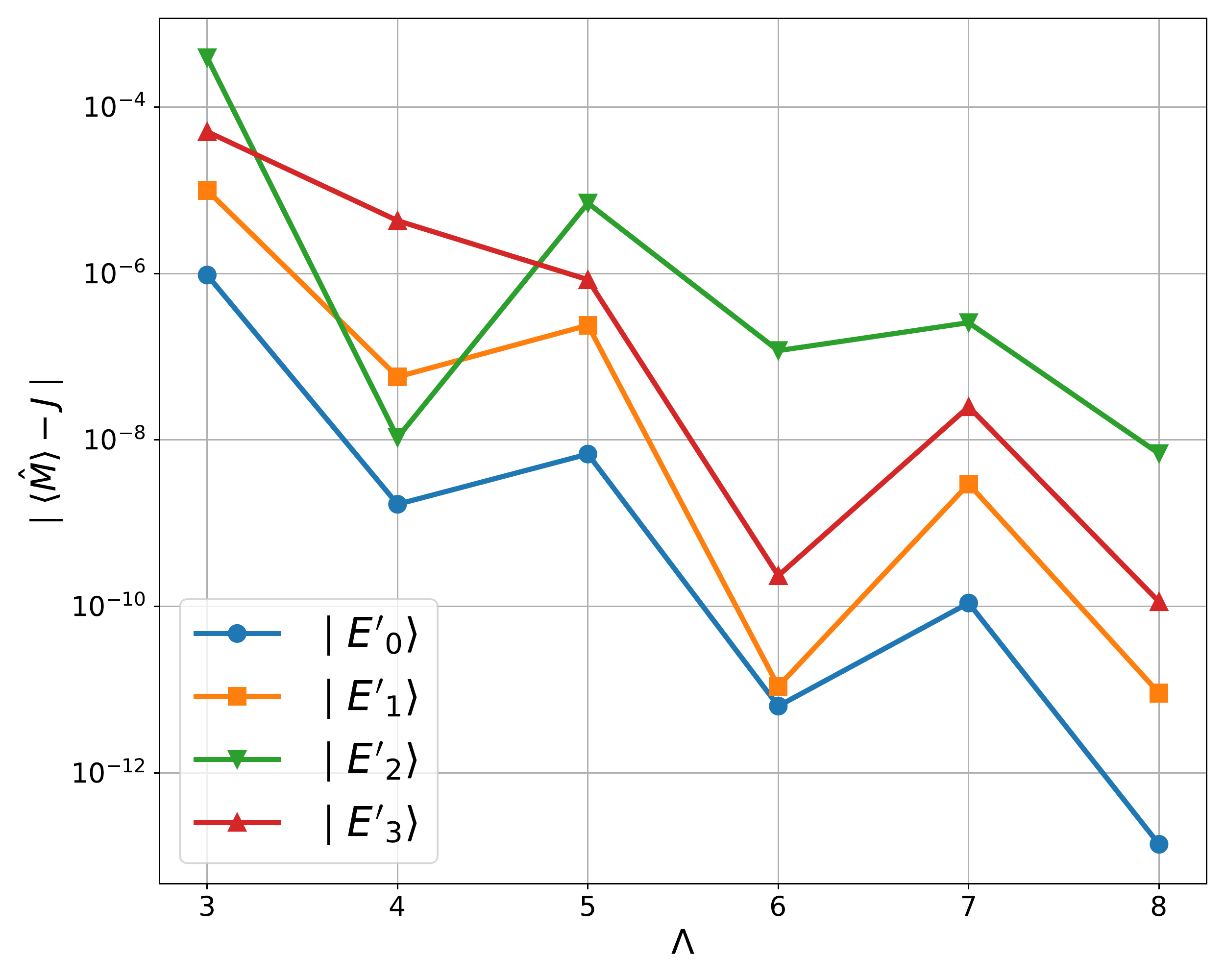} \\
      \includegraphics[width=0.30\textwidth]{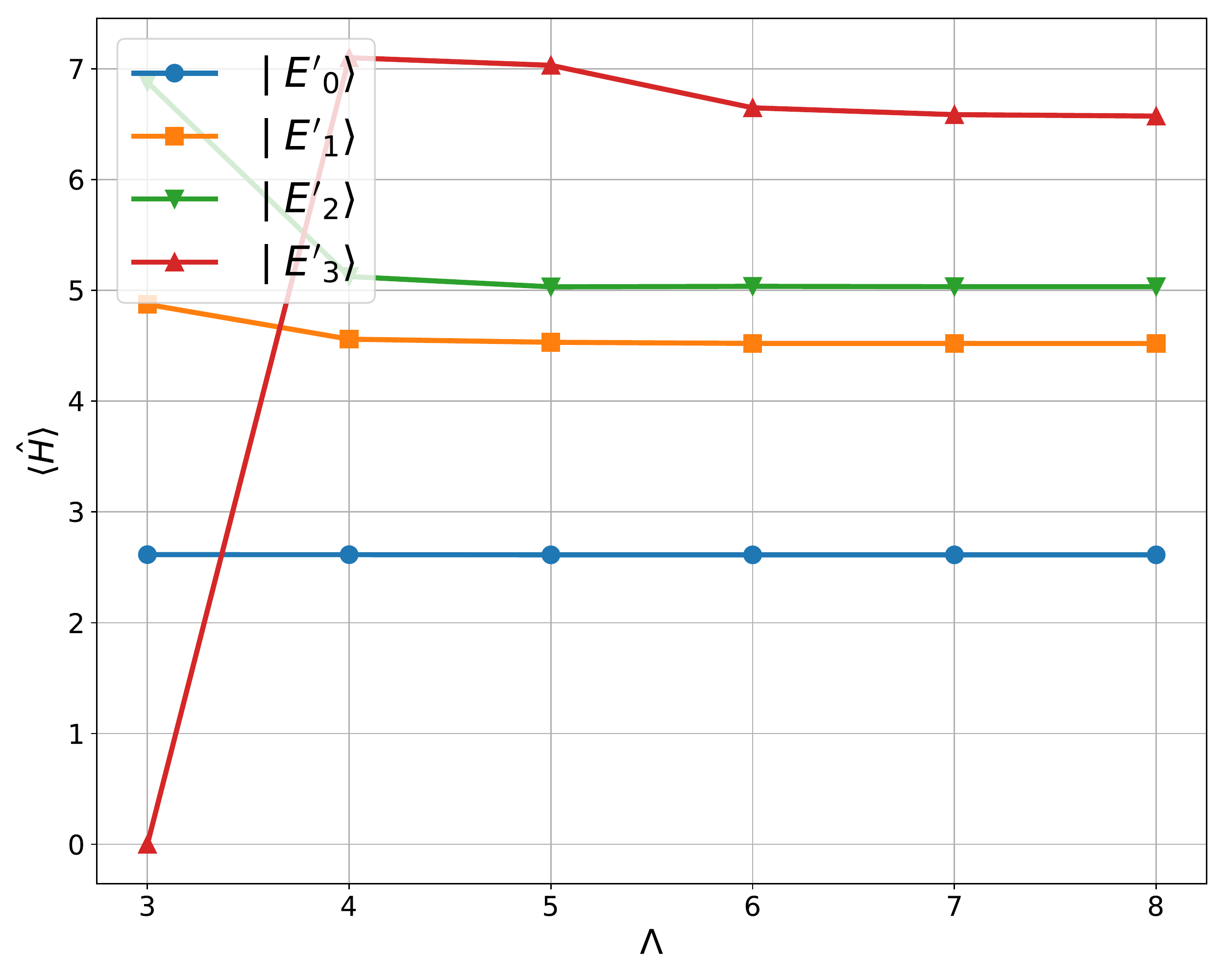} &
      \includegraphics[width=0.30\textwidth]{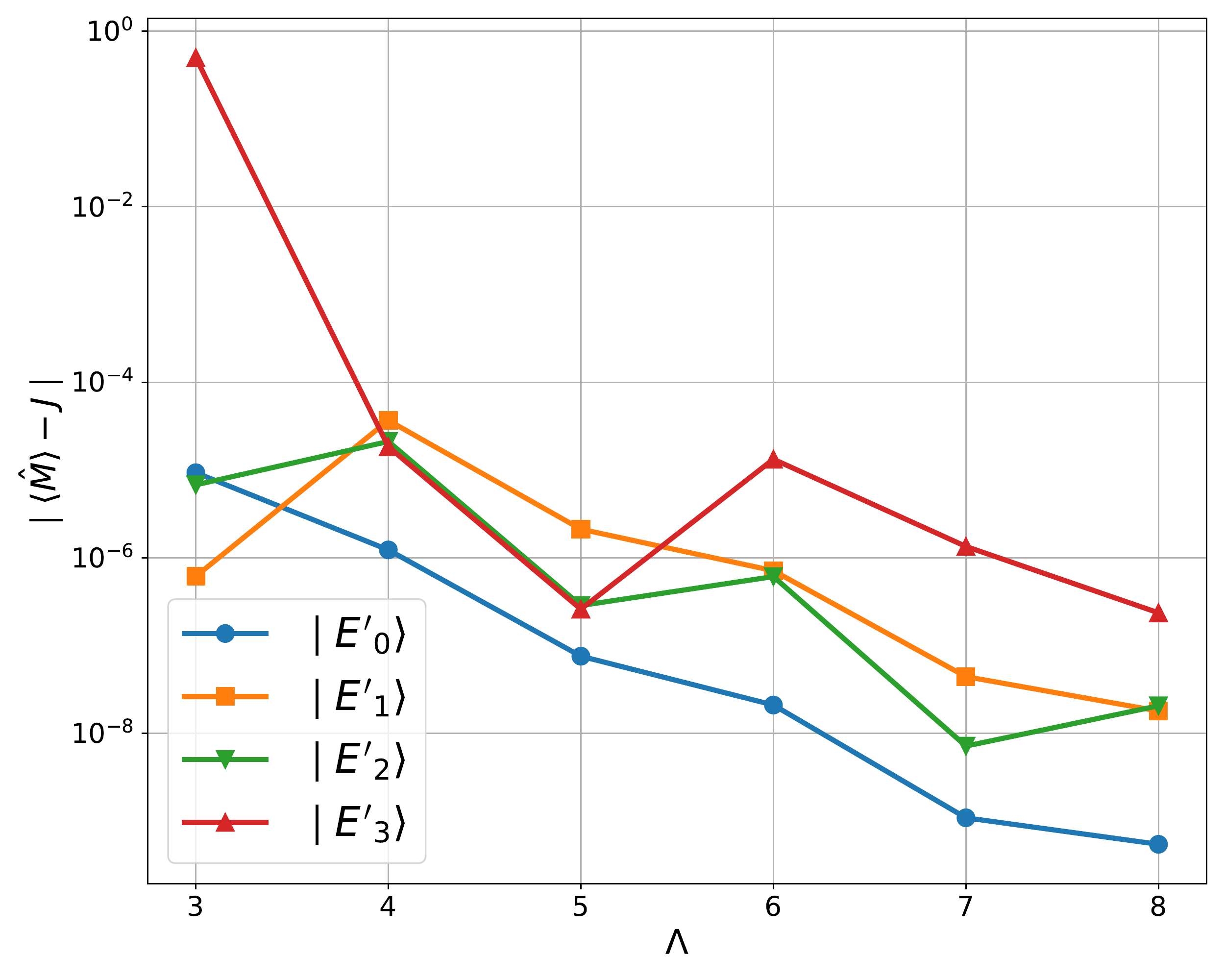}
  \end{tabular}
  \end{center}
  \caption{\label{fig:ham_su2_mini_gs_gm_e0123}
  $\langle E'_{i,J}|\hat{H}|E'_{i,J}\rangle$ and $|\langle E'_{i,J}|\hat{M} - J|E'_{i,J}\rangle|$ in SU(2) minimal BMN, for a few low-lying excited modes.
  The modes are ordered by their $\lambda_{E'}$ eigenvalues of $\hat{H}'$ and their order can change depending on the cutoff $\Lambda$, because the deformed Hamiltonian's parameters in $\hat{H}'$ are changing.
  The upper row is for the sector with $J=0$ and the lower row is for $J=1/2$.
  Other parameters are $\lambda=g^2N=0.2$, $m^2=1$, $c=\Lambda$, $c'=10\Lambda$.
  }
\end{figure}
Let the eigenstates of $\hat{H}'$ in the sector of angular momentum $J$ be $|E'_{0,J}\rangle,|E'_{1,J}\rangle,\cdots$.
The amount of breaking of gauge symmetry and SO(2) rotational symmetry due to the truncation can be seen from $\langle E'_{i,J}|\hat{G}^2|E'_{i,J}\rangle$ and $\langle E'_{i,J}|\hat{M}|E'_{i,J}\rangle-J$.
These quantities are plotted in Fig.~\ref{fig:ham_su2_mini_gv} and Fig.~\ref{fig:ham_su2_mini_gm_J0}.
We can see the quick restoration of these symmetries as the cutoff increases, for all couplings.

In Fig.~\ref{fig:ham_su2_mini_gs}, $\langle E'_{0,J=0}|\hat{H}|E'_{0,J=0}\rangle$ is plotted.
Given that the breaking of gauge symmetry and rotational symmetry is small, this is very close to the ground state energy $E_{0,J=0}$, which has to be zero at $\Lambda=\infty$ due to the BPS condition.
We can see an exponentially fast approach to zero as $\Lambda$ increases.
A few of the low-lying modes (up to the third excited state) for the SU(2) model with $\lambda=0.2$ are plotted in Fig.~\ref{fig:ham_su2_mini_gs_gm_e0123} for the $J=0$ and the $J=1/2$ sectors.
As the cutoff $\Lambda$ increases, the ground states reach their asymptotic energy value, which is zero for the $J=0$ sector where the ground states is BPS, but is different from zero for the $J=1/2$ sector.

\section{Quantum-classical hybrid algorithm (VQE)}\label{sec:VQE}
The Variational Quantum Eigensolver (VQE) algorithm is expected to be a practically useful tool in the NISQ era.
The algorithm is a hybrid quantum-classical algorithm.
The expectation value of the Hamiltonian is efficiently computed on quantum hardware for quantum states represented by parametrized quantum circuits which mimic wave functions with variational parameters.
The parameters of the quantum circuits, or, in other words, of the wave function, are optimized using classical algorithms (e.g. steepest descent) on classical hardware such that each step of the optimization is moving towards convergence to the lower bound of the energy expectation value.
The algorithm returns an upper bound for the ground-state energy $E_{\rm var}$:
\begin{equation}\label{eq:vqe-bound}
E_0 \le E_{\rm var} =
\frac{\left\langle \psi (\theta _i) \right|H\left| \psi (\theta _i) \right\rangle }
{\left\langle \psi (\theta _i) \right|\left. \psi (\theta _i) \right\rangle } \ ,
\end{equation}
where $E_0$ is the true ground-state energy and $\theta_i$ are variational parameters of the trial wave function, which is represented in terms of parametrized quantum gates.
More details about the VQE hybrid quantum-classical algorithm can be found in Ref.~\cite{Moll_2018}.
There is also a fully quantum algorithm for computing eigenvalues described in Ref.~\cite{wei2020quantum}.

On NISQ devices it is of paramount importance to test how well the VQE performs for different systems~\cite{2021arXiv210506208K}.
In this section, we test the potential performance of the VQE approach in estimating the ground states of matrix models for the first time.
First, we construct a truncated Hamiltonian which maps the Hamiltonian of the matrix model of interest into strings of Pauli matrices which can be implemented in terms of gates on quantum computers.
We use the IBM \qiskit~\cite{qiskit} software framework and use the VQE algorithm to place an upper bound on the ground-state energy of the matrix model.

The performance of the VQE can depend on the choice of the ansatz for the wave function and the optimizer used for the minimization of the energy.
We will try several different combinations of classical optimizers and ans\"{a}tze for the wave function.
Even if the same ansatz is used, different upper bounds might be obtained depending on the optimizers because the convergence to the true minimum can be sabotaged by restricting the maximum number of iterations or starting from an unfavorable point in parameter space, i.e. getting trapped in local minima.
The ans\"{a}tze we use in this section are not designed specifically for matrix models, and hence, there is no reason to expect good results.
A first step towards identifying the best ansatz within a certain class of parametrized quantum circuits is shown at the end of this section, but we leave a more comprehensive search to future work.
In fact, we find larger deviation from the exact result at stronger couplings, which suggests a need for finding a better ansatz for the wave function as we change the coupling of the matrix quantum mechanics model.
This is expected, since the ground state at weak coupling is expected to exhibit different properties than at strong coupling.

\begin{figure}[htbp]
    \centering
    \includegraphics[width=0.5\textwidth]{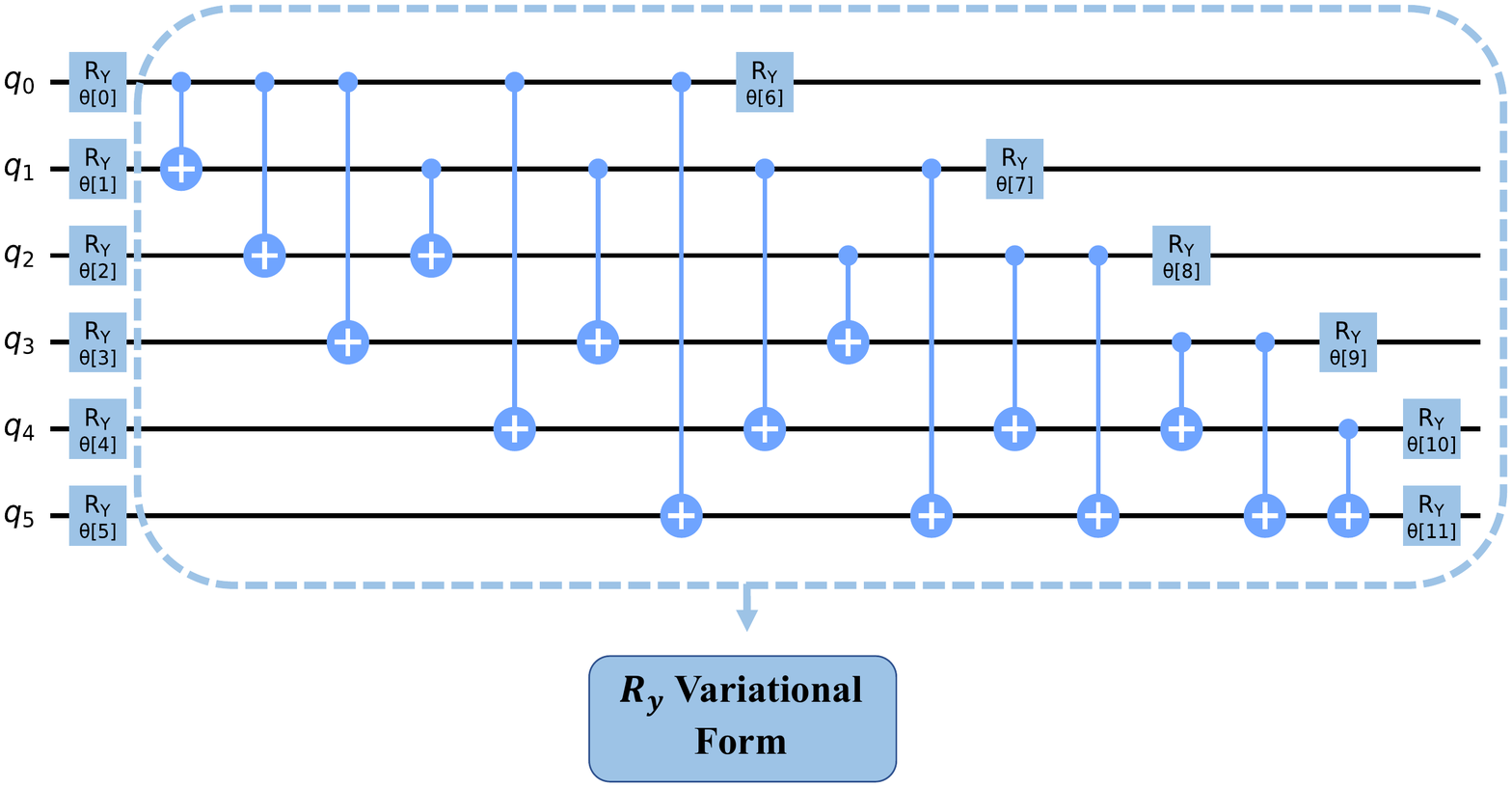}
    \caption{\label{fig:2x2_varForm_oneLayer}
    A single block of the variational wave function in terms of parametrized quantum circuits.
    A parameterized $R_y$ gate is applied to each qubit, and each qubit is entangled with every other qubit using CNOT gates.
    We also consider a similar block ($R_yR_z$ variational form) in which the $R_y$ gate is replaced with a sequence of $R_y$ and $ R_z$ gates.
    }
\end{figure}

\begin{figure}[htbp]
    \centering
    \includegraphics[width=0.5\textwidth]{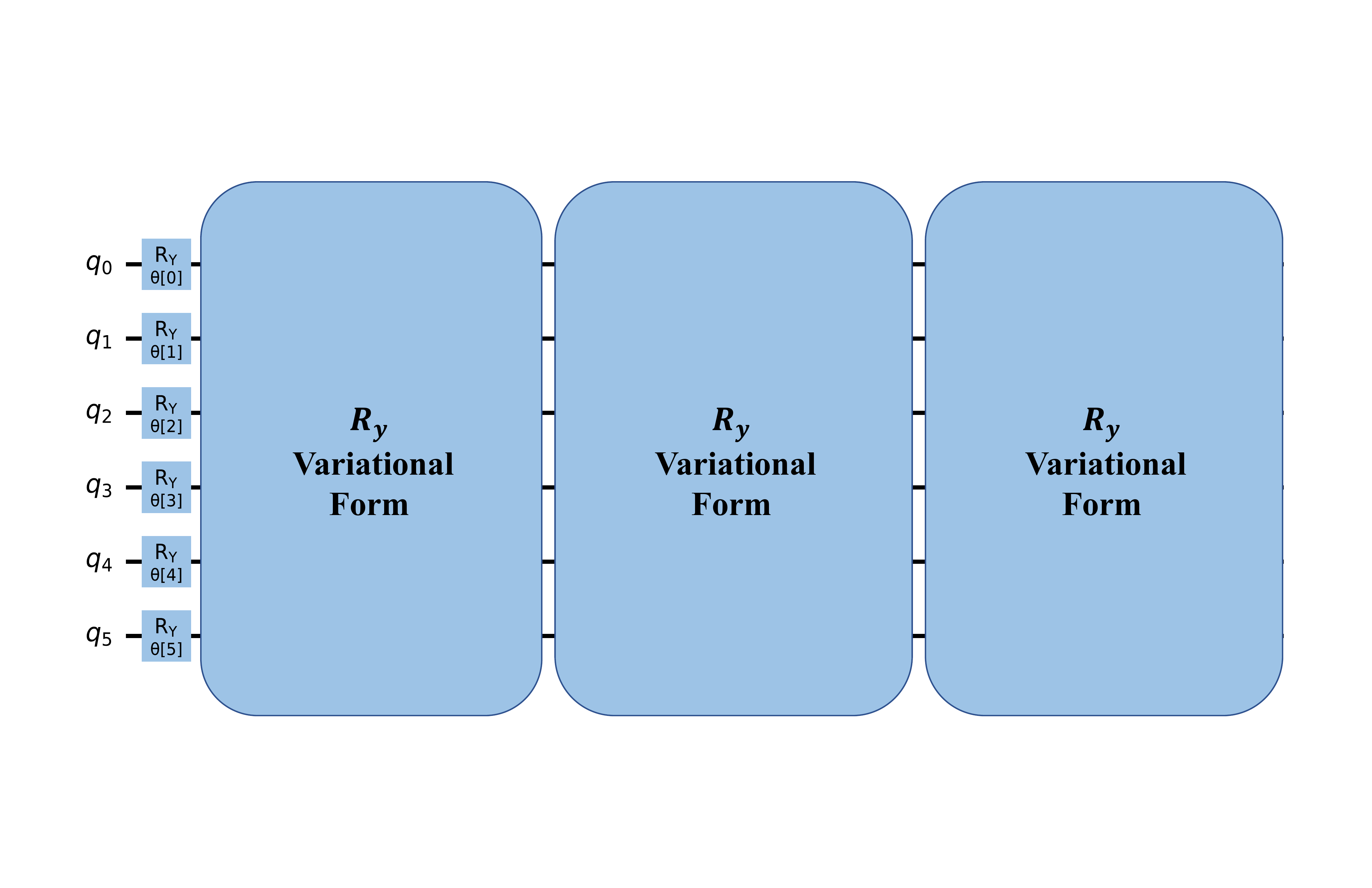}
    \caption{\label{fig:2x2_varForm_Full}
    The variational ansatz used in the VQE simulation.
    The circuit has a depth of 3 (i.e. the $R_y$ variational form is repeated three times).
    We also consider the $R_yR_z$ variational form.
    }
\end{figure}

A VQE solver obtains an upper bound for the ground state energy by starting from a parametrized wave function in the form of quantum circuits made of parametrized gates.
Such circuits can be chosen arbitrarily and, typically, the final energy will depend on the choice of the initial form of the wave function.
In Fig.~\ref{fig:2x2_varForm_oneLayer} we show a single block of the variational circuit of 6 qubits that we use in the next section, where the parameters are encoded in gates that rotate each qubit about a certain axis.
In the figure we show specifically $R_y$ gates which are single qubit rotations about the $Y$ axis, so that
\begin{equation}\label{eq:ry-gate}
{R_y}(\theta ) = \exp \left( { - i\frac{\theta }{2}} \right) = \left( {\begin{array}{*{20}{c}}
{\cos \frac{\theta }{2}}&{ - \sin \frac{\theta }{2}}\\
{\sin \frac{\theta }{2}}&{\cos \frac{\theta }{2}}
\end{array}} \right) \ .
\end{equation}
We will also consider a similar circuit obtained by replacing an $R_y$ with a sequence of $R_y$ and $R_z$ gates.

These blocks can be composed into multiple ``layers'' to make deeper circuits, which allow the representation of more expressive wave functions.
By repeating the same blocks multiple times in a sequence, we also increase the number of parameters to optimize.
For most optimizers in the following sections we use three repetitions of the $R_y$ fully-entangled block as shown in Fig.~\ref{fig:2x2_varForm_Full}, and we refer to this variational ansatz as having a depth of 3 (just counting the number of layers).
This amounts to $8 \times 3 = 24$ parameters in total, which are the angles $\theta[0],\theta[1],\cdots,\theta[23]$ for the rotation gates.
If the variational ansatz requires more qubits because, for example, we want to use a larger cutoff of the Hilbert space, then we would need a larger number of parameters.
In general, having access to more parameters to optimize will allow for a more expressive wave function and we can expect a better overlap with the true ground state wave function.
However, the optimization landscape in a high dimensional space (for a large number of parameters) becomes more complex and some optimizers might fail in finding a true minimum or might take too long to converge.

After choosing the variational form of the quantum circuit, we have to choose a classical optimizer that will be used to find the correct set of parameters to minimize the energy.
We use four different optimizers: a Sequential Least SQuares Programming optimizer (SLSQP), a Constrained Optimization By Linear Approximation optimizer (COBYLA), a Limited-memory BFGS Bound optimizer (L-BFGS-B), and a Nelder-Mead optimizer.
More information on these optimizers can be found in~\cite{numerical_python}.
For a series of $N_{r}$ runs starting from different initial parameter values $\theta[i]$, the least upper bound (minimum) gives the closest value to the ground state.
For our optimizers, we limit the maximum number of iterations to $10^4$, which is sufficient to reach convergence in most of the cases we study in the following sections, and we discuss the case of more iterations in a dedicated section featuring deeper parametrized circuits.
To give a summary of the performance of each optimizer for a fixed variational form of the quantum circuit, we report the minimum value of the energy, the maximum value, the mean value, and the standard deviation across $N_r$ runs.

\subsection{SU(2) bosonic matrix Model}
In this section, we consider the bosonic matrix model defined in Sec.~\ref{sec:def_bos_MM}.
We study the case of two SU(2) matrices ($N=2, d=2$), and different truncation levels: $\Lambda=2$ and $\Lambda=4$.
They require only 6 and 12 qubits, respectively, and hence, the emulation on a classical computer is straightforward.
IBM \qiskit provides several quantum simulators as well as access to fully-quantum hardware resources.
The QASM simulator can be used to simulate noise on an actual quantum device and the state vector simulator holds the value of the quantum state in computer memory through the computation.
Using a NISQ device with the current knowledge we have on the best variational wave function for matrix models would not be an efficient use of resources.
For this reason in this paper we use the state vector simulator on a classical computer.
We then focus on identifying good variational ansatze.
We would like to use the QASM simulator and hardware resources in upcoming works when we can make the most out of them.

\subsubsection{One qubit for each boson}\label{sec:one-qubit-bosonic}
When the cutoff for the truncated Hilbert space is $\Lambda=2$, we can represent each degree of freedom in the matrix model with one qubit.
In the SU(2) two-matrix model we have six degrees of freedom, $X_{I=1,2}^{\alpha=1,2,3}$.
We call $X_1^1$, $X_1^2$, $X_1^3$, $X_2^1$, $X_2^2$, $X_2^3$ in Eq.~\eqref{eq:bos-ham-reg} to be $x_1,x_2,\cdots,x_6$.
There are six annihilation operators $\hat{a}_{i=1,\cdots,6}$, which can be represented in terms of tensor products as
\begin{align}\label{eq:bosonic-annihilation-oneq}
\hat{a}_i =\hat{I}_1\otimes  \ldots \otimes\hat{I}_{i-1}
\otimes \left( {\begin{array}{*{20}{c}}
0&1\\
0&0
\end{array}} \right) \otimes \hat{I}_{i + 1} \otimes  \ldots \otimes\hat{I}_6
\ ,
\end{align}
where $\hat{I}_i$ is the $2\times 2 $ identity matrix for $\Lambda=2$.

The regularized Hamiltonian in Eq.~\eqref{eq:bos-ham-reg} for $\Lambda=2$ is a $2^6\times 2^6 = 64\times 64$ matrix whose smallest eigenvalue at coupling $\lambda = 0.2$ is $E_0 = 3.14808$.
This is the exact result from the Hamiltonian truncation (HT) approach.
We compare the VQE result for the truncated Hamiltonian, which is only an upper bound, to this exact HT result.
In Tab.~\ref{tab:6bosons_2x2_VQEResults_100runs}, the statistics of the results for the four different VQE solvers are shown for $N_r=100$ runs and variational forms $R_y$, $R_yR_z$ of depth 3.
The best results are $E_{\rm var} = 3.148972$ and $E_{\rm var} = 3.149157$ for $R_y$ and $R_yR_z$, respectively.
The $R_yR_z$ variational form is an extension of the $R_y$ form, i.e. one can set the $R_z$ gates angles to zero, but the high dimensionality of the space of parameters does not allow the optimizers to find an optimal point that is better unless the starting point is carefully chosen.

\begin{table}[ht]
    \centering
    \begin{tabular}{|l||l|l|l|l||l|l|l|l|}
    \hline
    Optimizer & \multicolumn{4}{|c||}{Var. form: $R_y$} & \multicolumn{4}{|c|}{Var. form: $R_yR_z$} \\ \hline
                & Min. &  Max. & Mean & Std. & Min. &  Max. & Mean & Std. \\ \hline
    COBYLA      & 3.149370	& 4.147156	& 3.159740	& 0.099739 & {\bf 3.149157}	& 3.150034	& 3.149862	& 0.000202 \\ \hline
    L-BFGS-B	& 3.149268	& 4.150000	& 3.159886	& 0.100012 & 3.149375	& 4.148751	& 3.159925	& 0.099882 \\ \hline
    SLSQP       & 3.149397	& 4.150000	& 3.164968	& 0.111340 & 3.149377	& 4.149946	& 3.164980	& 0.111349 \\ \hline
    NELDER-MEAD	& {\bf 3.148972} & 3.195922	& 3.150774	& 0.005065 & 3.149516	& 4.149891	& 3.171468	& 0.140469 \\ \hline
    \end{tabular}
    \caption{\label{tab:6bosons_2x2_VQEResults_100runs}
    VQE results for the bosonic BMN matrix model with SU(2) and $\lambda=0.2$, where each boson is represented by one qubit.
    We use four optimizers and two variational forms (both with depth 3).
    Each optimizer began from the same initial point.
    The minimum, maximum, average, and standard deviation of the best results across $N_r=100$ runs are reported.
    The exact result from the Hamiltonian truncation is $E_0 = 3.14808$ and the best results are in bold for each variational form.
    We limited the maximum number of iterations to $10^4$.
    }
\end{table}

\begin{figure}[htbp]
    \centering
    \includegraphics[width=0.8\textwidth]{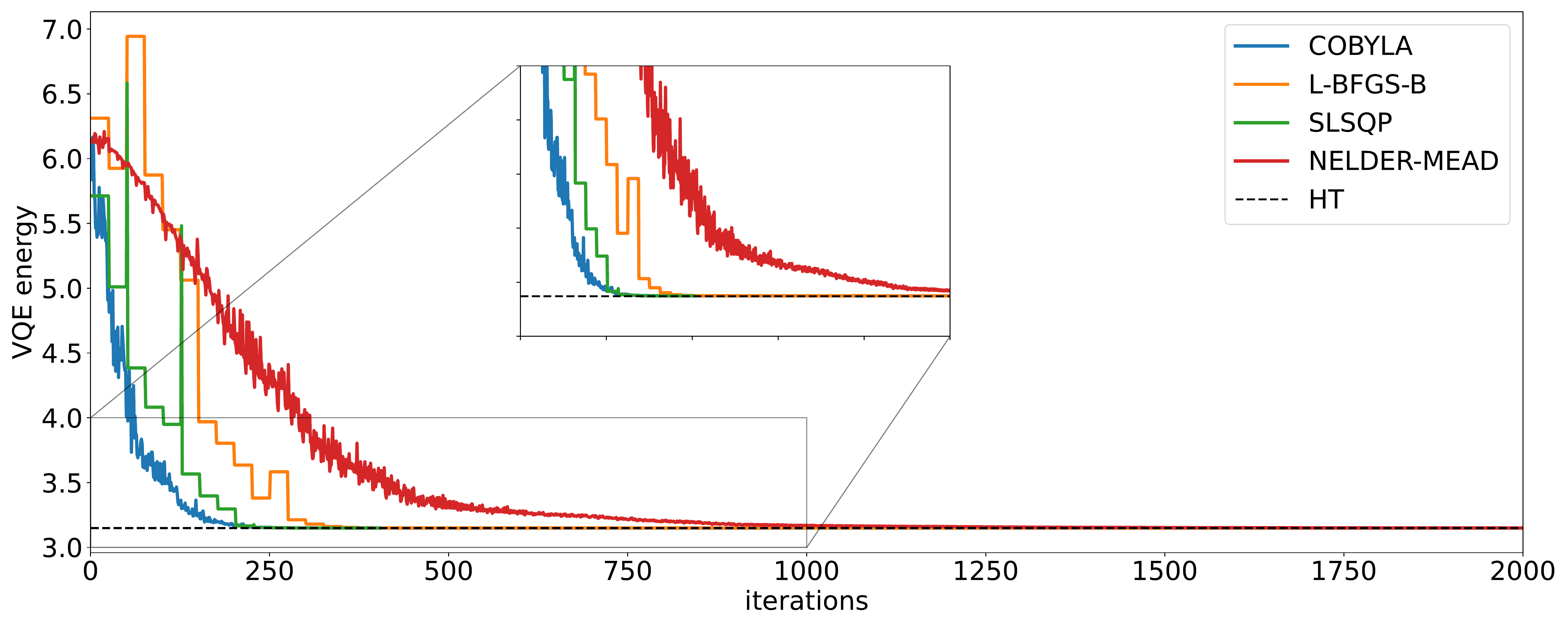}
    \caption{\label{fig:6bosons2x2_varOptimizers}
    Convergence of the VQE results for the bosonic matrix model with SU(2) and $\lambda=0.2$, where each boson is represented by one qubit.
    Each curve represents a different classical optimizer (SLSQP, COBYLA, L-BFGS-B, and Nelder-Mead) and we only show the first 2000 iterations out of 10000.
    The VQE result represents the least upper bound from 10 runs of each optimizer with an $R_y$ variational form of depth 3.
    The exact energy from the truncated Hamiltonian (HT) is represented by the dotted line at $E_0 = 3.14808$.
    }
\end{figure}

The upper bounds we find are very close to the exact value of $E_0$, even though these variational quantum circuits have not been specifically defined for the bosonic matrix models.
The number of steps needed for the convergence of the solvers depends on the optimizer, as shown in Fig.~\ref{fig:6bosons2x2_varOptimizers}.
Among the solvers we use, COBYLA, L-BFGS-B, and SLSQP exhibit faster convergence than NELDER-MEAD.

\subsubsection{Two qubits for each boson}\label{sec:two-qubits-bosonic}
When we use a cutoff $\Lambda=4$, we can represent each matrix degree of freedom with two qubits.
The SU(2) bosonic matrix model will then have each boson represented by a $4 \times 4 $ matrix (corresponding to two qubits).
This is similar to Sec.~\ref{sec:one-qubit-bosonic}, but with the annihilation operators represented by the tensor product
\begin{align}
\hat{a}_i =\hat{I}_1\otimes  \ldots \otimes\hat{I}_{i-1}
\otimes \left( {\begin{array}{*{20}{c}}
0&1&0&0\\
0&0&{\sqrt 2 }&0\\
0&0&0&{\sqrt 3 }\\
0&0&0&0
\end{array}} \right)
 \otimes \hat{I}_{i + 1} \otimes  \ldots \otimes\hat{I}_6
\ ,
\end{align}
where now the identity matrix $\hat{I}_i$ we use is $4\times 4$.
The truncated Hamiltonian will then be a $2^{12}\times 2^{12} = 4096 \times 4096$ matrix (on 12 qubits) and the exact ground state energy is $E_0 = 3.13406$.

\begin{table}[ht]
    \centering
    \begin{tabular}{|l||l|l|l|l||l|l|l|l|}
    \hline
    Optimizer & \multicolumn{4}{|c||}{Var. form: $R_y$} & \multicolumn{4}{|c|}{Var. form: $R_yR_z$} \\ \hline
                & Min. &  Max. & Mean & Std. & Min. &  Max. & Mean & Std. \\ \hline
    COBYLA   & {\bf 3.137059} & 4.769101 & 3.251414 & 0.347646 & 3.137237 & 4.782013 & 3.378628 & 0.472015 \\ \hline
    L-BFGS-B & {\bf 3.137059} & 5.769553 & 3.283462 & 0.434162 & {\bf 3.137050} & 4.286367 & 3.243110 & 0.307549 \\ \hline
    SLSQP        & 3.137060 & 5.769554 & 3.327706 & 0.471957 & 3.137059 & 4.232419 & 3.236925 & 0.290855 \\ \hline
    NELDER-MEAD  & 3.137471 & 5.713976 & 3.492673 & 0.478810 & 3.273614 & 6.443055 & 4.428032 & 0.758732 \\ \hline
    \end{tabular}
    \caption{\label{tab:6bosons_4x4_VQEResults_10runs}
    VQE results for the bosonic matrix model with SU(2) and $\lambda=0.2$, where each boson is represented by two qubits.
    We use four optimizers and two variational forms (both with depth 3).
    Each optimizer began from the same initial point.
    The minimum, maximum, average, and standard deviation of the best results across $N_r=100$ runs are reported.
    The exact result from the Hamiltonian truncation is $E_0 = 3.13406$ and the best results are in bold for each variational form.
    We limited the maximum number of iterations to $10^4$.
    }
\end{table}

\begin{figure}[htbp]
    \centering
    \includegraphics[width=0.8\textwidth]{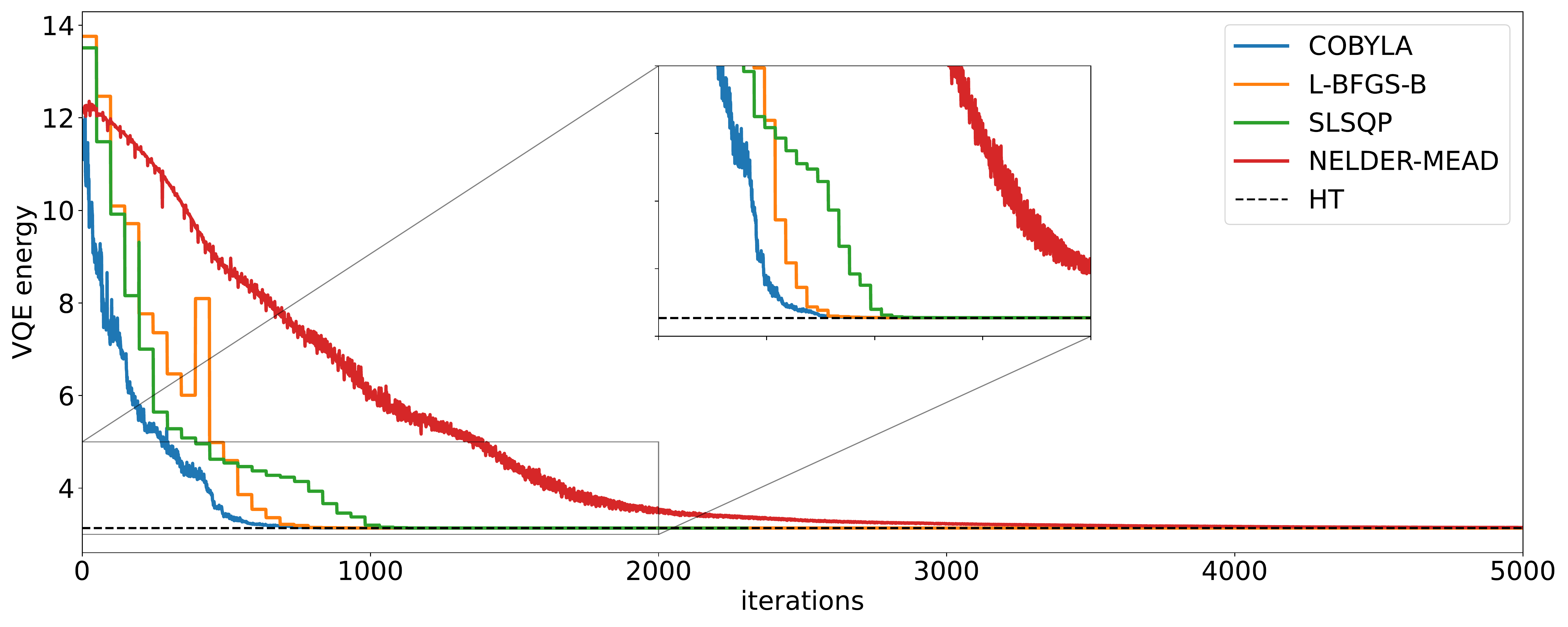}
    \caption{\label{fig:6bosons4x4_varOptimizers}
    Convergence of the VQE results for the SU(2) bosonic matrix model with $\lambda=0.2$, where each boson is represented by two qubits.
    Each curve represents a different classical optimizer (SLSQP, COBYLA, L-BFGS-B, and Nelder-Mead), and we only show the first 5000 iterations out of 10000.
    The VQE result represents the least upper bound from 10 runs of each optimizer with an $R_y$ variational form of depth 3.
    The exact energy from the truncated Hamiltonian (HT) is represented by the dotted line at $E_0 = 3.13406$.
    }
\end{figure}

We test the performance of the same four types of VQE solvers as before, and we use the 12-qubit equivalent of the parametrized variational form of Fig.~\ref{fig:2x2_varForm_Full} which has 48 parameters, $\theta[0],\theta[1],\cdots,\theta[47]$.
In Fig.~\ref{fig:6bosons4x4_varOptimizers} we plot the convergence of the best run out of $N_r=100$ for each optimizer.
After a certain number of optimization steps (which depends on the optimizer and the Hamiltonian) the estimate for the upper bound of the ground state energy flattens out and this yields the best value for that run.
In Tab.~\ref{tab:6bosons_4x4_VQEResults_10runs} we report the statistics for the VQE results with $N_r = 100$.
The best results are $E_{\rm var} = 3.137059$ and $E_{\rm var} = 3.137050$ for $R_y$ and $R_yR_z$, respectively and are again very close to the exact value of $E_0 = 3.13406$.

\subsubsection{VQE for different coupling constants}
\begin{figure}[htbp]
    \centering
    \begin{tabular}{cc}
        \includegraphics[width=0.4\textwidth]{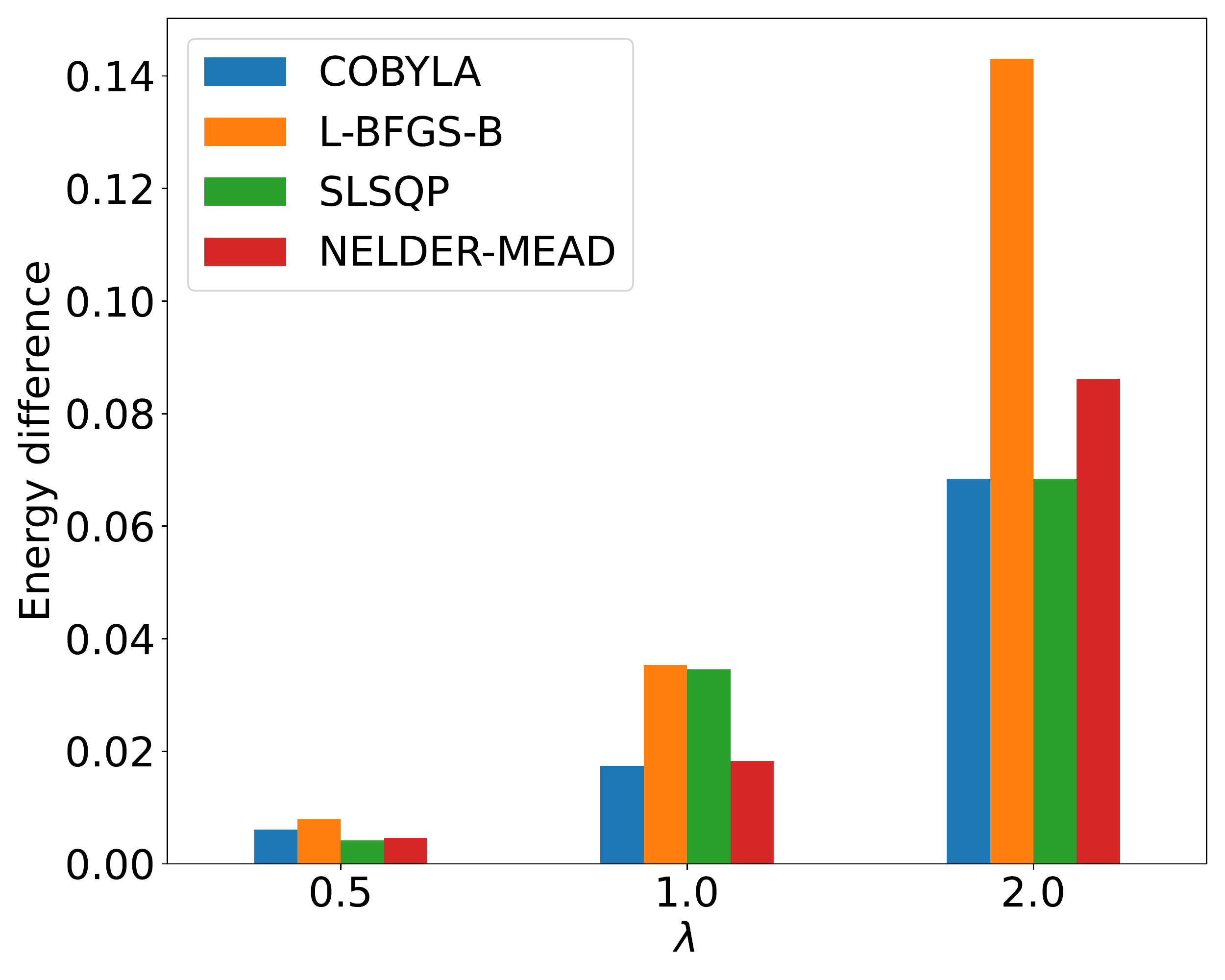} &
        \includegraphics[width=0.4\textwidth]{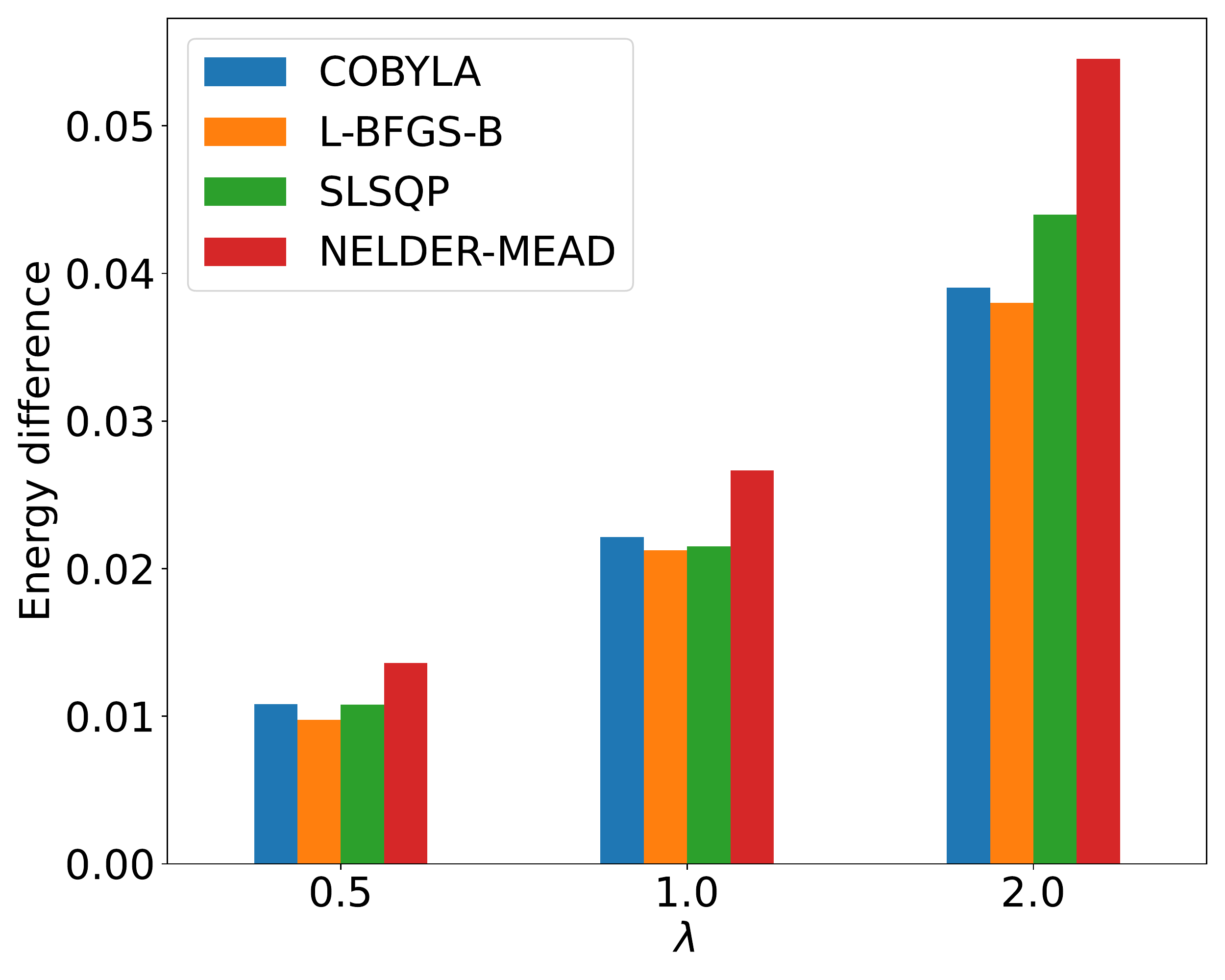} \\
    \end{tabular}
    \caption{\label{fig:6Boson_VaryCoupling_vqeANDexact}
    The energy difference between the VQE upper bound for the SU(2) bosonic matrix model using the three best optimizers for coupling $\lambda = 0.5$, 1.0, and 2.0.
    The left plot is for bosons represented by one qubit and the right plot is for bosons represented by two qubits.
    The $R_y$ ansatz with depth 3 is used and the best of $N_r=100$ runs is displayed.
    We limited the maximum number of iterations to $10^4$.
    }
\end{figure}

To test the accuracy of the VQE results with respect to the coupling, we ran the VQE while varying the coupling for the SU(2) Hamiltonian for both cases where the bosons are represented by one qubit or by two qubits.
A graph of the VQE results with respect to coupling is shown in Fig.~\ref{fig:6Boson_VaryCoupling_vqeANDexact}, with the values tabulated in Tab.~\ref{tab:6Bosons_VaryCoupling}, for both cases where bosons are represented by one qubit or two qubits.
In both cases we used the same variational forms as before and we compared them using all solvers.

\begin{table}[htbp]
    \centering
    \begin{tabular}{|l||l|l|l||l|l|l|}
    \hline
    Optimizer & \multicolumn{3}{|c||}{One qubit} & \multicolumn{3}{|c|}{Two qubits} \\ \hline
                & $\lambda=0.5$ &  $\lambda=1.0$ & $\lambda=2.0$ & $\lambda=0.5$ &  $\lambda=1.0$ & $\lambda=2.0$ \\ \hline
    COBYLA       &      3.36859  & {\bf 3.71463} &      4.33638  &      3.30975  &      3.54839  &      3.93452  \\ \hline
    L-BFGS-B     &      3.37043  &      3.73259  &      4.41103  & {\bf 3.30869} & {\bf 3.54748} & {\bf 3.93348} \\ \hline
    SLSQP        & {\bf 3.36675} &      3.73179  & {\bf 4.33636} &      3.30974  &      3.54776  &      3.93946  \\ \hline
    NELDER-MEAD  &      3.36718  &      3.71547  &      4.35411  &      3.31255  &      3.55292  &      3.95003  \\ \hline
    HT (exact)   &      3.36254  &      3.69722  &      4.26795  &      3.29894  &      3.52625  &      3.89548  \\ \hline
    \end{tabular}
    \caption{\label{tab:6Bosons_VaryCoupling}
    VQE results for the bosonic BMN matrix model with SU(2) group at different couplings, with each boson represented by one qubit and two qubits.
    The $R_y$ ansatz with depth 3 is used and the best of $N_r=100$ runs is reported.
    The last row is the exact result obtained from the Hamiltonian truncation (HT).
    Best results are reported in bold.
    We limited the maximum number of iterations to $10^4$.
    }
\end{table}

We conclude our VQE studies of the bosonic matrix model with a discussion of the dependence of the upper bound determined by the VQE for different couplings.
Clearly the gap between the upper bound determined by the VQE and the exact value grows with the coupling for both the one-qubit bosons and two-qubits bosons.
This is consistent with what was found in studying the an-harmonic oscillator and supersymmetric an-harmonic oscillator in Ref.~\cite{8538962}.
The most likely reason for the increase in the gap is that the ansatz of the variational wave function does not have as large of an overlap with the ground state wave function at strong coupling as it does at weak coupling.
Some possibilities to improve this include using specially designed ansatz tailored to the Hamiltonian as is done with coupled cluster ansatz for chemistry, adaptive methods to obtain better ansatz described in Ref.~\cite{Grimsley_2019}, and machine learning methods to improve overlap with the ground state in Ref.~\cite{Choo_2020}.

\subsection{SU(2) minimal BMN}

\subsubsection{One qubit for each boson}
For the minimal BMN matrix model we use the same representation as above with each boson represented by one qubit (a $2 \times 2$ matrix) and each fermion also represented by one qubit.
In total we will have 9 degrees of freedom to be described by 9 qubits.
The fermionic space operators will have to be tensored with the bosonic space which is $2^6=64$ dimensional.
Therefore we define three fermionic annihilation operators as:
\begin{align}\label{eq:fermionic-operators}
{c_1} &= {\hat{I}_{64}} \otimes \left( {\begin{array}{*{20}{c}}
0&1\\
0&0
\end{array}} \right) \otimes \left( {\begin{array}{*{20}{c}}
1&0\\
0&1
\end{array}} \right) \otimes \left( {\begin{array}{*{20}{c}}
1&0\\
0&1
\end{array}} \right) \ ,
\\
{c_2} &= {\hat{I}_{64}} \otimes \left( {\begin{array}{*{20}{c}}
1&0\\
0&{ - 1}
\end{array}} \right) \otimes \left( {\begin{array}{*{20}{c}}
0&1\\
0&0
\end{array}} \right) \otimes \left( {\begin{array}{*{20}{c}}
1&0\\
0&1
\end{array}} \right) \ ,
\\
{c_3} &= {\hat{I}_{64}} \otimes \left( {\begin{array}{*{20}{c}}
1&0\\
0&{ - 1}
\end{array}} \right) \otimes \left( {\begin{array}{*{20}{c}}
1&0\\
0&{ - 1}
\end{array}} \right) \otimes \left( {\begin{array}{*{20}{c}}
0&1\\
0&0
\end{array}} \right) \ ,
\end{align}
where $\hat{I}_{64}$ is the identity matrix for the bosonic space.
The bosonic annihilation operators are the same as those defined in Eq.~\ref{eq:bosonic-annihilation-oneq}, except that they are tensored with the identity matrix $\hat{I}_{8}$ for the $2^3=8$ dimensional fermionic space.

The exact ground state energy for the minimal BMN matrix model model is zero.
The truncated Hamiltonian for this model is represented by a $2^9 \times 2^9 = 512 \times 512$ matrix whose exact ground state energy for $\lambda=0.2$ is $E_0 = 0.003287$.
We repeat the same strategy used in the bosonic BMN model and report the performance statistics for various optimizers and variational forms in Tab.~\ref{tab:6bosons_3fermions_2x2_100Runs}.

\begin{table}[htbp]
    \centering
    \begin{tabular}{|l||l|l|l|l||l|l|l|l|}
    \hline
    Optimizer & \multicolumn{4}{|c||}{Var. form: $R_y$} & \multicolumn{4}{|c|}{Var. form: $R_yR_z$} \\ \hline
                & Min. &  Max. & Mean & Std. & Min. &  Max. & Mean & Std. \\ \hline
    COBYLA       & {\bf 0.10079} & 0.15002 & 0.13020 & 0.01116 & {\bf 0.07783} & 1.34775 & 0.20278 & 0.27819 \\ \hline
    L-BFGS-B     & 0.12452 & 1.35761 & 0.18785 & 0.19427 & 0.10068 & 1.38351 & 0.18251 & 0.18785 \\ \hline
    SLSQP        & 0.12456 & 1.65000 & 0.23570 & 0.28736 & 0.10068 & 1.36707 & 0.19168 & 0.21056 \\ \hline
    NELDER-MEAD  & 0.12496 & 1.35808 & 0.19479 & 0.20055 & 0.13218 & 1.88856 & 0.32012 & 0.31914 \\ \hline
    \end{tabular}
    \caption{\label{tab:6bosons_3fermions_2x2_100Runs}
    VQE results for the minimal BMN matrix model with SU(2) and $\lambda=0.2$, where each boson is represented by one qubit.
    The VQE optimizers were run $N_r=100$ times with two variational forms, $R_y$ and $R_yR_z$ (both with depth 3).
    We limited the maximum number of iterations to $10^4$.
    Each optimizer began from the same initial point.
    The minimum, maximum, average, and standard deviation of the best results across all runs are reported.
    The exact value is $E_0 = 0.003287$.
    Best results are in bold.
    }
\end{table}

\begin{figure}[htbp]
    \centering
    \includegraphics[width=0.8\textwidth]{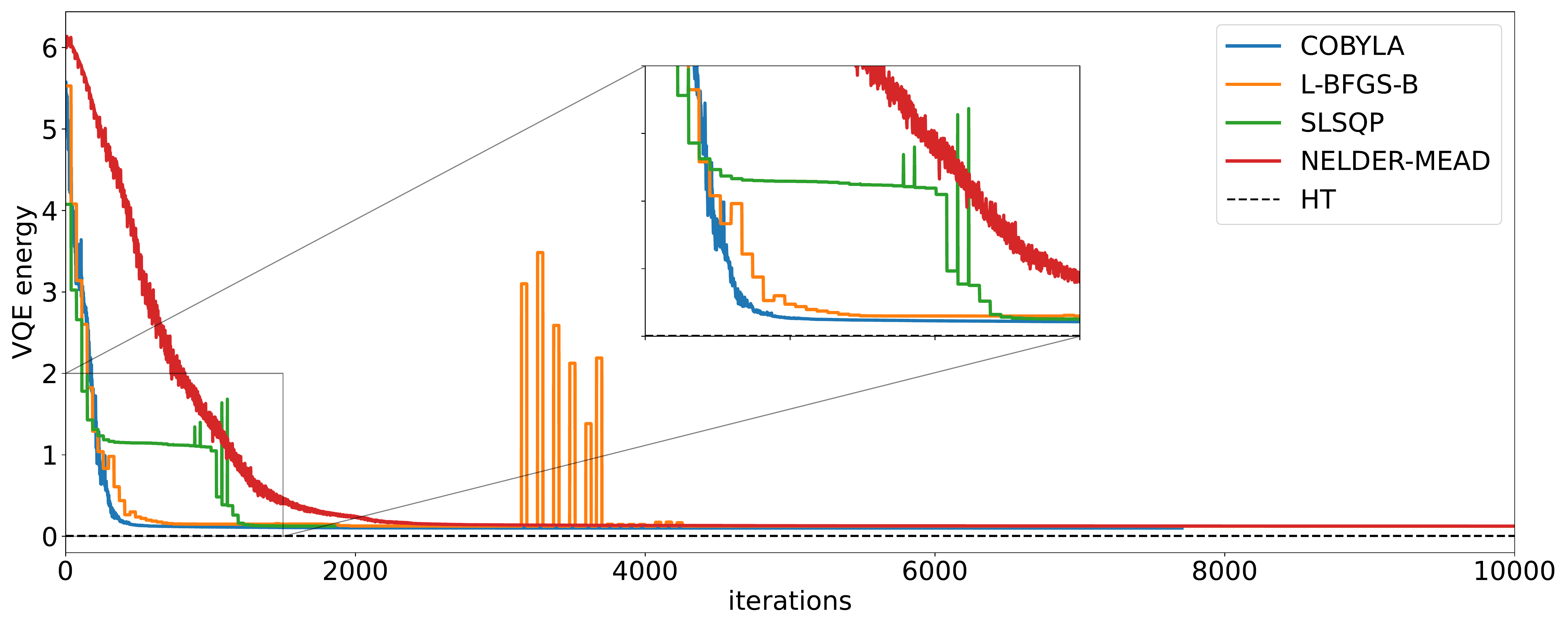}
    \caption{\label{fig:6bosons_3fermions_2x2_variousOptimizers}
    Convergence of the VQE results for the minimal BMN matrix model with SU(2) and $\lambda=0.2$, where each boson is represented by one qubit.
    Each curve represents a different classical optimizer that was used with the VQE algorithm (SLSQP, COBYLA, L-BFGS-B, and Nelder-Mead).
    The VQE result represents the least upper bound from 100 runs of each optimizer.
    The exact energy of the truncated Hamiltonian (HT) is represented by the dotted line at $E_0 = 0.003287$.
    We limited the maximum number of iterations to $10^4$.
    }
\end{figure}

For the minimal BMN model, we found that the least upper bound is further from the exact value than in the bosonic BMN if we used the same type of variational wave functions.
In this case, the parametrization of the variational wave function does not have as strong an overlap with the true ground state wave function as was found in the bosonic matrix model case.
It would be interesting to investigate the form of the variational ground states to further examine the differences between the two models and in the following we will show results for deeper parametrized quantum circuits.
A plot of the convergence for the different optimizers is shown in Fig.~\ref{fig:6bosons_3fermions_2x2_variousOptimizers}, where only the best out of $N_r=100$ is plotted for each classical optimizer.
In some cases the optimizers seem to have converged to a minimum energy, but during some iterations they briefly move away (larger energy) from the minimum before moving closer to it again.
This behavior affects the total number of iterations needed for convergence and it depends on the path taken by the optimizer in parameter space to navigate the energy function landscape and find the optimal point.

\subsubsection{VQE computations at different Depths}
Increasing the depth of a quantum circuit allows for access to a larger number of attainable states over which the expectation value of the Hamiltonian can be calculated.
To understand how the results change with respect to depth, we run the VQE algorithm for the minimal BMN model over various depths.
We vary the depth from 1 to 9, with an increment of 1.

\begin{figure}[htbp]
    \centering
    \includegraphics[width=0.5\textwidth]{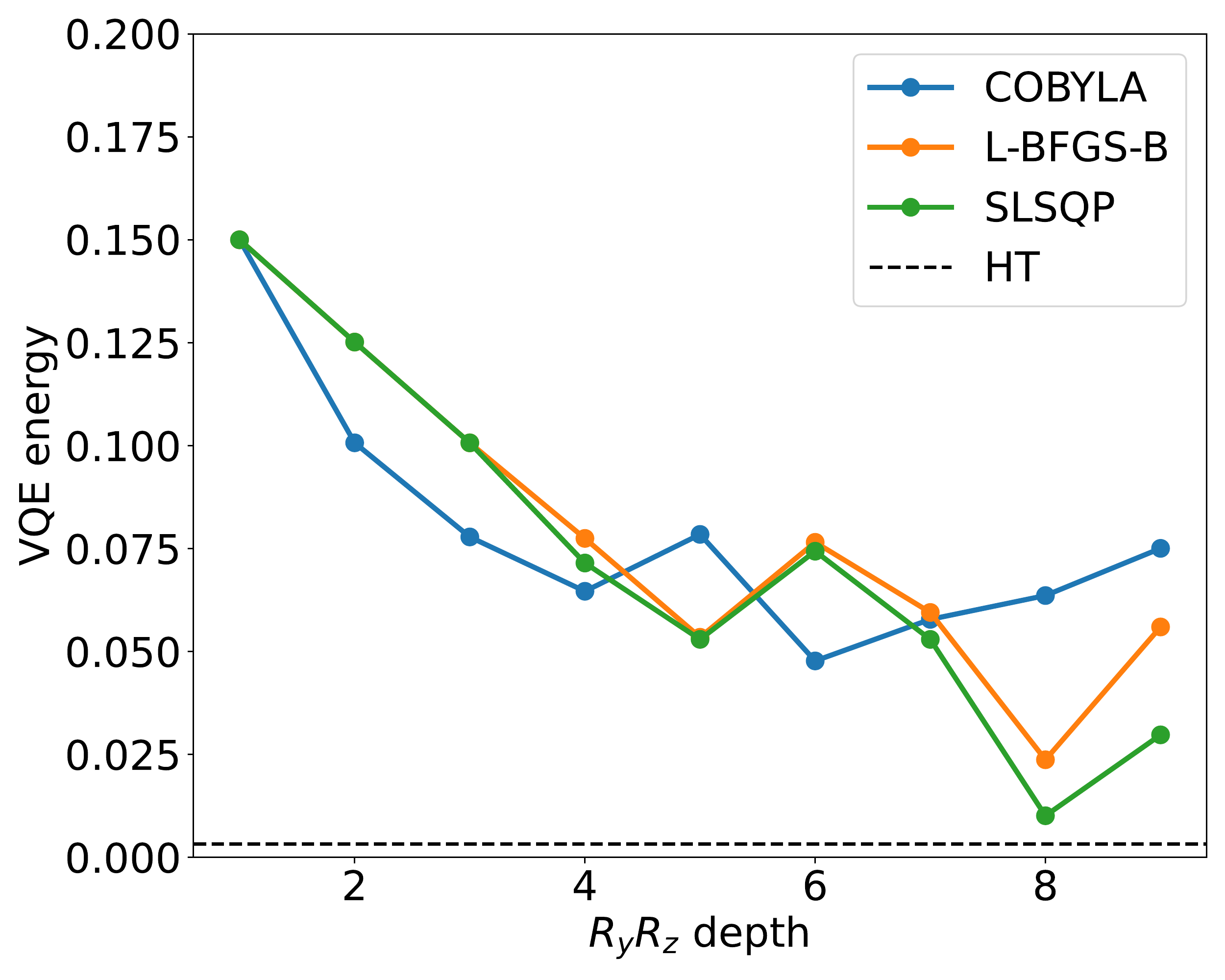}
    \caption{\label{fig:6bosons_3fermions_varyDepth}
    VQE results with respect to depth for the minimal BMN matrix model with SU(2) and $\lambda=0.2$.
    The VQE was run using various optimizers, for an $R_yR_z$ variational form of depth ranging from 1 to 9.
    The least upper bound across 100 different runs (with $10^4$ iterations) is reported.
    The exact ground state energy of the truncated Hamiltonian is represented by the dotted line.
    }
\end{figure}

\begin{table}[htbp]
    \centering
    \begin{tabular}{|l|l|l|l|}
    \hline
    Depth & COBYLA & L-BFGS-B & SLSQP \\ \hline
    1	& 0.150000	& 0.150000	& 0.150000 \\ \hline
    2	& 0.100689	& 0.125154	& 0.125158 \\ \hline
    3	& 0.077833	& 0.100683	& 0.100683 \\ \hline
    4	& 0.064637	& 0.077476	& 0.071500 \\ \hline
    5	& 0.078448	& 0.053471	& 0.052959 \\ \hline
    6	& 0.047723	& 0.076528	& 0.074366 \\ \hline
    7	& 0.057819	& 0.059495	& 0.052949 \\ \hline
    8	& 0.063603	& 0.023726	& 0.010126 \\ \hline
    9	& 0.075062	& 0.055973	& 0.029769 \\ \hline
    \end{tabular}
    \caption{\label{tab:6b3f_varyDepth_Table}
    VQE results for the minimal BMN model with SU(2) and $\lambda=0.2$.
    The VQE was run using various optimizers, for an $R_yR_z$ variational form of depth ranging from 1 to 9.
    The least upper bound across 100 different runs (with 10000 iterations) is reported.
    The exact value is $E_0 = 0.003287$.
    }
\end{table}

\begin{figure}[htbp]
    \centering
    \includegraphics[width=0.8\textwidth]{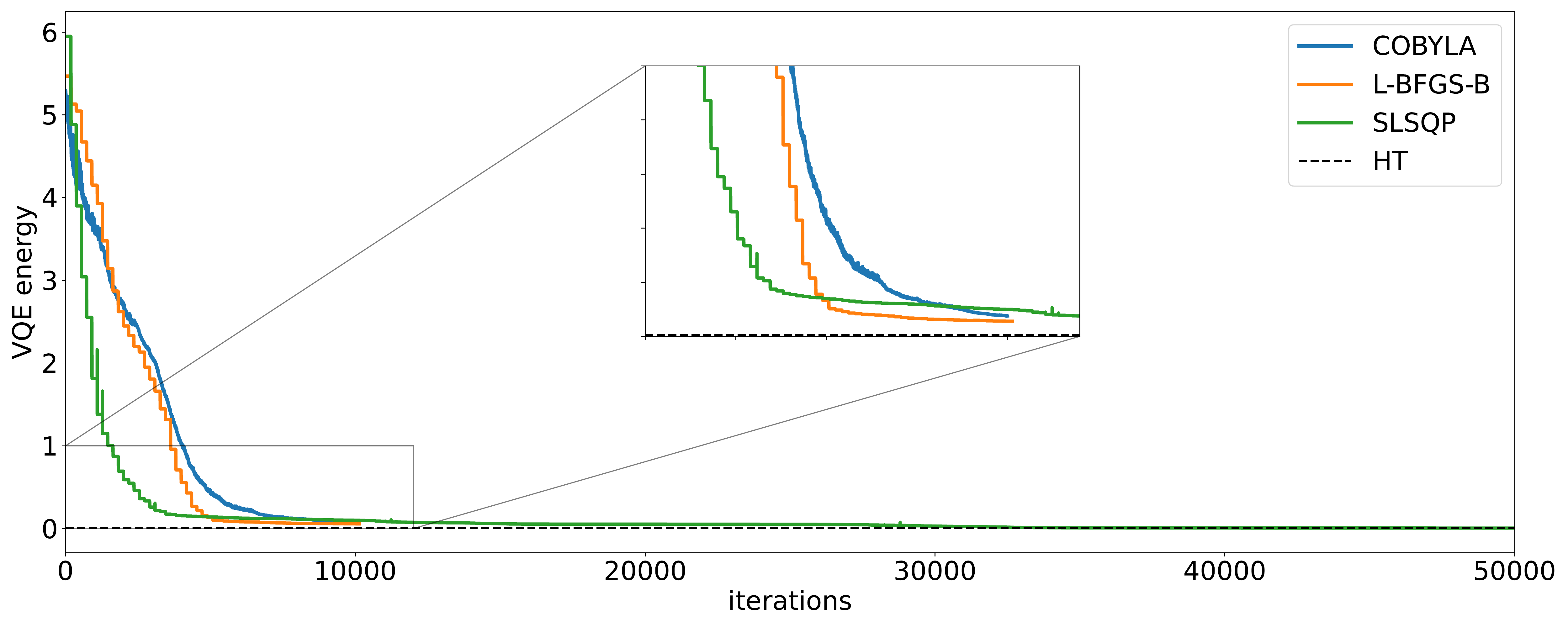}
    \caption{\label{fig:6bosons_3fermions_Depth9_convergence}
    Convergence of the VQE results for the minimal BMN matrix model with SU(2) and $\lambda=0.2$, where each boson is represented by a one qubit.
    The VQE results are the least upper bound from 100 runs of each optimizer, and an $R_yR_z$ variational form of depth 9 was used.
    The SLSQP optimizer was allowed to run for five times more iterations than the other two which were stopped at $10^4$ iterations.
    The exact energy of the truncated Hamiltonian (HT) is represented by the dotted line at $E_0 = 0.003287$.
    }
\end{figure}

The results are shown in Fig.~\ref{fig:6bosons_3fermions_varyDepth} and tabulated in Tab.~\ref{tab:6b3f_varyDepth_Table}.
For each depth, the initial point for each optimizer is the same, which ensures a fair comparison of each optimizer's ability to converge.
Additionally, we used an $R_yR_z$ variational form to further increase the number of attainable states by starting already with a larger number of parameters.
We use the least upper bound out of $N_r=100$ different runs.
As expected, a more expressive variational form allows us to get results which are increasingly closer to the exact diagonalization result for this truncated Hamiltonian.
For a depth of 8 the SLSQP optimizer gives the best result, a least upper bound of $E_{\rm var} = 0.010126$ to be compared with the exact result of $E_0 = 0.003287$.
We also note that the algorithms take more iterations to reach convergence when more parameters are included in the variational quantum circuits representing the wave function ansatz.
This is shown in Fig.~\ref{fig:6bosons_3fermions_Depth9_convergence} for an $R_yR_z$ variational form of depth 9.
These results imply that increasing the number of maximum iterations above $10^4$ for some of the solvers might allow them to find better optimal points.
For example, if we let the SLSQP optimizer run until convergence with no limit on the number of iterations at depth 9, it will get to a least upper bound of $E_{\rm var} =0.004755$, but it would need five times more iterations $\approx 5 \times 10^4$.

\subsubsection{VQE for different coupling constants}

\begin{table}[htbp]
    \centering
    \begin{tabular}{|l||l|l|l|l|l||l|}
    \hline
    $\lambda$ & COBYLA & L-BFGS-B & SLSQP & NELDER-MEAD & Best & HT (exact) \\ \hline
    0.5  & 0.088492 & 0.139702 & 0.134517 & 0.406003 & 0.02744 & 0.01690 \\ \hline
    1.0  & 0.135800 & 0.219268 & 0.308781 & 0.752459 & 0.07900 & 0.04829 \\ \hline
    2.0  & 0.387977 & 0.622704 & 0.522396 & 1.271939 & 0.17688 & 0.08385 \\ \hline
    \end{tabular}
    \caption{\label{tab:6Boson3fermion_VaryCoupling}
    VQE results for the minimal BMN matrix model, where each boson and fermion are represented by one qubit, at different couplings.
    The $R_yR_z$ ansatz with depth 5 is used and the least upper bound across $N_r=100$ runs is reported (with $10^4$ iterations).
    In the `Best' column we report the least upper bound we obtained from a depth 9 $R_yR_z$ ansatz using the best optimizer for ten times more iterations.
    The `HT' column is the exact energy of the truncated Hamiltonian.
    }
\end{table}

\begin{figure}[htbp]
    \centering
    \begin{tabular}{ccc}
        \includegraphics[width=0.3\textwidth]{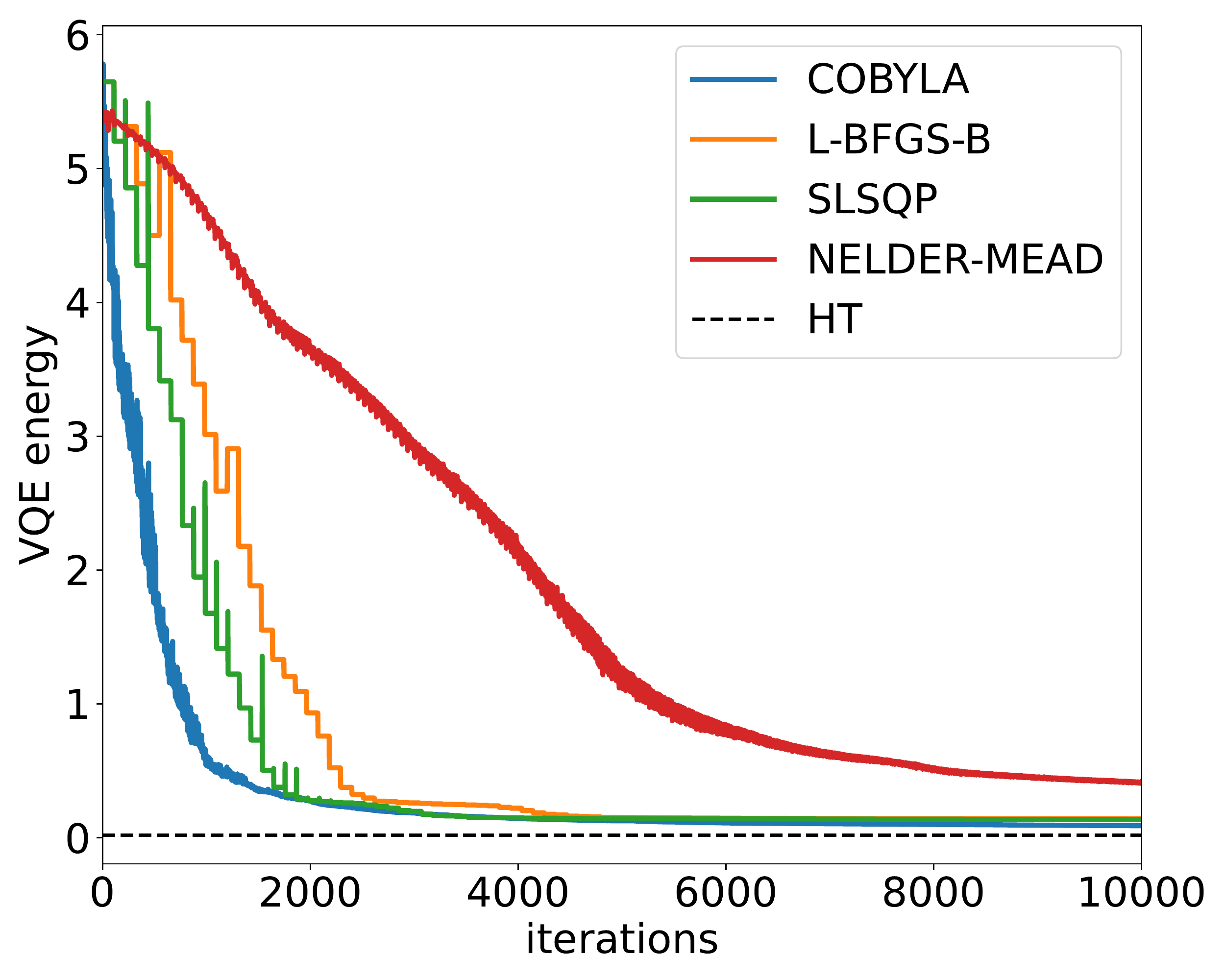} &
        \includegraphics[width=0.3\textwidth]{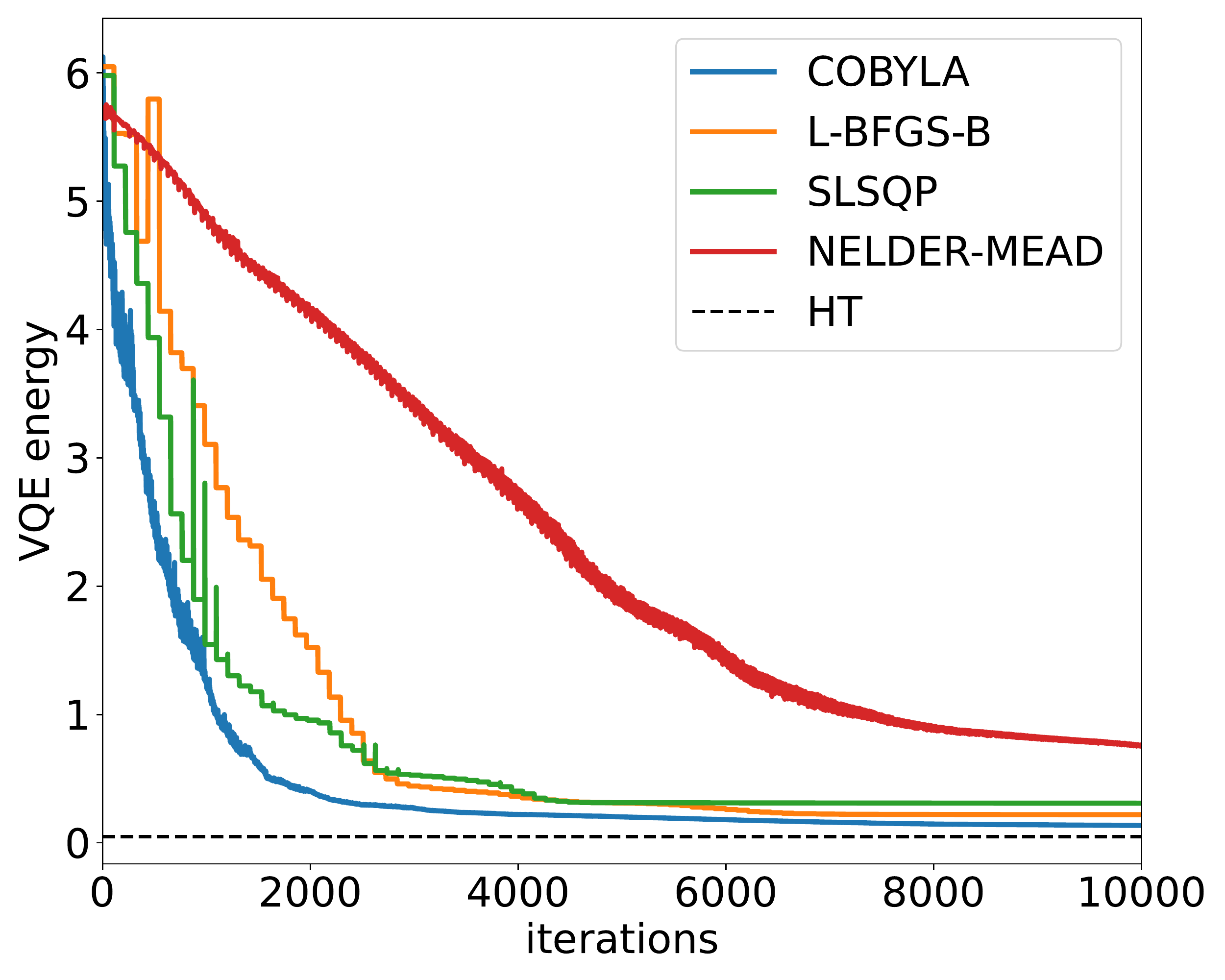} &
        \includegraphics[width=0.3\textwidth]{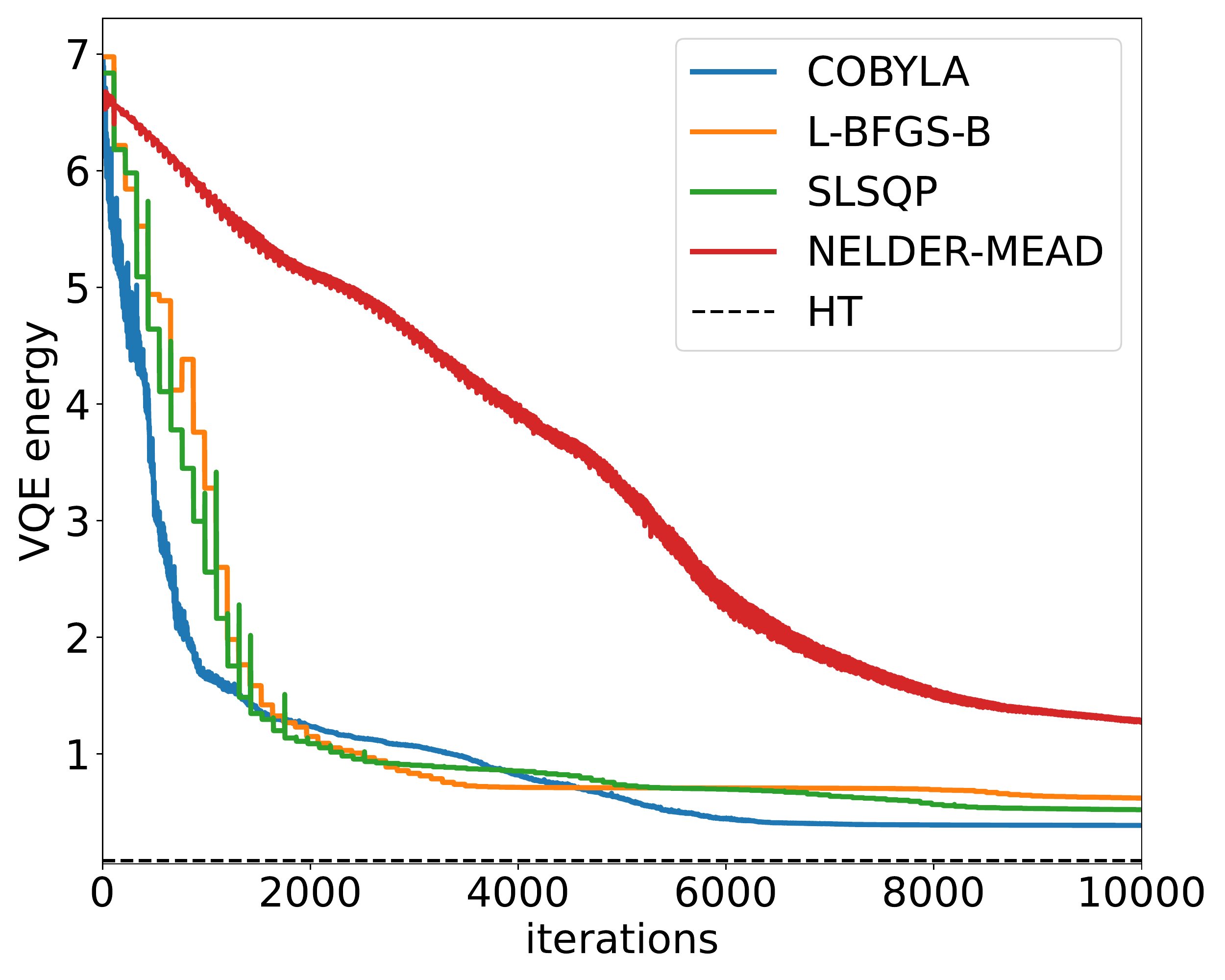} \\
    \end{tabular}
    \caption{\label{fig:6Boson3fermion_VaryCoupling_ConvergencePlots}
    Convergence plots of the results of the VQE computation for the minimal BMN matrix model with SU(2) for coupling (left) $\lambda = 0.5$ , (middle) $\lambda = 1.0$, and (right) $\lambda =2.0$.
    The $R_yR_z$ ansatz with depth 5 is used and the best of $N_r=100$ runs is displayed up to $10^4$ iterations.
    }
\end{figure}

As was done with the SU(2) bosonic model, we test the accuracy of the VQE results with respect to coupling for the minimal BMN matrix model by running the VQE for different coupling constants.
We used an $R_yR_z$ ansatz with depth 5 to expand the number of attainable states (by including more variational parameters) since we have seen this to be advantageous in the previous section at $\lambda=0.2$.
The VQE results at different coupling constants are tabulated in Tab.~\ref{tab:6Boson3fermion_VaryCoupling}.
The accompanying optimizer convergence plots are shown in Fig.~\ref{fig:6Boson3fermion_VaryCoupling_ConvergencePlots}.
It is evident from the convergence plots that the NELDER-MEAD optimizers needs more iterations to converge.

\section{Deep learning}\label{sec:deep-learning}
\subsection{Overview of variational quantum Monte Carlo}
Next we discuss the deep learning approach that was applied to the minimal BMN matrix model in Ref.~\cite{Han:2019wue}.
Specifically, neural networks are used as the wave function ansatz in the variational quantum Monte Carlo method for the ground state.
The algorithm can be roughly decomposed into four steps:

\begin{enumerate}
\item[{\it (i)}] Parametrize a quantum state $|\psi_\theta\rangle$ by a set of parameters $\theta$.
Specifically, we construct a function which is interpreted as the wave function $\psi_\theta(X)=\langle X|\psi_\theta\rangle$ by using a neural network.
The explicit form of $\psi_\theta(X)$ will be given later in Sec.~\ref{sec:neural-ansatz}.

\item[{\it (ii)}] Estimate the energy from Monte Carlo samples of the wave function.
The energy of the state $|\psi_\theta\rangle$ is defined by
\begin{align}\label{eq:variational-mc}
E_\theta \equiv
\langle\psi_\theta|\hat{H}|\psi_\theta\rangle
=
\int dX |\psi_\theta(X)|^2\cdot\frac{\langle X |\hat{H} |\psi_\theta\rangle}{\psi_\theta(X)} = \mathbf{E}_{X \sim |\psi_\theta|^2}[\epsilon_\theta(X)]
\ .
\end{align}

In the last expression, the samples of matrices are drawn from the probability distribution $|\psi_\theta(X)|^2$, $\epsilon_\theta(X)$ is defined as $\langle X | \hat{H} | \psi_\theta \rangle / \psi_\theta(X)$ and $E_\theta$ is estimated as the mean of $\epsilon_\theta(X)$ from these samples.
The generative nature of our wave function ansatz, to be discussed in detail later, allows for both efficient sampling from $|\psi_\theta|^2$ and evaluation of $\epsilon_\theta$.

\item[{\it (iii)}] Compute the gradient of the energy with respect to model parameters $\theta$.
Note that the sampling distribution depends on $\theta$ as well, and hence
\begin{align}\label{eq:variational-grad}
\nabla_\theta E_\theta = \mathbf{E}_{X \sim |\psi_\theta|^2}[\nabla_\theta \epsilon_\theta(X)] + \mathbf{E}_{X \sim |\psi_\theta|^2}[\epsilon_\theta(X) \nabla_\theta \ln |\psi_\theta|^2] \ .
\end{align}

\item[{\it (iv)}] Update the model parameters via the gradient descent as
\begin{align}
\theta' = \theta - \beta \nabla_\theta E_\theta \ ,
\end{align}
where the step size $\beta$ is the ``learning rate''.
The energy in Eq.~\eqref{eq:variational-mc} is then minimized within the family of wave functions parametrized by $\theta$.
The (local) minimum found is then a variational upper bound for the ground state energy.
\end{enumerate}

\subsection{Form of the wave function ansatz}\label{sec:neural-ansatz}
Firstly we focus on the wave function form in the SU($N$) bosonic matrix quantum mechanics.
The quantum wave function is a complex function $\psi(X) = |\psi(X)| e^{i \theta(X)}$ of Hermitian matrices $X$.
The wave function norm $|\psi(X)|$ and the phase $\theta(X)$ are modeled separately.

The wave function norm is parametrized as an autoregressive flow model.
Let $|\psi(X)| = \sqrt{p_\theta(X)}$, and $p_\theta(X)$, the wave function probability distribution, takes the following form:
\begin{align}\label{eq:ansatz}
    p_\theta(X) = p(x_1; F^0_\theta) p(x_2; F^1_\theta(x_1)) p(x_3; F^2_\theta(x_1, x_2)) \ldots,
\end{align}
where $p(x;F)$ is a function of $x$ specified by parameters $F$.
Here $x_i$ are the components of the Hermitian matrices expanded onto some basis.
For the SU($N$) bosonic matrix model, $i$ runs from 1 to $N^2 - 1$ times the number of matrices.
The $x_i$'s are ordered in some arbitrary way in this ansatz, and both choices of the ordering and of the basis of the $\mathfrak{su}(N)$ algebra used for expansion do not significantly affect the final results.
As an example, in Eq.~\eqref{eq:ansatz}, $p$'s on the right side could be normal distributions, which are parametrized by their locations and scales.
The location and the scale parameters of the coordinate $x_i$ depend on $x_j$ with $j < i$.
The dependence is parametrized by function $F^i_\theta$ which will be discussed later.
This autoregressive ordering of coordinates allows for easy sampling from $p_\theta(X)$, as we can firstly sample $x_1$ independent of any other variables, then sample $x_2$ dependent on $x_1$, $x_3$ dependent on $x_1$ and $x_2$ and so on.

The final component in Eq.~\eqref{eq:ansatz} is the form of $F^i_\theta$.
If the distributions $p$ in Eq.~\eqref{eq:ansatz} are normal with two parameters, $F^i_\theta$ is then a vector function from $i$ to two real numbers.
For each $i$, $F^i_\theta$ is given by a fully-connected neural network ($\circ$ denotes a concatenation of functions):
\begin{align}\label{eq:fc}
F_\theta^i = A_\theta^{i, m}\circ\tanh\circ A_\theta^{i, m-1}\circ\tanh\circ
\cdots \circ A_\theta^{i, 2}\circ\tanh\circ A_\theta^{i,1} \ ,
\end{align}
where for $a = 1, 2, \ldots, m$,
\begin{align}
A_\theta^{i, a}(\vec{x}) = M_\theta^{i, a}\vec{x}+\vec{b}_\theta^{i, a}
\end{align}
is an affine transformation.
The weight matrices $M_\theta^{i, a}$ and the bias vectors $\vec{b}_\theta^{i, a}$ are trainable parameters, not shared among $i$'s.
The input and output dimensions are determined ($i$ and $2$ in the Gaussian example), and dimensions of the intermediate layers are hyperparameters.

Of course, if the distribution $p$ in Eq.~\eqref{eq:ansatz} is only Gaussian, the ansatz cannot describe non-Gaussian conditional distributions, although it is still much more flexible than a free ansatz.
For more flexibility, in this work we have used the neural autoregressive flow (NAF)~\cite{huang2018neural}, which is an extension of the simple Gaussian example discussed.
We use $m = 3$ for the fully-connected network in NAF and denote the ratio of the intermediate (hidden) dimension to the input dimension as $\alpha$.

The phase of the wave function $\theta(X)$ is parametrized by another fully connected neural network as in Eq.~\eqref{eq:fc}.
The input dimension of the network is $N^2 - 1$ times the number of matrices.
The output dimension is one, because the phase is a scalar number.
For ground states of the bosonic models, the phase is not strictly necessary.

For the minimal BMN model, the quantum wave function $\psi(X, \xi)$ depends on fermionic matrices as well.
We expand fermionic matrices onto some SU($N$) basis and choose the fermion number basis as the basis for quantum states.
Then the coordinates $(X, \xi)$ for the wave function contain $2 (N^2 - 1)$ real numbers and $(N^2 - 1)$ binaries.
The wave function is parameterized as $\sqrt{p(X)} (f_\xi(X) + i g_\xi(X))$, where $p(X)$ is given by the NAF as previously discussed, and $f_\xi$ and $g_\xi$ are real functions given by fully-connected neural networks as in (\ref{eq:fc}).
The weights and biases in $f_\xi$ and $g_\xi$ depend on the binary vector $\xi$, via additional fully-connected networks as well.

\subsection{Gauge constraints}
Physically we may hope to keep only SU($N$) gauge invariant states.
In Ref.~\cite{Han:2019wue} this is achieved by gauge fixing.
Here we propose two more methods of imposing gauge invariance (cf. Sec.~\ref{sec:singlet-constraint-QC}).

The first alternative is to add a gauge Casimir penalty to the Hamiltonian:
\begin{align}
\hat{H}' = \hat{H} + c\sum_\alpha\hat{G}_\alpha^2
\ .
\end{align}
Here $c > 0$ is a hyperparameter.
We have introduced this modified Hamiltonian in Sec.~\ref{sec:singlet-constraint-QC} where we used exact diagonalization to find the eigenstates.
The penalty term favors gauge singlet states and imposes strict gauge invariance as $c \to \infty$.
However, in practice as our ansatz is not strictly gauge invariant, very large $c$ will interfere with minimization of the energy.

The second approach is to project the wave function onto the singlet sector.
Let $\psi(X)$ be a variational wave function, not necessarily gauge invariant.
The projection onto the singlet sector can be written as an average over SU($N$) gauge transformations:
\begin{align}
    \hat{P} |\psi\rangle = \int d U \, \hat{U} |\psi\rangle,
\end{align}
where the integration is done with the Haar measure.
For any gauge invariant observable $\hat{O}$, its expectation value in the projected (unnormalized) state is
\begin{align}\label{eq:singlet_obs}
    \langle \hat{O} \rangle_{\text{singlet}} = \frac{\langle \psi | \hat{P} \hat{O} | \psi \rangle}{\langle \psi | \hat{P} | \psi \rangle}.
\end{align}
Note that $\hat{U} |X\rangle = |U X U^\dagger \rangle$, so the numerator is
\begin{align}
    \langle \psi | \hat{P} \hat{O} | \psi \rangle &= \int dX\, \langle \psi | \hat{P} | X \rangle \langle X | \hat{O} |\psi \rangle = \int dU dX\, \psi^*(U X U^\dagger) \langle X | \hat{O} |\psi \rangle \nonumber \\
    &= \mathbf{E}_{U, X \sim |\psi|^2}\left[\frac{\langle X | \hat{O} |\psi \rangle}{\psi(X)} \frac{\psi^*(U X U^\dagger)}{\psi^*(X)}\right].
\end{align}
The denominator is simply $\hat{O} = 1$.
Hence the singlet expectation values can be estimated with Monte Carlo samples of $U$ and $X$ as well.

However, for ground states of bosonic models it is not necessary to impose gauge invariance, as the ground state is always a singlet.
This is numerically verified in Tab.~\ref{tab:ml_with_gauge_constraint}, where the gauge Casimir in the variational ground states is less than $10^{-3}$ even at $c = 0$.
With $c = 10$ the expectation value of $G^2$ is less than $10^{-4}$, and the variational ground state energy is not significantly increased.

\begin{table}[ht]
    \centering
    \begin{tabular}{|c||c c c c|}
    \hline
    &   &\qquad\quad  $c = 0$ & &\\
    \hline
    \hline
    $\lambda$ & $0.2$ & $0.5$ & $1.0$ & $2.0$  \\
    \hline
    $E_{0,\text{var}}$ & 3.137(2) & 3.299(2) & 3.518(2) & 3.856(3) \\ \hline
    $G^2_{\text{var}}$ & 0.0028(4) & 0.0059(6) & 0.0062(7) & 0.0122(9) \\ \hline
    $E_{0,\text{singlet}}$ & 3.135(2) & 3.297(2) & 3.520(2) & 3.859(3) \\ \hline
    \end{tabular}
    \\
    \vspace{5mm}
    \begin{tabular}{|c ||c c c c|}
    \hline
    & & \qquad\quad $c = 10$ & & \\
    \hline
    \hline
    $\lambda$ & $0.2$ & $0.5$ & $1.0$ & $2.0$  \\
    \hline
    $E_{0,\text{var}}$ & 3.137(2) & 3.309(2) & 3.545(3) & 3.912(3) \\ \hline
    $G^2_{\text{var}}$ & 0.00011(8) & 0.00019(7) & 0.00022(11) & 0.00021(8) \\ \hline
    $E_{0,\text{singlet}}$ & 3.139(2) & 3.307(2) & 3.544(3) & 3.908(3) \\ \hline
    \end{tabular}
    \caption{\label{tab:ml_with_gauge_constraint}
    Neural variational quantum Monte Carlo ground state energy and gauge casimir for the bosonic SU(2) matrix quantum mechanics, at coupling $\lambda = g^2 N = 0.2, 0.5, 1.0, 2.0$ and gauge penalty $c = 0, 10$ (cf. Eq.~\eqref{eq:deformed-Hamiltonian}).
    Here $E_{0, \text{var}} = \langle \hat{H} \rangle$ in the variational state for $N = 2$, $\hat{H}$ in~\eqref{eq:bosonic_Hamiltonian} and $E_{0,\text{singlet}} = \langle \hat{H} \rangle_{\text{singlet}}$ as in Eq.~\eqref{eq:singlet_obs}.
    The ratio of the hidden dimension and the input dimension $\alpha = 20$.
    }
\end{table}

\subsection{Ground state energy and observables}
\subsubsection{Bosonic matrix model}
We start with the ground state energy in the SU(2) bosonic two-matrix quantum mechanics.
The variational energies will be compared to Monte Carlo results and they agree within numerical uncertainties.
Accuracy of the variational energies can also be assessed by increasing the number of hidden units in the neural network ansatz.
With more parameters we expect that the ansatz is more flexible and accurate, and indeed convergence to exact (Hamiltonian truncation) results is observed in Tab.~\ref{tab:ml_bosonic_su2}.

\begin{table}[ht]
    \centering
    \begin{tabular}{|c||c|c|c|c|c|c|c|}
    \hline
     $\alpha$ & $1$ & $2$ & $5$ & $10$ & $20$ & $50$ & HT (exact) \\
    \hline
    \hline
    $\lambda = 0.2$ & 3.137(2) & 3.137(2) & 3.140(2) & 3.138(2) & 3.137(2) & 3.135(2) & 3.134 \\ \hline
    $\lambda = 0.5$ & 3.313(2) & 3.312(2) & 3.308(2) & 3.307(2) & 3.302(2) & 3.305(2) & 3.297\\ \hline
    $\lambda = 1.0$ & 3.544(3) & 3.544(2) & 3.541(3) & 3.528(2) & 3.519(2) & 3.520(2) & 3.516\\ \hline
    $\lambda = 2.0$ & 3.914(3) & 3.910(3) & 3.892(3) & 3.872(3) & 3.857(3) & 3.859(3) & 3.854\\ \hline
    \end{tabular}
    \caption{\label{tab:ml_bosonic_su2} Neural variational quantum Monte Carlo ground state energy for the bosonic SU(2) matrix quantum mechanics, at coupling $\lambda = g^2 N = 0.2$, 0.5, 1.0, and 2.0.
    The neural autoregressive flow ansatz contains a fully-connected network with an input and an output layer of $2 (N^2 - 1) = 6$ units, and a hidden layer of $\alpha \times 2 (N^2 - 1)$ units.
    Hence the $\alpha$ here is the ratio of the hidden dimension and the input dimension.
    Larger $\alpha$ means more parameters and a more flexible ansatz.
    The exact result of the Hamiltonian truncation (HT) at high truncation level $\Lambda=11$ is reported in the last column.
    }
\end{table}

\begin{table}[ht]
    \centering
    \begin{tabular}{|c||c|c|c|c|c|c|c|}
    \hline
     $\alpha$ & $1$ & $2$ & $5$ & $10$ & $20$ & $50$ & MC \\
    \hline
    \hline
    $\lambda = 0.5$ & 8.833(10) & 8.828(7) & 8.829(7) & 8.835(7) & 8.824(7) & 8.826(7) & 8.84(4) \\ \hline
    $\lambda = 1.0$ & 9.459(7) & 9.447(7) & 9.468(8) & 9.466(10) & 9.432(7) & 9.440(8) & 9.38(4) \\ \hline
    $\lambda = 2.0$ & 10.447(12) & 10.448(9) & 10.441(9) & 10.457(12) & 10.426(8) & 10.397(13) & 10.24(4) \\ \hline
    \end{tabular}
    \caption{\label{tab:ml_bosonic_su3}
    Neural variational quantum Monte Carlo ground state energy for the bosonic SU(3) matrix quantum mechanics, at coupling $\lambda = g^2 N = 0.5$, 1.0, and 2.0.
    The neural autoregressive flow ansatz contains a fully-connected network with an input and an output layer of $2 (N^2 - 1) = 16$ units, and a hidden layer of $\alpha \times 2 (N^2 - 1)$ units.
    Hence the $\alpha$ here is the ratio of the hidden dimension and the input dimension.
    Larger $\alpha$ means more parameters and a more flexible ansatz.
    The results of the lattice Monte Carlo (MC) simulations are reported in the last column.
    }
\end{table}

No significant improvement is observed from $\alpha = 20$ to $\alpha = 50$ in Tab.~\ref{tab:ml_bosonic_su2}.
Similarly for the SU(3) bosonic matrix model we summarize the results in Tab.~\ref{tab:ml_bosonic_su3}

\subsubsection{SU(2) minimal BMN}
In the supersymmetric model the ground state energy should be zero, and indeed we see vanishing variational energies.
In Tab.~\ref{tab:ml_susy_su2} for $N = 2$ and $\lambda = \mu = 1$, the variational energy converges to zero as we increase $\alpha$.
Other observables in the variational state are accurate as well.
Results at $\alpha = 20$ for other couplings are summarized in Tab.~\ref{tab:ml_susy_su2_couplings}.
Note that a natural energy scale of the theory is of order one.
For example, the ground-state energy of bosonic and fermionic oscillators are $\pm(N^2-1)=\pm 3$, and they cancel out.
The results at $\alpha = 20$ means the cancellation is reproduced up to a-percent-order error.

\begin{table}[ht]
    \centering
    \begin{tabular}{|c||c|c|c|c|c|c|c|}
    \hline
     $\alpha$ & $1$ & $2$ & $5$ & $10$ & $20$ & $50$ & HT (exact)\\
    \hline
    \hline
    $H$   & 0.058(6) & 0.053(6) & 0.041(6) & 0.031(6) & 0.014(6) & 0.005(6) & 0.000 \\ \hline
    $G^2$ & 0.007(8) & $-0.008(8)$ & 0.014(8) & 0.007(9) & 0.022(9) & 0.012(9) & 0.000 \\ \hline
    $M$   & $-0.0003(3)$ & $-0.0004(3)$ & $-0.0001(4)$ & 0.0001(4) & $-0.0003(5)$ & $-0.0001(4)$ & 0.0000 \\ \hline
    $F$   & 0.1844(6) & 0.1833(6) & 0.1895(6) & 0.1922(6) & 0.1946(7) & 0.1935(7) & 0.2034 \\ \hline
    \end{tabular}
    \caption{\label{tab:ml_susy_su2}
    Neural variational quantum Monte Carlo ground state observables for the SU(2) minimal BMN model, at coupling $\lambda = g^2 N = 1.0$ and $\mu = 1$.
    The observables are the Hamiltonian $H$, the gauge Casimir $G^2$, the SO(2) angular momentum $M$, and the fermion number $F$.
    Larger $\alpha$ means more parameters and a more flexible ansatz.
    The last column contains the exact results from the the Hamiltonian truncation (HT) at $\Lambda=8$.
    }
\end{table}

\begin{table}[ht]
    \centering
    \begin{tabular}{|c||c|c|c|}
    \hline
    & $\lambda = 0.5$ & $\lambda = 1.0$ & $\lambda = 2.0$ \\
    \hline
    \hline
    $H$ & 0.009(5) & 0.014(6) & 0.034(7) \\ \hline
    $G^2$ & 0.010(6) & 0.022(9) & 0.038(14) \\ \hline
    $M$ & $-0.0002(3)$ & $-0.0003(5)$ & 0.0006(7) \\ \hline
    $F$ & 0.1224(4) & 0.1946(7) & 0.2729(9) \\ \hline
    \end{tabular}
    \caption{\label{tab:ml_susy_su2_couplings}
    Neural variational quantum Monte Carlo ground state observables for the SU(2) minimal BMN, at coupling $\lambda = g^2 N = 0.5$, 1.0, and 2.0, using $\mu = 1$.}
\end{table}

\section{Monte Carlo Simulation on a Euclidean Lattice}\label{sec:lattice-simulation}
In this section, we employ the lattice Monte Carlo method to study the bosonic matrix model and compare its results to the methods described in the previous sections.
For the minimal BMN model, since several exact relations are known about the ground state due to supersymmetry, we do not repeat the study using lattice Monte Carlo simulations.
For an introductory review, see e.g. Ref.~\cite{Hanada:2018fnp}.

We use the following Euclidean action on the circle:
\begin{align}
S = N\int_0^\beta dt {\rm Tr}\left(
\frac{1}{2}(D_tX_I)^2
+
\frac{m^2}{2}X_I^2
-
\frac{\lambda}{4}[X_I,X_J]^2
\right) \ ,
\end{align}
where $\lambda=g^2N$.
The circumference of the temporal circle, $\beta$, is the inverse temperature $\beta=1/T$ of the system.
The matrices $X_I$ are traceless Hermitian.

The lattice regularization we use is the tree-level improved action which is essentially the same as the one used in Ref.~\cite{Berkowitz:2018qhn}.
The discretized lattice action can be written as
\begin{align}
S_{\rm lattice} = Na\sum_{t=1}^{n_t}{\rm Tr}
\left[
\frac{1}{2}(D_tX)_{I,t}^2
+
\frac{m^2}{2}X_{I,t}^2
-
\frac{\lambda}{4}[X_{I,t},X_{J,t}]^2
\right] \ ,
\end{align}
where $n_t$ is the number of lattice points used for discretizing the circle and $a=\beta/n_t$ is the lattice spacing.
The lattice spacing has to be removed because it introduces systematic effects and we are interested in the so-called ``continuum limit'' system, when $a \rightarrow 0$.

The covariant derivative $D_t$ is defined by
\begin{align}
(D_tX)_{I,t} = \frac{1}{a}
\left[
-\frac{1}{2}U_t U_{t+a}X_{I,t+2a}U_{t+a}^\dagger U_{t}^\dagger
+
2U_t X_{I,t+a}U_{t}^\dagger
-
\frac{3}{2}X_{I,t}
\right] \ .
\end{align}
Here we adopt periodic boundary conditions for the matrices, $X_{I,n_t+1}=X_{I,1}$.
If we set the unitary link variable $U_t$ to be the identity, the ungauged version of the model can be studied as well~\cite{Berkowitz:2018qhn}.
Note that $\beta=an_t$.

In lattice Monte Carlo, a sequence of lattice configurations
$\{X^{(1)},U^{(1)}\}\to \{X^{(2)},U^{(2)}\}
\to\cdots\to
\{X^{(k)},U^{(k)}\}\to\cdots$
is generated, such that their probability distribution converges to $e^{-S_{\rm lattice}[X,U]}$.
Then the average over lattice configurations converges to the expectation value in the Euclidean path integral:
\begin{align}\label{eq:path-integral}
\langle f\rangle \equiv
\frac{\int dX dU f(X,U)e^{-S_{\rm lattice}[X,U]}}{\int dX dU e^{-S_{\rm lattice}[X,U]}}
=
\lim_{K\to\infty} \frac{1}{K}\sum_{k=1}^K f(X^{(k)},U^{(k)}) \ .
\end{align}
To generate configurations with the correct probability distribution we use the Hybrid Monte Carlo (HMC) algorithm~\cite{Duane:1987de}, which allows us to sample configurations along molecular dynamics trajectories in a fictitious Monte Carlo time.
The configurations sampled along these HMC trajectories are autocorrelated because of the dynamics used to select the next step in a trajectory and successive configurations must be discarded to reduce such autocorrelations.

To determine the energy of the system in the ground state using this path integral formulation, it is convenient to use the virial theorem, which relates the kinetic ($K$) and potential terms ($V$) in the Hamiltonian:
\begin{align}
\langle K\rangle = \left\langle \frac{1}{2}\sum_i x_i\frac{\partial V}{\partial x_i} \right \rangle \ .
\end{align}

Hence the total energy $E$ at temperature $T=\beta^{-1}$ is evaluated as
\begin{align}
E &= \left\langle \frac{1}{\beta} \int_0^\beta dt \left(V + \frac{1}{2}\sum_i x_i\frac{\partial V}{\partial x_i} \right)
\right\rangle
\nonumber\\
&= \left\langle \frac{N}{\beta} \int_0^\beta dt \left( m^2X_{I}^2- \frac{3\lambda}{4}[X_{I},X_{J}]^2 \right)
\right\rangle \ .
\end{align}

On the discretized lattice circle, this becomes
\begin{align}
E_{\rm lattice} = \left\langle \frac{N}{n_t}\sum_{t=1}^{n_t} \left( m^2X_{I,t}^2 - \frac{3\lambda}{4}[X_{I,t},X_{J,t}]^2
\right) \right\rangle \ .
\end{align}

We compute the energy for the bosonic 2-matrix model with gauge group SU(2) and SU(3) at three values of the coupling $\lambda=0.5$, 1.0, and 2.0, with fixed $m^2 = 1$.
The simulations are performed at temperatures $0.025\le T\le 0.4$ and lattice size $n_t$ ranging from $16$ to $192$.
For each parameter set we simulate about one million HMC trajectories of length between 0.5 and 1 (a length of one corresponds to one molecular dynamics time unit MDTU).
We discard 1000 MDTUs at the beginning of each trajectory to remove thermalization (burn-in) effects and have configurations that are from the correct distribution obtained when the HMC algorithm converges.
We also save configuration every 10 to 50 MDTUs to reduce autocorrelation times.
\emph{A posteriori} we check that the integrated autocorrelation time of the energy is always around unity.

\begin{figure}[htbp]
  \begin{center}
   \includegraphics[width=0.8\textwidth]{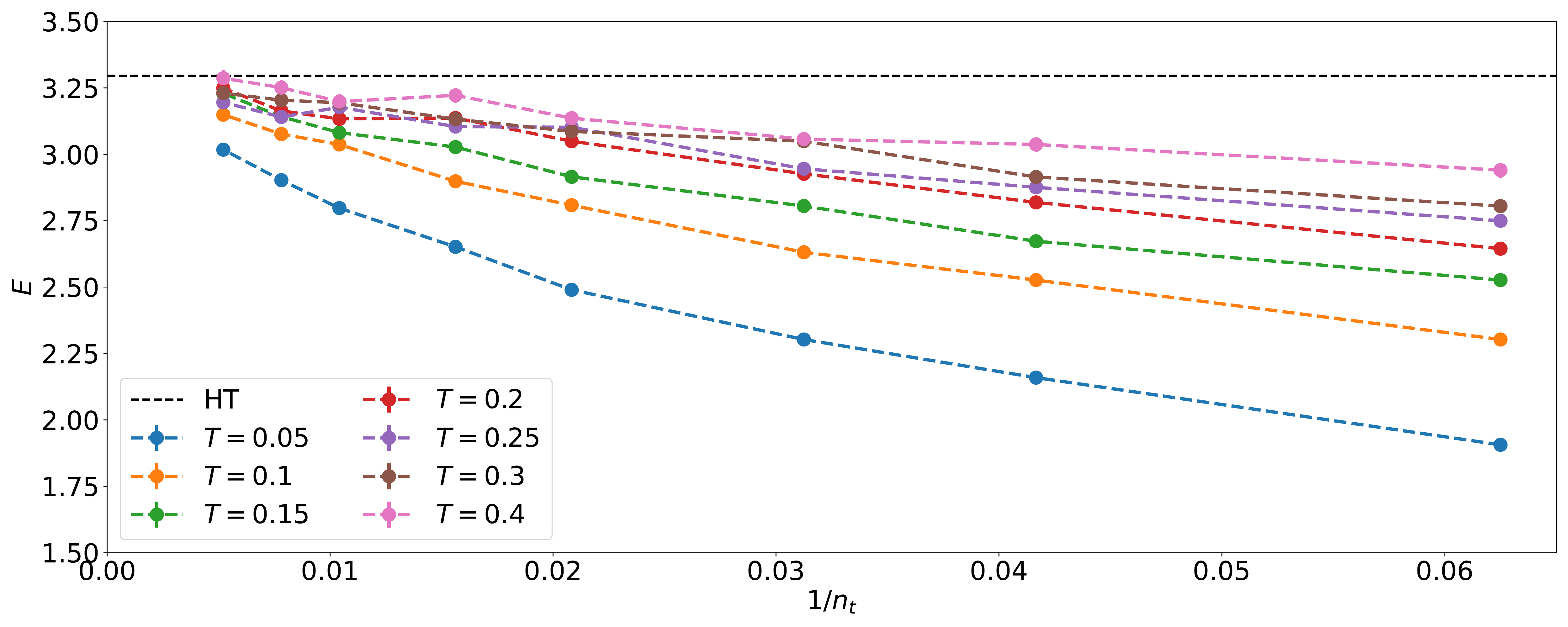}
  \end{center}
  \caption{\label{fig:lattice-SU2-energy_L}
  The energy of the SU(2) bosonic model at $m^2=1$ and $\lambda = g^2N = 0.5$ for various temperatures $T$ and lattice sizes $n_t$.
  The black dashed line is the exact result from the Hamiltonian truncation (HT) approach at cutoff $\Lambda=14$, $E_0=3.297$.
  Larger temperatures approach it from below.
  }
\end{figure}

\begin{figure}[htbp]
  \begin{center}
   \includegraphics[width=0.8\textwidth]{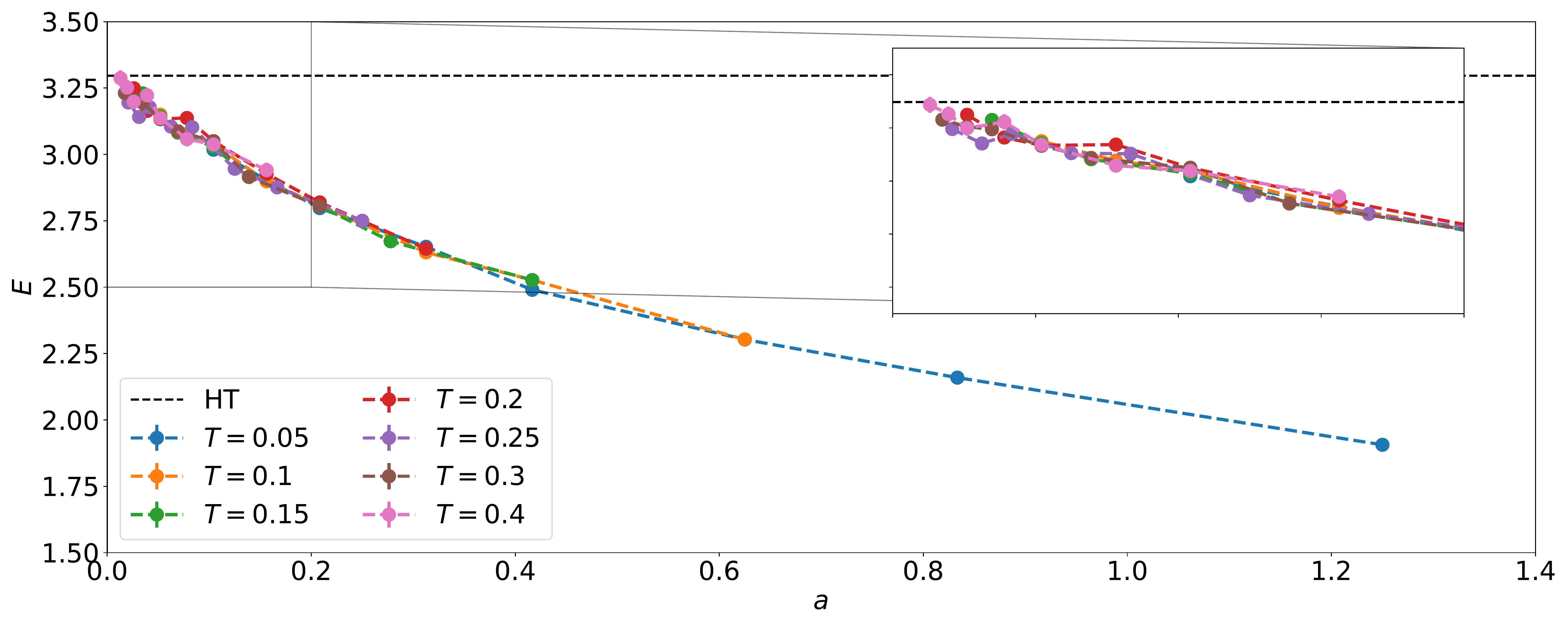}
  \end{center}
  \caption{\label{fig:lattice-SU2-energy}
  The energy of the SU(2) bosonic model at $m^2=1$ and $\lambda = g^2N = 0.5$ for various temperatures $T$ and lattice sizes $n_t$.
  The horizontal axis is the lattice spacing $a=1/(Tn_t)$.
  The black dashed line is the exact result from the Hamiltonian truncation (HT) approach at cutoff $\Lambda=14$, $E_0=3.297$.
  }
\end{figure}

In Fig.~\ref{fig:lattice-SU2-energy_L}, we show the energy of the SU(2) theory with $m^2=1$ and $\lambda=g^2N=0.5$ at several values of temperature between $T=0.4$ and $T=0.05$.
The horizontal axis is the inverse lattice size $1/n_t$.
In Fig.~\ref{fig:lattice-SU2-energy} we change the horizontal axis to be proportional to the lattice spacing $a=1/(Tn_t)$.
In this representation, the data does not show a significant $T$-dependence and, at fixed lattice spacing $a$, different temperatures mostly agree within statistical uncertainties.
The black horizontal line is the value coming from the Hamiltonian truncation (HT) of Sec.~\ref{sec:Hamiltonian-truncation} computed at a cutoff of $\Lambda=14$.
We can see that the lattice data towards the continuum limit $a \rightarrow 0$ is consistent with this value but the comparison requires a careful extrapolation of the lattice data.
The continuum limit is taken using energy data in the range $a=1/(Tn_t) \in [0,0.25]$ and extrapolated with simple polynomial functions of the lattice spacing.

We fit the data using Bayesian least-squares fits of gaussian-distributed variables implemented in \textrm{lsqfit}~\cite{peter_lepage_2021_4568470} and using uniformative priors for the parameters.
A way to assess systematic effects of the continuum limit extrapolation is to try cutting the data used at different maximum lattice spacing, that is using $a \le a_\textrm{max}$, and also increase the order $n_p$ of the fitting functions (polynomials) to include higher powers of $a$.
We use the fit function
\begin{equation}\label{eq:lattice_fit_gen}
F(T,n_t) = E + \sum_{i=1}^{n_p} a_i \left(\frac{1}{Tn_t}\right)^i \ ,
\end{equation}
where $E$ is the energy value in the continuum limit that we are after.
We repeat the fit to Eq.~\eqref{eq:lattice_fit_gen} for $n_p=1$, 2, and 3, while sliding $a_{\rm max}$ from values closer to the continuum limit all the way to the largest we have for each matrix model.
The results for the SU(2) bosonic matrix model at $\lambda=0.5$ are summarized in Tab.~\ref{tab:lat_su2_g05_fit_sys_allT}.

\begin{table}[htbp]
    \centering
    \begin{tabular}{rrlr||rrlr}
    \hline
       $a_\textrm{max}$ &   $n_p$ & $E$ &   $\chi^2$/dof & $a_\textrm{max}$ &   $n_p$ & $E$ &   $\chi^2$/dof \\
    \hline
    0.05 & 1 & 3.281(24)  & 1.70 &  0.30 & 1 & 3.2590(53) & 1.43 \\
    0.05 & 2 & 3.317(57)  & 1.67 &  0.30 & 2 & 3.2841(88) & 1.15 \\
    0.05 & 3 & 3.317(57)  & 1.67 &  0.30 & 3 & 3.285(14)  & 1.15 \\
    0.10 & 1 & 3.278(11)  & 1.46 &  0.40 & 1 & 3.2416(46) & 2.31 \\
    0.10 & 2 & 3.312(26)  & 1.37 &  0.40 & 2 & 3.2878(76) & 1.12 \\
    0.10 & 3 & 3.311(27)  & 1.37 &  0.40 & 3 & 3.282(12)  & 1.12 \\
    0.20 & 1 & 3.2725(66) & 1.29 &  0.50 & 1 & 3.2202(41) & 4.13 \\
    0.20 & 2 & 3.280(12)  & 1.28 &  0.50 & 2 & 3.2823(65) & 1.23 \\
    0.20 & 3 & 3.287(22)  & 1.27 &  0.50 & 3 & 3.2900(98) & 1.21 \\
    \hline
    \end{tabular}
    \caption{\label{tab:lat_su2_g05_fit_sys_allT}
    Results of systematic fitting for the SU(2) bosonic model with $m^2=1$ and $\lambda=g^2N=0.5$.
    }
\end{table}

\begin{figure}[htbp]
    \centering
    \includegraphics[width=0.45\textwidth]{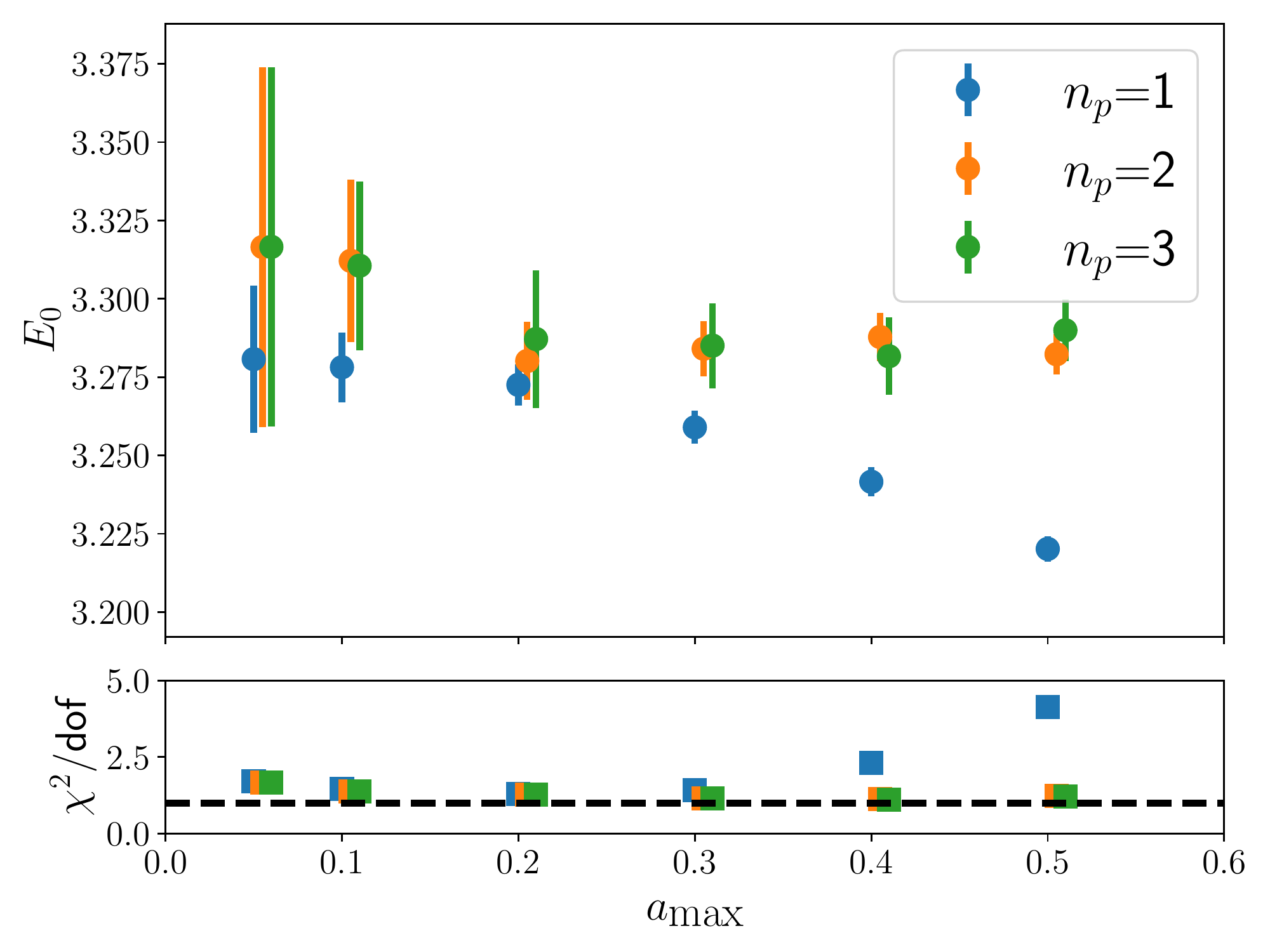}
    \caption{\label{fig:lat_su2_g02_fit_sys_allT}
    Results of systematic fitting for the SU(2) bosonic model with $m^2=1$ and $\lambda=g^2N=0.5$ using different data portions with polynomials of different order $n_p$.
    the lower panel shows the reduced $\chi^2$/dof which becomes very large for the low-order polynomials when larger lattice spacing are included, going well above the black dashed line representing 1.
    }
\end{figure}

We choose the final result based on where the results stabilize and become consistent across models with polynomials of different orders.
This can be seen in Fig.~\ref{fig:lat_su2_g02_fit_sys_allT}, where a linear function fit only works very close to the continuum limit, while higher-order polynomials can model the data at larger lattice spacing.
We repeat the procedure above for all the couplings.

Results from this systematic procedure, choosing the $n_p=2$ polynomial model for the continuum limit and cutting the data at $a_{\rm max}=0.1$, are summarized in Tab.~\ref{tab:lat_su2} and Tab.~\ref{tab:lat_su3} for the SU(2) and the SU(3) bosonic matrix model, respectively.

\begin{table}[htbp]
    \centering
    \begin{tabular}{|c||c|c|c|}
    \hline
    &  $\lambda = 0.5$ & $\lambda = 1.0$ & $\lambda = 2.0$ \\
    \hline
    \hline
    $E_{0, \text{MC}}$ &  3.312(26) & 3.497(33) & 3.847(30) \\ \hline
    \end{tabular}
    \caption{\label{tab:lat_su2}
    Ground state energy for the SU(2) bosonic model with $m^2=1$ at coupling $\lambda = g^2 N = 0.5$, 1.0, and 2.0.
    These are selected with an order $n_p=2$ polynomial and removing lattice spacing larger than 0.1.
    }
\end{table}

\begin{table}[htbp]
    \centering
    \begin{tabular}{|c||c|c|c|}
    \hline
    &  $\lambda = 0.5$ & $\lambda = 1.0$ & $\lambda = 2.0$ \\
    \hline
    \hline
    $E_{0, \text{MC}}$ &  8.836(38) & 9.381(38) & 10.236(41) \\ \hline
    \end{tabular}
    \caption{\label{tab:lat_su3}
    Ground state energy for the SU(3) bosonic model with $m^2=1$ at coupling $\lambda = g^2 N = 0.5$, 1.0, and 2.0.
    These are selected with an order $n_p=2$ polynomial and removing lattice spacing larger than 0.1.
    }
\end{table}

\begin{table}[htbp]
    \centering
    \begin{tabular}{rrrlr}
    \hline
       $T$ &   $n_t^{\rm cut}$ &   $n_p$ & $E$       &   $\chi^2$/dof \\
    \hline
      0.05 &     16 &       2 & 3.219(13) &           1.54 \\
      0.10 &     16 &       2 & 3.294(18) &           0.33 \\
      0.15 &     16 &       2 & 3.315(23) &           0.99 \\
      0.20 &     16 &       2 & 3.274(27) &           1.21 \\
      0.25 &     16 &       2 & 3.221(29) &           1.05 \\
      0.30 &     16 &       2 & 3.263(31) &           0.41 \\
      0.40 &     16 &       2 & 3.338(38) &           0.52 \\
    \hline
    \end{tabular}
    \caption{\label{tab:lat_su2_g05_fit_sys_eachT}
    Results of systematic fitting for the SU(2) matrix model with $m^2=1$ and $\lambda=g^2N=0.5$ using a $T$-dependent energy function.
    }
\end{table}

\begin{figure}[htbp]
    \centering
    \includegraphics[width=0.45\textwidth]{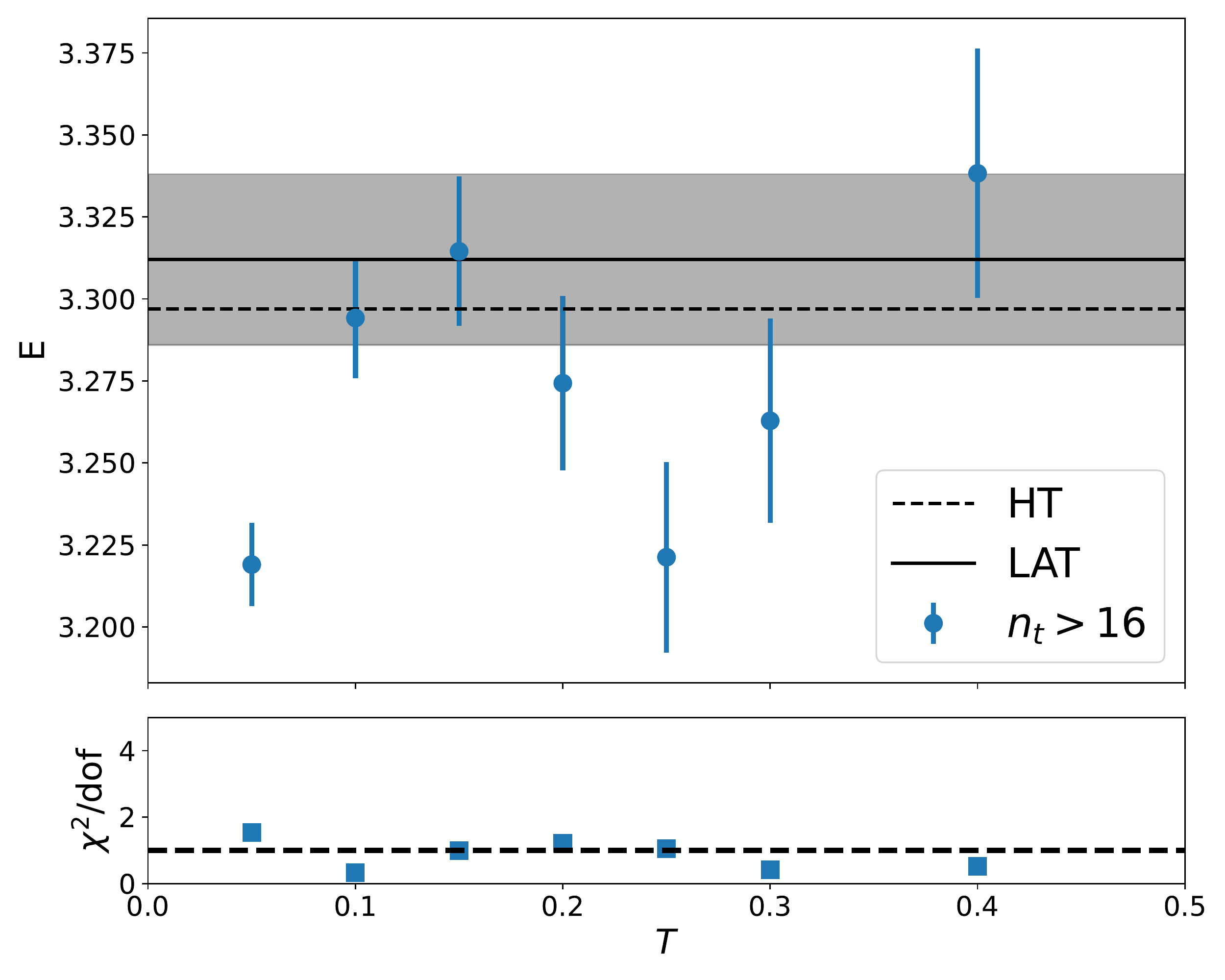}
    \caption{\label{fig:lat_su2_g05_fit_sys_c_eachT}
    Results of fitting different temperatures with a second order polynomial and including all values of $n_t>16$ for the SU(2) matrix model with $m^2=1$ and $\lambda=g^2N=0.5$.
    The result from the Hamiltonian truncation (HT) is the black dashed line.
    The result from the original fitting strategy in Tab.~\ref{tab:lat_su2} with the corresponding errorbar is the black line and the gray shaded area.
    the lower panel shows the reduced $\chi^2$/dof with the dashed line representing 1.
    }
\end{figure}

As an alternative analysis to study fitting systematic effects, we can also fit the energy data at each temperature $T$ with
the function
\begin{equation}
    F(T, n_t) = E(T) + \sum_i^{n_p} a_i(T) \left( \frac{1}{n_t} \right)^i \ ,
\end{equation}
where the $E(T)$ energy in the large-$n_t$ limit has $T$ dependence.
Fig.~\ref{fig:lat_su2_g05_fit_sys_c_eachT} shows that the fit for the lowest temperature is of poor quality and that makes the result for $E(T)$ deviate from the expected result.
This is due to smaller temperatures requiring much larger values of $n_t$ to achieve small discretization effects.
We summarize the results of the fits (including only lattice sizes $n_t> n_t^{\rm cut} = 16$) in Tab.~\ref{tab:lat_su2_g05_fit_sys_eachT}.
\clearpage
\section{Comparison of different approaches}\label{sec:comparison}
In this section, we compare the approaches investigated in this paper.

\subsection{Computational cost}\label{sec:comparison-cost}
In principle, the Hamiltonian truncation method on a classical computer is the most straightforward and most reliable approach, because we are guaranteed to obtain the precise value at large cutoffs $\Lambda$.
As we have seen for the SU(2) matrix models, the cutoff-dependence of the ground-state energy is exponentially suppressed.
However, this method is not tractable for SU($N$) models with $N \ge 3$, because the dimension of the Hilbert space is $\Lambda^{2(N^2-1)}$ for bosonic two-matrix model and $2^{N^2-1}\Lambda^{2(N^2-1)}$ for the minimal BMN model.
This is too large for current technology when $N \ge 3$, but the situation could be different if we could use an error-corrected quantum computer on which the adiabatic state preparation and quantum phase estimation can be used reliably.
The VQE can circumvent the quantum-noise issue if a quantum computer with sufficiently many qubits becomes available.

On the other hand, the deep learning method does not suffer from the quick growth of the dimension of the Hilbert space, similarly to the lattice Monte Carlo method.
The deep learning approach we have adopted requires computational resources to study hyperparameters for the neural network parameterizations of the wave function which will give the best (in a variational sense) energy.
One of the advantages is that the information on the entire wave function can be obtained instead of just a few observables.
The computational cost scales with the dimension of the neural network (the number of training parameters): the input dimension of the neural networks scales as $O(N^2)$ for SU($N$) matrices and the number of parameters in hidden layers scales as $O(N^4)$.
For example, in bosonic models $N = 20 $ (about 400 bosons), and in the supersymmetric case $N = 12$ (about 150 bosons with 150 fermions) are readily accessible on a laptop\footnote{The training process of the deep neural network might be affected by slow down if some hyperparameters are not chosen correctly, such as the learning rate or the optimizer.}.

The lattice Monte Carlo simulation can be used even when $N$ is very large with the computational time\footnote{Molecular dynamics time units might require more steps as a function of the temperature or lattice size, therefore increasing the computational time to obtain uncorrelated configurations along a Markov Chain.} scaling roughly with $N^3$.
However, we can only obtain expectation values of observables related to the ground state or a few low-lying state.
For a reliable estimation of these observables we have to study sufficiently low temperatures at various (large) lattice sizes, such that discretization effects become small or can be modeled accurately.
Technically, the most complicated part is the uncertainty of the continuum extrapolation.
Note that it is difficult to determine the spectrum of the excited modes precisely when using lattice Monte Carlo methods.
Moreover, we should remark that typical lattice field configurations are not related to the wave function.

\subsection{Results}\label{sec:comparison-results}
\subsubsection{SU(2) bosonic model}
We compare the ground state energy obtained with different methods for SU(2) bosonic matrix model.
A very good agreement can be seen between the Hamiltonian truncation method (at large cutoff $\Lambda=14$), the VQE method, the deep learning (DL) method and the lattice Monte Carlo (MC) method.
The level of agreement deteriorates for the VQE as the model is explored at larger gauge couplings $\lambda$, because the VQE has a large truncation error due to the small cutoff $\Lambda=4$ used.
The agreement might improve if a larger cutoff (larger number of qubits) is used or if more expressive variational wave functions are employed for the ground state.
Both these directions require an increased computational time.

The values for the ground-state energy coming from all the approaches are compared in Tab.~\ref{tab:comparison-su2}.
Note the errobars in the DL and MC approaches because of their stochastic Monte Carlo estimation nature.
The DL method errorbars are statistical and they can be reduced with more Monte Carlo samples.
On the other hand, the MC method errorbars are mostly from a parameter inference procedure to extract the continuum limit extrapolated energy.
These errorbars can be reduced but there will be a lower limit given by the systematic effects of finding the right extrapolation model.

\begin{table}[ht]
    \centering
    \begin{tabular}{|r||l|l|l|}
    \hline
    &  $\lambda = 0.5$ & $\lambda = 1.0$ & $\lambda = 2.0$ \\
    \hline
    \hline
    $E_{0, \text{HT}}$ &  3.297 & 3.516  & 3.855 \\
    \hline
    $E_{0, \text{DL}}$ &  3.302(2) & 3.519(2) & 3.857(3) \\
    \hline
    $E_{0, \text{MC}}$ &  3.312(26) & 3.497(33) & 3.847(30) \\
    \hline
    \hline
    $E_{0, \text{VQE}}$ & 3.309 & 3.547 & 3.933 \\
    \hline
    \end{tabular}
    \caption{\label{tab:comparison-su2}
    The ground-state energy in the SU(2) bosonic two-matrix model obtained from the Hamiltonian truncation ($E_{0, \text{HT}}$), the VQE ($E_{0, \text{VQE}}$), deep learning ($E_{0, \text{DL}}$), and lattice Monte Carlo ($E_{0, \text{MC}}$) at coupling $\lambda = g^2 N = 0.5$, 1.0, and 2.0.
    The apparent poor performance of the VQE may come from using a small cutoff $\Lambda=4$ compared to the Hamiltonian truncation.
    The Hamiltonian truncation results at $\Lambda=14$ have a negligible truncation error at this number of digits.
    }
\end{table}

\subsubsection{SU(3) bosonic model}
We can not treat the truncated Hamiltonian exactly (or with the VQE approach) due to the growth of the Hilbert space and the corresponding increase in the computational resources for the SU(3) gauge group.
However, the deep learning (DL) and lattice Monte Carlo (MC) techniques do not suffer from this unfavorable scaling.
We can actually see very good agreement between these two approaches for the bosonic SU(3) matrix model.
At the large coupling $\lambda=2.0$, the ML bound is higher than the MC estimation in the continuum limit.
At smaller coupling they are compatible within errors.
The results are summarized in Tab.~\ref{tab:comparison-su3}.

\begin{table}[ht]
    \centering
    \begin{tabular}{|r||l|l|l|}
    \hline
    & $\lambda = 0.5$ & $\lambda = 1.0$ & $\lambda = 2.0$ \\
    \hline
    \hline
    $E_{0, \text{DL}}$ & 8.824(7) & 9.432(7) & 10.426(8) \\
    \hline
    $E_{0, \text{MC}}$ &  8.836(38) & 9.381(38) & 10.236(41) \\
    \hline
    \end{tabular}
    \caption{\label{tab:comparison-su3}
    The ground-state energy in the SU(3) bosonic two-matrix model at coupling $\lambda = g^2 N = 0.5$, 1.0, and 2.0.
    Neural variational quantum Monte Carlo (DL) and lattice Monte Carlo (MC) are compared.
    }
\end{table}

\subsubsection{SU(2) minimal BMN}
For the minimal BMN matrix model with SU(2) group we know the exact value for the ground state energy at each coupling $\lambda$.
This energy is zero due to supersymmetry, hence this case is perfectly suitable to benchmark our computational approaches.
The VQE approach is limited by the size of the Hilbert space that can be explored with a small number of qubits and the ground-state energy variational upper bound is close to zero for small coupling but much higher for larger couplings.
Similarly, the DL method is compatible with the exact energy at small coupling and it becomes less compatible at larger couplings, due to the difficulty in determining a good variational ansatz for a strongly-coupled system.
We summarize the results in Tab.~\ref{tab:comparison-su2-bmn}.

\begin{table}[ht]
    \centering
    \begin{tabular}{|r||l|l|l|}
    \hline
    &  $\lambda = 0.5$ & $\lambda = 1.0$ & $\lambda = 2.0$ \\
    \hline
    \hline
    $E_{0, \text{HT}}$ &  0.000 & 0.000 & 0.000 \\
    \hline
    $E_{0, \text{DL}}$ &  0.009(5) & 0.014(6) & 0.034(7) \\
    \hline
    \hline
    $E_{0, \text{VQE}}$ & 0.027 & 0.079 & 0.177 \\
    \hline
    \end{tabular}
    \caption{\label{tab:comparison-su2-bmn}
    The ground-state energy in the SU(2) minimal BMN model obtained from the Hamiltonian truncation ($E_{0, \text{HT}}$), the VQE ($E_{0, \text{VQE}}$), and deep learning ($E_{0, \text{DL}}$) at coupling $\lambda = g^2 N = 0.5$, 1.0, and 2.0.
    The VQE results are obtained at cutoff $\Lambda=2$ with the best optimizer and variational form, but they may improve at larger cutoff.
    The Hamiltonian truncation results at $\Lambda=11$ can not be distinguished from the exact supersymmetric results of zero, at this number of digits.
    }
\end{table}

\section{Conclusions and discussions}\label{sec:conclusions}
In this paper, we conducted a systematic survey of quantum-simulation and deep-learning approaches to matrix models.
Specifically, we considered a bosonic two-matrix model and its supersymmetric extension (minimal BMN model), and computed the spectrum of the low-lying modes.
We used the Hamiltonian truncation method (in the Fock basis), the Variational Quantum Eigensolver (VQE) method, and variational quantum Monte Carlo with a neural network ansatz for the wave function.
We also performed lattice Monte Carlo simulations based on the Euclidean path integral for the bosonic model as a comparison.

With the Hamiltonian truncation method, we considered the matrix model with a fixed number of modes for each boson up to a cutoff value that is slowly taken to infinity in order to remove truncation effects.
The spectrum of the truncated Hamiltonian is obtained using \qutip on classical computers.
We explored how quickly the correct answer at infinite cutoff can be recovered.
With this setup we only studied the bosonic matrix model with SU(2) gauge group, because the SU(3) model contains more bosonic degrees of freedom (16 compared to 6) resulting in a much larger Hilbert space: this would have required computational resources not available to us at the moment.
We found that the truncation effects to the low-energy spectrum are suppressed exponentially with respect to the cutoff $\Lambda$.
This is a favorable scaling and we are able to prove it empirically for the first time in this work.
Note that the number of qubits in quantum simulation scales as $\log_2\Lambda$; hence the truncation error can be suppressed double-exponentially with respect to the number of qubits.
It would be natural to assume the same scaling of the truncation errors for larger systems (larger gauge group, or more matrices).
Such a good scaling would be a good news for quantum simulation of quantum gravity via holography.
Quantum simulations of matrix models can naturally lead to ``quantum gravity in the lab''~\cite{Danshita:2016xbo,Brown:2019hmk,Nezami:2021yaq}.
So far toy models such as the SYK models were considered (regarding quantum simulation protocols of the SYK model on universal quantum computers, see Refs.~\cite{Garcia-Alvarez:2016wem,Babbush:2018mlj}), but matrix model can be a much better, theoretically well-controlled setup for quantum gravity.
Moreover, understanding these matrix models through quantum simulations could help in the context of quantum error correction studies through their connection with the holographic conjecture~\cite{Faist:2019ahr}.

We also wanted to understand how quantum hardware can be utilized efficiently to study matrix models in the NISQ era.
For this reason we resorted to a hybrid quantum-classical algorithm, the Variational Quantum Eigensolver, to extract the ground state of the models.
We used the IBM \qiskit software to estimate the ground-state energy at small system sizes, and compared the numbers with those from exact diagonalization of the truncated Hamiltonian using classical hardware.
In the future we plan to use quantum hardware for the simulations, but first we needed to investigate the performance of the variational quantum circuits and the classical optimizers.
Our results for the bosonic model show very good agreement at weak coupling, but there is a growing discrepancy between the least upper bound for the energy from the VQE and the exact diagonalization value of the truncated Hamiltonian as the coupling becomes larger.
This is not surprising because the quantum circuits we used to approximate the ground-state wave function are not crafted for matrix models.
We need to find better circuits, and for that purpose, the deep learning approach may give us some hints.
Moreover, we noticed that the upper bound given by the VQE for the bosonic model is usually better than the one in the minimal BMN model.
In the case of the minimal BMN model, the energy of the ground state is zero up to exponentially small truncation effects.
However, we noticed that the VQE is giving an upper bound that is non-zero and the difference with the exact value grows with the coupling.

To explore what deep learning can do for the matrix models, we used a variational quantum Monte Carlo approach with an autoregressive neural flow as the parametrization for the wave function of the ground state.
This approach can be readily applicable to larger system sizes and larger gauge groups.
We studied the SU(2) bosonic model and the SU(2) minimal BMN model and we also looked at the SU(3) bosonic model.
Because we did not study the SU(3) bosonic model via the Hamiltonian truncation due to the large Hilbert space, a comparison with lattice Monte Carlo was crucial to establish a baseline for the results.
This neural network based variational quantum Monte Carlo approach is able to obtain a good approximation of the ground state energy for all the parameters we studied in the bosonic and minimal BMN matrix models.
This is a very promising outcome, which is demonstrated here for the first time.
We believe that this deep learning approach is readily applicable to various problems such as real-time evolution or complex models such as the full BMN or BFSS matrix models.

\acknowledgments
We wish to thank David Berenstein for useful suggestions and discussions on Matrix Models and Fabrizio Minganti for discussions about \qutip.
We also thank Antonio Mezzacapo for useful suggestions on quantum computing applied to gauge theories.
This material is based upon work supported in part by the U.S. Department of Energy, Office of Science, National Quantum Information Science Research Centers, Co-design Center for Quantum Advantage (C2QA) under contract number DE-SC0012704. This project was supported in part by the U.S. Department of Energy, Office of Science, Office of Workforce Development for Teachers and Scientists (WDTS) under the Science Undergraduate Laboratory Internships Program (SULI).
F.N. is supported in part by: Nippon Telegraph and Telephone Corporation (NTT) Research, the Japan Science and Technology Agency (JST) [via the Quantum Leap Flagship Program (Q-LEAP), the Moonshot R\&D Grant Number JPMJMS2061, and the Centers of Research Excellence in Science and Technology (CREST) Grant No. JPMJCR1676], the Japan Society for the Promotion of Science (JSPS) [via the Grants-in-Aid for Scientific Research (KAKENHI) Grant No. JP20H00134 and the JSPS–RFBR Grant No. JPJSBP120194828], the Army Research Office (ARO) (Grant No. W911NF-18-1-0358), the Asian Office of Aerospace Research and Development (AOARD) (via Grant No. FA2386-20-1-4069), and the Foundational Questions Institute Fund (FQXi) via Grant No. FQXi-IAF19-06.
E.R. is supported by Nippon Telegraph and Telephone Corporation (NTT) Research.

\bibliography{main}

\appendix

\section{Operator formalism and path-integral formalism}\label{appendix:operator-formalism-vs-path-integral}
\subsection{Hilbert space with and without singlet constraint}
When quantum gauge theory is treated in the Hamiltonian formulation in the temporal gauge ($A_t=0$ gauge), it is often said that ``physical states are gauge singlets''.
The canonical partition function can be written as a trace over the space of gauge-invariant states ${\cal H}_{\rm inv}$,
\begin{align}\label{eq:Z-H-inv}
Z(T) &= {\rm Tr}_{{\cal H}_{\rm inv}}\left( e^{-\hat{H}/T} \right) \ .
\end{align}

There is another, but equivalent, formulation with non-singlet Hilbert space (``extended'' Hilbert space) ${\cal H}_{\rm ext}$, in which the quantum states that mapped to each other via gauge transformation are identified.
The partition function is written as
\begin{align}\label{eq:Z-H-ext}
Z(T) &= \frac{1}{{\rm vol}(G)}\int_G dg {\rm Tr}_{{\cal H}_{\rm ext}}\left( \hat{g} e^{-\hat{H}/T} \right) \ .
\end{align}

Here $G={\rm SU}(N)$, and the integral is taken with the Haar measure.
Note that
\begin{align}
\hat{{\cal P}} \equiv \frac{1}{{\rm vol}(G)}\int_G dg \hat{g}
\end{align}
is a projection operator from ${\cal H}_{\rm ext}$ to ${\cal H}_{\rm inv}$ that satisfies $\hat{{\cal P}}^2=\hat{{\cal P}}$.

We show the equivalence of \eqref{eq:Z-H-inv} and \eqref{eq:Z-H-ext}.
This is a well-known fact, but we present the derivation to make the paper self-contained.
Essentially, we repeat the materials in Refs.~\cite{Hanada:2020uvt,Hanada:2021ipb}.

Suppose that an energy eigenstate $|\Phi\rangle\in{\cal H}_{\rm ext}$ is invariant only under $G_\Phi\subset G$.
From $|\Phi\rangle$, we can obtain $|\Phi\rangle_{\rm inv}\in{\cal H}_{\rm inv}$ as
\begin{align}
|\Phi\rangle_{\rm inv} = \frac{1}{\sqrt{C_\Phi}}\times\frac{1}{{\rm vol}(G)}\int_G dg (\hat{g}|\Phi\rangle) \ .
\end{align}

The normalization factor $C_\Phi$ is determined such that the norm of $|\Phi\rangle_{\rm inv}$ is 1.
Namely,
\begin{align}
C_\Phi &=
\frac{1}{({\rm vol}(G))^2}\int_G dg\int_G dg' \langle\Phi|\hat{g}^{-1}\hat{g}'|\Phi\rangle
\nonumber\\
&= \frac{1}{{\rm vol}(G)}\int_G dg \langle\Phi|\hat{g}|\Phi\rangle
\nonumber\\
&= \frac{{\rm vol}(G_\Phi)}{{\rm vol}(G)} \ .
\end{align}

Again, this is the inverse of the over-counting factor.
Therefore,
\begin{align}
{\rm Tr}_{{\cal H}_{\rm inv}}\left(
e^{-\hat{H}/T}
\right)
&=
\sum_\Phi
\frac{{\rm vol}(G_\Phi)}{{\rm vol}(G)}
{}_{\rm inv}\langle\Phi|
e^{-\hat{H}/T}
|\Phi\rangle_{\rm inv}
\nonumber\\
&=
\sum_\Phi
\frac{{\rm vol}(G_\Phi)}{{\rm vol}(G)}
\times
\frac{1}{C_\Phi}\frac{1}{({\rm vol}(G))^2}
\int_G dg\int_G dg'
\langle\Phi|\hat{g}^{-1}
e^{-\hat{H}/T}\hat{g}'
|\Phi\rangle
\nonumber\\
&=
\sum_\Phi
\frac{1}{{\rm vol}(G)}
\int_G dg
\langle\Phi|\hat{g}
e^{-\hat{H}/T}
|\Phi\rangle
\nonumber\\
&=
\frac{1}{{\rm vol}(G)}
\int_Gdg
{\rm Tr}_{{\cal H}_{\rm ext}}\left(
\hat{g}
e^{-\hat{H}/T}
\right) \ .
\end{align}
Here the sum over $\Phi$ is taken over the extended Hilbert space.

\subsection{Relation between canonical quantization and path integral formalism}\label{sec:H-to-L}
In the canonical partition function defined on the extended Hilbert space, the group element $g$ is interpreted as the Polyakov loop.
It can be seen as follows.
The expression \eqref{eq:Z-H-ext} can be rewritten as
\begin{align}
Z(T)
&=
\frac{1}{[{\rm vol}(G)]^K}
\int\left(\prod_{k=1}^KdU_{(k)}\right)
{\rm Tr}_{{\cal H}_{\rm ext}}
\Bigl(
\hat{U}_{(K)}
e^{-\frac{H(\hat{P},\hat{X})}{TK}}
\hat{U}_{(K-1)}^{-1}\hat{U}_{(K-1)}
\nonumber\\
&
\quad
e^{-\frac{H(\hat{P},\hat{X})}{TK}}
\hat{U}_{(K-2)}^{-1}\hat{U}_{(K-2)}
\cdots
\hat{U}_{(1)}^{-1}\hat{U}_{(1)}
e^{-\frac{H(\hat{P},\hat{X})}{TK}}
\Bigl)
\nonumber\\
&=
\frac{1}{[{\rm vol}(G)]^K}
\int\left(\prod_{k=1}^KdU_{(k)}\right)
\int\left(\prod_{k=1}^KdX_{(k)}\right)
\nonumber\\
&
\qquad
\langle X_{(K)}|
\hat{U}_{(K)}
e^{-\frac{H(\hat{P},\hat{X})}{TK}}
\hat{U}_{(K-1)}^{-1}
|X_{(K-1)}\rangle
\nonumber\\
&
\qquad
\times
\langle X_{(K-1)}|
\hat{U}_{(K-1)}
e^{-\frac{H(\hat{P},\hat{X})}{TK}}
\hat{U}_{(K-2)}^{-1}
|X_{(K-2)}\rangle
\nonumber\\
&
\qquad
\times
\cdots
\times
\langle X_{(1)}|
\hat{U}_{(1)}
e^{-\frac{H(\hat{P},\hat{X})}{TK}}
|X_{(K)}\rangle \ .
\end{align}
At the first line, we simply inserted $\hat{U}^{-1}\hat{U}=\hat{1}$ at many places.
$\hat{U}_{(K)}$ corresponds to $\hat{g}$.
At the next line, we inserted $\int dX_{(k)}|X_{(k)}\rangle\langle X_{(k)}|=\hat{1}$ at many places.
\begin{align}
&
\langle X_{(k)}|
\hat{U}_{(k)}
e^{-\frac{H(\hat{P},\hat{X})}{TK}}
\hat{U}_{(k-1)}^{-1}
|X_{(k-1)}\rangle
\nonumber\\
&\quad=
\langle U_{(k)}X_{(k)}U_{(k)}^{-1}|
e^{-\frac{H(\hat{P},\hat{X})}{TK}}
|U_{(k-1)}X_{(k-1)}U_{(k-1)}^{-1}\rangle
\nonumber\\
&\quad=
\int dP
\langle U_{(k)}X_{(k)}U_{(k)}^{-1}|
e^{-\frac{H(\hat{P},\hat{X})}{TK}}
|P\rangle\langle P|
U_{(k-1)}X_{(k-1)}U_{(k-1)}^{-1}\rangle
\nonumber\\
&\quad=
\int dP
e^{i{\rm Tr}[P(U_{(k)}X_{(k)}U_{(k)}^{-1}-U_{(k-1)}X_{(k-1)}U_{(k-1)}^{-1})]}
e^{-H(P,U_{(k)}X_{(k)}U_{(k)}^{-1})/(TK)}
\nonumber\\
&\quad=
e^{-KT{\rm Tr}[(U_{(k)}X_{(k)}U_{(k)}^{-1}-U_{(k-1)}X_{(k-1)}U_{(k-1)}^{-1})^2]}
e^{-V(U_{(k)}X_{(k)}U_{(k)}^{-1})/(TK)}
\nonumber\\
&\quad\simeq
e^{-L[D_t(U_{(k)}X_{(k)}U_{(k)}^{-1}),(U_{(k)}X_{(k)}U_{(k)}^{-1})]/(TK)}
\nonumber\\
&\quad=
e^{-L[D_tX_{(k)},X_{(k)})]/(TK)} \ .
\end{align}

Here we used
\begin{align}
U_{(k-1)}^{-1}U_{(k)}\equiv e^{iA_{(K)}/(KT)}
\end{align}
and
\begin{align}
X_{(k)}-(U_{(k-1)}U_{(k)}^{-1})^{-1}X_{(k-1)}(U_{(k-1)}U_{(k)}^{-1})
\simeq
\frac{D_tX_{(k)}}{KT}.
\end{align}
$L[D_tX,X]$ is the Lagrangian with the Euclidean signature.
By taking $K\to\infty$ limit, we obtain
\begin{align}
Z(T) &= \int [dA] [dX]e^{-\int dt L[D_tX,X]} \ .
\end{align}
The temporal direction is compactified with the circumference $\frac{1}{T}$, and the periodic boundary condition is imposed to $X_I$ and $A_t$.
Therefore,
\begin{align}
g &= U_K
\nonumber\\
&=
U_K(U_{K-1}^{-1}U_{K-1})(U_{K-2}^{-1}U_{K-2})\cdots(U_{1}^{-1}U_{1})
\nonumber\\
&=
(U_KU_{K-1}^{-1})(U_{K-1}U_{K-2}^{-1})\cdots(U_{2}U_{1}^{-1})U_1
\nonumber\\
&=
{\rm P}e^{i\int_0^{1/T} dt A_t}
\end{align}
is the Polyakov loop.
(Here ${\rm P}$ stands for the path-ordered product along time direction.)

Note that, unless the time direction is noncompact or a special boundary condition is taken, the condition $A_t=0$ cannot be imposed in the literal sense.
In the case of the canonical partition function, we cannot eliminate the degrees of freedom of the Polyakov loop.
The integration over Polyakov loop leads to the projector $\frac{1}{{\rm vol}(G)}\int dg\hat{g}$ in the operator formalism.

\section{Table of results of Hamiltonian truncation}\label{appendix:Ham_truncation}
\subsection{SU(2) bosonic matrix model}
\begin{table}[H]
\scriptsize
    \centering
    \begin{tabular}{c|c|c|c|c}
         & \multicolumn{4}{|c}{$\langle E_0|\hat{H}|E_0\rangle$} \\ \hline
    $\Lambda$ & $\lambda=0.2$ & $\lambda=0.5$ & $\lambda=1.0$ & $\lambda=2.0$ \\ \hline
     3 & 3.13230465 & 3.28285159 & 3.45847992 & 3.66904008 \\
     4 & 3.13406307 & 3.29894363 & 3.52625444 & 3.89547837 \\
     5 & 3.13390803 & 3.29649279 & 3.51211772 & 3.83339247 \\
     6 & 3.13392706 & 3.29702515 & 3.51650758 & 3.85950613 \\
     7 & 3.13392487 & 3.29692058 & 3.51533549 & 3.85081424 \\
     8 & 3.13392519 & 3.29694663 & 3.51573342 & 3.85457565 \\
     9 & 3.13392514 & 3.29694061 & 3.51561221 & 3.85316716 \\
    10 & 3.13392515 & 3.29694226 & 3.51565635 & 3.85380864 \\
    11 & 3.13392515 & 3.29694183 & 3.51564151 & 3.85354647 \\
    12 & 3.13392515 & 3.29694196 & 3.51564726 & 3.85367079 \\
    13 & 3.13392515 & 3.29694192 & 3.51564518 & 3.85361668 \\
    14 & 3.13392515 & 3.29694194 & 3.51564602 & 3.85364330 \\
    \end{tabular}
    \caption{\label{tab:ham_su2_c0_gs}
    The ground state energy in the SU(2) bosonic model at different coupling constants $\lambda$ for various cutoff $\Lambda$.
    The precision of the iterative sparse eigensolver is set to $10^{-8}$.
    }
\end{table}

\begin{table}[H]
\scriptsize
    \centering
    \begin{tabular}{c|c|c|c|c}
     & \multicolumn{4}{|c}{$\sum_\alpha\langle E_0|\hat{G}^2_\alpha|E_0\rangle$} \\ \hline
    $\Lambda$ & $\lambda=0.2$ & $\lambda=0.5$ & $\lambda=1.0$ & $\lambda=2.0$  \\ \hline
     3 & 0.000224293831 & 0.003688212672 & 0.018964932725 & 0.060510041162 \\
     4 & 0.000184221658 & 0.002400096378 & 0.011889840764 & 0.045924094777 \\
     5 & 0.000004511062 & 0.000178633922 & 0.001732981158 & 0.011493185261 \\
     6 & 0.000002506324 & 0.000101373665 & 0.000986724060 & 0.006352557866 \\
     7 & 0.000000087650 & 0.000009947253 & 0.000176500569 & 0.001788911061 \\
     8 & 0.000000049542 & 0.000005629387 & 0.000099441023 & 0.001011302266 \\
     9 & 0.000000002327 & 0.000000683314 & 0.000020545750 & 0.000315815031 \\
    10 & 0.000000001312 & 0.000000389452 & 0.000011932039 & 0.000185582780 \\
    11 & 0.000000000077 & 0.000000055646 & 0.000002785207 & 0.000062685672 \\
    12 & 0.000000000044 & 0.000000032055 & 0.000001643162 & 0.000037949122 \\
    13 & 0.000000000003 & 0.000000005217 & 0.000000421954 & 0.000013712553 \\
    14 & 0.000000000002 & 0.000000003031 & 0.000000252683 & 0.000008471522 \\
    \end{tabular}
    \caption{\label{tab:ham_su2_c0_gv}
    The ground state violation of the singlet constraint in the SU(2) bosonic model at different coupling constants $\lambda$ for various cutoff $\Lambda$.
    The precision of the iterative sparse eigensolver is set to $10^{-8}$.
    }
\end{table}

\subsection{SU(2) minimal BMN}
\begin{table}[H]
\scriptsize
    \centering
    \begin{tabular}{c|c|c|c|c}
     & \multicolumn{4}{|c}{$\langle E'_0|\hat{H}|E'_0\rangle$} \\ \hline
     $\Lambda$ & $\lambda=0.2$ & $\lambda=0.5$ & $\lambda=1.0$ & $\lambda=2.0$ \\ \hline
     3 & -0.000348435200 & -0.003873948083 & -0.019907205965 & -0.084936973789 \\
     4 &  0.000114126215 &  0.002116374610 &  0.013418689187 &  0.060446205687 \\
     5 & -0.000003216759 & -0.000102255010 & -0.000909125958 & -0.005803433350 \\
     6 &  0.000002429833 &  0.000134468904 &  0.001628985778 &  0.012489096584 \\
     7 & -0.000000036383 & -0.000003214703 & -0.000051505094 & -0.000504975266 \\
     8 &  0.000000063447 &  0.000010165342 &  0.000231428987 &  0.002895837534 \\
     9 & -0.000000000464 & -0.000000099555 & -0.000002210416 & -0.000018843848 \\
    10 &  0.000000002099 &  0.000000910719 &  0.000036818848 &  0.000724187898 \\
    11 & -0.000000000002 &  0.000000001411 &  0.000000259748 &  0.000012480441 \\
    \end{tabular}
    \caption{\label{tab:ham_su2_mini_gs}
    The energy expectation value $\langle E'_0|\hat{H}|E'_0\rangle$ for the ground state of the $\hat{H}'$ Hamiltonians of the SU(2) minimal BMN model at different coupling constants $\lambda$ for various cutoff $\Lambda$.
    Only the first 12 digits are shown.
    }
\end{table}

\begin{table}[H]
\scriptsize
    \centering
    \begin{tabular}{c|c|c|c|c}
     & \multicolumn{4}{|c}{$\sum_\alpha\langle E'_0|\hat{G}^2_\alpha|E'_0\rangle$} \\ \hline
     $\Lambda$ & $\lambda=0.2$ & $\lambda=0.5$ & $\lambda=1.0$ & $\lambda=2.0$ \\ \hline
      3 &  0.000027144384 &  0.000270665767 &  0.001217678752 &  0.004391570468 \\
      4 &  0.000003466155 &  0.000071748567 &  0.000483864489 &  0.002211074147 \\
      5 &  0.000000377832 &  0.000013303385 &  0.000129198647 &  0.000837166225 \\
      6 &  0.000000087652 &  0.000005300120 &  0.000065635237 &  0.000487463732 \\
      7 &  0.000000007701 &  0.000000831344 &  0.000015273650 &  0.000161843291 \\
      8 &  0.000000002422 &  0.000000413367 &  0.000009393689 &  0.000110264245 \\
      9 &  0.000000000211 &  0.000000063891 &  0.000002115537 &  0.000035041177 \\
     10 &  0.000000000080 &  0.000000036382 &  0.000001444227 &  0.000026392551 \\
     11 &  0.000000000007 &  0.000000005736 &  0.000000326893 &  0.000008241117 \\
    \end{tabular}
    \caption{\label{tab:ham_su2_mini_gv}
    The gauge constraint violation $\sum_{\alpha} \langle E'_0|\hat{G}^2_{\alpha}|E'_0\rangle$ for the ground state of the $\hat{H}'$ Hamiltonians of the SU(2) minimal BMN model at different coupling constants $\lambda$ for various cutoff $\Lambda$.
    Only the first 12 digits are shown.
    }
\end{table}

\begin{table}[H]
\scriptsize
    \centering
    \begin{tabular}{c|c|c|c|c}
     & \multicolumn{4}{|c}{$|\sum_\alpha\langle E'_0|\hat{M}_\alpha|E'_0\rangle - J|$} \\ \hline
     $\Lambda$ & $\lambda=0.2$ & $\lambda=0.5$ & $\lambda=1.0$ & $\lambda=2.0$ \\ \hline
     3 & 0.000016801483 & 0.000182372511 & 0.000913116520 & 0.003744814441 \\
     4 & 0.000000040458 & 0.000001248745 & 0.000008311970 & 0.000017373811 \\
     5 & 0.000000225739 & 0.000007305553 & 0.000064178889 & 0.000381297761 \\
     6 & 0.000000000342 & 0.000000015301 & 0.000000143203 & 0.000002510586 \\
     7 & 0.000000004147 & 0.000000371546 & 0.000005900379 & 0.000056563759 \\
     8 & 0.000000000002 & 0.000000000600 & 0.000000031637 & 0.000000714895 \\
     9 & 0.000000000099 & 0.000000023729 & 0.000000662279 & 0.000009762673 \\
    10 & 0.000000000000 & 0.000000000077 & 0.000000005308 & 0.000000175686 \\
    11 & 0.000000000003 & 0.000000001820 & 0.000000087067 & 0.000001946922 \\
    \end{tabular}
    \caption{\label{tab:ham_su2_mini_gm_J0}
    The angular momentum constraint violation $|\sum_\alpha\langle E'_0|\hat{M}_\alpha|E'_0\rangle - J|$ for the ground state of the $\hat{H}'$ Hamiltonians of the SU(2) minimal BMN model at different coupling constants $\lambda$ for various cutoff $\Lambda$.
    Only the first 12 digits are shown and $J=0$.
    }
\end{table}

\section{Table of results of lattice Monte Carlo simulations}

\subsection{Tables for the results of the ground state energy}
An example of raw results from the analysis of the Markov Chain Monte Carlo histories is shown in Table~\ref{tab:lat_su3_g05}.
For each parameter set of $T$ and $n_t$ we thermalize the MCMC chain and remove a burn-in subset of trajectories (in units of molecular dynamics time (MDTU)) of 1000 MDTU at the beginning of each chain.
We save the energy observable every $N_\textrm{drop}$ MDTU and we collect the observables on a total of $N_\textrm{cfgs}$ configurations.
We also measure the integrated autocorrelation time $\tau$  of the energy observable using the Madras-Sokal windowing algorithm.

\begin{table}[H]
\scriptsize
\centering
\begin{tabular}{rrlrrr}
\hline
   $T$ &   $n_t$ &  $E$  &   $N_\textrm{cfgs}$ &   $N_\textrm{drop}$ &   $\tau$ \\
\hline
 0.400 &      16 & 7.845(70) &      980 &       50 &    0.810 \\
 0.350 &      16 & 7.639(41) &     1980 &       50 &    0.950 \\
 0.300 &      16 & 7.481(54) &      980 &       50 &    1.040 \\
 0.250 &      16 & 7.354(52) &      990 &      100 &    0.900 \\
 0.200 &      16 & 7.072(44) &      990 &      100 &    0.850 \\
 0.150 &      16 & 6.710(38) &      988 &       80 &    1.130 \\
 0.100 &      16 & 6.131(29) &      988 &       80 &    0.750 \\
 0.050 &      16 & 5.071(20) &      990 &      100 &    0.980 \\
 0.025 &      16 & 3.873(13) &      990 &      100 &    1.020 \\
 0.400 &      24 & 7.912(60) &      990 &      100 &    0.990 \\
 0.350 &      24 & 8.036(43) &     1980 &       50 &    1.000 \\
 0.300 &      24 & 7.759(57) &      990 &      100 &    0.990 \\
 0.250 &      24 & 7.773(52) &      990 &      100 &    1.120 \\
 0.200 &      24 & 7.558(44) &      990 &      100 &    1.450 \\
 0.150 &      24 & 7.134(38) &      990 &      100 &    0.800 \\
 0.100 &      24 & 6.669(31) &      990 &      100 &    0.930 \\
 0.050 &      24 & 5.670(20) &      990 &      100 &    1.170 \\
 0.025 &      24 & 4.613(13) &      990 &      100 &    0.910 \\
 0.400 &      32 & 8.284(69) &      990 &      100 &    0.860 \\
 0.350 &      32 & 8.162(44) &     1980 &       50 &    0.940 \\
 0.300 &      32 & 8.113(57) &      990 &      100 &    1.000 \\
 0.250 &      32 & 8.001(49) &      990 &      100 &    1.030 \\
 0.200 &      32 & 7.720(45) &      990 &      100 &    0.990 \\
 0.150 &      32 & 7.513(39) &      990 &      100 &    0.770 \\
 0.100 &      32 & 7.053(31) &      990 &      100 &    0.820 \\
 0.050 &      32 & 6.137(22) &      990 &      100 &    1.130 \\
 0.025 &      32 & 5.064(14) &      990 &      100 &    0.820 \\
 0.400 &      48 & 8.430(68) &      980 &       50 &    0.980 \\
 0.350 &      48 & 8.334(44) &     1980 &       50 &    0.860 \\
 0.300 &      48 & 8.191(58) &      980 &       50 &    0.790 \\
 0.250 &      48 & 8.219(51) &      988 &       80 &    0.990 \\
 0.200 &      48 & 7.986(43) &      988 &       80 &    0.680 \\
 0.150 &      48 & 7.809(41) &      988 &       80 &    1.020 \\
 0.100 &      48 & 7.489(30) &      990 &      100 &    0.960 \\
 0.050 &      48 & 6.697(21) &      990 &      100 &    1.130 \\
 0.025 &      48 & 5.723(15) &      991 &      102 &    1.030 \\
 0.400 &      64 & 8.455(62) &      980 &       50 &    0.930 \\
 0.350 &      64 & 8.456(43) &     1980 &       50 &    1.150 \\
 0.300 &      64 & 8.418(57) &      980 &       50 &    0.790 \\
 0.250 &      64 & 8.258(50) &      980 &       50 &    1.020 \\
 0.200 &      64 & 8.192(45) &      980 &       50 &    1.020 \\
 0.150 &      64 & 8.024(38) &      980 &       50 &    0.790 \\
 0.100 &      64 & 7.810(31) &      980 &       50 &    0.830 \\
 0.050 &      64 & 7.036(22) &      980 &       50 &    1.030 \\
 0.025 &      64 & 6.125(15) &      980 &       50 &    1.250 \\
 0.400 &      96 & 8.654(68) &      980 &       50 &    1.140 \\
 0.350 &      96 & 8.518(43) &     1980 &       50 &    0.940 \\
 0.300 &      96 & 8.630(56) &      980 &       50 &    1.250 \\
 0.250 &      96 & 8.405(30) &     2760 &       25 &    1.380 \\
 0.200 &      96 & 8.458(48) &      980 &       50 &    0.870 \\
 0.150 &      96 & 8.215(39) &      980 &       50 &    0.930 \\
 0.100 &      96 & 8.009(31) &      980 &       50 &    0.900 \\
 0.050 &      96 & 7.475(22) &      980 &       50 &    0.930 \\
 0.025 &      96 & 6.672(15) &      980 &       50 &    0.860 \\
 0.400 &     128 & 8.740(54) &     1267 &       12 &    5.510 \\
 0.350 &     128 & 8.568(46) &     1980 &       50 &    0.990 \\
 0.300 &     128 & 8.636(59) &      925 &        8 &    4.250 \\
 0.250 &     128 & 8.520(52) &     1025 &        8 &    3.150 \\
 0.200 &     128 & 8.481(32) &     2000 &       20 &    1.350 \\
 0.150 &     128 & 8.413(29) &     1950 &       20 &    1.700 \\
 0.100 &     128 & 8.222(14) &     4900 &       10 &    1.300 \\
 0.050 &     128 & 7.761(16) &     2000 &       20 &    1.140 \\
 0.025 &     128 & 7.058(11) &     2000 &       20 &    1.340 \\
 0.400 &     192 & 8.754(47) &     1980 &       50 &    1.470 \\
 0.350 &     192 & 8.622(43) &     1980 &       50 &    0.980 \\
 0.300 &     192 & 8.654(41) &     1980 &       50 &    1.250 \\
 0.250 &     192 & 8.651(37) &     1980 &       50 &    1.180 \\
 0.200 &     192 & 8.560(33) &     1980 &       50 &    0.950 \\
 0.150 &     192 & 8.512(28) &     1980 &       50 &    1.020 \\
 0.100 &     192 & 8.369(24) &     1980 &       50 &    0.920 \\
 0.050 &     192 & 8.028(16) &     1980 &       50 &    0.980 \\
 0.025 &     192 & 7.478(11) &     1980 &       50 &    1.120 \\
\hline
\end{tabular}
\caption{\label{tab:lat_su3_g05}
Results for the internal energy of the 2-matrix bosonic model with gauge group SU(3) and $\lambda=0.5$.
$N_\textrm{cfgs}$ is the number of configurations used in the analysis and $N_\textrm{drop}$ is the number of molecular dynamics time units (MDTU) between successive configurations.
$\tau$ is the integrated autocorrelation time (in units of $N_\textrm{drop}$) and it is always close to unity, indicating that there are no strong autocorrelations.}
\end{table}
\begin{table}[H]
\scriptsize
\centering
\begin{tabular}{rrlrrr}
\hline
   $T$ &   $n_t$ &  $E$   &   $N_\textrm{cfgs}$ &   $N_\textrm{drop}$ &   $\tau$ \\
\hline
 0.400 &      16 & 8.084(49)  &     1980 &       50 &    0.930 \\
 0.350 &      16 & 7.994(45)  &     1980 &       50 &    0.930 \\
 0.300 &      16 & 7.880(41)  &     1980 &       50 &    0.850 \\
 0.250 &      16 & 7.608(37)  &     1980 &       50 &    0.920 \\
 0.200 &      16 & 7.291(32)  &     1980 &       50 &    0.980 \\
 0.150 &      16 & 6.936(28)  &     1980 &       50 &    0.950 \\
 0.100 &      16 & 6.352(22)  &     1980 &       50 &    1.280 \\
 0.050 &      16 & 5.170(15)  &     1980 &       50 &    1.070 \\
 0.025 &      16 & 3.9119(91) &     1980 &       50 &    1.040 \\
 0.400 &      24 & 8.433(49)  &     1980 &       50 &    1.260 \\
 0.350 &      24 & 8.310(45)  &     1980 &       50 &    1.030 \\
 0.300 &      24 & 8.214(42)  &     1980 &       50 &    0.950 \\
 0.250 &      24 & 8.052(38)  &     1980 &       50 &    0.820 \\
 0.200 &      24 & 7.807(33)  &     1980 &       50 &    0.840 \\
 0.150 &      24 & 7.470(29)  &     1980 &       50 &    0.980 \\
 0.100 &      24 & 6.943(22)  &     1980 &       50 &    1.130 \\
 0.050 &      24 & 5.880(15)  &     1980 &       50 &    1.130 \\
 0.025 &      24 & 4.6610(99) &     1980 &       50 &    0.910 \\
 0.400 &      32 & 8.759(50)  &     1980 &       50 &    0.840 \\
 0.350 &      32 & 8.572(46)  &     1980 &       50 &    1.130 \\
 0.300 &      32 & 8.521(42)  &     1980 &       50 &    0.980 \\
 0.250 &      32 & 8.247(38)  &     1980 &       50 &    1.060 \\
 0.200 &      32 & 8.151(34)  &     1980 &       50 &    0.890 \\
 0.150 &      32 & 7.857(29)  &     1980 &       50 &    0.870 \\
 0.100 &      32 & 7.328(23)  &     1980 &       50 &    1.160 \\
 0.050 &      32 & 6.311(15)  &     1980 &       50 &    1.000 \\
 0.025 &      32 & 5.182(10)  &     1980 &       50 &    0.850 \\
 0.400 &      48 & 8.834(50)  &     1980 &       50 &    0.770 \\
 0.350 &      48 & 8.808(48)  &     1980 &       50 &    0.960 \\
 0.300 &      48 & 8.654(42)  &     1980 &       50 &    1.090 \\
 0.250 &      48 & 8.654(39)  &     1980 &       50 &    0.930 \\
 0.200 &      48 & 8.568(35)  &     1980 &       50 &    0.950 \\
 0.150 &      48 & 8.251(29)  &     1980 &       50 &    0.820 \\
 0.100 &      48 & 7.825(24)  &     1980 &       50 &    1.090 \\
 0.050 &      48 & 6.972(16)  &     1980 &       50 &    0.900 \\
 0.025 &      48 & 5.877(11)  &     1980 &       50 &    1.170 \\
 0.400 &      64 & 9.033(50)  &     1980 &       50 &    0.870 \\
 0.350 &      64 & 8.958(48)  &     1980 &       50 &    1.020 \\
 0.300 &      64 & 8.934(43)  &     1980 &       50 &    0.910 \\
 0.250 &      64 & 8.768(39)  &     1980 &       50 &    0.840 \\
 0.200 &      64 & 8.605(35)  &     1980 &       50 &    0.970 \\
 0.150 &      64 & 8.482(30)  &     1980 &       50 &    0.960 \\
 0.100 &      64 & 8.139(25)  &     1980 &       50 &    0.970 \\
 0.050 &      64 & 7.362(17)  &     1980 &       50 &    0.930 \\
 0.025 &      64 & 6.324(11)  &     1980 &       50 &    0.970 \\
 0.400 &      96 & 9.173(49)  &     1980 &       50 &    1.080 \\
 0.350 &      96 & 9.092(46)  &     1980 &       50 &    0.810 \\
 0.300 &      96 & 8.997(43)  &     1980 &       50 &    1.340 \\
 0.250 &      96 & 8.981(38)  &     1980 &       50 &    0.900 \\
 0.200 &      96 & 8.851(35)  &     1980 &       50 &    1.350 \\
 0.150 &      96 & 8.748(30)  &     1980 &       50 &    0.910 \\
 0.100 &      96 & 8.476(25)  &     1980 &       50 &    1.010 \\
 0.050 &      96 & 7.844(17)  &     1980 &       50 &    0.930 \\
 0.025 &      96 & 6.928(12)  &     1980 &       50 &    0.970 \\
 0.400 &     128 & 9.200(50)  &     1980 &       50 &    0.940 \\
 0.350 &     128 & 9.104(47)  &     1980 &       50 &    0.840 \\
 0.300 &     128 & 9.143(43)  &     1980 &       50 &    0.990 \\
 0.250 &     128 & 9.112(40)  &     1980 &       50 &    0.960 \\
 0.200 &     128 & 8.962(35)  &     1980 &       50 &    0.850 \\
 0.150 &     128 & 8.866(31)  &     1980 &       50 &    0.950 \\
 0.100 &     128 & 8.715(24)  &     1980 &       50 &    0.990 \\
 0.050 &     128 & 8.138(17)  &     1980 &       50 &    1.130 \\
 0.025 &     128 & 7.326(12)  &     1980 &       50 &    1.100 \\
 0.400 &     192 & 9.238(51)  &     1980 &       50 &    1.380 \\
 0.350 &     192 & 9.252(48)  &     1980 &       50 &    1.110 \\
 0.300 &     192 & 9.182(43)  &     1980 &       50 &    0.980 \\
 0.250 &     192 & 9.148(40)  &     1980 &       50 &    0.790 \\
 0.200 &     192 & 9.068(35)  &     1980 &       50 &    1.130 \\
 0.150 &     192 & 9.012(30)  &     1980 &       50 &    0.940 \\
 0.100 &     192 & 8.905(26)  &     1820 &       50 &    0.790 \\
 0.050 &     192 & 8.469(17)  &     1980 &       50 &    1.100 \\
 0.025 &     192 & 7.834(12)  &     1980 &       50 &    1.050 \\
\hline
\end{tabular}
\caption{\label{tab:lat_su3_g10}
Results for the internal energy of the 2-matrix bosonic model with gauge group SU(3) and $\lambda=1.0$.
$N_\textrm{cfgs}$ is the number of configurations used in the analysis and $N_\textrm{drop}$ is the number of molecular dynamics time units (MDTU) between successive configurations.
$\tau$ is the integrated autocorrelation time (in units of $N_\textrm{drop}$) and it is always close to unity, indicating that there are no strong autocorrelations.}
\end{table}
\begin{table}[H]
\scriptsize
\centering
\begin{tabular}{rrlrrr}
\hline
   $T$ &   $n_t$ &  $E$   &   $N_\textrm{cfgs}$ &   $N_\textrm{drop}$ &   $\tau$ \\
\hline
 0.400 &      16 & 8.831(54)  &     1980 &       50 &    1.120 \\
 0.350 &      16 & 8.510(48)  &     1980 &       50 &    0.900 \\
 0.300 &      16 & 8.275(44)  &     1980 &       50 &    0.990 \\
 0.250 &      16 & 8.104(40)  &     1980 &       50 &    0.960 \\
 0.200 &      16 & 7.780(35)  &     1980 &       50 &    0.830 \\
 0.150 &      16 & 7.322(29)  &     1980 &       50 &    1.060 \\
 0.100 &      16 & 6.614(24)  &     1980 &       50 &    0.950 \\
 0.050 &      16 & 5.316(15)  &     1980 &       50 &    0.840 \\
 0.025 &      16 & 3.9537(95) &     1980 &       50 &    0.960 \\
 0.400 &      24 & 9.186(54)  &     1980 &       50 &    1.100 \\
 0.350 &      24 & 9.069(51)  &     1980 &       50 &    0.910 \\
 0.300 &      24 & 8.864(46)  &     1980 &       50 &    0.910 \\
 0.250 &      24 & 8.606(41)  &     1980 &       50 &    0.840 \\
 0.200 &      24 & 8.312(38)  &     1980 &       50 &    0.730 \\
 0.150 &      24 & 7.901(32)  &     1980 &       50 &    1.050 \\
 0.100 &      24 & 7.349(24)  &     1980 &       50 &    0.960 \\
 0.050 &      24 & 6.064(16)  &     1980 &       50 &    0.950 \\
 0.025 &      24 & 4.762(10)  &     1980 &       50 &    1.140 \\
 0.400 &      32 & 9.394(54)  &     1980 &       50 &    0.980 \\
 0.350 &      32 & 9.097(48)  &     1980 &       50 &    0.870 \\
 0.300 &      32 & 9.238(47)  &     1980 &       50 &    1.150 \\
 0.250 &      32 & 8.954(40)  &     1980 &       50 &    1.070 \\
 0.200 &      32 & 8.714(37)  &     1980 &       50 &    1.120 \\
 0.150 &      32 & 8.337(31)  &     1980 &       50 &    0.910 \\
 0.100 &      32 & 7.761(25)  &     1980 &       50 &    1.050 \\
 0.050 &      32 & 6.604(17)  &     1980 &       50 &    1.150 \\
 0.025 &      32 & 5.321(11)  &     1980 &       50 &    1.000 \\
 0.400 &      48 & 9.627(54)  &     1980 &       50 &    1.000 \\
 0.350 &      48 & 9.552(52)  &     1980 &       50 &    0.870 \\
 0.300 &      48 & 9.385(47)  &     1980 &       50 &    0.980 \\
 0.250 &      48 & 9.348(42)  &     1980 &       50 &    0.960 \\
 0.200 &      48 & 9.184(38)  &     1980 &       50 &    1.000 \\
 0.150 &      48 & 8.813(32)  &     1980 &       50 &    0.890 \\
 0.100 &      48 & 8.331(25)  &     1980 &       50 &    1.130 \\
 0.050 &      48 & 7.315(18)  &     1980 &       50 &    0.950 \\
 0.025 &      48 & 6.083(11)  &     1980 &       50 &    1.050 \\
 0.400 &      64 & 9.829(55)  &     1980 &       50 &    0.990 \\
 0.350 &      64 & 9.711(51)  &     1980 &       50 &    0.860 \\
 0.300 &      64 & 9.570(47)  &     1980 &       50 &    0.950 \\
 0.250 &      64 & 9.470(42)  &     1980 &       50 &    0.940 \\
 0.200 &      64 & 9.325(37)  &     1980 &       50 &    0.890 \\
 0.150 &      64 & 9.112(33)  &     1980 &       50 &    0.910 \\
 0.100 &      64 & 8.669(26)  &     1980 &       50 &    1.080 \\
 0.050 &      64 & 7.756(18)  &     1980 &       50 &    0.920 \\
 0.025 &      64 & 6.605(12)  &     1980 &       50 &    0.980 \\
 0.400 &      96 & 9.882(54)  &     1980 &       50 &    1.130 \\
 0.350 &      96 & 9.927(52)  &     1980 &       50 &    0.980 \\
 0.300 &      96 & 9.822(46)  &     1980 &       50 &    1.040 \\
 0.250 &      96 & 9.778(43)  &     1980 &       50 &    1.120 \\
 0.200 &      96 & 9.647(39)  &     1980 &       50 &    0.990 \\
 0.150 &      96 & 9.414(33)  &     1980 &       50 &    0.940 \\
 0.100 &      96 & 9.114(26)  &     1980 &       49 &    1.110 \\
 0.050 &      96 & 8.312(18)  &     1980 &       49 &    1.010 \\
 0.025 &      96 & 7.290(12)  &     1980 &       50 &    1.130 \\
 0.400 &     128 & 9.920(53)  &     1980 &       50 &    1.160 \\
 0.350 &     128 & 10.017(51) &     1980 &       50 &    1.150 \\
 0.300 &     128 & 10.047(49) &     1980 &       50 &    0.920 \\
 0.250 &     128 & 9.864(42)  &     1980 &       50 &    0.880 \\
 0.200 &     128 & 9.797(39)  &     1980 &       50 &    0.990 \\
 0.150 &     128 & 9.656(34)  &     1980 &       50 &    0.950 \\
 0.100 &     128 & 9.338(27)  &     1980 &       50 &    1.200 \\
 0.050 &     128 & 8.677(18)  &     1980 &       50 &    0.960 \\
 0.025 &     128 & 7.748(12)  &     1980 &       50 &    0.940 \\
 0.400 &     192 & 10.129(55) &     1980 &       50 &    0.980 \\
 0.350 &     192 & 10.124(52) &     1980 &       50 &    0.970 \\
 0.300 &     192 & 10.074(46) &     1980 &       50 &    0.980 \\
 0.250 &     192 & 9.915(44)  &     1980 &       50 &    0.940 \\
 0.200 &     192 & 9.891(40)  &     1980 &       50 &    1.120 \\
 0.150 &     192 & 9.848(34)  &     1980 &       50 &    1.090 \\
 0.100 &     192 & 9.659(27)  &     1980 &       50 &    1.120 \\
 0.050 &     192 & 9.128(19)  &     1980 &       50 &    0.970 \\
 0.025 &     192 & 8.335(13)  &     1980 &       50 &    1.080 \\
\hline
\end{tabular}
\caption{\label{tab:lat_su3_g20}
Results for the internal energy of the 2-matrix bosonic model with gauge group SU(3) and $\lambda=2.0$.
$N_\textrm{cfgs}$ is the number of configurations used in the analysis and $N_\textrm{drop}$ is the number of molecular dynamics time units (MDTU) between successive configurations.
$\tau$ is the integrated autocorrelation time (in units of $N_\textrm{drop}$) and it is always close to unity, indicating that there are no strong autocorrelations.}
\end{table}
\begin{table}[H]
\scriptsize
\centering
\begin{tabular}{rrlrrr}
\hline
   $T$ &   $n_t$ &  $E$   &   $N_\textrm{cfgs}$ &   $N_\textrm{drop}$ &   $\tau$ \\
\hline
 0.400 &      16 & 3.296(32)  &     1980 &       50 &    0.960 \\
 0.300 &      16 & 3.121(26)  &     1980 &       50 &    0.930 \\
 0.250 &      16 & 3.017(24)  &     1980 &       50 &    0.820 \\
 0.200 &      16 & 2.924(22)  &     1980 &       50 &    0.920 \\
 0.150 &      16 & 2.750(18)  &     1980 &       50 &    0.780 \\
 0.100 &      16 & 2.477(15)  &     1980 &       50 &    1.120 \\
 0.050 &      16 & 1.9920(94) &     1980 &       50 &    1.080 \\
 0.400 &      24 & 3.428(32)  &     1980 &       50 &    1.120 \\
 0.300 &      24 & 3.289(28)  &     1980 &       50 &    0.920 \\
 0.250 &      24 & 3.237(25)  &     1980 &       50 &    1.240 \\
 0.200 &      24 & 3.101(22)  &     1980 &       50 &    1.060 \\
 0.150 &      24 & 2.935(19)  &     1980 &       50 &    1.120 \\
 0.100 &      24 & 2.722(14)  &     1980 &       50 &    0.990 \\
 0.050 &      24 & 2.2931(99) &     1980 &       50 &    1.050 \\
 0.400 &      32 & 3.531(32)  &     1980 &       50 &    0.940 \\
 0.300 &      32 & 3.373(28)  &     1980 &       50 &    1.000 \\
 0.250 &      32 & 3.394(27)  &     1980 &       50 &    1.150 \\
 0.200 &      32 & 3.258(22)  &     1980 &       50 &    0.990 \\
 0.150 &      32 & 3.157(20)  &     1980 &       50 &    1.000 \\
 0.100 &      32 & 2.922(16)  &     1980 &       50 &    0.880 \\
 0.050 &      32 & 2.464(10)  &     1980 &       50 &    0.990 \\
 0.400 &      48 & 3.649(34)  &     1980 &       50 &    1.160 \\
 0.300 &      48 & 3.524(28)  &     1980 &       50 &    0.980 \\
 0.250 &      48 & 3.518(26)  &     1980 &       50 &    1.040 \\
 0.200 &      48 & 3.446(24)  &     1980 &       50 &    1.120 \\
 0.150 &      48 & 3.328(20)  &     1980 &       50 &    1.140 \\
 0.100 &      48 & 3.129(16)  &     1980 &       50 &    0.930 \\
 0.050 &      48 & 2.721(10)  &     1980 &       50 &    0.970 \\
 0.400 &      64 & 3.666(34)  &     1980 &       50 &    1.040 \\
 0.300 &      64 & 3.697(30)  &     1980 &       50 &    0.930 \\
 0.250 &      64 & 3.560(25)  &     1980 &       50 &    0.960 \\
 0.200 &      64 & 3.512(23)  &     1980 &       50 &    1.180 \\
 0.150 &      64 & 3.398(19)  &     1980 &       50 &    0.820 \\
 0.100 &      64 & 3.246(16)  &     1980 &       50 &    1.010 \\
 0.050 &      64 & 2.916(11)  &     1980 &       50 &    1.100 \\
 0.400 &      96 & 3.736(34)  &     1980 &       50 &    1.170 \\
 0.300 &      96 & 3.733(29)  &     1980 &       50 &    1.100 \\
 0.250 &      96 & 3.658(27)  &     1980 &       50 &    0.800 \\
 0.200 &      96 & 3.630(24)  &     1980 &       50 &    1.040 \\
 0.150 &      96 & 3.530(20)  &     1980 &       50 &    0.870 \\
 0.100 &      96 & 3.379(16)  &     1980 &       50 &    0.880 \\
 0.050 &      96 & 3.122(11)  &     1980 &       50 &    0.950 \\
 0.400 &     128 & 3.796(34)  &     1980 &       50 &    0.870 \\
 0.300 &     128 & 3.725(28)  &     1980 &       50 &    1.170 \\
 0.250 &     128 & 3.700(26)  &     1980 &       50 &    1.030 \\
 0.200 &     128 & 3.640(23)  &     1980 &       50 &    1.050 \\
 0.150 &     128 & 3.617(20)  &     1980 &       50 &    0.950 \\
 0.100 &     128 & 3.507(16)  &     1980 &       50 &    0.990 \\
 0.050 &     128 & 3.262(12)  &     1980 &       50 &    1.170 \\
 0.400 &     192 & 3.741(34)  &     1980 &       50 &    0.940 \\
 0.300 &     192 & 3.740(28)  &     1980 &       50 &    1.060 \\
 0.250 &     192 & 3.762(26)  &     1980 &       50 &    0.990 \\
 0.200 &     192 & 3.737(23)  &     1980 &       50 &    0.900 \\
 0.150 &     192 & 3.705(20)  &     1980 &       50 &    0.960 \\
 0.100 &     192 & 3.581(16)  &     1980 &       50 &    0.950 \\
 0.050 &     192 & 3.425(12)  &     1980 &       50 &    1.170 \\
\hline
\end{tabular}
\caption{\label{tab:lat_su2_g20}
Results for the internal energy of the 2-matrix bosonic model with gauge group SU(2) and $\lambda=2.0$.
$N_\textrm{cfgs}$ is the number of configurations used in the analysis and $N_\textrm{drop}$ is the number of molecular dynamics time units (MDTU) between successive configurations.
$\tau$ is the integrated autocorrelation time (in units of $N_\textrm{drop}$) and it is always close to unity, indicating that there are no strong autocorrelations.}
\end{table}
\begin{table}[H]
\scriptsize
\centering
\begin{tabular}{rrlrrr}
\hline
   $T$ &   $n_t$ &  $E$   &   $N_\textrm{cfgs}$ &   $N_\textrm{drop}$ &   $\tau$ \\
\hline
 0.400 &      16 & 3.070(31)  &     1980 &       50 &    1.050 \\
 0.300 &      16 & 2.973(26)  &     1980 &       50 &    1.010 \\
 0.250 &      16 & 2.880(23)  &     1980 &       50 &    0.980 \\
 0.200 &      16 & 2.799(20)  &     1980 &       50 &    0.920 \\
 0.150 &      16 & 2.615(17)  &     1980 &       50 &    1.090 \\
 0.100 &      16 & 2.359(13)  &     1980 &       50 &    0.760 \\
 0.050 &      16 & 1.9534(87) &     1980 &       50 &    0.950 \\
 0.400 &      24 & 3.169(30)  &     1980 &       50 &    1.180 \\
 0.300 &      24 & 3.103(26)  &     1980 &       50 &    0.880 \\
 0.250 &      24 & 3.017(22)  &     1980 &       51 &    0.970 \\
 0.200 &      24 & 2.927(20)  &     1981 &       51 &    0.890 \\
 0.150 &      24 & 2.819(18)  &     1980 &       51 &    1.050 \\
 0.100 &      24 & 2.592(14)  &     1980 &       50 &    1.000 \\
 0.050 &      24 & 2.2188(93) &     1980 &       50 &    1.380 \\
 0.400 &      32 & 3.273(30)  &     1980 &       50 &    1.090 \\
 0.300 &      32 & 3.160(26)  &     1980 &       50 &    1.150 \\
 0.250 &      32 & 3.111(24)  &     1980 &       50 &    1.040 \\
 0.200 &      32 & 3.059(21)  &     1980 &       50 &    1.050 \\
 0.150 &      32 & 2.918(17)  &     1981 &       51 &    0.980 \\
 0.100 &      32 & 2.738(14)  &     1981 &       51 &    0.970 \\
 0.050 &      32 & 2.3711(94) &     1980 &       50 &    1.100 \\
 0.400 &      48 & 3.374(32)  &     1980 &       50 &    1.000 \\
 0.300 &      48 & 3.281(27)  &     1980 &       50 &    0.970 \\
 0.250 &      48 & 3.209(23)  &     1980 &       50 &    1.020 \\
 0.200 &      48 & 3.194(21)  &     1980 &       50 &    1.000 \\
 0.150 &      48 & 3.081(18)  &     1980 &       50 &    0.980 \\
 0.100 &      48 & 2.910(14)  &     1980 &       51 &    0.970 \\
 0.050 &      48 & 2.5936(95) &     1980 &       50 &    0.870 \\
 0.400 &      64 & 3.393(32)  &     1980 &       50 &    0.980 \\
 0.300 &      64 & 3.354(26)  &     1980 &       50 &    0.920 \\
 0.250 &      64 & 3.279(24)  &     1980 &       50 &    1.160 \\
 0.200 &      64 & 3.262(21)  &     1980 &       50 &    0.950 \\
 0.150 &      64 & 3.163(18)  &     1980 &       50 &    0.880 \\
 0.100 &      64 & 3.040(15)  &     1980 &       50 &    1.000 \\
 0.050 &      64 & 2.7500(99) &     1980 &       50 &    0.950 \\
 0.400 &      96 & 3.430(22)  &     3980 &       50 &    1.030 \\
 0.300 &      96 & 3.341(27)  &     1980 &       50 &    1.080 \\
 0.250 &      96 & 3.373(24)  &     1980 &       50 &    1.040 \\
 0.200 &      96 & 3.339(22)  &     1980 &       50 &    1.380 \\
 0.150 &      96 & 3.288(19)  &     1980 &       50 &    1.250 \\
 0.100 &      96 & 3.179(15)  &     1980 &       50 &    0.870 \\
 0.050 &      96 & 2.941(10)  &     1981 &       51 &    0.980 \\
 0.300 &     128 & 3.420(27)  &     1980 &       50 &    1.070 \\
 0.250 &     128 & 3.418(24)  &     1980 &       50 &    0.990 \\
 0.200 &     128 & 3.367(22)  &     1980 &       50 &    1.270 \\
 0.150 &     128 & 3.334(18)  &     1980 &       50 &    0.990 \\
 0.100 &     128 & 3.237(15)  &     1980 &       50 &    0.950 \\
 0.050 &     128 & 3.029(10)  &     1980 &       50 &    0.980 \\
 0.300 &     192 & 3.461(27)  &     1980 &       50 &    1.390 \\
 0.250 &     192 & 3.417(24)  &     1980 &       50 &    1.090 \\
 0.200 &     192 & 3.437(22)  &     1980 &       50 &    1.130 \\
 0.150 &     192 & 3.378(18)  &     1980 &       50 &    1.110 \\
 0.100 &     192 & 3.332(15)  &     1980 &       50 &    0.940 \\
 0.050 &     192 & 3.176(11)  &     1980 &       50 &    0.940 \\
\hline
\end{tabular}
\caption{\label{tab:lat_su2_g10}
Results for the internal energy of the 2-matrix bosonic model with gauge group SU(2) and $\lambda=1.0$.
$N_\textrm{cfgs}$ is the number of configurations used in the analysis and $N_\textrm{drop}$ is the number of molecular dynamics time units (MDTU) between successive configurations.
$\tau$ is the integrated autocorrelation time (in units of $N_\textrm{drop}$) and it is always close to unity, indicating that there are no strong autocorrelations.}
\end{table}
\begin{table}[H]
\scriptsize
\centering
\begin{tabular}{rrlrrr}
\hline
   $T$ &   $n_t$ &  $E$   &   $N_\textrm{cfgs}$ &   $N_\textrm{drop}$ &   $\tau$ \\
\hline
 0.400 &      16 & 2.941(28)  &     1980 &       50 &    1.000 \\
 0.300 &      16 & 2.805(24)  &     1980 &       50 &    0.870 \\
 0.250 &      16 & 2.750(21)  &     1980 &       50 &    1.000 \\
 0.200 &      16 & 2.645(19)  &     1980 &       50 &    0.850 \\
 0.150 &      16 & 2.527(16)  &     1980 &       50 &    1.070 \\
 0.100 &      16 & 2.303(13)  &     1980 &       50 &    1.080 \\
 0.050 &      16 & 1.9063(85) &     1980 &       50 &    1.010 \\
 0.400 &      24 & 3.038(28)  &     1980 &       50 &    1.080 \\
 0.300 &      24 & 2.915(24)  &     1980 &       50 &    0.770 \\
 0.250 &      24 & 2.876(22)  &     1980 &       50 &    0.990 \\
 0.200 &      24 & 2.819(20)  &     1980 &       50 &    1.100 \\
 0.150 &      24 & 2.673(16)  &     1980 &       50 &    1.130 \\
 0.100 &      24 & 2.527(14)  &     1980 &       50 &    1.160 \\
 0.050 &      24 & 2.1593(88) &     1980 &       50 &    1.080 \\
 0.400 &      32 & 3.058(27)  &     1980 &       50 &    1.150 \\
 0.300 &      32 & 3.050(24)  &     1980 &       50 &    0.940 \\
 0.250 &      32 & 2.946(22)  &     1980 &       50 &    0.980 \\
 0.200 &      32 & 2.927(19)  &     1980 &       50 &    0.930 \\
 0.150 &      32 & 2.806(17)  &     1980 &       50 &    1.040 \\
 0.100 &      32 & 2.631(13)  &     1980 &       50 &    1.190 \\
 0.050 &      32 & 2.3029(89) &     1980 &       50 &    0.930 \\
 0.400 &      48 & 3.137(28)  &     1980 &       50 &    1.180 \\
 0.300 &      48 & 3.087(25)  &     1980 &       50 &    1.150 \\
 0.250 &      48 & 3.102(23)  &     1980 &       50 &    0.940 \\
 0.200 &      48 & 3.050(20)  &     1980 &       50 &    0.970 \\
 0.150 &      48 & 2.916(17)  &     1980 &       50 &    1.100 \\
 0.100 &      48 & 2.809(14)  &     1980 &       50 &    1.050 \\
 0.050 &      48 & 2.4901(91) &     1980 &       50 &    0.990 \\
 0.400 &      64 & 3.223(29)  &     1980 &       50 &    0.820 \\
 0.300 &      64 & 3.133(24)  &     1980 &       50 &    1.000 \\
 0.250 &      64 & 3.105(22)  &     1980 &       50 &    1.110 \\
 0.200 &      64 & 3.137(20)  &     1980 &       50 &    0.940 \\
 0.150 &      64 & 3.028(17)  &     1980 &       50 &    1.050 \\
 0.100 &      64 & 2.899(14)  &     1980 &       50 &    1.060 \\
 0.050 &      64 & 2.6519(93) &     1980 &       50 &    1.020 \\
 0.400 &      96 & 3.199(28)  &     1980 &       50 &    1.130 \\
 0.300 &      96 & 3.195(24)  &     1980 &       49 &    0.730 \\
 0.250 &      96 & 3.178(22)  &     1980 &       50 &    1.060 \\
 0.200 &      96 & 3.134(20)  &     1980 &       50 &    1.090 \\
 0.150 &      96 & 3.082(17)  &     1980 &       50 &    0.860 \\
 0.100 &      96 & 3.037(14)  &     1980 &       50 &    1.150 \\
 0.050 &      96 & 2.7978(97) &     1980 &       50 &    1.190 \\
 0.400 &     128 & 3.252(29)  &     1980 &       50 &    1.040 \\
 0.300 &     128 & 3.204(24)  &     1980 &       50 &    0.970 \\
 0.250 &     128 & 3.141(22)  &     1980 &       49 &    0.710 \\
 0.200 &     128 & 3.164(20)  &     1980 &       50 &    1.100 \\
 0.150 &     128 & 3.142(18)  &     1980 &       50 &    0.940 \\
 0.100 &     128 & 3.077(14)  &     1980 &       50 &    0.900 \\
 0.050 &     128 & 2.903(10)  &     1980 &       50 &    0.990 \\
 0.400 &     192 & 3.287(30)  &     1980 &       50 &    1.310 \\
 0.300 &     192 & 3.231(23)  &     1980 &       50 &    1.200 \\
 0.250 &     192 & 3.196(22)  &     1980 &       50 &    1.070 \\
 0.200 &     192 & 3.249(21)  &     1980 &       50 &    1.070 \\
 0.150 &     192 & 3.230(18)  &     1980 &       50 &    0.940 \\
 0.100 &     192 & 3.150(14)  &     1980 &       50 &    0.790 \\
 0.050 &     192 & 3.018(10)  &     1980 &       50 &    0.910 \\
\hline
\end{tabular}
\caption{\label{tab:lat_su2_g05}
Results for the internal energy of the 2-matrix bosonic model with gauge group SU(2) and $\lambda=0.5$.
$N_\textrm{cfgs}$ is the number of configurations used in the analysis and $N_\textrm{drop}$ is the number of molecular dynamics time units (MDTU) between successive configurations.
$\tau$ is the integrated autocorrelation time (in units of $N_\textrm{drop}$) and it is always close to unity, indicating that there are no strong autocorrelations.}
\end{table}

\subsection{Tables and figures for the results of the systematic fitting of the continuum limit}
\begin{table}[H]
\scriptsize
    \centering
    \begin{tabular}{rrlr||rrlr}
    \hline
       $a_\textrm{max}$ &   $n_p$ & $E$ &   $\chi^2$/dof & $a_\textrm{max}$ &   $n_p$ & $E$ &   $\chi^2$/dof \\
    \hline
    0.05 &  1 & 8.823(33)  &   1.11 & 0.30 &  1 & 8.6919(87) &   2.13 \\
    0.05 &  2 & 8.822(68)  &   1.11 & 0.30 &  2 & 8.773(15)  &   1.30 \\
    0.05 &  3 & 8.822(68)  &   1.11 & 0.30 &  3 & 8.809(22)  &   1.20 \\
    0.10 &  1 & 8.768(17)  &   1.12 & 0.40 &  1 & 8.6369(74) &   4.49 \\
    0.10 &  2 & 8.836(38)  &   0.99 & 0.40 &  2 & 8.767(12)  &   1.24 \\
    0.10 &  3 & 8.836(38)  &   0.99 & 0.40 &  3 & 8.794(20)  &   1.19 \\
    0.20 &  1 & 8.722(11)  &   1.82 & 0.50 &  1 & 8.5880(67) &   7.94 \\
    0.20 &  2 & 8.807(19)  &   1.24 & 0.50 &  2 & 8.747(10)  &   1.36 \\
    0.20 &  3 & 8.809(30)  &   1.24 & 0.50 &  3 & 8.789(15)  &   1.14 \\
    \hline
    \end{tabular}
    \caption{Systematic fitting for the bosonic SU(3) model with  $\lambda=0.5$}
    \label{tab:lat_su3_g05_fit_sys}
\end{table}
\begin{table}[H]
\scriptsize
    \centering
    \begin{tabular}{rrlr||rrlr}
    \hline
       $a_\textrm{max}$ &   $n_p$ & $E$ &   $\chi^2$/dof & $a_\textrm{max}$ &   $n_p$ & $E$ &   $\chi^2$/dof \\
    \hline
    0.05 & 1 & 9.359(35)  & 0.57 & 0.30 &  1 & 9.2273(86) &   2.72 \\
    0.05 & 2 & 9.375(70)  & 0.57 & 0.30 &  2 & 9.337(14)  &   0.99 \\
    0.05 & 3 & 9.375(70)  & 0.57 & 0.30 &  3 & 9.367(21)  &   0.92 \\
    0.10 & 1 & 9.328(17)  & 1.00 & 0.40 &  1 & 9.1622(73) &   6.30 \\
    0.10 & 2 & 9.381(38)  & 0.93 & 0.40 &  2 & 9.326(12)  &   1.04 \\
    0.10 & 3 & 9.381(39)  & 0.93 & 0.40 &  3 & 9.361(19)  &   0.95 \\
    0.20 & 1 & 9.281(11)  & 1.45 & 0.50 &  1 & 9.0787(64) &  14.87 \\
    0.20 & 2 & 9.355(19)  & 1.01 & 0.50 &  2 & 9.307(10)  &   1.21 \\
    0.20 & 3 & 9.359(31)  & 1.01 & 0.50 &  3 & 9.348(15)  &   0.99 \\
    \hline
    \end{tabular}
    \caption{Systematic fitting for the bosonic SU(3) model with  $\lambda=1.0$}
    \label{tab:lat_su3_g10_fit_sys}
\end{table}
\begin{table}[H]
\scriptsize
    \centering
    \begin{tabular}{rrlr||rrlr}
    \hline
       $a_\textrm{max}$ &   $n_p$ & $E$ &   $\chi^2$/dof & $a_\textrm{max}$ &   $n_p$ & $E$ &   $\chi^2$/dof \\
    \hline
    0.05 &  1 & 10.235(38)  & 1.09 & 0.30 &  1 & 10.0660(93) &   4.43 \\
    0.05 &  2 & 10.265(73)  & 1.08 & 0.30 &  2 & 10.227(15)  &   1.34 \\
    0.05 &  3 & 10.265(73)  & 1.08 & 0.30 &  3 & 10.264(23)  &   1.25 \\
    0.10 &  1 & 10.247(19)  & 1.10 & 0.40 &  1 & 9.9657(80)  &  11.43 \\
    0.10 &  2 & 10.236(41)  & 1.09 & 0.40 &  2 & 10.213(13)  &   1.31 \\
    0.10 &  3 & 10.237(42)  & 1.09 & 0.40 &  3 & 10.256(21)  &   1.19 \\
    0.20 &  1 & 10.142(12)  & 2.21 & 0.50 &  1 & 9.8484(70)  &  25.52 \\
    0.20 &  2 & 10.251(21)  & 1.41 & 0.50 &  2 & 10.176(11)  &   1.78 \\
    0.20 &  3 & 10.288(33)  & 1.36 & 0.50 &  3 & 10.249(16)  &   1.21 \\
    \hline
    \end{tabular}
    \caption{Systematic fitting for the bosonic SU(3) model with  $\lambda=2.0$}
    \label{tab:lat_su3_g20_fit_sys}
\end{table}
\begin{table}[H]
\scriptsize
    \centering
    \begin{tabular}{rrlr||rrlr}
    \hline
       $a_\textrm{max}$ &   $n_p$ & $E$ &   $\chi^2$/dof & $a_\textrm{max}$ &   $n_p$ & $E$ &   $\chi^2$/dof \\
    \hline
    0.05 &  1 & 3.519(31)  &  0.57 & 0.30 &  1 & 3.4579(58) &   1.80 \\
    0.05 &  2 & 3.526(77)  &  0.57 & 0.30 &  2 & 3.5159(98) &   0.58 \\
    0.05 &  3 & 3.526(77)  &  0.57 & 0.30 &  3 & 3.516(16)  &   0.58 \\
    0.10 &  1 & 3.507(13)  &  0.55 & 0.40 &  1 & 3.4351(50) &   3.17 \\
    0.10 &  2 & 3.497(33)  &  0.55 & 0.40 &  2 & 3.5074(84) &   0.74 \\
    0.10 &  3 & 3.498(34)  &  0.55 & 0.40 &  3 & 3.523(14)  &   0.70 \\
    0.20 &  1 & 3.4854(72) &  0.64 & 0.50 &  1 & 3.4080(44) &   5.46 \\
    0.20 &  2 & 3.514(14)  &  0.50 & 0.50 &  2 & 3.4923(71) &   0.95 \\
    0.20 &  3 & 3.513(26)  &  0.50 & 0.50 &  3 & 3.522(11)  &   0.68 \\
    \hline
    \end{tabular}
    \caption{Systematic fitting for the bosonic SU(2) model with  $\lambda=1.0$}
    \label{tab:lat_su2_g10_fit_sys}
\end{table}
\begin{table}[H]
\scriptsize
    \centering
    \begin{tabular}{rrlr||rrlr}
    \hline
       $a_\textrm{max}$ &   $n_p$ & $E$ &   $\chi^2$/dof & $a_\textrm{max}$ &   $n_p$ & $E$ &   $\chi^2$/dof \\
    \hline
    0.05 &  1 & 3.836(28)  & 0.68 & 0.30 &  1 & 3.7745(62) &  2.25 \\
    0.05 &  2 & 3.775(62)  & 0.59 & 0.30 &  2 & 3.831(10)  &  1.19 \\
    0.05 &  3 & 3.775(62)  & 0.59 & 0.30 &  3 & 3.864(16)  &  1.03 \\
    0.10 &  1 & 3.835(13)  & 1.02 & 0.40 &  1 & 3.7414(54) &  4.63 \\
    0.10 &  2 & 3.847(30)  & 1.01 & 0.40 &  2 & 3.8341(89) &  1.12 \\
    0.10 &  3 & 3.845(31)  & 1.01 & 0.40 &  3 & 3.848(14)  &  1.09 \\
    0.20 &  1 & 3.8028(77) & 1.55 & 0.50 &  1 & 3.7056(47) &  7.96 \\
    0.20 &  2 & 3.854(14)  & 1.10 & 0.50 &  2 & 3.8147(76) &  1.43 \\
    0.20 &  3 & 3.849(25)  & 1.10 & 0.50 &  3 & 3.851(11)  &  1.07 \\
    \hline
    \end{tabular}
    \caption{Systematic fitting for the bosonic SU(2) model with  $\lambda=2.0$}
    \label{tab:lat_su2_g20_fit_sys}
\end{table}
\clearpage

\begin{figure}
    \centering
    \includegraphics[width=0.45\textwidth]{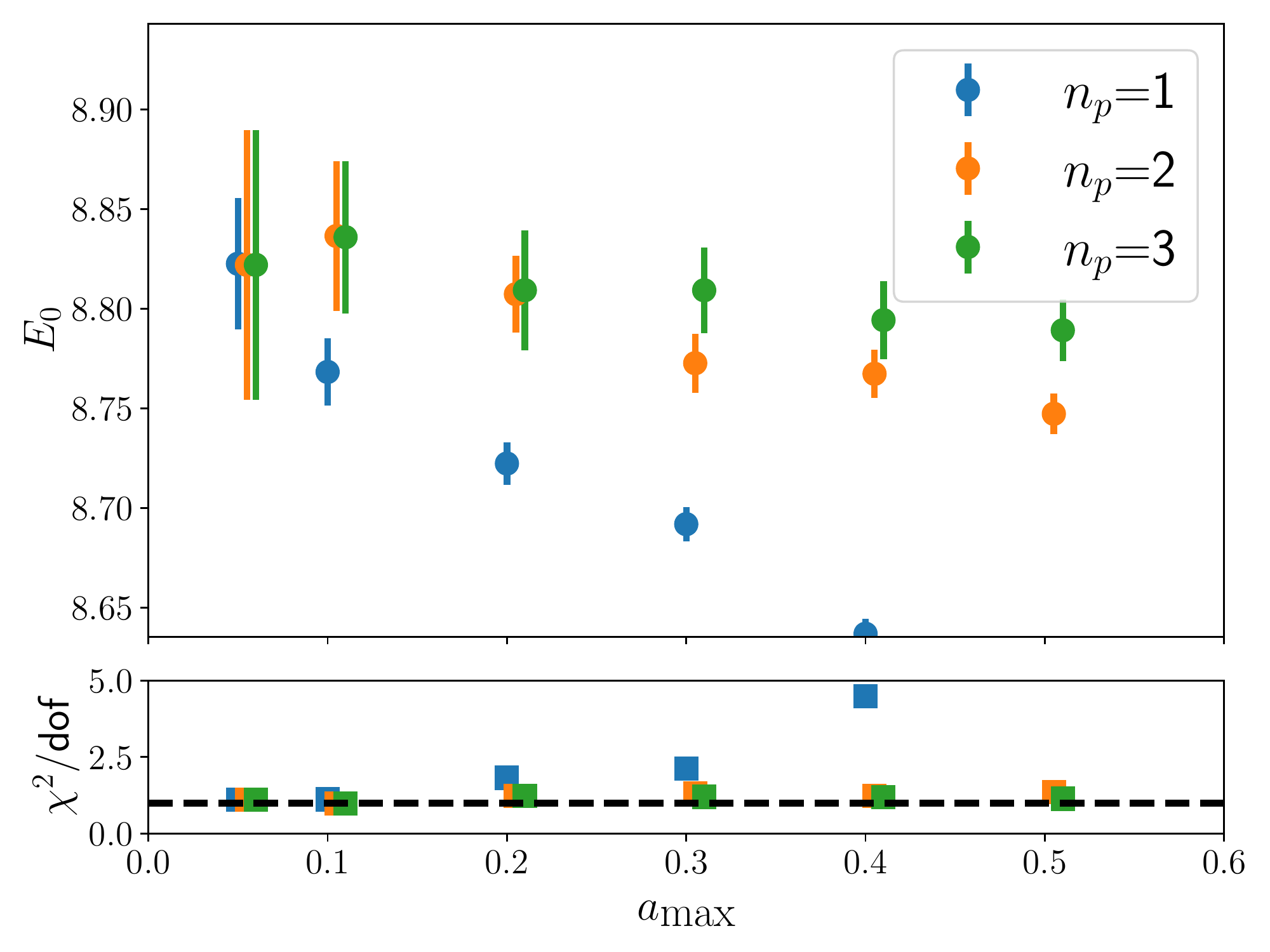}
    \caption{\label{fig:lat_su3_g05_fit_sys}
    Bosonic SU(3) model with $\lambda=0.5$.
    Results of fitting different data portions with polynomials of different order $n_p$.
    The lower panel shows the reduced $\chi^2$ which becomes very large for the low order polynomials when larger lattice spacings are included.
    }
\end{figure}
\begin{figure}
    \centering
    \includegraphics[width=0.45\textwidth]{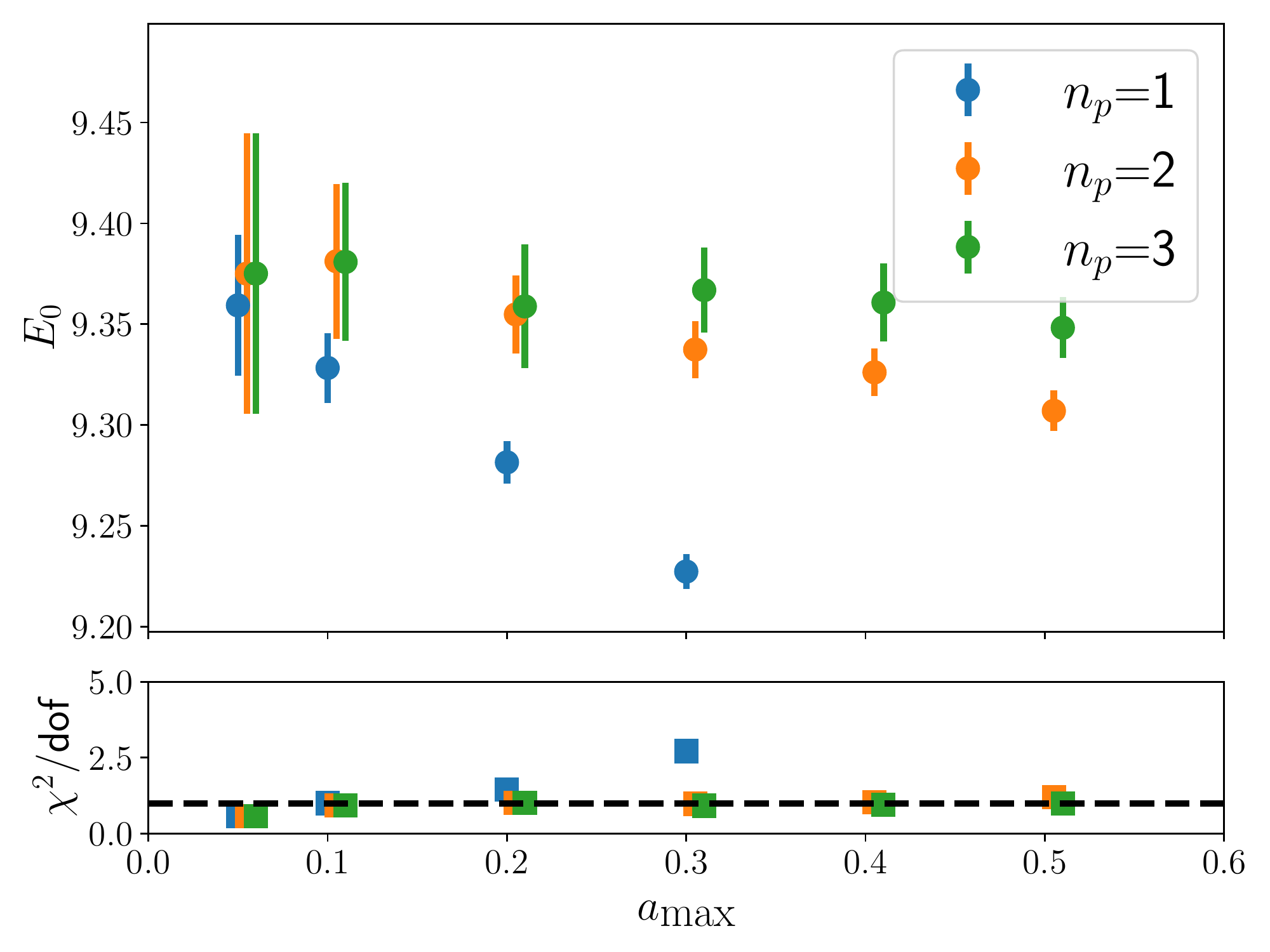}
    \caption{\label{fig:lat_su3_g10_fit_sys}
    Bosonic SU(3) model with $\lambda=1.0$.
    Results of fitting different data portions with polynomials of different order $n_p$.
    The lower panel shows the reduced $\chi^2$ which becomes very large for the low order polynomials when larger lattice spacings are included.
    }
\end{figure}
\begin{figure}
    \centering
    \includegraphics[width=0.45\textwidth]{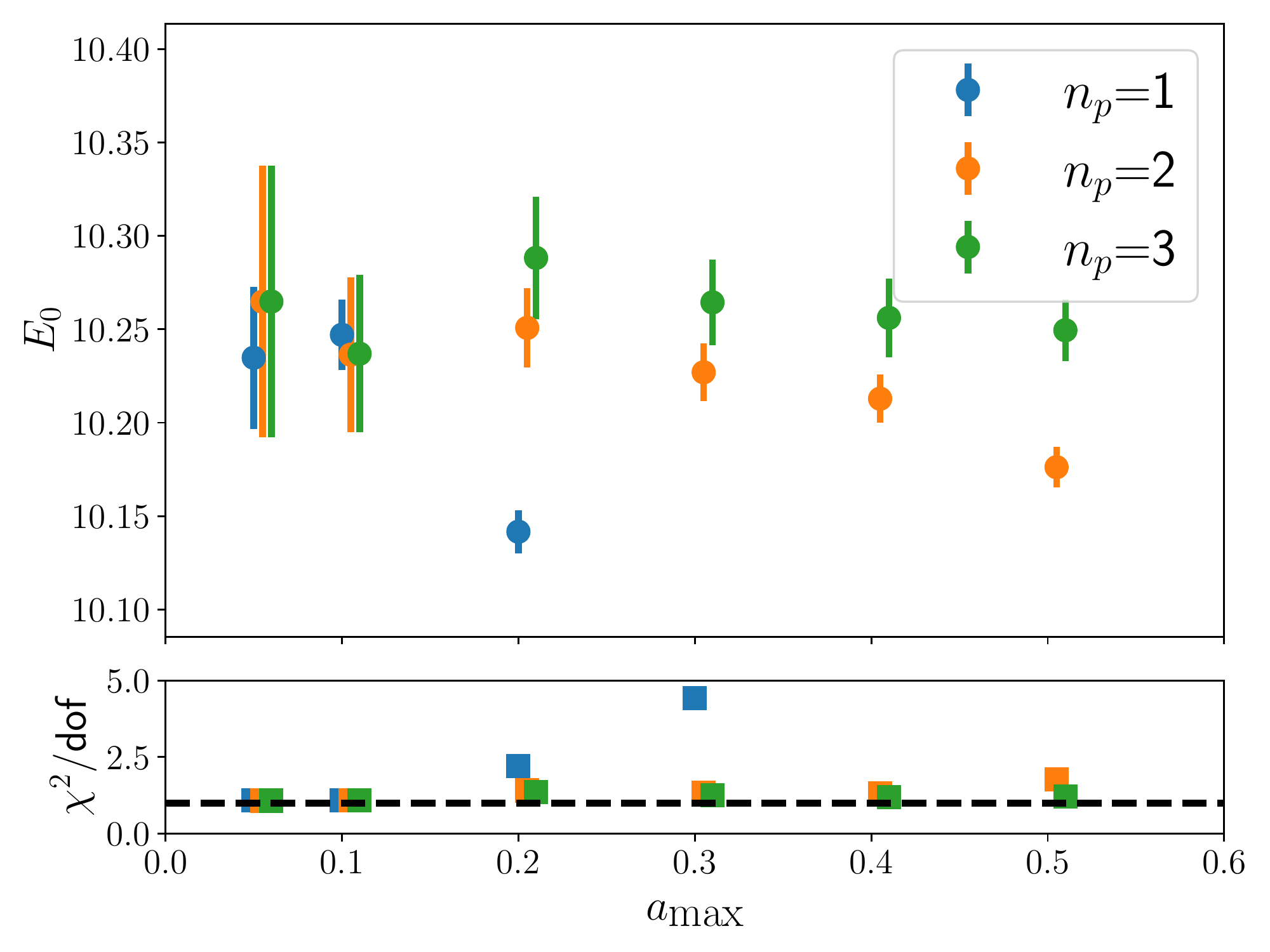}
    \caption{\label{fig:lat_su3_g20_fit_sys}
    Bosonic SU(3) model with $\lambda=2.0$.
    Results of fitting different data portions with polynomials of different order $n_p$.
    The lower panel shows the reduced $\chi^2$ which becomes very large for the low order polynomials when larger lattice spacings are included.
    }
\end{figure}
\begin{figure}
    \centering
    \includegraphics[width=0.45\textwidth]{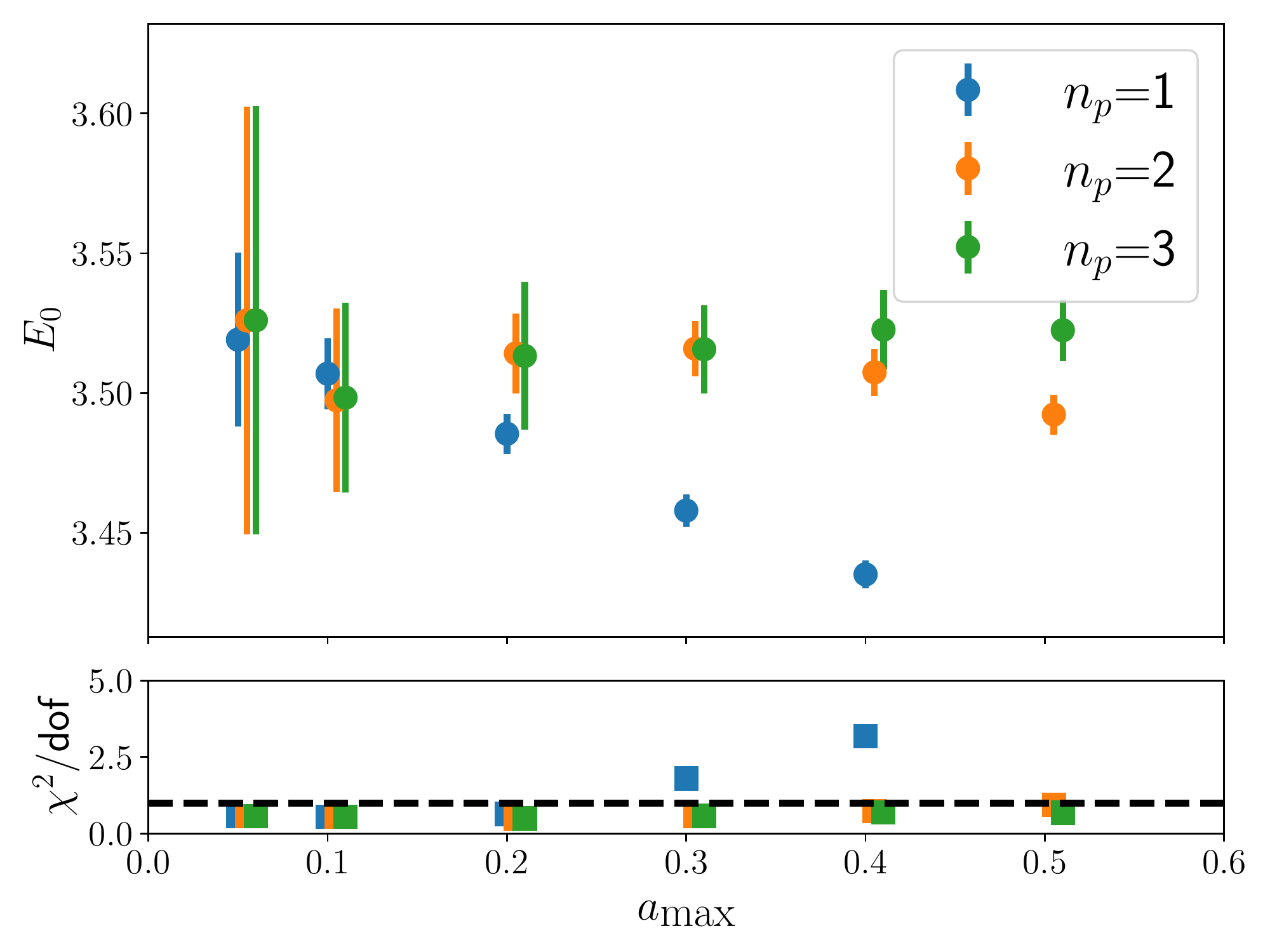}
    \caption{\label{fig:lat_su2_g10_fit_sys}
    Bosonic SU(2) model with $\lambda=1.0$.
    Results of fitting different data portions with polynomials of different order $n_p$.
    The lower panel shows the reduced $\chi^2$ which becomes very large for the low order polynomials when larger lattice spacings are included.
    }
\end{figure}
\begin{figure}
    \centering
    \includegraphics[width=0.45\textwidth]{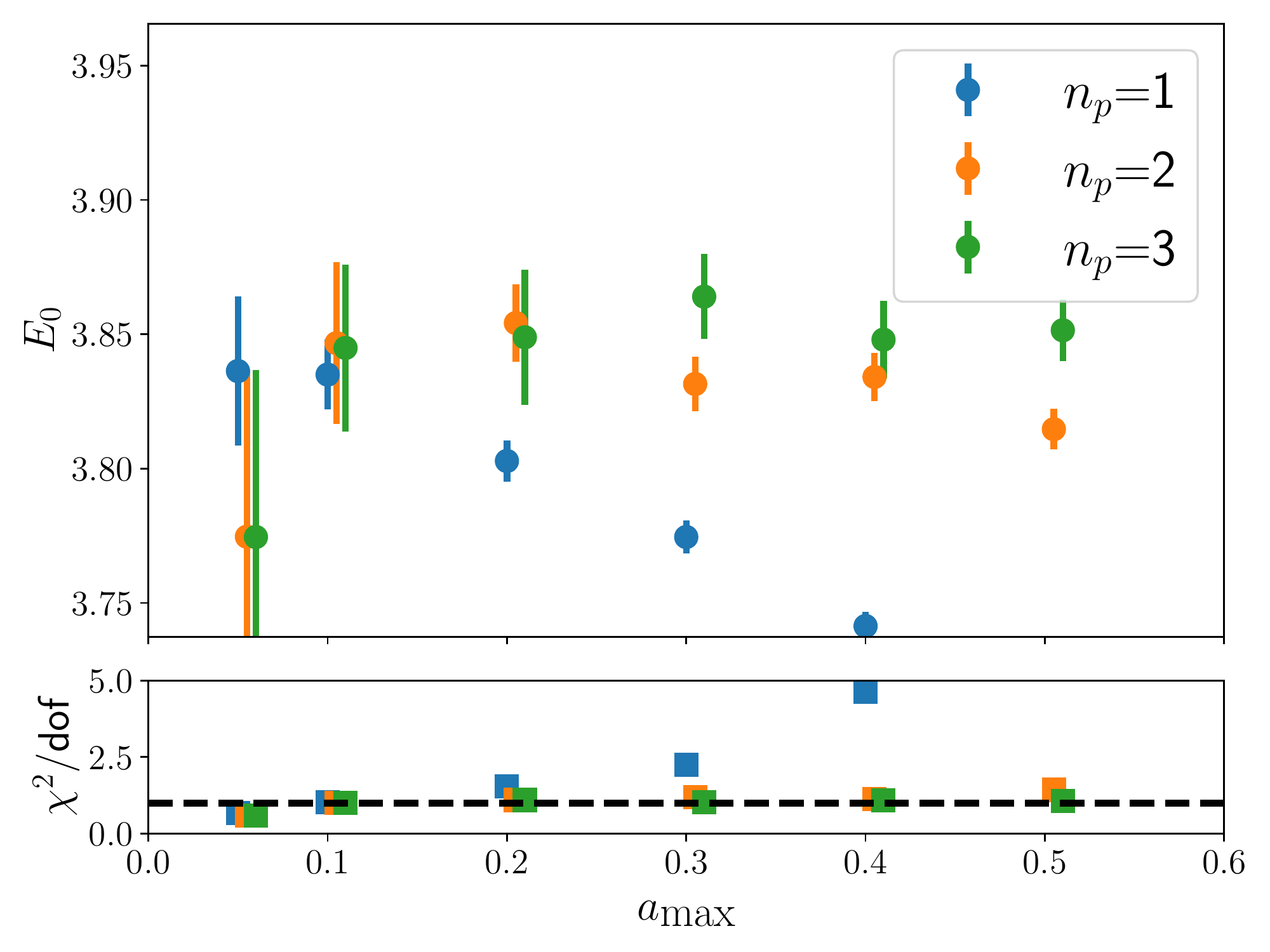}
    \caption{\label{fig:lat_su2_g20_fit_sys}
    Bosonic SU(2) model with $\lambda=2.0$.
    Results of fitting different data portions with polynomials of different order $n_p$.
    The lower panel shows the reduced $\chi^2$ which becomes very large for the low order polynomials when larger lattice spacings are included.
    }
\end{figure}

\end{document}